\documentclass{llncs}
\usepackage{llncsdoc}

\usepackage[linesnumbered,ruled,vlined]{algorithm2e}
\usepackage{mathtools}
\usepackage[space]{cite}
\usepackage{url}

\usepackage{subfig}
\usepackage{graphicx}
\usepackage{multirow}
\usepackage{verbatim}
\usepackage{mathtools}
\usepackage{booktabs,makecell,tabularx}
\usepackage{url}
\usepackage{multirow}
\usepackage{amssymb}
\usepackage{slashbox}
\usepackage{float}
\usepackage{amsmath}
\usepackage{arydshln}
\usepackage{array}
\newcolumntype{C}[1]{>{\centering\arraybackslash$}p{#1}<{$}}

\begin{document}
\title{Balanced Encoding of Near-Zero Correlation for an AES Implementation}

\author{Seungkwang Lee \inst{1} \and Jeong-Nyeo Kim \inst{2,*}}
\institute{Dept. Industrial Security, Dankook University (\email{sk.cryptographic@dankook.ac.kr}) \and Dept. Information Security Research Division, ETRI (\email{jnkim@etri.re.kr})}

\maketitle 

\begin{abstract}

Power analysis poses a significant threat to the security of cryptographic algorithms, as it can be leveraged to recover secret keys. While various software-based countermeasures exist to mitigate this non-invasive attack, they often involve a trade-off between time and space constraints. Techniques such as masking and shuffling, while effective, can noticeably impact execution speed and rely heavily on run-time random number generators. On the contrary, internally encoded implementations of block ciphers offer an alternative approach that does not rely on run-time random sources, but it comes with the drawback of requiring substantial memory space to accommodate lookup tables. Internal encoding, commonly employed in white-box cryptography, suffers from a significant security limitation as it does not effectively protect the secret key against statistical analysis.
To overcome this weakness, this paper introduces a secure internal encoding method for an AES implementation. By addressing the root cause of vulnerabilities found in previous encoding methods, we propose a balanced encoding technique that aims to minimize the problematic correlation with key-dependent intermediate values. We analyze the potential weaknesses associated with the balanced encoding and present a method that utilizes complementary sets of lookup tables. In this approach, the size of the lookup tables is approximately 512KB, and the  number of table lookups is 1,024. This is comparable to the table size of non-protected white-box AES-128 implementations, while requiring only half the number of lookups. By adopting this method, our aim is to introduce a non-masking technique that mitigates the vulnerability to statistical analysis present in current internally-encoded AES implementations.

\end{abstract}


\section{Introduction}
\label{sec:introduction}

In the context of the gray-box model, when software cryptographic implementations run in untrusted environments, they become vulnerable to attacks exploiting side-channel information like timing or power consumption. This type of information often exposes secret keys, allowing attackers to retrieve them without the need for laborious reverse engineering efforts. Among the various sources of side-channel information leakage, one widely exploited aspect is the correlation with key-dependent intermediate values.

For instance, power analysis techniques, including Differential Power Analysis (DPA)\cite{Kocher:DPA:1999} and Correlation Power Analysis (CPA)\cite{Brier:CPA:2004}, analyze the correlation between the power consumption during cryptographic computations and hypothetical values associated with those computations. In such attacks, the attacker interacts with the software implementation of a cryptographic primitive by providing arbitrary inputs and recording power traces using an oscilloscope. For each pair of input and subkey candidate relevant to the target cryptographic function (e.g., SubBytes), the attacker computes the hypothetical value. The correct candidate yields hypothetical values that strongly correlate with specific points in the power traces because the power consumption is directly related to the data processed in the circuit.

In DPA, power traces are grouped into distinct sets based on these hypothetical values, and average traces are generated for each set. If the subkey candidate is correct, a distinctive peak will appear in the differential trace at the points associated with the target function. In contrast, in CPA, the subkey is deduced by calculating the correlation between hypotheses and individual points in the power traces.

The number of power traces necessary for a successful power analysis primarily depends on the power model and electronic noise. A power model, such as a bit model or Hamming Weight (HW) model, defines how the power consumption of the target function is represented. On the other hand, electronic noise results from power measurement and constant components like leakage current and transistor switching~\cite{PA:book}. Generally, as noise levels increase, the required number of power traces for key recovery also increases.

In recent years, several studies~\cite{Sasdrich:Walsh:2016,Bos:DCA:2015,Lee:maskedWB:2018} have introduced a novel approach to computational analysis. This approach involves providing a security verifier or attacker access to either the source code or the executable of the target primitive. In this model, computational traces can be collected by observing the computations in memory without the interference of electronic noise. Consequently, key-dependent intermediate values and memory addresses related to read and write access are readily exposed.

The concept of computational traces was initially introduced by Differential Computation Analysis (DCA)\cite{Bos:DCA:2015}. DCA is a non-invasive attack that can extract secret keys from a wide range of different white-box implementations\cite{Chow:WB-AES:2002,Chow:WB-DES:2002} without necessitating a detailed reverse engineering of the target implementation. From the attacker's perspective, noise-free computational traces have also facilitated DCA-variant attacks, such as zero difference enumeration~\cite{Banik:ZDE:2017}, collision attacks~\cite{Rivain:Collision:2019}, and bucketing attacks~\cite{Mohamed:Bucketing:2019}. From the standpoint of cryptographic engineers, noise-free traces can be employed to assess the protection of secret keys against power analysis with a reduced number of traces compared to classical power analysis and without the need for an expensive oscilloscope.

To mitigate problematic correlations in side-channel attacks, common software countermeasures include the use of masking and operation shuffling techniques. Masking relies on the concept of splitting a sensitive variable $s$ into $d+1$ shares, denoted as $s_0, s_1, \ldots, s_d$. These shares are combined through a group operation $\circ$ (typically XOR or modular addition) to recover the original value $s$. It is crucial that any subset of fewer than $d+1$ shares or leakage signals remains statistically independent of $s$. To achieve this, the masks $s_1, \ldots, s_d$ are randomly selected, and the masked variable $s_0$ is computed as $s_0 = s \circ s_1 \circ \ldots \circ s_d$. The parameter $d$ is referred to as the masking order.

The adoption of higher-order masking schemes in the physical security of block-cipher implementations has gained significant attention. This is primarily because higher-order DPA attacks become exponentially more expensive~\cite{PA:book}. Notably, Rivain and Prouff introduced a practical $d$th-order masking scheme for AES, which incurs reasonable software implementation overhead~\cite{Rivain:HO-Masking:2010}. Their approach builds upon the hardware-oriented masking scheme proposed by Ishai \textit{et al.}~\cite{Ishai:masking:2003}.
Additionally, Coron extended the classical randomized table countermeasure against first-order attacks in their work~\cite{Coron:HO-Mask-LUT:2014}. The author presented a technique for masking lookup tables of block ciphers at any order, thereby serving as an effective countermeasure against side-channel attacks.

However, higher-order countermeasures place a significant demand on randomness during execution, which can be costly in practice, especially on resource-constrained embedded devices. To tackle this challenge, Coron \textit{et al.} introduced a construction that minimizes the total amount of required randomness by utilizing multiple pseudo-random generators instead of relying solely on a True Random Number Generator (TRNG)~\cite{Coron:HO-Mask-PRG:2020}.

Remarkably, real-world experiments have demonstrated that straightforward software implementations of theoretically secure runtime masking schemes are unlikely to achieve their anticipated level of security~\cite{Beckers:TVLA:2022}. This is primarily attributed to certain sections of the code unintentionally exposing unmasked data, thus making it susceptible to side-channel analysis. In their evaluation of leakage, Beckers \textit{et al.} illustrated that Test Vector Leakage Assessment (TVLA) uncovered substantial leakage, rendering nearly all the assessed implementations vulnerable to basic CPA attacks.

To further strengthen the resistance against key leakage, a commonly employed approach by industrial practitioners is to combine higher-order masking with shuffling techniques~\cite{Herbst:smartcard:2006}. Shuffling involves randomizing the sequence of independent operations for each encryption execution, effectively dispersing the leakage points of key-sensitive variables across $t$ different locations. This strategy mitigates the concentration of leakage at specific points, enhancing the overall security of the implementation. 
Provided that the sequence is shuffled uniformly at random, then the signal-to-noise ratio of the instantaneous leakage on the sensitive variable is reduced by a factor of $t$. 
Shuffling is easy to implement and less costly than higher-order masking especially when applied to the S-box. 
However, the combination of first-order masking and shuffling is not sufficiently secure against advanced power analysis~\cite{Stefan:protectingAES:2007,Stefan:attackingAES:2008}.

It's worth mentioning that higher-order DCA~\cite{Bogdanov:HODCA:2019} bears some resemblance to higher-order DPA attacks. This method was devised to retrieve secret keys from implementations that use masking and shuffling techniques by analyzing computational traces. Additionally, an algebraic DCA attack\cite{Biryukov:AGB-DCA:2018} has been proposed, which can break the linear masking in white-box implementations independently of the masking orders. To combat both computational and algebraic attacks, a white-box masking scheme has been introduced, combining both linear and nonlinear components~\cite{Seker:WB-Masking:2021}. However, this masking technique is a hardware-based countermeasure that is largely dependent on a run-time random number generator. In general, masking schemes are costly in terms of run time and often lead to the joint leakage. While it's feasible to statically embed a masking technique into the internally-encoded lookup tables of white-box instances~\cite{Lee:maskedWB:2018,Lee:ImproveMaskedWB:2020}, this approach comes with a significant memory overhead, sometimes exceeding 10MB.

Given these challenges, there is a clear demand for a software-based, non-masking countermeasure against non-invasive statistical analysis in the gray-box setting. Such an approach would enable the effective protection of secret keys in low-cost devices like IC cards with limited resources. To attain this objective, we revisit the concept of internal encoding for table-based AES implementations. Initially used in white-box cryptography to enhance table diversity and ambiguity~\cite{Chow:WB-AES:2002,Chow:WB-DES:2002}, internal encoding has faced vulnerabilities in white-box cryptography due to the remaining high correlation in encoded intermediate values~\cite{Sasdrich:Walsh:2016,Bos:DCA:2015}. This vulnerability has emphasized the necessity for balanced encoding methods.

In this paper, our primary focus is on discovering a balanced encoding method. Notably, we tailor an 8-bit linear transformation to eliminate problematic correlations. It's worth mentioning that our linear transformation is designed to employ both nonsingular and singular matrices. This improvement substantially increases table diversity and ambiguity. However, since the linear transformation alone cannot effectively conceal zeros, we incorporate a nonlinear transformation to obscure them. By implementing our encoding method in the AES algorithm, the security of the key can be bolstered against various statistical analysis. 

Our research introduces several significant contributions, which can be summarized as follows:
\begin{itemize}
    \item We propose a novel 8-bit linear transformation technique that achieves a balanced encoding. In contrast to the limited table diversity and ambiguity caused by using only 8$\times$8 block invertible matrices, we present an innovative approach that incorporates non-invertible matrices within the linear transformation. This allows us to enhance the security while maintaining table diversity.
    \item To hide the values of zeros in intermediate computations, we employ a simple nibble encoding method. Since multiplying with zeros always yields zeros, our nibble encoding replaces zeros with other nibbles in a way that preserves the balance.
    \item We apply our proposed balanced encoding technique to an AES implementation, achieving the lowest correlation coefficients with the correct hypothetical values. To enhance security further, we employ complementary sets of lookup tables to thwart DPA-like variants targeting the lowest correlations. This AES implementation requires approximately 512KB of memory space for lookup tables and involves 1,024 table lookups.
    \item Our experimental results underscore the effectiveness of our AES implementation in defending against various statistical analysis, including DPA-like variants, collision-based approaches, mutual information analysis, and TVLA attacks.
    \item Importantly, our approach does not rely on masking and shuffling techniques and can be implemented without having to utilize run-time random sources in the device. This eliminates the risk of unintended unmasking during run-time and mitigates the performance overhead associated with heavyweight random number generators.
\end{itemize}

The remaining sections of the paper are organized as follows:
Section~\ref{sec:preliminaries} provides an overview of the fundamental concepts, including non-invasive statistical analysis, the existing method of internal encoding, and its vulnerabilities.
Section~\ref{sec:proposed_encoding} presents our proposed encoding method.
In Section~\ref{sec:protected_aes}, we introduce our AES implementation that incorporates balanced encoding. We analyze its performance and cost in detail.
Section~\ref{sec:experiment} presents the experimental results, focusing on the evaluation of key protection against a range of non-invasive attacks.
Section~\ref{sec:conclusion} provides a summary of the key findings and engages in a discussion of this research.

\section{Preliminaries}
\label{sec:preliminaries}

In this section, we delve into the realm of gray-box attacks, which are widely recognized as one of the primary methods used to extract secret keys from low-cost devices. These attacks typically rely on side-channel information. For instance, in the case of IC cards, power analysis leverages oscilloscopes to capture a series of power traces, which inevitably contain measurement noise.

However, when assessing the resilience against gray-box attacks with access to the source code or executable of the target cipher, it becomes possible to directly collect computational traces from memory without any measurement noise. It is important to note that if the key cannot be extracted from these noise-free computational traces, it can be inferred that the key remains secure against power analysis.
In this section, we provide an explanation of the fundamental concepts related to well-known techniques of side-channel analysis and highlight the vulnerability associated with internal encoding.

\subsection{Gray-box attacks}

\noindent \textbf{Power analysis} is a non-invasive attack to extract the key from low-\section{Preliminaries}
\label{sec:preliminaries}

In this section, we delve into the realm of gray-box attacks, which are widely recognized as one of the primary methods used to extract secret keys from low-cost devices. These attacks typically rely on side-channel information. For instance, in the case of IC cards, power analysis leverages oscilloscopes to capture a series of power traces, which inevitably contain measurement noise.

However, when assessing the resilience against gray-box attacks with access to the source code or executable of the target cipher, it is possible to directly collect computational traces from memory without any measurement noise. It is important to note that if the key is hard to extract from these noise-free computational traces, the key remains secure against power analysis. 
In this section, we provide an explanation of the fundamental concepts related to well-known techniques of side-channel analysis and highlight the vulnerability associated with internal encoding.

\subsection{Gray-box attacks}

\noindent \textbf{Power analysis} is a non-invasive attack to extract the key from low-cost devices such as IC cards. 
In the gray-box model of power analysis, an attacker can select a plaintext and access the ciphertext. 
The attacker can also collect timing  or power consumption information related to the encryption operation.
However, it is not possible to access the internal resources of the computing environment to observe or manipulate memory.
Power consumption of IC cards is proportional (or inversely proportional) to the HW of data processed in the circuit~\cite{PA:book}. 
Suppose that an attacker knows the correct subkey of the AES algorithm. 
For a target operation like SubBytes, power consumption of a circuit is then strongly correlated to  the attacker's hypothetical value at a point of power traces. 
The following explains two fundamental techniques of power analysis, DPA and CPA.\\

\noindent \textbf{DPA} is a statistical method for analyzing power consumption.
A selection function $D$($b$, $C$, $k^{\ast}$) is defined to compute a  key-sensitive variable $b$, where $C$ is a ciphertext and $k^{\ast}$ is a subkey candidate. 
If $k^{\ast}$ is a wrong key candidate, $D$ will evaluate $b$ correctly with a $1/2$ probability.
DPA records $N$ power traces $\textbf{V}_{1..N}[1..\kappa]$ consisting of $\kappa$ points.
Then the differential trace $\Delta_D[j]$, where $j$ $\in [1, \kappa]$, is generated by computing the difference between the average of traces for $D$($\cdot$) = 1 and $D$($\cdot$) = 0 as follows:

\footnotesize
\begin{align}
\begin{split}
\Delta_D[j] = {}& \frac{\sum_{n=1}^{N}D(b, C_{n},  k^{\ast})\textnormal{V}_{n}[j]}{\sum_{n=1}^{N}D(b, C_{n}, k^{\ast})}  - \frac{\sum_{n=1}^{N}(1 - D(b, C_{n}, k^{\ast}))\textnormal{V}_{n}[j]}{\sum_{n=1}^{N}(1 - D(b, C_{n}, k^{\ast}))} \\
\approx {}& 2\left (\frac{\sum_{n=1}^{N}D(b, C_{n}, k^{\ast})\textnormal{V}_{n}[j]}{\sum_{n=1}^{M}D(b, C_{n}, k^{\ast})} - \frac{\sum_{n=1}^{N}\textnormal{V}_{n}[j]}{N} \right ) \nonumber
\end{split}
\end{align}
\normalsize

If $k^{\ast}$ is correct, $D$ will be proportional to $\textnormal{V}_{m}[j]$, a sample point in the trace.
That point indicates power consumption of computing $b$.
For this reason, a correct subkey will result in a noticeable peak in the differential trace, but it is unlikely that the wrong key candidate will produce a spike. \\

\noindent \textbf{CPA} improves on DPA by taking advantage of leakage models such as the HW.
For $K$ different key candidates, let $h_{n, k^{\ast}}$ be the hypothetical value for the power estimation, where $n$ $\in [1, N]$, and $k^{\ast}$ $[0, K)$. 
A correlation between hypothetical values and power traces can be obtained by the estimator $r$ as follows~\cite{PA:book}:
\begin{align}
\begin{split}
r_{k^{\ast},j} = {}& \frac{\sum_{n=1}^N(h_{n,k^{\ast}} - \overline{h_k^{\ast}}) \cdot (V_n[j] - \overline{V[j]})}{\sqrt{\sum_{n=1}^N (h_{n,k^{\ast}} - \overline{h_k^{\ast}})^2 \cdot \sum_{n=1}^N(V_n[j] - \overline{V[j]})^2}},  \nonumber
\end{split}
\end{align}
where $\overline{h_k^{\ast}}$ and $\overline{V[j]}$ are the sample means of $h_k^{\ast}$ and $V[j]$, respectively.
Similar to the case of DPA, a correct subkey exhibits a noticeable spike in the correlation plot during CPA attacks. Throughout this paper, we specifically focus on CPA attacks instead of DPA attacks due to their efficiency~\cite{Brier:CPA:2004}.\\

\noindent \textbf{DCA} is an alternative approach to power analysis that leverages computational traces instead of power traces. Unlike power traces, which are susceptible to electronic noise and constant components like leakage currents and transistor switching, computational traces offer a noise-free representation of the software's read-write data or memory addresses accessed during execution.

When evaluating the resistance to key leakage in cryptographic implementations, it is not always necessary to collect power traces. If the source code or execution binary is available, DCA can be employed to gather and analyze computational traces. Since computational traces are devoid of noise, they provide an efficient basis for CPA techniques. In other words, the number of traces required to recover the secret key from the implementations is reduced.


To capture computational traces during the encryption process, one can utilize dynamic binary instrumentation (DBI) tools such as Intel Pin~\cite{Luk:Pin} and Valgrind~\cite{Nethercote:Valgrind}. These tools allow for the collection and analysis of computational traces. It is important to note that DBI tools can monitor binaries as they execute and dynamically insert new instructions into the instruction stream, rendering DCA a technique often categorized as an invasive attack.
The source code for DCA is available in a public repository~\footnote{https://github.com/SideChannelMarvels/Deadpool}, providing further resources for understanding and implementing this approach.
By utilizing DCA and computational traces, researchers can mitigate the impact of noise and enhance the efficiency of power analysis techniques. \\

\noindent \textbf{Collision-based attacks}, which fall under the category of side-channel analyses, have been previously explored in research works~\cite{Schramm:Collision:2003,Rivain:Collision:2019}. In a block cipher algorithm, a collision occurs when two or more distinct inputs produce the same output from a function within the algorithm. Such collisions indicate that the traces exhibit significant similarity during the period when the internal collision persists. In this collision-based attack against internal encodings, the following observation holds true: if a sensitive variable collides for a pair of inputs, the corresponding encoded variable in the trace also experiences a collision. Conversely, if a sensitive variable does not collide for a pair of inputs, the corresponding encoded variable in the trace remains collision-free. Differential cluster analysis~\cite{Batina:ClusterAnalysis:2009} also utilizes this property to create clusters and evaluate   cluster criteria like the sum-of-squared-error.   \\
It is noteworthy that a collision attack succeeds whenever MIA succeeds, but the reverse is not true~\cite{Rivain:Collision:2019}. \\

\noindent \textbf{Walsh transform} can also quantify a correlation using only simple XOR and addition 
if the intermediate values or computational traces can be collected~\cite{Sasdrich:Walsh:2016}.
To evaluate the security of software cryptographic implementations, the Walsh transform can be inserted in the source code to calculate the correlation without having to use DBI tools. 
By doing so, scanning every sample point of the traces is not necessary.
Compared to Pearson correlation coefficients, the Walsh transform consists of lightweight operations and shows the correlation as a natural number.

\begin{definition}
Let $x$ = $\langle$$x_1$, $\ldots$, $x_n$$\rangle$, $\omega$ = $\langle$$\omega_1$, $\ldots$, $\omega_n$$\rangle$
be elements of $\{0,1\}^n$ and $x\cdot\omega$ = $x_1\omega_1$$\oplus$$\ldots$$\oplus$$x_n\omega_n$.
Let $f(x)$ be a Boolean function of $n$ variables.
Then the Walsh transform of the function $f(x)$ is a real valued function over $\{0, 1\}^n$ that can be defined as $W_f(\omega)$ = $\Sigma_{x\in\{0,1\}^n}(-1)^{f(x)\oplus x\cdot\omega}$.\\
\label{def:walsh}
\end{definition}
\begin{definition}
Iff the Walsh transform $W_f$ of a Boolean function $f(x_1, \ldots, x_n)$
satisfies $W_f(\omega)$ = 0, for 0 $\leq$ $HW(\omega)$ $\leq$ $d$,
it is called a balanced $d$-th order correlation immune function or
an $d$-resilient function.\\
\label{def:walsh_balance}
\end{definition}

In Definition~\ref{def:walsh}, $W_f(\omega)$ quantifies a correlation between $f(x)$ and $x\cdot\omega$, where $f(x)$ is the intermediate value and $x\cdot\omega$ is the attacker's hypothetical value; if HW($\omega$) = 1, $\omega$ selects a particular bit of $x$. 
For each key candidate $k^\ast$, for every $\omega \in \{0,1\}^8$, and for $n$ Boolean functions $f_i$, we can also calculate the Walsh transforms $W_{f_i}$ and accumulate the imbalances for each key candidate using the following formula:
\begin{equation}
\label{eq:delta}
    \Delta^{f}_{k^{\ast} \in \{0,1\}^8} = \sum_{\forall\omega \in \{0,1\}^8}\sum_{i \in [1, n]}|W_{f_i}(\omega)| 
\end{equation}
While it is possible to analyze and quantify the risk of key leakage at $f_i$ using individual $W_{f_i}$ values, a comprehensive understanding of key leakage can be achieved by considering the collective sum of imbalances $\Delta^{f}_{k^{\ast}}$. This sum provides an overall indication of the correlation with the correct hypothetical values for each subkey candidate.
In the event of key leakage, the sum of imbalances $\Delta^{f}_k$ calculated using the correct subkey will exhibit a prominent spike, making it easily distinguishable from the sum of imbalances of other key candidates.

In this context, the computational trace $f(x)$ represents the noise-free intermediate value. Definition~\ref{def:walsh_balance} outlines the criterion for a balanced first-order correlation immune function, ensuring that each bit of the intermediate values has no correlation with any bit of the hypothetical values.



\subsection{High Correlation in the Existing Encoding Method}

In the context of white-box cryptographic implementations of block ciphers, an internal encoding scheme comprising linear and nonlinear transformations has been utilized to obfuscate key-dependent intermediate values. However, it has been observed that the current encoding method is not entirely effective in concealing the keys from power analysis~\cite{Bock:nonlinear:2018, Lee:Linear:2020}.

Firstly, the concatenation of two 4-bit nonlinear transformations demonstrates limited confusion effects on 8-bit values. This limits its ability to provide strong protection against power analysis attacks.

Secondly, the linear combinations of intermediate values exhibit high correlation with the intermediate values themselves due to their dependency on the encryption key. 
For example, suppose that an 8-bit linear transformation is applied to $3 \cdot x$, where $x$ is the SubBytes output. 
Then the linear combination of the first to fifth bits of $3 \cdot x$ is always correlated to the first bit of $x$. 

Let $f_i(\cdot)$ denote the encoding on the MixColumns multiplication in the first round of a white-box implementation of the AES-128 algorithm (WB-AES)~\cite{Chow:WB-AES:2002}. To be specific, the MixColumns multiplication is performed between the SubBytes output and the first column of MixColumn matrix.  Given the SubBytes output $x$, $f_{i \in [0, 31] }(x)$ is an encoded 32-bit value of MixColumns multiplication.  Fig.~\ref{fig:chow_wbaes_walsh} illustrates the Walsh transform with $\omega = 8$. 
Although the correct subkey has $W_{fi}$ = 0 at 30 bits out of 32 bits, the 6th and the 7th bits of $f(\cdot)$ are highly correlated to the 4th LSB of $x$. Their Walsh transform scores are 128.  In contrary, the average of $|W_{fi}|$ of the wrong key candidates was 13.07 (max = 52 and s.d = 9.44). This means that the subkey can be recovered by power analysis with an overwhelming probability.

\begin{figure}
\centering
\includegraphics[width=0.9\linewidth]{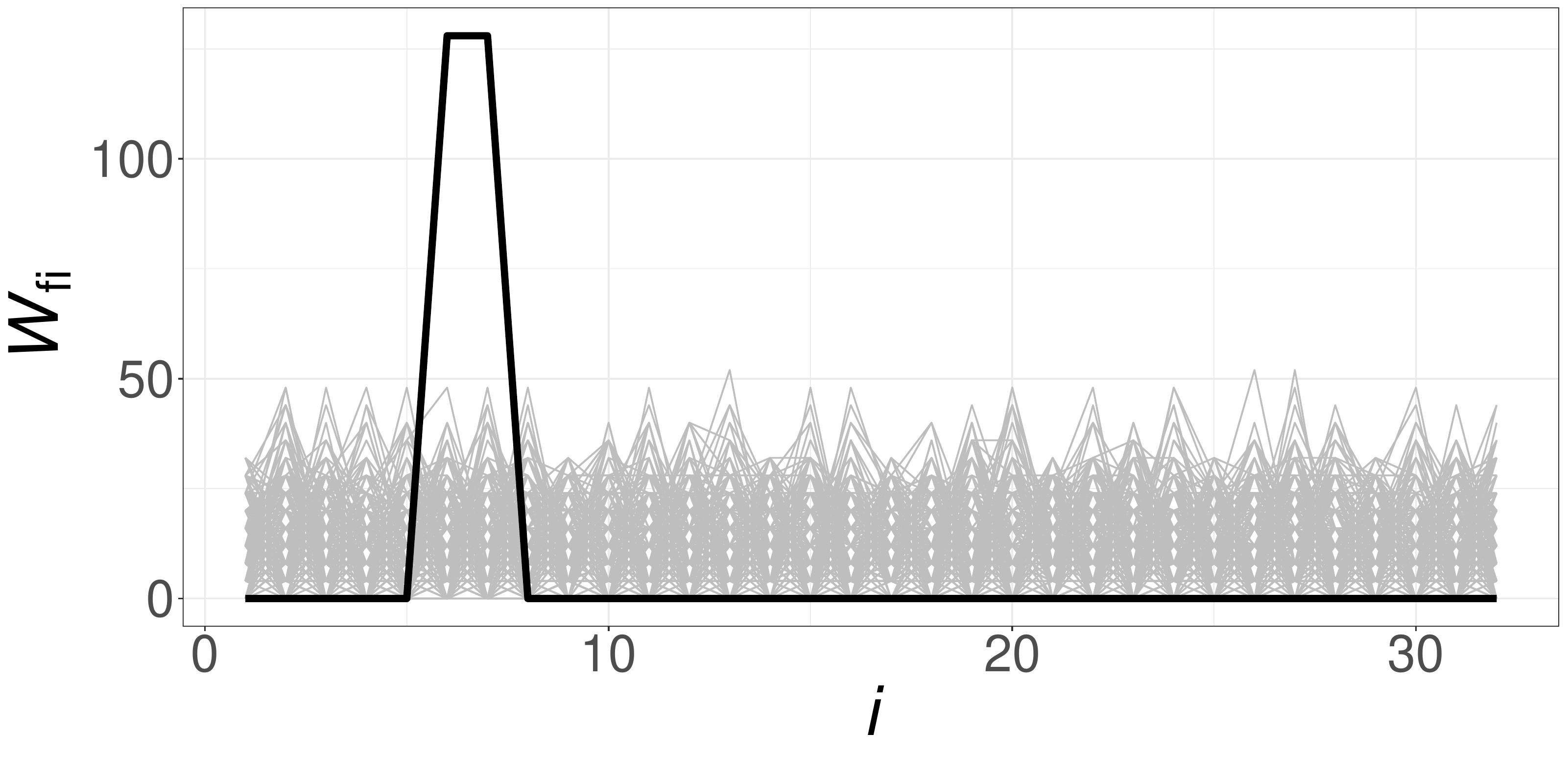}
\caption{Walsh transform on the MixColumns multiplication in the first round of WB-AES. Black: correct subkey. Gray: wrong key candidates.}
\label{fig:chow_wbaes_walsh}
\end{figure}

Various approaches have been proposed in the past to tackle the issue of internal encoding, such as classical rum-time and customized static masking techniques~\cite{Bogdanov:HODCA:2019,Lee:maskedWB:2018,Lee:ImproveMaskedWB:2020}.
However, rum-time masking is computationally expensive and susceptible to higher-order DPA attacks.
Meanwhile, static masking requires significant memory resources, typically in the order of tens of megabytes, making it unsuitable for low-cost devices. To overcome these limitations, we propose a non-masking encoding method that achieves efficient memory utilization while maintaining security for the AES-128 algorithm.

\subsection{Notations} 
For 8-bit binary vectors X, Y, Z, the superscripts H and L represent their upper 4 bits and lower 4 bits, respectively.  
Thus, X = $\textnormal{X}^{\textnormal{H}} || \textnormal{X}^{\textnormal{L}}$.
The encoding is denoted by $\mathcal{E}$, consisting of the linear and nonlinear transformations, denoted by $\mathcal{L}$ and $\mathcal{N}$, respectively. 
By abuse of notation, $\mathcal{N}^H$ denotes the nibble encoding for the upper 4 bits whereas $\mathcal{N}^L$ denotes the lower 4 bits of the input. 
The subscripts to $\mathcal{N}$ are used to indicate different nibble encodings. 
The decoding is denoted by $\mathcal{D}$. 
Let denote two sets of $4\times4$ binary matrices by $\mathcal{F}$ and $\mathcal{G}$, which are
chosen under the certain conditions explained in the following section.
$f \xleftarrow[]{\$} \mathcal{F}$ means a random sampling from $\mathcal{F}$.    
\textit{Idx}($v$) is defined to be a function:  \textit{Idx}($v$) = $\{i \mid  v_{i} = 1 \textnormal{ for } i \in [1, 8]\}$, where $v$ is an 8-bit binary vector.
For example, $\{1, 5, 6\} \leftarrow $ \textit{Idx}([1, 0, 0, 0, 1, 1, 0, 0]).
\\

cost devices such as IC cards. 
In the gray-box model of power analysis, an attacker can select a plaintext and access the ciphertext. 
The attacker can also collect timing  or power consumption information related to the encryption operation.
However, it is not possible to access the internal resources of the computing environment to observe or manipulate memory.
Power consumption of IC cards is proportional (or inversely proportional) to the HW of data processed in the circuit~\cite{PA:book}. 
Suppose that an attacker knows the correct subkey of the AES algorithm. 
For a target operation like SubBytes, power consumption of a circuit is then strongly correlated to  the attacker's hypothetical value at a point of power traces. 
The following explains two fundamental techniques of power analysis, DPA and CPA.\\

\noindent \textbf{DPA} is a statistical method for analyzing power consumption.
A selection function $D$($b$, $C$, $k^{\ast}$) is defined to compute a  key-sensitive variable $b$, where $C$ is a ciphertext and $k^{\ast}$ is a subkey candidate. 
If $k^{\ast}$ is a wrong key candidate, $D$ will evaluate $b$ correctly with a $1/2$ probability.
DPA records $N$ power traces $\textbf{V}_{1..N}[1..\kappa]$ consisting of $\kappa$ points.
Then the differential trace $\Delta_D[j]$, where $j$ $\in [1, \kappa]$, is generated by computing the difference between the average of traces for $D$($\cdot$) = 1 and $D$($\cdot$) = 0 as follows:

\footnotesize
\begin{align}
\begin{split}
\Delta_D[j] = {}& \frac{\sum_{n=1}^{N}D(b, C_{n},  k^{\ast})\textnormal{V}_{n}[j]}{\sum_{n=1}^{N}D(b, C_{n}, k^{\ast})}  - \frac{\sum_{n=1}^{N}(1 - D(b, C_{n}, k^{\ast}))\textnormal{V}_{n}[j]}{\sum_{n=1}^{N}(1 - D(b, C_{n}, k^{\ast}))} \\
\approx {}& 2\left (\frac{\sum_{n=1}^{N}D(b, C_{n}, k^{\ast})\textnormal{V}_{n}[j]}{\sum_{n=1}^{M}D(b, C_{n}, k^{\ast})} - \frac{\sum_{n=1}^{N}\textnormal{V}_{n}[j]}{N} \right ) \nonumber
\end{split}
\end{align}
\normalsize

If $k^{\ast}$ is correct, $D$ will be proportional to $\textnormal{V}_{m}[j]$, a sample point in the trace.
That point indicates power consumption of computing $b$.
For this reason, a correct subkey will result in a noticeable peak in the differential trace, but it is unlikely that the wrong key candidate will produce a spike. \\

\noindent \textbf{CPA} improves on DPA by taking advantage of leakage models such as the HW.
For $K$ different key candidates, let $h_{n, k^{\ast}}$ be the hypothetical value for the power estimation, where $n$ $\in [1, N]$, and $k^{\ast}$ $[0, K)$. 
A correlation between hypothetical values and power traces can be obtained by the estimator $r$ as follows~\cite{PA:book}:
\begin{align}
\begin{split}
r_{k^{\ast},j} = {}& \frac{\sum_{n=1}^N(h_{n,k^{\ast}} - \overline{h_k^{\ast}}) \cdot (V_n[j] - \overline{V[j]})}{\sqrt{\sum_{n=1}^N (h_{n,k^{\ast}} - \overline{h_k^{\ast}})^2 \cdot \sum_{n=1}^N(V_n[j] - \overline{V[j]})^2}},  \nonumber
\end{split}
\end{align}
where $\overline{h_k^{\ast}}$ and $\overline{V[j]}$ are the sample means of $h_k^{\ast}$ and $V[j]$, respectively.
Similar to the case of DPA, a correct subkey exhibits a noticeable spike in the correlation plot during CPA attacks. Throughout this paper, we specifically focus on CPA attacks instead of DPA attacks due to their efficiency~\cite{Brier:CPA:2004}.\\

\noindent \textbf{DCA} is an alternative approach to power analysis that leverages computational traces instead of power traces. Unlike power traces, which are susceptible to electronic noise and constant components like leakage currents and transistor switching, computational traces offer a noise-free representation of the software's read-write data or memory addresses accessed during execution.

When evaluating the resistance to key leakage in cryptographic implementations, it is not always necessary to collect power traces. If the source code or execution binary is available, DCA can be employed to gather and analyze computational traces. Since computational traces are devoid of noise, they provide an efficient basis for CPA techniques. In other words, the number of traces required to recover the secret key from the implementations is reduced.


To capture computational traces during the encryption process, one can utilize dynamic binary instrumentation (DBI) tools such as Intel Pin~\cite{Luk:Pin} and Valgrind~\cite{Nethercote:Valgrind}. These tools allow for the collection and analysis of computational traces. It is important to note that DBI tools can monitor binaries as they execute and dynamically insert new instructions into the instruction stream, rendering DCA a technique often categorized as an invasive attack.
In this paper, we assume that the attacker does not employ DBI tools to perform invasive attacks. 
The source code for DCA is available in a public repository~\footnote{https://github.com/SideChannelMarvels/Deadpool}, providing further resources for understanding and implementing this approach.
By utilizing DCA and computational traces, researchers can mitigate the impact of noise and enhance the efficiency of power analysis techniques. \\

\noindent \textbf{Collision-based attacks}, which fall under the category of side-channel analyses, have been previously explored in research works~\cite{Schramm:Collision:2003,Rivain:Collision:2019}. In a block cipher algorithm, a collision occurs when two or more distinct inputs produce the same output from a function within the algorithm. Such collisions indicate that the traces exhibit significant similarity during the period when the internal collision persists. In this collision-based attack against internal encodings, the following observation holds true: if a sensitive variable collides for a pair of inputs, the corresponding encoded variable in the trace also experiences a collision. Conversely, if a sensitive variable does not collide for a pair of inputs, the corresponding encoded variable in the trace remains collision-free. Differential cluster analysis~\cite{Batina:ClusterAnalysis:2009} also utilizes this property to create clusters and evaluate   cluster criteria like the sum-of-squared-error.  

In the side-channel attack context, Mutual Information Analysis (MIA) was introduced to address scenarios where the adversary has limited knowledge about the distribution of the leakage and its relationship with computed data~\cite{Gierlichs:MIA:2008}. It is noteworthy that a collision attack succeeds whenever MIA succeeds, but the reverse is not true~\cite{Rivain:Collision:2019}. \\

\noindent \textbf{Walsh transform} can also quantify a correlation using only simple XOR and addition 
if the intermediate values or computational traces can be collected~\cite{Sasdrich:Walsh:2016}.
To evaluate the security of software cryptographic implementations, the Walsh transform can be inserted in the source code to calculate the correlation without having to use DBI tools. 
By doing so, scanning every sample point of the traces is not necessary.
Compared to Pearson correlation coefficients, the Walsh transform consists of lightweight operations and shows the correlation as a natural number.

\begin{definition}
Let $x$ = $\langle$$x_1$, $\ldots$, $x_n$$\rangle$, $\omega$ = $\langle$$\omega_1$, $\ldots$, $\omega_n$$\rangle$
be elements of $\{0,1\}^n$ and $x\cdot\omega$ = $x_1\omega_1$$\oplus$$\ldots$$\oplus$$x_n\omega_n$.
Let $f(x)$ be a Boolean function of $n$ variables.
Then the Walsh transform of the function $f(x)$ is a real valued function over $\{0, 1\}^n$ that can be defined as $W_f(\omega)$ = $\Sigma_{x\in\{0,1\}^n}(-1)^{f(x)\oplus x\cdot\omega}$.\\
\label{def:walsh}
\end{definition}
\begin{definition}
Iff the Walsh transform $W_f$ of a Boolean function $f(x_1, \ldots, x_n)$
satisfies $W_f(\omega)$ = 0, for 0 $\leq$ $HW(\omega)$ $\leq$ $d$,
it is called a balanced $d$-th order correlation immune function or
an $d$-resilient function.\\
\label{def:walsh_balance}
\end{definition}

In Definition~\ref{def:walsh}, $W_f(\omega)$ quantifies a correlation between $f(x)$ and $x\cdot\omega$, where $f(x)$ is the intermediate value and $x\cdot\omega$ is the attacker's hypothetical value; if HW($\omega$) = 1, $\omega$ selects a particular bit of $x$. 
For each key candidate $k^\ast$, for every $\omega \in \{0,1\}^8$, and for $n$ Boolean functions $f_i$, we can also calculate the Walsh transforms $W_{f_i}$ and accumulate the imbalances for each key candidate using the following formula:
\begin{equation}
\label{eq:delta}
    \Delta^{f}_{k^{\ast} \in \{0,1\}^8} = \sum_{\forall\omega \in \{0,1\}^8}\sum_{i \in [1, n]}|W_{f_i}(\omega)| 
\end{equation}
While it is possible to analyze and quantify the risk of key leakage at $f_i$ using individual $W_{f_i}$ values, a comprehensive understanding of key leakage can be achieved by considering the collective sum of imbalances $\Delta^{f}_{k^{\ast}}$. This sum provides an overall indication of the correlation with the correct hypothetical values for each subkey candidate.
In the event of key leakage, the sum of imbalances $\Delta^{f}_k$ calculated using the correct subkey will exhibit a prominent spike, making it easily distinguishable from the sum of imbalances of other key candidates.

In this context, the computational trace $f(x)$ represents the noise-free intermediate value. Definition~\ref{def:walsh_balance} outlines the criterion for a balanced first-order correlation immune function, ensuring that each bit of the intermediate values has no correlation with any bit of the hypothetical values.



\subsection{High Correlation in the Existing Encoding Method}

In the context of white-box cryptographic implementations of block ciphers, an internal encoding scheme comprising linear and nonlinear transformations has been utilized to obfuscate key-dependent intermediate values. However, it has been observed that the current encoding method is not entirely effective in concealing the keys from power analysis~\cite{Bock:nonlinear:2018, Lee:Linear:2020}.

Firstly, the concatenation of two 4-bit nonlinear transformations demonstrates limited confusion effects on 8-bit values. This limits its ability to provide strong protection against power analysis attacks.

Secondly, the linear combinations of intermediate values exhibit high correlation with the intermediate values themselves due to their dependency on the encryption key. 
For example, suppose that an 8-bit linear transformation is applied to $3 \cdot x$, where $x$ is the SubBytes output. 
Then the linear combination of the first to fifth bits of $3 \cdot x$ is always correlated to the first bit of $x$. 

Let $f_i(\cdot)$ denote the encoding on the MixColumns multiplication in the first round of a white-box implementation of the AES-128 algorithm (WB-AES)~\cite{Chow:WB-AES:2002}. To be specific, the MixColumns multiplication is performed between the SubBytes output and the first column of MixColumn matrix.  Given the SubBytes output $x$, $f_{i \in [0, 31] }(x)$ is an encoded 32-bit value of MixColumns multiplication.  Fig.~\ref{fig:chow_wbaes_walsh} illustrates the Walsh transform with $\omega = 8$. 
Although the correct subkey has $W_{fi}$ = 0 at 30 bits out of 32 bits, the 6th and the 7th bits of $f(\cdot)$ are highly correlated to the 4th LSB of $x$. Their Walsh transform scores are 128.  In contrary, the average of $|W_{fi}|$ of the wrong key candidates was 13.07 (max = 52 and s.d = 9.44). This means that the subkey can be recovered by power analysis with an overwhelming probability.

\begin{figure}
\centering
\includegraphics[width=0.9\linewidth]{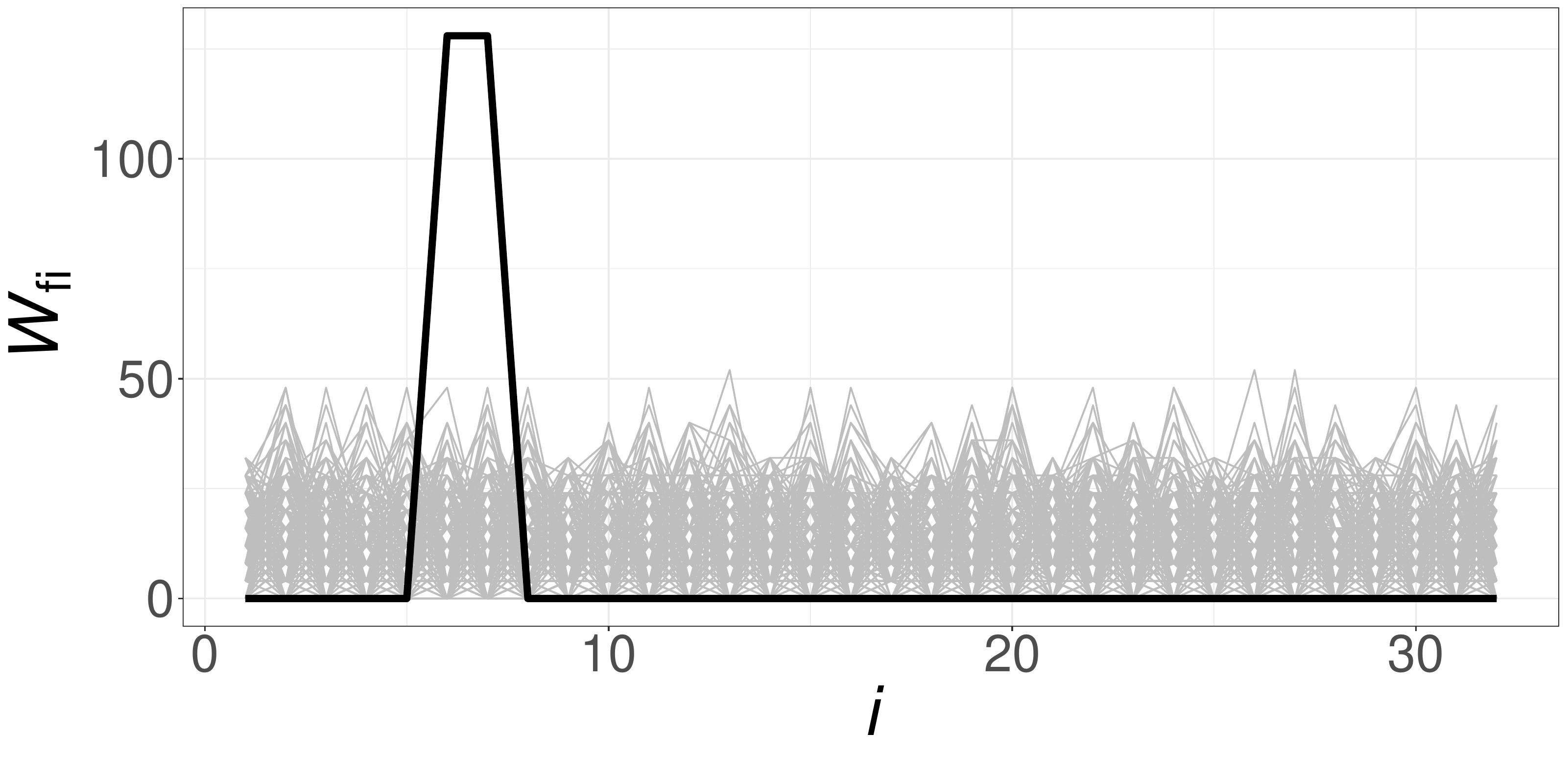}
\caption{Walsh transform on the MixColumns multiplication in the first round of WB-AES. Black: correct subkey. Gray: wrong key candidates.}
\label{fig:chow_wbaes_walsh}
\end{figure}

Various approaches have been proposed in the past to tackle the issue of internal encoding, such as classical rum-time and customized static masking techniques~\cite{Bogdanov:HODCA:2019,Lee:maskedWB:2018,Lee:ImproveMaskedWB:2020}.
However, rum-time masking is computationally expensive and susceptible to higher-order DPA attacks.
Meanwhile, static masking requires significant memory resources, typically in the order of tens of megabytes, making it unsuitable for low-cost devices. To overcome these limitations, we propose a non-masking encoding method that achieves efficient memory utilization while maintaining security for the AES-128 algorithm.

\subsection{Notations} 
For 8-bit binary vectors X, Y, Z, the superscripts H and L represent their upper 4 bits and lower 4 bits, respectively.  
Thus, X = $\textnormal{X}^{\textnormal{H}} || \textnormal{X}^{\textnormal{L}}$.
The encoding is denoted by $\mathcal{E}$, consisting of the linear and nonlinear transformations, denoted by $\mathcal{L}$ and $\mathcal{N}$, respectively. 
By abuse of notation, $\mathcal{N}^H$ denotes the nibble encoding for the upper 4 bits whereas $\mathcal{N}^L$ denotes the lower 4 bits of the input. 
The subscripts to $\mathcal{N}$ are used to indicate different nibble encodings. 
The decoding is denoted by $\mathcal{D}$. 
Let denote two sets of $4\times4$ binary matrices by $\mathcal{F}$ and $\mathcal{G}$, which are
chosen under the certain conditions explained in the following section.
$f \xleftarrow[]{\$} \mathcal{F}$ means a random sampling from $\mathcal{F}$.    
\textit{Idx}($v$) is defined to be a function:  \textit{Idx}($v$) = $\{i \mid  v_{i} = 1 \textnormal{ for } i \in [1, 8]\}$, where $v$ is an 8-bit binary vector.
For example, $\{1, 5, 6\} \leftarrow $ \textit{Idx}([1, 0, 0, 0, 1, 1, 0, 0]).
\\

\section{Proposed Encoding}
\label{sec:proposed_encoding}


Before going into depth on our balanced encoding, let's start by explaining the imbalances in the existing 8-bit linear transformation adopted in white-box AES implementations.  
For $x \in$ GF$(2^8)$, let $S^{\ell}(x)$ denote the SubBytes output multiplied by $\ell \in \{1, 2, 3\}$, the elements in the MixColumns matrix. 
Here, $S^{\ell}(x, y)$ refers to the $y$-th bit of $S^{\ell}(x)$, where $y \in [1, 8]$ (the MSB is the first bit).
The following matrix $\textbf{S}^{\ell}$ is defined as
\begin{align}  
\bf{S}^{\ell} & = 
\begin{bmatrix}
    \textbf{S}^{\ell}_{1,1} ~~   & \textbf{S}^{\ell}_{1,2}  ~~ & \dots   ~~ & \textbf{S}^{\ell}_{1,256} \\
    \textbf{S}^{\ell}_{2,1} ~~   & \textbf{S}^{\ell}_{2,2}  ~~ & \dots   ~~ & \textbf{S}^{\ell}_{2,256} \\    
    \vdots         ~~   & \vdots          ~~ & \ddots  ~~ & \vdots   \\
    \textbf{S}^{\ell}_{8,1} ~~   & \textbf{S}^{\ell}_{8,2}  ~~ & \cdots  ~~ & \textbf{S}^{\ell}_{8,256}
\end{bmatrix} \notag \\
& = 
\begin{bmatrix}
    S^{\ell}(0,1) ~~   & S^{\ell}(1,1)  ~~ & \dots   ~~ & S^{\ell}(255,1) \\
    S^{\ell}(0,2) ~~   & S^{\ell}(1,2)  ~~ & \dots   ~~ & S^{\ell}(255,2) \\    
    \vdots         ~~   & \vdots          ~~ & \ddots  ~~ & \vdots   \\
    S^{\ell}(0,8) ~~   & S^{\ell}(1,8)  ~~ & \cdots  ~~ & S^{\ell}(255,8)
\end{bmatrix}.  \notag
\end{align}
An asterisk often refers to either a row or a column. 
For example, $\textbf{S}^{\ell}_{i, \ast}$ refers to the $i$-th row, and $\textbf{S}^{\ell}_{\ast, j}$ refers to the $j$-th column of $\textbf{S}^{\ell}$. 
If $M_{i,j}$ is an 8 $\times$ 8 binary invertible matrix, an 8-bit linear transformation with $\textbf{S}^{\ell}$ is given by $\textbf{R}^{\ell} = M \cdot \textbf{S}^{\ell}$.
Each row in $\textbf{R}^{\ell}_{i, \ast}$ is then computed by the XOR operations between the selected rows in $\textbf{S}^{\ell}$; if $M_{i,j}$ = 1, $\textbf{S}^{\ell}_{j, \ast}$ is XORed to compute  $\textbf{R}^{\ell}_{i, \ast}$

By Definition~\ref{def:walsh_balance}, for random integers $i$, $i'$ $\in$ [1,8] and $\ell$, $\ell'$ $\in$ [1,3], 
the balanced linear transformation must satisfy the Walsh transform as follows:
\begin{equation}
\sum_{j = 0}^{255}(-1)^{\textbf{R}^{\ell}_{i,j} \oplus \textbf{S}^{\ell'}_{i',j}} = 0.
\label{eq:R-S-0}
\end{equation}
In other words, this implies HW($\textbf{R}^{\ell}_{i,\ast} \oplus \textbf{S}^{\ell'}_{i',\ast}$) = 128.
The key-dependent distribution of $\textbf{S}^{\ell}$, however,  leads to HW($\textbf{R}^{\ell}_{i,\ast} \oplus \textbf{S}^{\ell'}_{i',\ast}$) = 0, which results in the Walsh transform value 256, with an overwhelming probability.
This has been proven by the following lemma, which can be found in~\cite{Lee:Linear:2020}. (Note that we have transposed matrix $\mathbf{H}$.) \\

\begin{lemma} 
\label{lem_1}
Assume that a 8$\times$256 binary matrix $\bf{H}$ is defined as
\[
\bf{H} = \begin{bmatrix}
    h_{1,1} ~~   & h_{1,2}  ~~ & \dots   ~~ & h_{1,256} \\
    h_{2,1} ~~   & h_{2,2}  ~~ & \dots   ~~ & h_{2,256} \\    
    \vdots         ~~   & \vdots          ~~ & \ddots  ~~ & \vdots   \\
    h_{8,1} ~~   & h_{8,2}  ~~ & \cdots  ~~ & h_{8,256}
\end{bmatrix},
\]
where the $i$-th column vector ${\bf h}_{*,i} = \langle h_{1, i}, h_{2, i}, \dots, h_{8, i} \rangle$
is an element of $GF(2^8)$ and ${\bf h}_{*, i} \neq {\bf h}_{*, j}$ for all $i \neq j$.
Then the HW of XORs of arbitrarily chosen row vectors from $\mathbf{H}$ is either 0 or 128.
In other words,
$HW({\bf h}_{j_1, *} \oplus {\bf h}_{j_2, *} \oplus \dots \oplus {\bf h}_{j_n, *})$ = 0 or 128,
where $n$ is a randomly chosen positive integer and $j_i \in \{1, 2, \dots, 8 \}$. \\
\end{lemma}

The proof can be simplified as follows, with more details provided in~\cite{Lee:Linear:2020}.
Let $\mathcal{J}$ be a set of randomly chosen indices from $\{1, 2, \dots, 8 \}$.
We can assume without loss of generality that $\mathcal{J}$ contains no duplicated indices and
$\big| \mathcal{J} \big| = n \leq 8$.
Define partitions of indices as
$$\mathcal{I}_{b_{1}, b_{2}, \dots, b_{n}} = \{ \ell \in \mathcal{I} | h_{j_i, \ell } = b_{i} \mbox{~for all~} j_i \in \mathcal{J} \},$$
where $\mathcal{I} = \{1, 2, \dots, 256 \}$, and $b_i \in \{0, 1\}$.
Here all $\mathcal{I}_{b_{1}, b_{2}, \dots, b_{n}}$ are disjoint to the others and
$\cup \mathcal{I}_{b_{1}, b_{2}, \dots, b_{n}} = \mathcal{I}$.
For any choice of $b_i$'s, we know that 
$$
\big| \mathcal{I}_{b_{1}, b_{2}, \dots, b_{n}} \big| = 256 / 2^n = 2^{8-n}.
$$
Using the definition of the HW, it follows that $\textnormal{HW}( \oplus_{j \in \mathcal{J}} {h}_{j,*})$ is summation of $\big| \mathcal{I}_{b_{1}, b_{2}, \dots, b_{n}} \big|$ where $\oplus_{i=1, \dots, n} b_i = 1$. 
$$\textnormal{HW}( \oplus_{j \in \mathcal{J}} {h}_{j,*}) = \Sigma_{\oplus_{i=1, \dots, n} b_i = 1} \big| \mathcal{I}_{b_{1}, b_{2}, \dots, b_{n}} \big| ~~~~~~~~$$ $$~~~~~~~~~~~~~~~~~~~~~~ = \Sigma_{\oplus_{i=1, \dots, n} b_i = 1} 2^{8-n} = \Sigma_{2^{n-1}} 2^{8-n}$$ $$~~~~~~~~~~~~~~~~~~~~~~~~~~~~~~~~ = {2^{n-1}} \cdot 2^{8-n} = 2^7 = 128.$$
If $\mathcal{J}$ is empty after removing duplicates, then the final HW becomes 0.

Equation~(\ref{eq:R-S-0}) can be simply re-written as 
\begin{align*}
\begin{split}
& ~~ 256 - (2 \times \textnormal{HW}( (M \cdot \textbf{S}^{\ell})_{i,*} \oplus \textbf{S}^{\ell'}_{i',*})) \\
 = & ~~ 256 - (2 \times \textnormal{HW}( \textbf{R}^{\ell}_{i,*} \oplus \textbf{S}^{\ell'}_{i',*})). 
 \label{eq:R-S-0-1}   
\end{split}
\end{align*}
It is important to note that for any $\ell$ and $\ell'$ in the range [1, 3], all row vectors of $\textbf{S}^{\ell}$ can be represented by XORing the row vectors of $\textbf{S}^{\ell'}$, and vice versa. 
By utilizing Lemma~\ref{lem_1} and considering the properties of GF($2^8$), it can be inferred that HW$((M \cdot \textbf{S}^{\ell})_{i,*} \oplus \textbf{S}^{\ell'}_{i',*})$ will result in either 0 or 128, depending on the specific rows of $\textbf{S}^{\ell}_{i,*}$ that are chosen by $M$.


Table~\ref{tab:blacklist} lists $\mathcal{V}$ and $\mathcal{W}$, the row index of $\textbf{S}^{\ell'}$ and the sets  of the row indexes of $\textbf{S}^{\ell}$, respectively,  producing the Walsh transform value of 256 for each pair of ($\ell$, $\ell'$).
The first row of the table means that if $\ell = \ell'$, the Walsh transform value would be 256 only if the $j$-th row from $\textbf{S}^{\ell}$ and $\textbf{S}^{\ell'}$ was selected, where $j \in [1, 8]$.
For ($\ell, \ell'$) = (2, 1), for example, suppose that $\textbf{R}^{2}$ contains a row computed by a linear combination of the 7- and 8-th rows of  $\textbf{S}^{2}$. 
This row is then identical to the 8-th bit of $\textbf{S}^{1}$. 
In other words, this linear transformation is unable to protect $\textbf{R}^{2}$ from an attacker calculating a correlation using the 8th bit of $\textbf{S}^{1}$. 

Considering AddRoundKey performed before SubBytes, a subkey $k$ will be added to the input $x$ of $S(\cdot)$.
Here, it is easy to know that $S(x \oplus k)$ can by expressed by a permutation of columns in $\textbf{S}^{\ell}$ and $\textbf{S}^{\ell'}$. 
Thus, the same row indexes listed in Table~\ref{tab:blacklist} will lead to the same results even after $k$ is added to $x$. 

\begin{table}[]
\centering
\caption{The row indexes of $\textbf{S}^{\ell}$ of which the XOR result becomes a row of $\textbf{S}^{\ell'}$. }
\label{tab:blacklist}
\begin{tabular}{@{}c@{\hskip 0.2in}c@{\hskip 0.2in}c@{\hskip 0.2in}c@{}}
\toprule
$\ell$ & $\ell'$ & $\mathcal{V}$                                                                         & $\mathcal{W}$                                                                         \\ 
\midrule
$i$ & $i$ & \begin{tabular}[c]{@{}c@{}} $j$ \end{tabular} & \begin{tabular}[c]{@{}c@{}}\{$j$\}\end{tabular} \\
\midrule

1 & 2 & \begin{tabular}[c]{@{}c@{}}1\\ 2\\ 3\\ 4\\ 5\\  6\\ 7\\ 8\end{tabular} & \begin{tabular}[c]{@{}c@{}}\{2\}\\ \{3\}\\ \{4\}\\ \{1, 5\}\\ \{1, 6\}\\ \{7\}\\ \{1, 8\}\\ \{1\}\end{tabular} \\
\midrule
1 & 3 & \begin{tabular}[c]{@{}c@{}}1\\ 2\\ 3\\ 4\\ 5\\ 6\\ 7\\ 8\end{tabular}     & \begin{tabular}[c]{@{}c@{}}\{1, 2\}\\ \{2, 3\}\\ \{3, 4\}\\ \{1, 4, 5\}\\ \{1, 5, 6\}\\ \{6, 7\}\\ \{1, 7, 8\}\\ \{1, 8\}\end{tabular}     \\
\midrule
2 & 1 & \begin{tabular}[c]{@{}c@{}}1\\ 2\\ 3\\ 4\\ 5\\ 6\\ 7\\ 8\end{tabular}     & \begin{tabular}[c]{@{}c@{}}\{8\}\\ \{1\}\\ \{2\}\\ \{3\}\\ \{4, 8\}\\ \{5, 8\}\\ \{6\}\\ \{7, 8\}\end{tabular}     \\
\midrule
2 & 3 & \begin{tabular}[c]{@{}c@{}}1\\ 2\\ 3\\ 4\\ 5\\ 6\\ 7\\ 8\end{tabular}     & \begin{tabular}[c]{@{}c@{}}\{1, 8\}\\ \{1, 2\}\\ \{2, 3\}\\ \{3, 4\}\\ \{4, 5, 8\}\\ \{5, 6, 8\}\\ \{6, 7\}\\ \{7\}\end{tabular}     \\
\midrule
3 & 1 & \begin{tabular}[c]{@{}c@{}}1\\ 2\\ 3\\ 4\\ 5\\ 6\\ 7\\ 8\end{tabular}     & \begin{tabular}[c]{@{}c@{}}\{1, 2, 3, 4, 5, 6, 7, 8\}\\ \{2, 3, 4, 5, 6, 7, 8\}\\ \{3, 4, 5, 6, 7, 8\}\\ \{4, 5, 6, 7, 8\}\\ \{1, 2, 3, 4\}\\ \{6, 7, 8\}\\ \{7, 8\}\\ \{1, 2, 3, 4, 5, 6, 7\} \end{tabular}     \\
\midrule
3 & 2 & \begin{tabular}[c]{@{}c@{}}1\\ 2\\ 3\\ 4\\ 5\\ 6\\ 7\\ 8\end{tabular}     & \begin{tabular}[c]{@{}c@{}}\{2, 3, 4, 5, 6, 7, 8\}\\ \{3, 4, 5, 6, 7, 8\}\\ \{4, 5, 6, 7, 8\}\\ \{5, 6, 7, 8\}\\ \{1, 2, 3, 4, 5\}\\ \{7, 8\}\\ \{8\}\\ \{1, 2, 3, 4, 5, 6, 7, 8\}\end{tabular}     \\ \bottomrule
\end{tabular}
\end{table}

\subsection{Basic Idea}

Our encoding scheme employs a customized 8-bit linear transformation and a simple nibble encoding to protect a key-dependent intermediate byte. To maintain balance within the linear transformation, it is crucial to construct a matrix $M$ that excludes certain row indexes, as specified in Table~\ref{tab:blacklist}. In order to achieve this, we introduce $\mathcal{W}$ as the set of index combinations (set) satisfying the condition:
$$
\mathcal{W} = \{\mathcal{J} | \oplus_{j \in \mathcal{J}} \textbf{S}^{\ell}_{j,*} = \textbf{S}^{\ell'}_{i,*} \mbox{~for all~} \ell, \ell' \in [1, 3] \},  
$$
where $i \in [1, 8]$. The set $\mathcal{W}$ is constructed by gathering the index sets from Table~\ref{tab:blacklist}. To ensure the balance requirement of the linear transformation, we enforce the condition that \textit{Idx}($M_{i,*}$) $\notin \mathcal{W}$, where $i$ represents each row in matrix $M$. This condition ensures that the selected indexes for each row of $M$ do not coincide with the sets in $\mathcal{W}$, thereby maintaining the desired balance within the linear transformation.
By doing so, we can achieve 
\begin{equation}
\textnormal{Pr}[\textbf{R}^{\ell}_{i,j} = \textbf{S}^{\ell'}_{i',j}] = 1/2,
\label{eq:Prob_R_S}
\end{equation}
which means that, from Equation~(\ref{eq:R-S-0}), the observed information and the correct hypothetical value is identical with a 1/2 probability.

Furthermore, as mentioned previously, the choice of different keys only affects the order of column vectors in $\textbf{S}^{\ell}$ and does not impact the HW of each row vector. Therefore, our encoding scheme remains independent of the key selection.


Previously, the linear transformation relied exclusively on invertible matrices to ensure its inverse transformation. However, this approach, which focused on binary invertible matrices and excluded specific row indexes, resulted in reduced table diversity.

To overcome this limitation, we have introduced a novel linear transformation that incorporates both singular and nonsingular matrices. Unlike before, where decoding (i.e., inverse transformation) was not feasible when using singular matrices, our new approach successfully addresses this challenge. By incorporating both singular and nonsingular matrices, we have developed a method that enables decoding even when singular matrices are utilized. This breakthrough enhances the versatility and effectiveness of the linear transformation, significantly expanding its capabilities.

Now, let's delve into the proposed 8-bit linear transformation, which is composed of 4-bit transformations. Additionally, we utilize 4-bit nonlinear transformations, known as nibble encoding, to partially conceal the value of 0. This is necessary because multiplying any value with 0 always yields 0, which can introduce problematic correlations with correct hypothetical values. By utilizing nibble encoding, we mitigate this issue and improve the robustness and security of the linear transformation scheme.\\

\subsection{Balanced Linear Transformation}
\label{sec:linear-transformation}

\begin{figure}[t]
\centering
\includegraphics[width=0.30\linewidth]{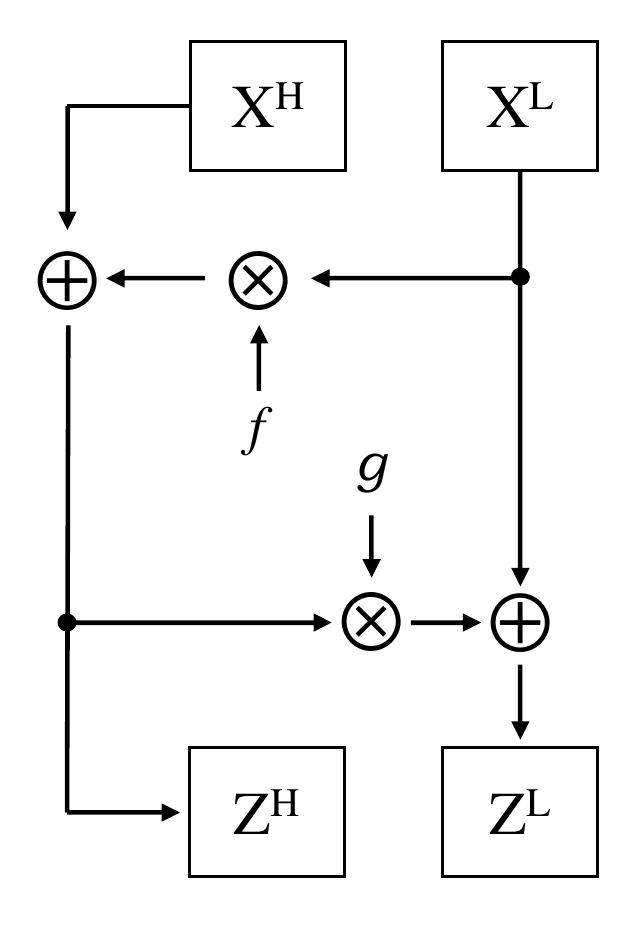}
\caption{Balanced encoding of an 8-bit vector $\textnormal{X}$ $(= \textnormal{X}^\textnormal{H} || \textnormal{X}^\textnormal{L})$ using 4$\times$4 binary matrices $f$ and $g$. }
\label{fig:balanced_encoding}
\end{figure}

\noindent \textbf{Proposed transformation $\mathcal{L}$: }
Suppose $f$ and $g$ are the elements randomly chosen from $\mathcal{F}$ and $\mathcal{G}$, respectively. 
For an 8-bit vector $\textnormal{X} = \textnormal{X}^\textnormal{H} || \textnormal{X}^\textnormal{L}$, the proposed linear transformation  $\textnormal{Z} = \mathcal{L}(\textnormal{X}, f, g)$ is defined as follows: 
\begin{equation*} \label{eq1}
\begin{split}
\textnormal{Z}^\textnormal{H} & = \textnormal{X}^\textnormal{H} \oplus (f \cdot \textnormal{X}^\textnormal{L})  \\
\textnormal{Z}^\textnormal{L} & = \textnormal{X}^\textnormal{L} \oplus (g \cdot \textnormal{Z}^\textnormal{H}), 
\end{split}
\end{equation*}
where $\cdot$ implies multiplication. 
Fig~\ref{fig:balanced_encoding} illustrates a graphical representation of it. 
This can be simply represented by $\textnormal{M} \cdot \textnormal{X}$, using an 8$\times$8 matrix $M$ defined as follows: \\
\begin{equation*}
\left[
    \begin{array}{c; {2pt/2pt}c}
        I_4 ~~~~& f \\ \hdashline[2pt/2pt]
        g ~~~~& I_4 \oplus g \cdot f 
    \end{array}
\right].
\end{equation*}
During the computation of $\mathcal{L}$, if $\textnormal{M}_{\ast, i}$ =1, the $i$-th row of $X$ is to be XORed.  
To ensure balance in the encoding, the initial step is to select random binary vectors $f_{i, \ast}$ 
such that \textit{Idx}($M_{i, \ast}$) $\not \in \mathcal{W}$, where $i \in [1, 4]$.\\

\noindent \textbf{Generation of \textit{f}: } Let $\mathcal{B}^{\mathcal{F}}_i$ denote the blacklist, which represents the set of binary vectors that must not appear in the $i$-th row of $f \in \mathcal{F}$. The random generation of $f$ can be described as follows:
\begin{enumerate}
    \item Initialize $f$ with all elements set to 0. 
    \item For each row $i$ in $f$:
    \begin{enumerate}
        \item Generate a 4-bit random binary vector $b_i$ = [$b_{i1}, b_{i2}, b_{i3}, b_{i4}$]. 
        \item While $b_i$ $\in$ $\mathcal{B}^{F}_{i}$, generate a new 4-bit random binary vector $b_i$.
        \item Set the row $i$ of $f$ as $b_i$.
    \end{enumerate}
    \item Return the generated matrix $f$.
\end{enumerate}
There exist 27,000 (=10$\times$15$\times$15$\times$12) binary matrices in $\mathcal{F}$ such that $\mathcal{F}_{i, \ast} \not \in \mathcal{B}^{\mathcal{F}}_i$. \\

\begin{table}[b]
\centering
\caption{The set of vectors that should be filtered out at the $i$-th row of $F$.}
\label{tab:blacklist_F}
\begin{tabular}{@{}c@{\hskip 0.2in}c@{}}
\toprule
$i$ & $\mathcal{B}^{F}_{i}$                                                                          \\ \midrule
1 & \{{[}0, 0, 0, 0{]}, {[}1, 0, 0, 0{]}, {[}0, 1, 0, 0 {]},                   \\
  &   {[}0, 0, 0, 1 {]},{[}1, 1, 0, 0 {]},{[}0, 0, 1, 1 {]}   \} \\
2 & \{{[}0, 0, 0, 0{]}\}                                                       \\
3 & \{{[}0, 0, 0, 0{]}\}                                     \\
4 & \{{[}0, 0, 0, 0{]}, {[}0, 0, 0, 1{]}, {[}1, 0, 0, 1{]}, {[}1, 1, 1, 1{]}\} \\ \bottomrule
\end{tabular}
\end{table}

\noindent \textbf{Generation of \textit{g}: } After generating $f$ which satisfies the condition, let $g$ be randomly chosen from $\mathcal{G}$. The pair ($g, f$) is checked to ensure that \textit{Idx}($g_{i, \ast} || (I_4 \oplus g \cdot f)_{i, \ast}$) does not belong to the set $\mathcal{W}$, where $i$ $\in$ [1, 4]. Our exhaustive search has revealed that there exist a total of 1,098,661,500 ($> 2^{30}$) valid pairs of ($g, f$) such that
$$
\textit{Idx} ( g_{i, \ast} || (I_4 \oplus g \cdot f)_{i, \ast}) \not \in \mathcal{W}. 
$$

In other words, our linear transformation guarantees the table diversity consisting of more than $2^{30}$ pairs of ($g, f$) producing the Walsh transform value of 0.  
On average, for each $f$, the number of row vectors at $g_{1, \ast}, g_{2, \ast}, g_{3, \ast}$, and $g_{4, \ast}$ is approximately 14, 13.6, 14.2, and 14.6, respectively. \\

\noindent \textbf{$\mathcal{L}^{-1}$ and XORs: }  For four 8-bit values $X_1$, $X_2$, $X_3$, and $X_4$, 
let $Y$ and $Z$ stand for the followings:
$$
Y =  \bigoplus_{i=1}^4 X_i \textnormal{ and } 
Z =  \bigoplus_{i=1}^4 \mathcal{L}(X_i, f, g), 
$$
where  $Z = Z^H || Z^L = \mathcal{L}(Y^H || Y^L, f, g)$.
Then, the inverse of the linear transformation is accomplished by $Y$ = $\mathcal{L}^{-1}(Z, f, g)$: 
\begin{align}
\begin{split}
Y^L = & Z^L \oplus (g \cdot Z^H) \\
Y^H = & Z^H \oplus (f \cdot Y^L) \\
Y   = & Y^H || Y^L. \nonumber
\end{split}
\end{align}

\noindent \textbf{Resistance to joint leakage: } 
One of the key features of our linear transformation $\mathcal{L}$ is its ability to maintain balance during the XOR operations of MixColumns between the encoded bytes. This property ensures resistance to joint leakage, which is a necessary condition for higher-order attacks. Therefore, our proposed implementation of AES, as described in the next section, demonstrates resilience against joint leakage, providing a robust defense against potential higher-order attacks.

Let $\ell_1, \ell_2, \ell_3, \ell_4$ represent the four coefficients of a row in the MixColumns matrix. For instance, in the first row, the coefficients are 2, 3, 1, and 1, respectively. During the MixColumns operation, each byte in the resulting state matrix is obtained by performing an XOR operation on four bytes selected from $\textbf{S}^{\ell_1}$, $\textbf{S}^{\ell_2}$, $\textbf{S}^{\ell_3}$, and $\textbf{S}^{\ell_4}$.

To demonstrate how $\mathcal{L}$ maintains balance during these XOR computations, we introduce an 8 $\times$ 256 lookup table denoted as $\overline{\textbf{S}}$. This table is constructed as follows:
$$
\overline{\textbf{S}}_{\ast, j} = \textbf{S}^{\ell_4}_{\ast, j} \oplus v^T, \textnormal{ for } 1 \leq j \leq 256, 
$$
where $v$ is a vector used for the XOR operation. It is worth noting that $\overline{\textbf{S}}$ can be considered as a permutation of columns in $\textbf{S}^{\ell'}$. Furthermore, we assume $\overline{\textbf{R}} = M \cdot \overline{\textbf{S}}$, without loss of generality.
If Equation~(\ref{eq:R-S-0}) is satisfied, then we have:
\begin{equation}
\sum_{j = 0}^{255}(-1)^{\overline{\textbf{R}}_{i,j} \oplus \overline{\textbf{S}}_{i',j}} = 0.
\label{eq:barR-barS-0}
\end{equation}
This equation implies that the HW($\overline{\textbf{R}}_{i,\ast} \oplus \overline{\textbf{S}}_{i',\ast}$) = 128. It provides evidence that the protection of $\mathcal{L}$ on MixColumns does not yield intermediate values that are highly correlated with the correct hypothetical value.

\subsection{Nibble Encoding for Hiding Zeros}
\label{sec:nibble-encoding}

The inclusion of nibble encoding serves a specific purpose in our scheme. Prior to applying nibble encodings, the balance provided by $\mathcal{L}$ ensures that all intermediate values exhibit no problematic correlation with the correct hypothetical values. However, it is necessary to address the challenge of hiding the value of 0, which cannot be concealed through simple multiplication.

To address this issue, we employ nibble encoding, which specifically targets the hiding of 0. Each nibble encoding operation involves swapping the value 0 with a candidate value $e \in [0, \texttt{0xF}]$, while leaving all other values unchanged. The primary objective is to identify an appropriate candidate $e$ that maintains the desired balance within the encoding process.

In summary, nibble encoding is utilized to address the requirement of concealing the value 0, which cannot be effectively hidden through standard multiplication operations. By selectively swapping 0 with a candidate value $e$, the scheme ensures that the overall balance is maintained while effectively hiding the presence of 0 within the encoding process.

\begin{figure}
\centering
\includegraphics[width=0.65\linewidth]{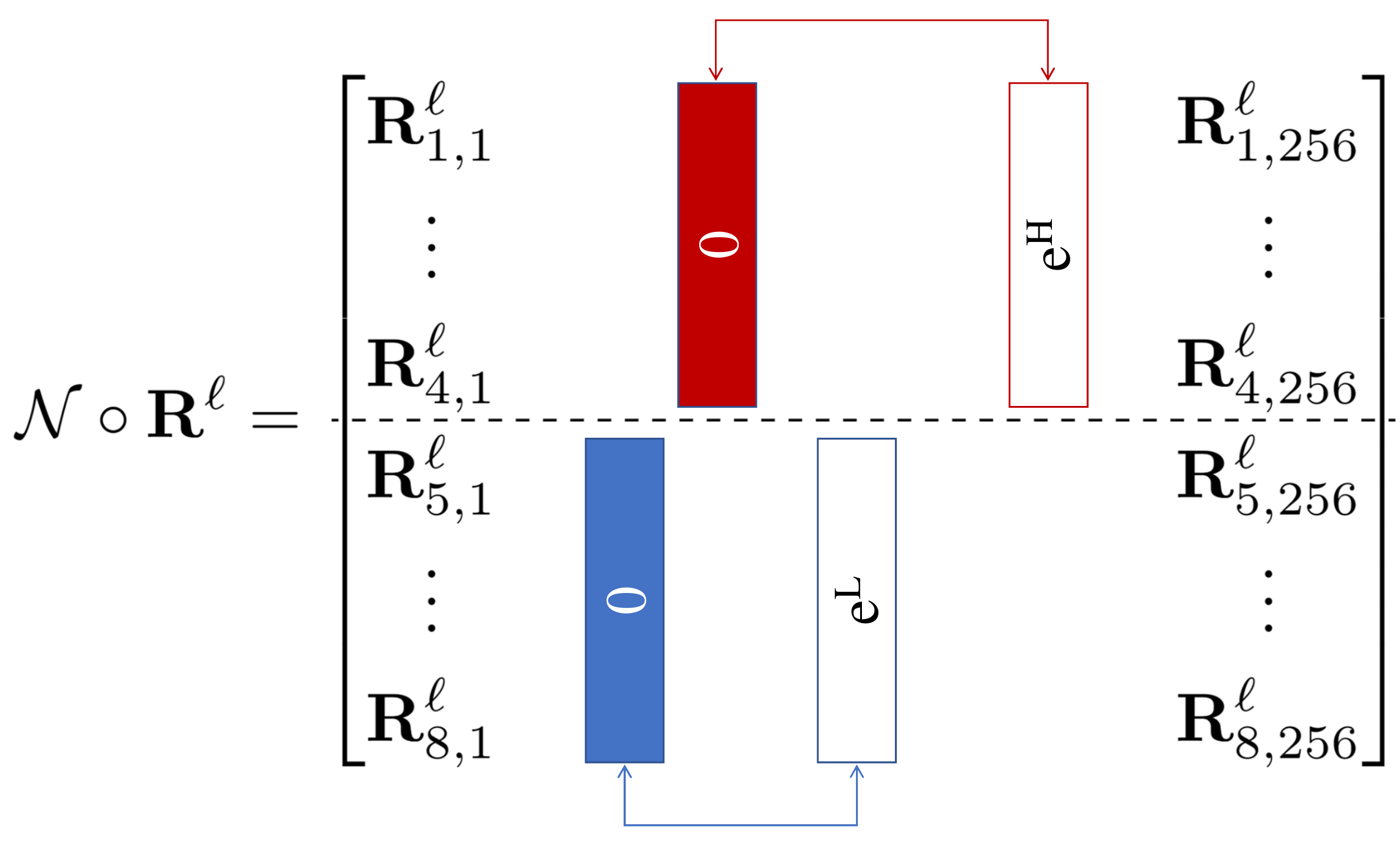}
\caption{Swapping zeros in $\mathcal{N}$ on $\textbf{R}^{\ell}$. }
\label{fig:nibble_encoding}
\end{figure}

\begin{figure}
\centering
\includegraphics[width=0.8\linewidth]{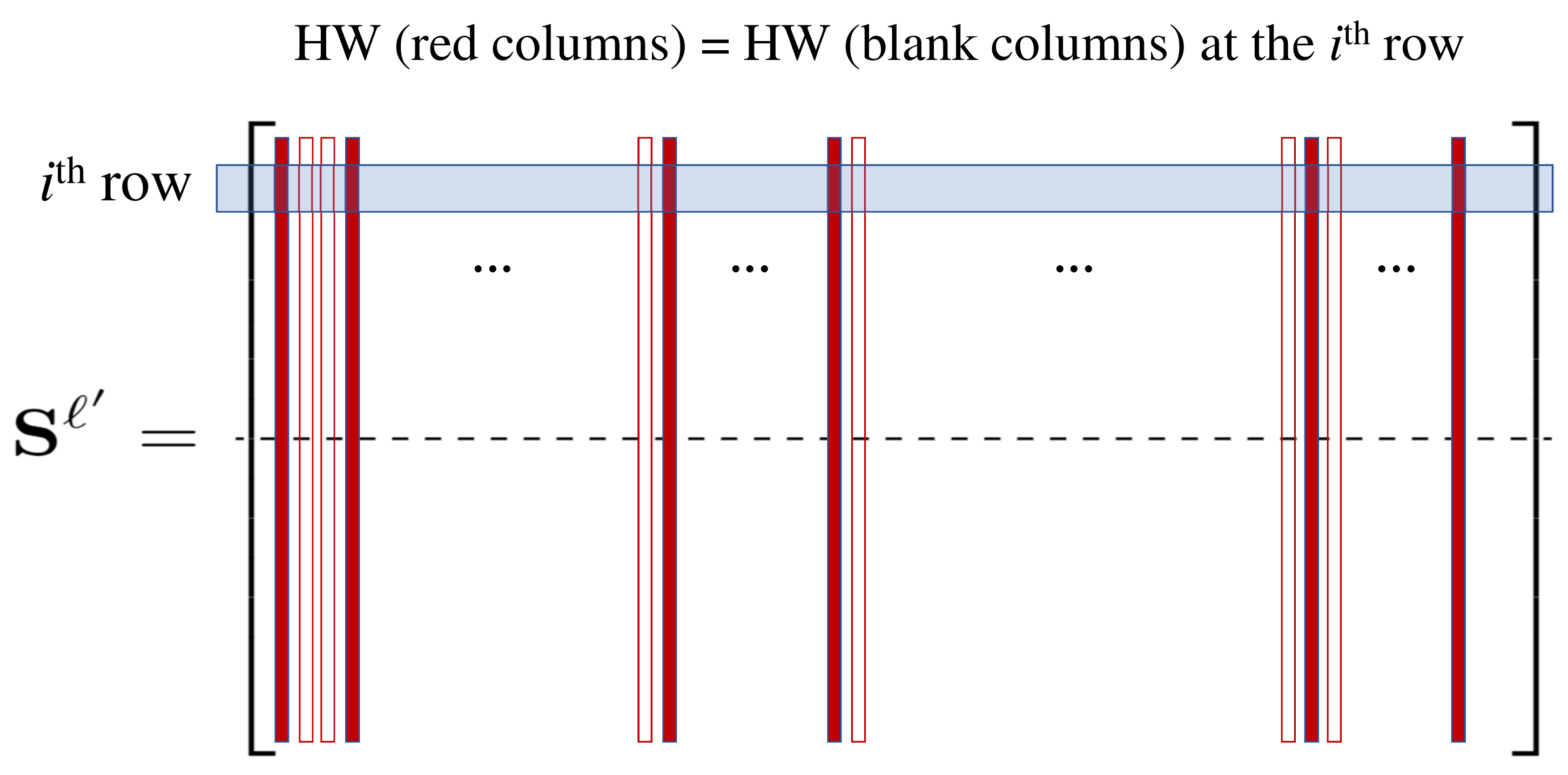}
\caption{$e^H$ can be a candidate if the same HW holds at every row after swaps. }
\label{fig:candidate_choice}
\end{figure}

Fig.~\ref{fig:nibble_encoding} shows graphical representation of two nibble encodings applied to $\textbf{R}^{\ell}$. 
In the upper and lower 4 bits, every four-bit chunk from 0 to \texttt{0xF} appears exactly 16 times because of the balance by $\mathcal{L}$.
Let denote two candidates to be swapped in the upper and the lower four bits by $e^H$ and $e^L$, respectively. 
In other words, the upper (resp. the lower) nibble encoding performs 16 swaps between 0 and $e^H$ (resp. $e^L$).  
Let $\mathcal{J}^H_0$ be a set of column indices with the upper four bits of $\textbf{R}^{\ell}$ equal to zero, and 
let $\mathcal{J}^H_e$ be a set of column indices with the upper four bits equal to $e$ as follows:
\begin{align*}
\begin{split}
\mathcal{J}^H_0  = & \{j | R^{\ell}_{1,j} || R^{\ell}_{2,j} || R^{\ell}_{3,j} || R^{\ell}_{4,j} = 0 \} \\
\mathcal{J}^H_e  = & \{j' | R^{\ell}_{1,j'} || R^{\ell}_{2,j'} || R^{\ell}_{3,j'} || R^{\ell}_{4,j'} = e \}
\end{split}
\end{align*}
For the upper 4-bit nibble encoding $\mathcal{N}^H$, a candidate $e^H$ must satisfy the following conditions to provide the first-order balanced encoding: 
$$
\sum_{j \in \mathcal{J}^H_0} S^{\ell'}_{i,j} = \sum_{j' \in \mathcal{J}^H_e} S^{\ell'}_{i,j'} \textnormal{  ~for all } \ell' \in \{1,2,3\} \textnormal{ and } 
1 \leq i \leq 8. 
$$
Here, it is worthy noting that the balance must be investigated for all $\ell'$ $\in$ \{1,2,3\} since a single matrix M will be used to protect $\textbf{S}^{1}$, $\textbf{S}^{2}$, and $\textbf{S}^{3}$ (see Section~\ref{sec:protected_aes}).
Similarly, we define 
\begin{align*}
\begin{split}
\mathcal{J}^L_0  = & \{j | R_{5,j} || R_{6,j} || R_{7,j} || R_{8,j} = 0 \} \\
\mathcal{J}^L_e  = & \{j' | R_{5,j'} || R_{6,j'} || R_{7,j'} || R_{8,j'} = e \}
\end{split}
\end{align*}
For the lower 4-bit nibble encoding $\mathcal{N}^L$, a candidate $e^L$ must satisfy the following conditions:
$$
\sum_{j \in \mathcal{J}^L_0} S^{\ell'}_{i,j} = \sum_{j' \in \mathcal{J}^L_e} S^{\ell'}_{i,j'} \textnormal{  ~for all } \ell' \in \{1,2,3\} \textnormal{ and } 
1 \leq i \leq 8. 
$$
A more graphical way of representing the implication is illustrated in Fig.~\ref{fig:candidate_choice}. 
Zeros can be swapped with $e^H$ if the HW of every column in $\mathcal{J}^H_0$ and $\mathcal{J}^H_e$ are the same for each row of $\textbf{S}^{\ell'}$. 
Also, $e^L$ can be found in the same manner using column indices belonging to $\mathcal{J}^L_0$ and $\mathcal{J}^L_e$.
By using 100 different pairs of ($f, g$), $\textbf{R}^{\ell \in \{1,2,3\}}$ was constructed and was then encoded by a pair of nibble encodings. 
We found that an average of 12.48 candidates (min: 2, max: 16) provides the balance for each nibble encoding.

\section{Protected AES-128}
\label{sec:protected_aes}

From now on, we describe a secure design of AES with a 128-bit key protected by our balanced encoding. 
Overall, this is inspired by a white-box cryptographic implementation of AES~\cite{Chow:WB-AES:2002}, which is mainly composed of a series of internally encoded lookup tables. 
Our focus is, however, on protecting against DPA-like attacks which take advantage of correlation to the key-dependent intermediate values.  
In other words, the rest of gray- and white-box attacks such as fault injection and cryptanalysis are not taken into account. 

\subsection{Design}
\label{sec:design}

\noindent\textbf{Rearrangement of AES:} The following describes a rearrangement of AES-128 by which a series of lookup tables is generated.   
By shifting the initial AddRoundKey into the first round and by applying ShiftRows to the round key (except the final round key), AES-128 can be described  concisely as follows: 

\setlength{\parindent}{25pt}
\indent  state $\leftarrow$ $plaintext$ \\
\indent  for $r$ = 1 $\cdots$ 9 \\
\indent\indent  ShiftRows(state)\\
\indent\indent  AddRoundKey(state, $\hat{k}^{r-1}$)\\
\indent\indent  SubBytes(state)\\
\indent\indent  MixColumns(state)\\
\indent  ShiftRows(state)\\
\indent  AddRoundKey (state, $\hat{k}^9$) \\
\indent  SubBytes(state)\\
\indent  AddRoundKey(state, $k^{10}$)\\
\indent  $ciphertext$ $\leftarrow$ state,\\
\setlength{\parindent}{0pt}

where ${k}^{r}$ is a 4 $\times$ 4 matrix of the $r$-th round key, and $\hat{k}^{r}$ is the result of applying ShiftRows to ${k}^{r}$.
By doing so, AddRoundKey can combine with SubBytes before multiplying each column of the MixColumns matrix for the first - ninth rounds. \\

\noindent\textbf{Generating lookup tables up to MixColumns multiplication:} We define an 8$\times$8 lookup table, denoted as \textit{T-boxes}, with the following expressions:
$$
\setlength\arraycolsep{0.1em}
\begin{array}{lcll}
  T^{r}_{i, j}(p)  & = & S(p \oplus \hat{k}^{r-1}_{i,j}), & \textnormal{for } 
  i, j \in [1, 4] \textnormal{ and } r \in [1, 9], \\
  T^{10}_{i, j}(p) & = & S(p \oplus \hat{k}^{9}_{i,j})\oplus k^{10}_{i,j}   & \textnormal{for } i, j \in [1, 4],
\end{array}
$$
where $p$ represents a subbyte of the \textit{state}.

Let $[x_1 ~ x_2 ~ x_3 ~ x_4]^T$ represent a column vector of the \textit{state} after performing the lookup in the \textit{T-boxes}. The subscript of each $x$ denotes the row index of the subbyte. 
To precompute the multiplication of $x_i$ with a column vector of the MixColumns matrix, we use the $U^{r}_{i,j}$ tables, which can be expressed as follows:

\begin{equation}
\begin{multlined}
\begin{array}{ccl}
U^r_{1,j}(x_1) &=& x_1 \cdot [\textnormal{02 01 01 03}]^{T}  \nonumber \\
U^r_{2,j}(x_2) &=& x_2 \cdot [\textnormal{03 02 01 01}]^{T}  \nonumber \\
U^r_{3,j}(x_3) &=& x_3 \cdot [\textnormal{01 03 02 01}]^{T}  \nonumber \\
U^r_{4,j}(x_4) &=& x_4 \cdot [\textnormal{01 01 03 02}]^{T}, \nonumber \\
\end{array}
\end{multlined}
\end{equation}
Here, $j$ represents the column index of the vector.

\begin{figure}
\centering
\subfloat[The first round]{
\centering
\includegraphics[width=0.5\linewidth]{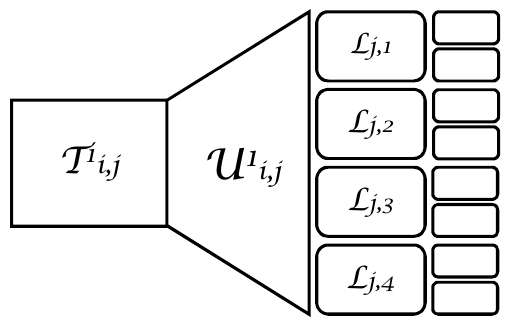}
\label{fig:tab-r1}} \\
\subfloat[The second round]{
\includegraphics[width=0.65\linewidth]{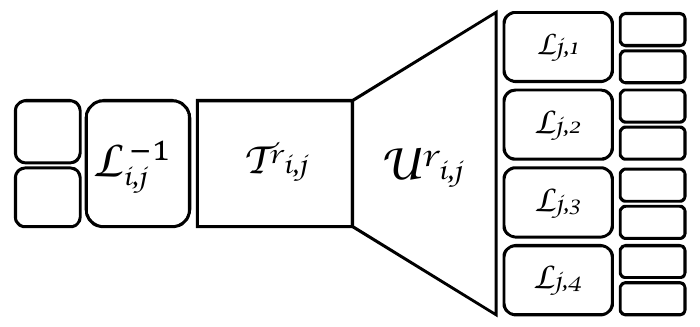}
\label{fig:tab-r2}} \\
\caption{Internal structure of $UT$ computing up to MixColumns multiplication. A blank squre is a nibble encoding.}
\label{fig:tab-mix-mul}
\end{figure}

The subsequent step involves composing $U$ and $T$ to generate the $UT$ tables, which map each subbyte of the plaintext to the MixColumns multiplication. It is important to note that the \textit{T-boxes}, serving as the precomputation tables for SubBytes, provide the inputs for the $U$ tables. Since the first key-dependent intermediate value, obtained by performing the encryption and looking up the $UT$ table, is exposed, it needs to be encoded. Let us denote these values as:
$$
UT^{r}_{i,j}(x_i) = [y_{i,1} ~~ y_{i,2} ~~ y_{i,3} ~~ y_{i,4}]^T.
$$ 
\\
\noindent\textbf{Encoding lookup tables:} In order to apply $\mathcal{L}$ to the four bytes of a column vector above, four sets of ($f, g$) are required; therefore each round uses 16 sets in total.  
The subscripts $_{j,k}$ are used to index a set pertaining to $\mathcal{L}$, where $j$ is a column index and $k \in \{1,2,3,4\}$.
$y_{i,k}$ is then encoded by $\mathcal{L}_{j,k}(y_{i,k}, f_{j,k}, g_{j,k})$. 
The structure of lookup tables precomputing the operations up to the MixColumns multiplication are simply depicted in Fig.~\ref{fig:tab-mix-mul}. 
Note that the second round begins with the decoding of the first round's output whereas the first round does not have to decode the plaintext. 
Through out this paper, we assume that the nibble encoding (blank squares in Fig.~\ref{fig:tab-mix-mul}) is always applied to every boundary of lookup tables.

\begin{figure}
\centering
\includegraphics[width = 0.8\linewidth]{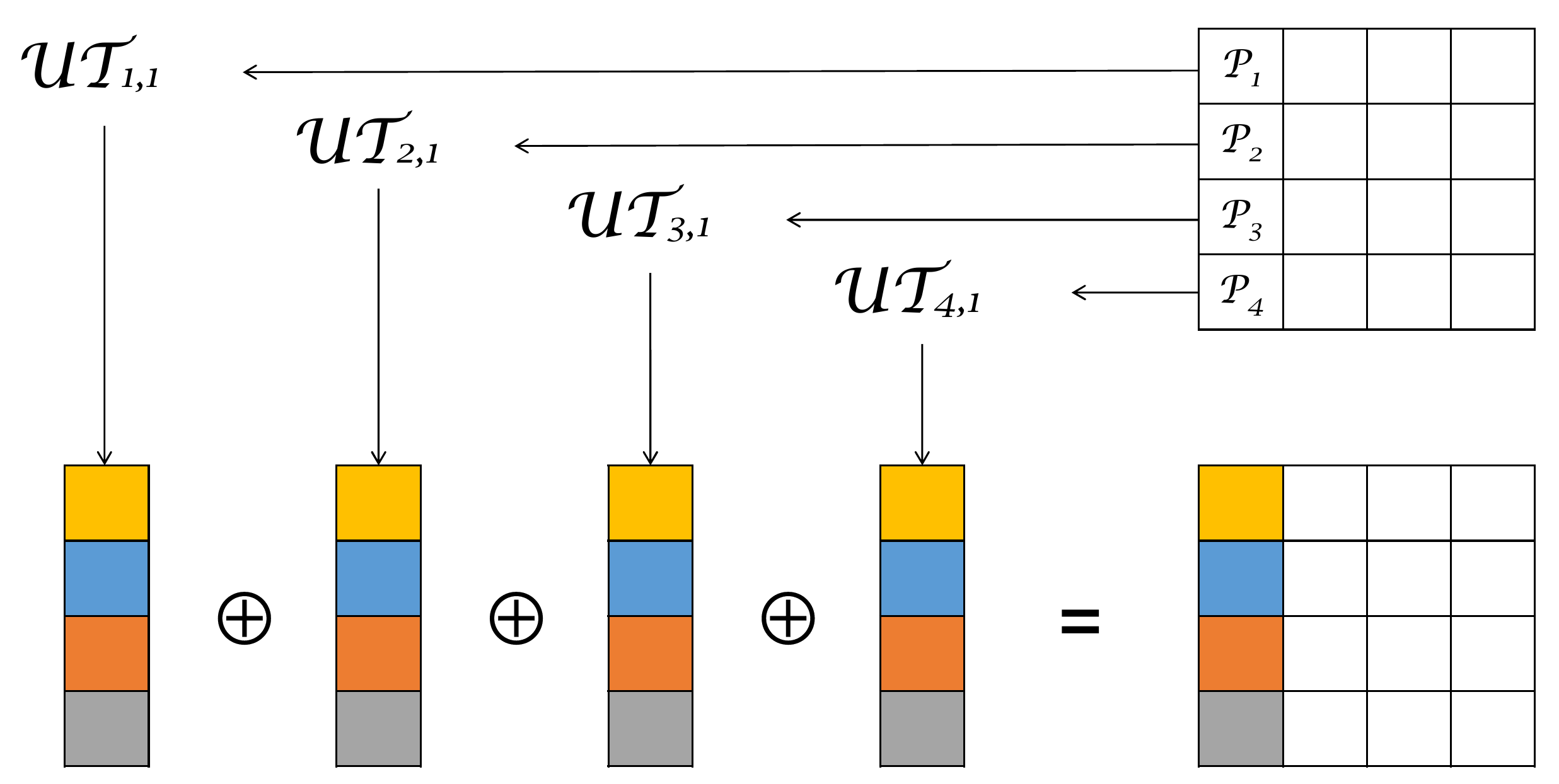}
\caption{Overall table lookups of a column vector in \textit{state}. 
A different color of the box means a different set of ($f, g$).  
$\oplus$ will be conducted by looking up the XOR tables $\mathcal{T}^x$. }
\label{fig:tab-lookup-round}
\end{figure}

Fig.~\ref{fig:tab-lookup-round} illustrates a simple description of lookups from \textit{state} to the first round output. 
For each column of \textit{state}, $UT$ takes a byte and provides a 4-byte vector of the MixColumns multiplication. 
The intermediate values after looking up $UT$  can be placed in a 4$\times$4$\times$4 array. \\

\noindent\textbf{XOR lookup tables:} The next step is to combine the encoded results of MixColumns multiplication into the round output by conducting XOR operations. 
Since the nibble encoding swaps zeros with unknown values, every XOR operation depicted by $\oplus$ in Fig.~\ref{fig:tab-lookup-round} must be performed by looking up the XOR tables, denoted by $\mathcal{T}^x$.
An instance of $\mathcal{T}^x$ is generated by using three nibble encodings to decode two 4-bit inputs and to encode a 4-bit output.
Due to the distributive property of multiplication over XOR, $\mathcal{L}$ does not have to be decoded in this process. 
The final round does not involve MixColumns, and thus $T^{10}$ is not composed with $U$. 
Because its outputs make a ciphertext, $T^{10}$ does not encode the output.  \\

\noindent\textbf{Generating the complementary set of the balanced lookup tables:}  
Up to this point, we have proposed a balanced encoding scheme for the AES implementation, aimed at reducing the correlation with hypothetical intermediate values that can be exploited. However, this balanced encoding approach gives rise to two vulnerabilities.
The first vulnerability stems from the lowest correlation, which increases the likelihood of the correct key being identified as the lowest-ranked key in CPA attacks. 
The second vulnerability arise from the bijectiveness of the encoding, making it susceptible to collision-based attacks. 

To address these vulnerabilities, we propose a method that involves the random selection of one set from multiple sets, each comprising pairs of complementary lookup tables. In the subsequent discussion, we focus on the utilization of a single pair denoted as ($Q_0$, $Q_1$).
Let $Q_0$ represent the set of lookup tables protected by the proposed encoding, as mentioned earlier. To ensure that the correct key does not exhibit the lowest correlation or perfect collision with the correct hypothetical values, we require another set of lookup tables, denoted as $Q_1$. In $Q_1$, the input and output bits of $Q_0$ are simply flipped. As a result, $Q_1$ also provides a balanced encoding for encrypting the plaintext.
In order to effectively protect the key, it is necessary to introduce a degree of imbalance. Consequently, a plaintext is encrypted using $Q_0$ with a probability of $\alpha$, while $Q_1$ is utilized with a probability of $1 - \alpha$.


Let $\{b_1, b_2 \cdots b_n \}$ be a sequence of binary numbers where  $b_i$ = 0 or 1 such that $\sum_{i = 1}^n b_i = n \times \alpha$. 
If a random number generator is available in the device, each encryption begins by picking up $i \in [1, n]$ at random and encrypts a plaintext using $Q_{b_i}$.
If $\alpha = 1/2$, a binary number, say $b$, may be simply generated to choose $Q_b$.   
In this case, this is the only operation for which the encryption depends on the run-time random number generator.
Otherwise, based on the fact that statistical analysis uses uniformly-distributed plaintexts, $i$ can be derived from plaintexts.
For example, if $n$ is unknown and less than 256, an XOR sum of every subbyte of the plaintext (\textit{mod} $n$) can be used to choose $i$.
By doing so, the encryption becomes independent of run-time random sources. 

\subsection{Costs}

Given the potential vulnerability of the AES algorithm's inner rounds to side-channel-assisted middle-round differential cryptanalysis~\cite{Bhasin:SITM:2020}, we expanded  our proposed encoding technique to encompass all rounds. Through a comparison between our single set, referred to as $Q_0$, of lookup tables and the unprotected WB-AES implementations introduced by Chow et al.~\cite{Chow:WB-AES:2002}, we observed reductions in both table size and the number of necessary table lookups achieved with $Q_0$.  This enhancement stems from our choice to employ four 8-bit transformations in place of a 32-bit transformation.

Table~\ref{tab:cost} provides an overview of the table sizes and the corresponding number of lookups. Our balanced encoding method, as previously explained, requires a complementary set of lookup tables, where both $Q_0$ and $Q_1$ possess equal sizes, resulting in a combined table size of approximately 512KB. While there is a slight increase in size compared to Chow's WB-AES implementation~\cite{Chow:WB-AES:2002}, which boasts  a table size of around 508KB without utilizing external encoding, our approach offers a significant advantage in terms of reducing the number of required lookups. In contrast to their implementation, which demands approximately 2,032 lookups, our method achieves a notable reduction of roughly 50\%. This improvement is feasible because each encryption operation only relies on just one of the table sets ($Q_0$ or $Q_1$), thereby minimizing the necessary lookup count.

It's crucial to emphasize that the threat model addressed by WB-AES specifically targets white-box attackers with the capability to access and modify computing resources. In this context, the use of a 32-bit linear transformation in WB-AES serves to enhance table diversity and introduce ambiguity as effective countermeasures against such attackers.

In our case, the choice to employ an 8-bit linear transformation is substantiated by the specific characteristics of our threat model. Given that our threat model primarily concentrates on non-invasive attacks, where adversaries lack the capability to modify computing resources, the use of an 8-bit linear transformation proves highly effective in providing the required protection and security measures within our particular context. Additionally, it's noteworthy that while the previous encoding method used in WB-AES does not offer protection against gray-box attacks, our approach provides defense against such attacks, thus offering supplementary security advantages.

In addition, Table~\ref{tab:elapsed_time} presents a comparison of the elapsed time and overall memory requirement required for encrypting a single block using various AES-128 implementations. 
This comparison encompasses Chow's WB-AES and Lee's WB-AES with static masking~\cite{Lee:ImproveMaskedWB:2020}, along with run-time masking countermeasures~\footnote{https://github.com/coron/htable} such as Rivain and Prouff's masking~\cite{Rivain:HO-Masking:2010} and Coron \textit{et al.}'s higher-order masking of lookup tables with common shares~\cite{Coron:LUT:2018}.

The experiments were carried out on a guest OS Ubuntu 16.04, running on a 3.4GHz Dual-core CPU with 8GB of RAM.
Please note that the elapsed time reported for run-time masking countermeasures excludes key scheduling. To ensure a fair comparison, we did not specifically measure the elapsed times for loading the lookup tables of both WB-AES and our implementations. Our approach demonstrates a performance improvement in comparison to existing countermeasures relying on either white-box implementations or run-time masking techniques.

From a practical standpoint, it's essential to acknowledge that our implementation does impose slightly higher memory resource demands, which could be a drawback for low-cost devices. The memory requirements of our proposed method significantly contrast with the relatively lower memory demands of higher-order masking techniques.
However, our implementation eliminates the requirement to generate random numbers or generate new sets of masked tables for each encryption execution. This streamlined approach greatly enhances efficiency, leading to faster encryption operations.

\begin{table}[]
\centering
\caption{Table size and lookups of the five outer rounds protected by the proposed encoding.}
\label{tab:cost}
\begin{tabular}{c@{\hskip 5mm}r@{\hskip 5mm}r}
\toprule
                          & \multicolumn{1}{c}{ Size (bytes)} & \multicolumn{1}{c}{\# of lookups} \\
\midrule
$UT$                      & $9\times4^3\times256$ = 147,456                & $9\times4^2$ = 144 \\
$\mathcal{T}^x$           & $9\times4^2\times3\times2\times128$ = 110,592  & $9\times4^2\times3\times2$ = 864 \\
$\mathcal{T}^{10}$        & $4^2\times256$ = 4,096                 & $4^2$ = 16 \\
\midrule
Total & 262,114 (approx. 256KB)  & 1,024 \\
\bottomrule
\end{tabular}
\end{table}

\begin{table}[]
\centering
\caption{Comparison of elapsed time and memory requirement for software implementations of encryption of one block. $n$: the number of shares.}
\label{tab:elapsed_time}
\begin{tabular}{lrr}
\toprule
                          & \multicolumn{1}{c}{ Elapsed time ($\mu$s)} & Mem (KB)\\
\midrule
AES-128 without countermeasure                                    & 0.7                & -\\
Chow's WB-AES~\cite{Chow:WB-AES:2002}                             & 26                 & 508\\
Lee's WB-AES~\cite{Lee:ImproveMaskedWB:2020}                      & 40                 & 17,964\\
Rivain-Prouff ($n$ = 3 / $n$ = 4)~\cite{Rivain:HO-Masking:2010}   & 38 / 45            & 4 / 5\\
Coron \textit{et al.}($n$ = 3 / $n$ = 4)~\cite{Coron:LUT:2018}    & 293 / 604          & 120 / 160 \\
This paper                                                        & 19                 & 512 \\ 
\bottomrule
\end{tabular}
\end{table}

\section{Experiments and Results}
\label{sec:experiment}

In this section, we evaluate the effectiveness of various non-invasive statistical analysis techniques using power traces. Building on this, we assume that the cryptographic binary cannot be extracted from the low-cost device, mitigating the risk of sophisticated invasive (white-box) attacks. Our primary focus is to assess the protection provided against well-known statistical analysis, as discussed in Section~\ref{sec:preliminaries}. It should be noted that the CPA attack using the HW model is ineffective against internally-encoded lookup tables~\cite{Lee:maskedWB:2018}. For the sake of convenience and efficiency in our analysis, we consider that the attacker employs mono-bit CPA, its variants, and collision-based attacks.
From a security verification standpoint, we either utilize the Walsh transform or collect noise-free computational traces directly from memory during binary execution. These traces are then subjected to the statistical analysis techniques mentioned above, demonstrating our effectiveness in defending against adversaries who rely on power traces collected using oscilloscopes.

\subsection{Methods}


To evaluate the protection against mono-bit CPA attacks, we analyze the correlation between the input and output of the encoded lookup tables, quantifying this correlation using the Walsh transform, as described in detail in Section~\ref{sec:preliminaries}. Additionally, we incorporate an open-source implementation of DCA from a public repository, which adds an extra layer of analysis to reinforce the security evaluation of our approach. For trace collection, we utilize Valgrind, which provides DBI capabilities. We then perform CPA attacks using the collected computational traces, aided by Daredevil~\cite{Paul:Daredevil:2015}. By employing noise-free computational traces, we can simultaneously evaluate resistance to CPA attacks using power traces. Furthermore, to demonstrate resistance against collision-based attacks, we utilize the Walsh transform as well.

The most frequently considered hypothetical value pertains to the output of the MixColumns multiplication. More specifically, it represents the outcome obtained by multiplying the SubBytes output with one of three coefficients (\texttt{0x01, 0x02, 0x03}) in the MixColumns matrix. It is worth noting that multiplication by \texttt{0x01} yields the SubBytes output itself.
Moreover, attacks can be targeted at the partial or full XOR operations performed on four bytes obtained from MixColumns multiplication. These XOR operations compute a subbyte of the round output. By analyzing the key leakage that occurs during these XOR operations, we can assess the resistance to joint leakage, which implies protection against higher-order attacks.

In our analysis, we first focus on the \textit{UT} outputs in the first round. To evaluate the impact of the balanced encoding, we utilize a single set of lookup tables, denoted as $Q_0$, and measure the correlation between the \textit{UT} outputs and the correct hypothetical values. 

Next, we investigate the influence of using either $Q_0$ or $Q_1$ with a probability of 1/2 on the key protection. By adapting this variation, we introduce an imbalance that affects the correlation between the \textit{UT} outputs and the hypothetical values. Consequently, the lowest correlation no longer reliably indicates the correct subkey, 
showcasing enhanced resistance against attacks.

Subsequently, we replicate the experiments targeting the first round's output. Initially, we utilize $Q_0$ and confirm that the balance in the encoding is maintained even when combining multiple intermediate values, thereby demonstrating resistance to joint leakage. Additionally, we assess the protection of the key when using either $Q_0$ or $Q_1$ with a probability of 1/2 for each encryption.
Furthermore, we demonstrate protection of collision and cluster analysis on the first round's output when using either $Q_0$ or $Q_1$. 

\subsection{Analysis of the UT Outputs}
\label{sec:analysis_UT_output}

\begin{figure}
\centering
\subfloat[With respect to the SubBytes output]{
\centering
\includegraphics[width=0.9\linewidth]{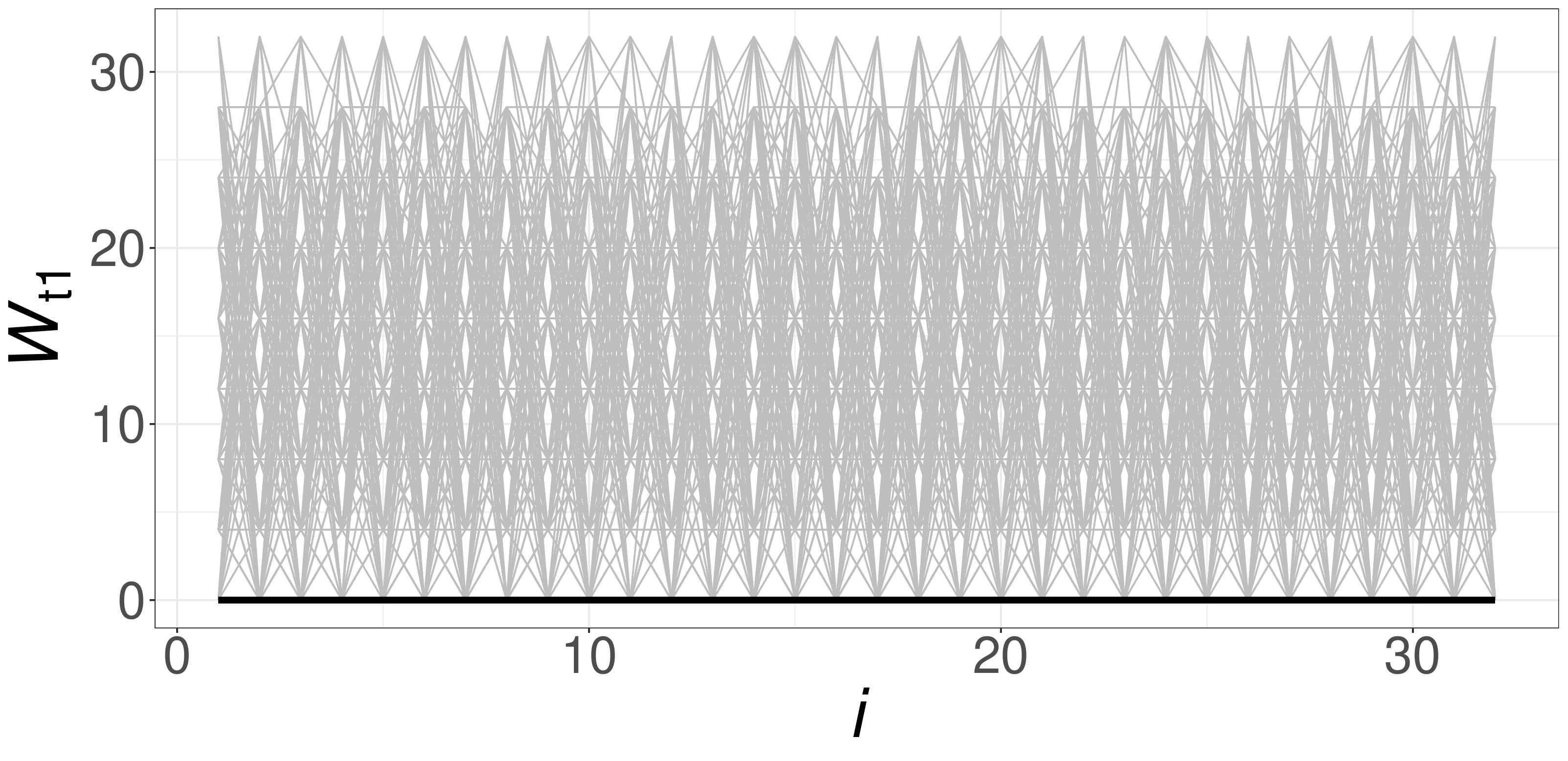}
} \\
\subfloat[With respect to the SubBytes output multiplied by 2]{
\includegraphics[width=0.9\linewidth]{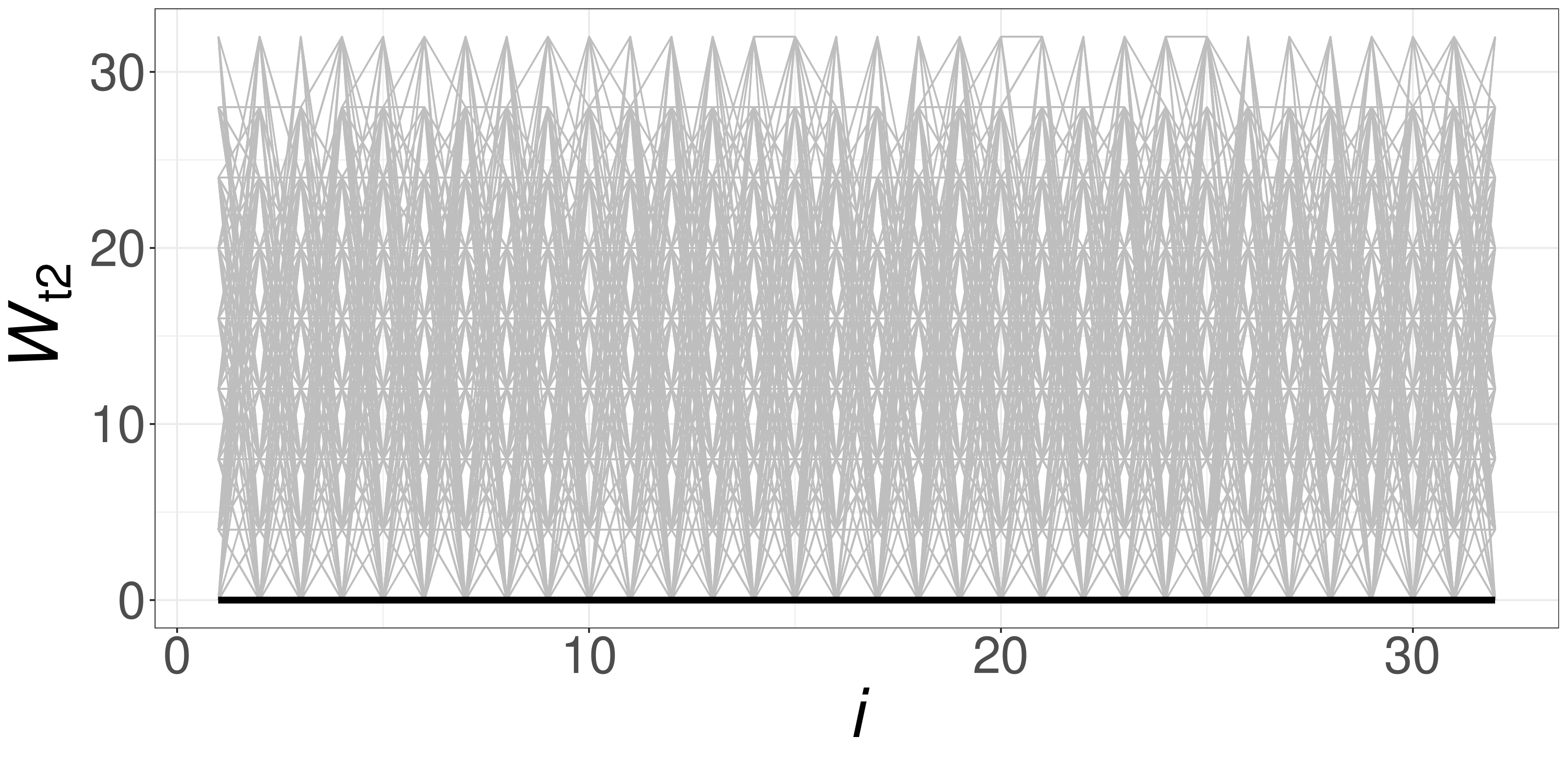}
} \\
\subfloat[With respect to the SubBytes output multiplied by 3]{
\includegraphics[width=0.9\linewidth]{fig/wt3_Q0_bit1.pdf}
} \\
\caption{Walsh transforms on the MixColumns multiplication obtained by $Q_0$. Black: correct subkey. Gray: wrong key candidates.}
\label{fig:Q0_UT_Walsh_bit1}
\end{figure}

To demonstrate the effect of the balanced encoding on the \textit{UT} outputs, we provide the following experimental results. 
First, the correlation of the \textit{UT} outputs to the hypothetical values was investigated in the first round.
Specifically, the hypothetical values guess the results of MixColumns multiplication, i.e., the SubBytes outputs multiplied by 1, 2, or 3.
In the following, we take a close look at  $UT^{1}_{0,0}$, a slice of the \textit{UT} table mapping the first subbyte of a plaintext to the MixColumns multiplication in the first round.
The Walsh transforms $W_{t\ell'}$ is defined:

\begin{equation}
W_{t\ell'}(i') = \sum_{j = 0}^{255}(-1)^{UT^{1}_{0,0}(j)_i \oplus \textbf{S}^{\ell'}_{i',j}}.
\label{eq:walsh_ut_1_0_0}
\end{equation}
Here, $W_{t\ell'}$ tells us how much the lookup values of $UT^{1}_{0,0}$ and the hypothetical SubBytes output multiplied by $\ell' \in \{1,2,3\}$ are correlated. 
For every single bit of the correct hypothetical values, all of the Walsh transforms give 0 as a result of the balanced encoding.
Due to limited space, Fig.~\ref{fig:Q0_UT_Walsh_bit1} shows a small part of the results obtained by the hypothetical values --- the first bit of the SubBytes output multiplied by 1, 2, and 3. 
The full experimental results are available in Appendix~\ref{sec:appendix_walsh}.
As outlined in Section~\ref{sec:protected_aes}, our encoding scheme utilizes a total of 16 sets of ($f$, $g$) in each round. By ensuring that every subkey is protected, we establish that the random selection of ($f$, $g$) maintains the desired balance within the encoding.


This balanced encoding has an influence on the result of DCA attacks using the hypothetical SubBytes output.
In order to illustrate it, 10,000 computational traces of encryption using 10,000 random plaintexts were collected by using Valgrind.
Table~\ref{tab:Q0_UT_DCA} shows the DCA result based on the SubBytes output. 
All correct subkeys ranked the lowest out of eight attacks at least once. 
Table~\ref{tab:Q0_UT_DCA_coefficients} compares the highest correlation coefficient from all key candidates and the coefficient from the correct key. 
When the computational traces are composed of only the \textit{UT} outputs, the DCA ranking of the correct subkey is always the lowest and the corresponding coefficients are much lower than the coefficients shown in Table~\ref{tab:Q0_UT_DCA_coefficients} (see also Appendix~\ref{sec:appendix_only_ut}).
This shows the proposed encoding addresses the correlation problem caused by the existing encoding method. 
These two experiments were carried out to show  there is no problematic correlation between the encoded lookup values and the correct values of $\textbf{S}^{\ell' \in \{1,2,3\}}$.

\begin{table*}
\centering
\scriptsize
\caption{DCA ranking on the AES encryption using only $Q_0$ when conducting mono-bit CPA on the SubBytes output with 10,000 computational traces.}
\label{tab:Q0_UT_DCA}
\begin{tabular}{c|*{16}{{c}@{\hskip 2mm}}}
\\
\backslashbox{TargetBit}{Subkey}   &       1   &       2   &    3   &    4   &   5    &    6   &      7   &    8   &   9   &  10   & 11   & 12   & 13  & 14  & 15  & 16  \\\hline
1 & 161	& 254	 & 247	 & 250	 & 256	 & 249	 & 250	 & 256	 & 167	 & 236	 & 250	 & 166	 & 211	 & 251	 & 253	 & 189	 \\
2 & 253	& 249	 & 256	 & 249	 & 203	 & 256	 & 251	 & 230	 & 253	 & 256	 & 255	 & 256	 & 230	 & 256	 & 256	 & 256	 \\
3 & 21	&  185	 & 136	 & 225	 & 199	 & 185	 & 24	 & 209	 & 183	 & 59	 & 29	 & 210	 & 18	 & 168	 & 235	 & 255	 \\
4 & 75	 & 76	 & 249	 & 13	 & 70	 & 158	 & 251	 & 52	 & 248	 & 190	 & 63	 & 124	 & 108	 & 171	 & 136	 & 239	 \\
5 & 252	 & 253	 & 256	 & 243	 & 256	 & 256	 & 256	 & 256	 & 256	 & 245	 & 228	 & 216	 & 249	 & 251	 & 253	 & 255	 \\
6 & 210	 & 138	 & 253	 & 241	 & 206	 & 161	 & 243	 & 256	 & 125	 & 219	 & 256	 & 247	 & 256	 & 120	 & 100	 & 90	 \\
7 & 256	 & 118	 & 249	 & 192	 & 68	 & 162	 & 235	 & 158	 & 167	 & 158	 & 119	 & 252	 & 118	 & 180	 & 230	 & 241	 \\
8 & 213	 & 256	 & 256	 & 256	 & 254	 & 248	 & 252	 & 254	 & 252	 & 234	 & 253	 & 256	 & 256	 & 239	 & 207	 & 125	 
\normalsize
\end{tabular}
\end{table*}

\begin{table*}
\centering
\scriptsize
\caption{Highest correlation of all key candidates vs. highest correlation of the correct subkeys when conducting the DCA attacks on the UT output obtained by $Q_0$. }
\label{tab:Q0_UT_DCA_coefficients}
\begin{tabular}{c|*{16}{{c}@{\hskip 1mm}}}
\\
Subkey    &       1   &       2   &    3   &    4   &   5    &    6   &      7   &    8   &   9   &  10   & 11   & 12   & 13  & 14  & 15  & 16  \\\hline
Highest   &  0.166	& 0.163 & 0.164 & 0.159 & 0.155 & 0.170 & 0.154 & 0.155 & 0.156 & 0.166 & 0.165 & 0.160	& 0.170  & 0.160 & 0.156 & 0.158	 		 \\
Key         &  0.141	& 0.132 & 0.129 & 0.142 & 0.132 & 0.127 & 0.139 & 0.135 & 0.129 & 0.134 & 0.139 & 0.130	& 0.144  & 0.130 & 0.132 & 0.132 		 
\normalsize
\end{tabular}
\end{table*}

With $\alpha$ = 1/2, the probability that a plaintext is encrypted with $Q_0$ or $Q_1$ is the same.
The next experiments are almost the same with the previous ones, except that $Q_0$ and $Q_1$ were selected with a 1/2 probability for each encryption.
In other words, the Walsh transforms  were calculated by using the \textit{UT} lookup values from $Q_0$ and $Q_1$.
All of the results are provided in Appendix~\ref{sec:appendix_walsh}. 
Here, Fig.~\ref{fig:Q0_Q1_UT_Walsh_bit1} shows one of them due to limited space.
The hypothetical value was, like the previous experiments, the first bit of the SubBytes output multiplied by 1, 2, and 3.
There was an increase in the encoding imbalance by which the correct subkey does not lead to  the lowest correlation to the correct hypothetical values.  

\begin{figure}
\centering
\subfloat[With respect to the SubBytes output]{
\centering
\includegraphics[width=0.9\linewidth]{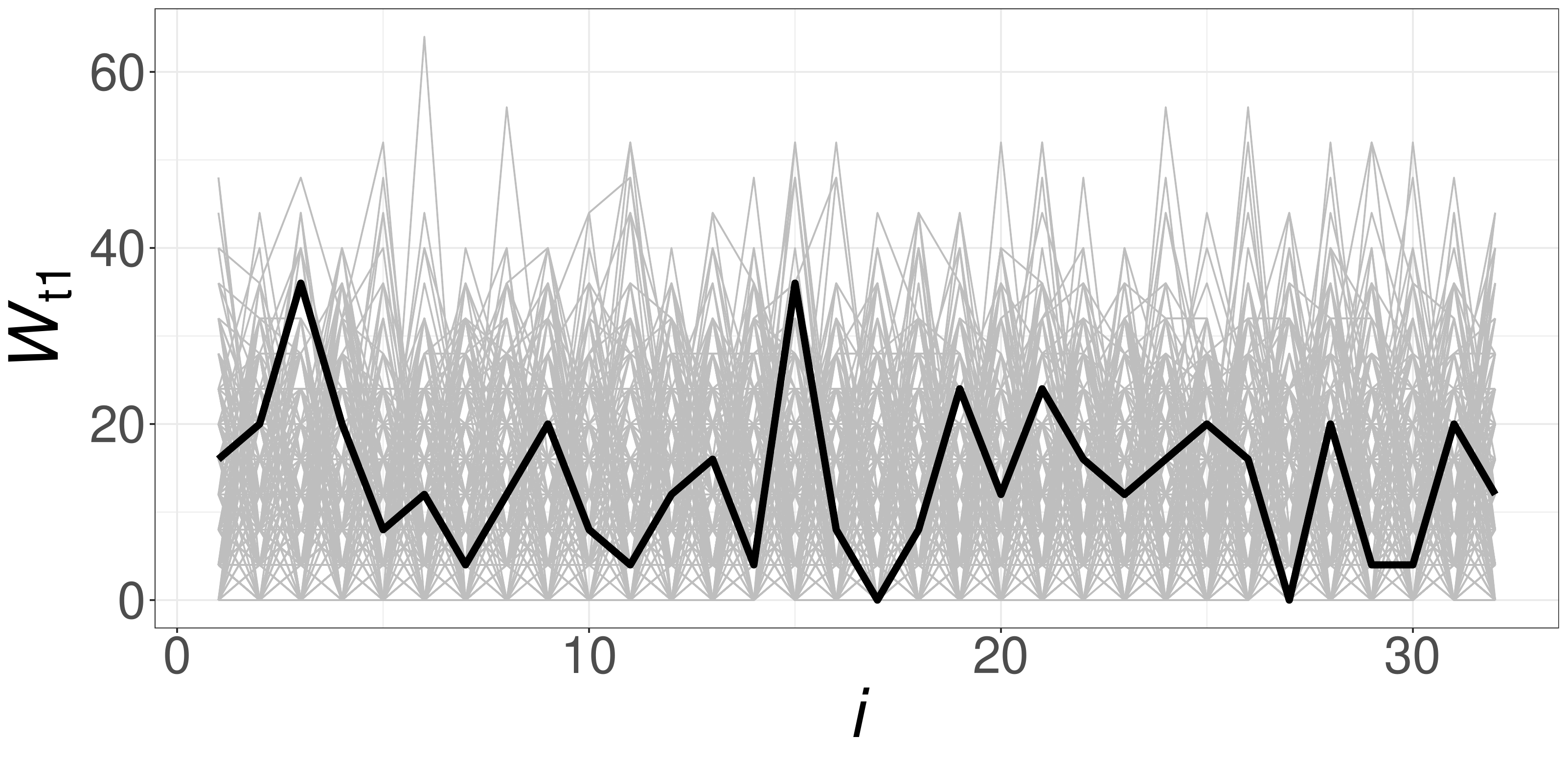}
} \\
\subfloat[With respect to the SubBytes output multiplied by 2]{
\includegraphics[width=0.9\linewidth]{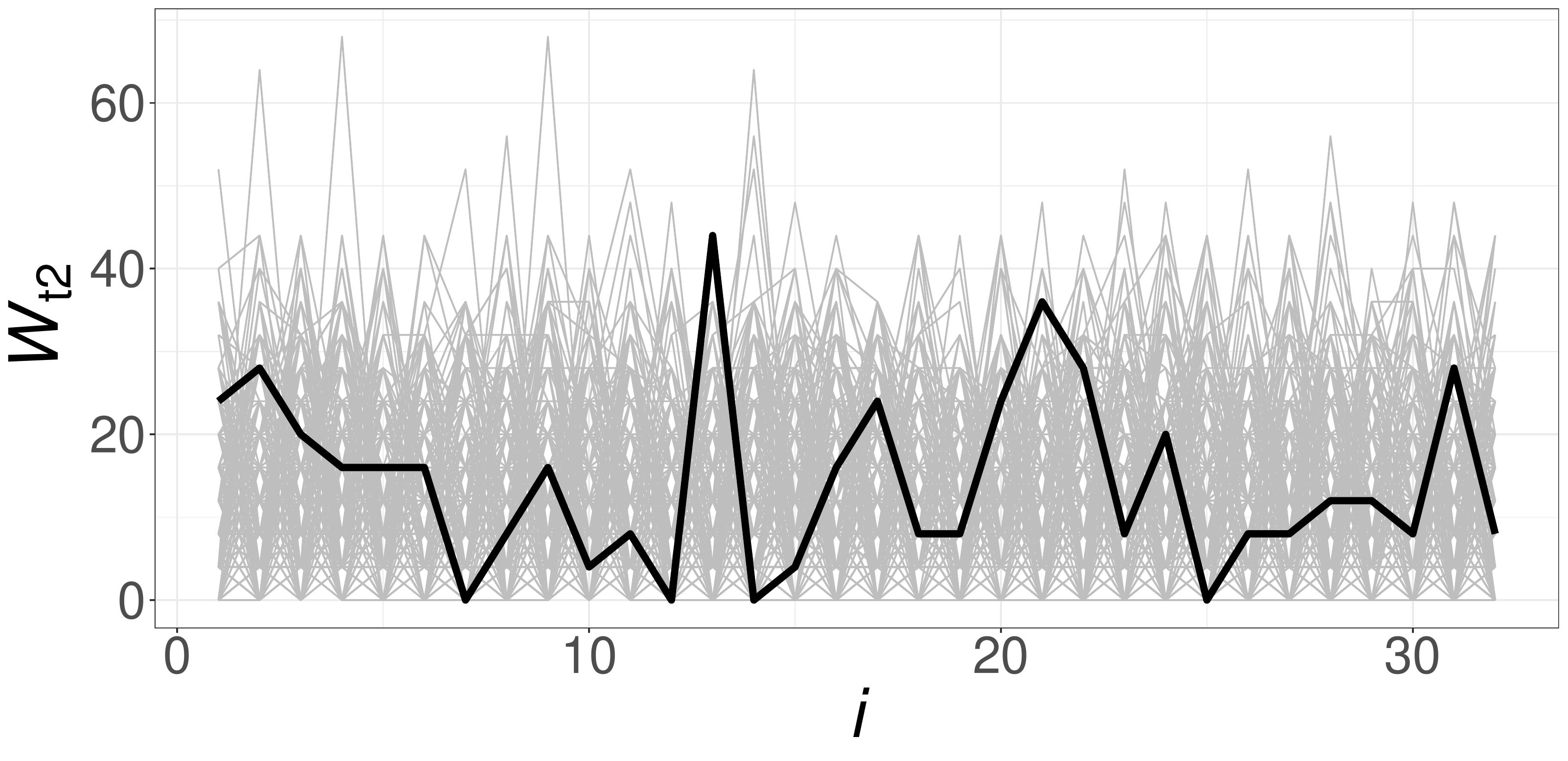}
} \\
\subfloat[With respect to the SubBytes output multiplied by 3]{
\includegraphics[width=0.9\linewidth]{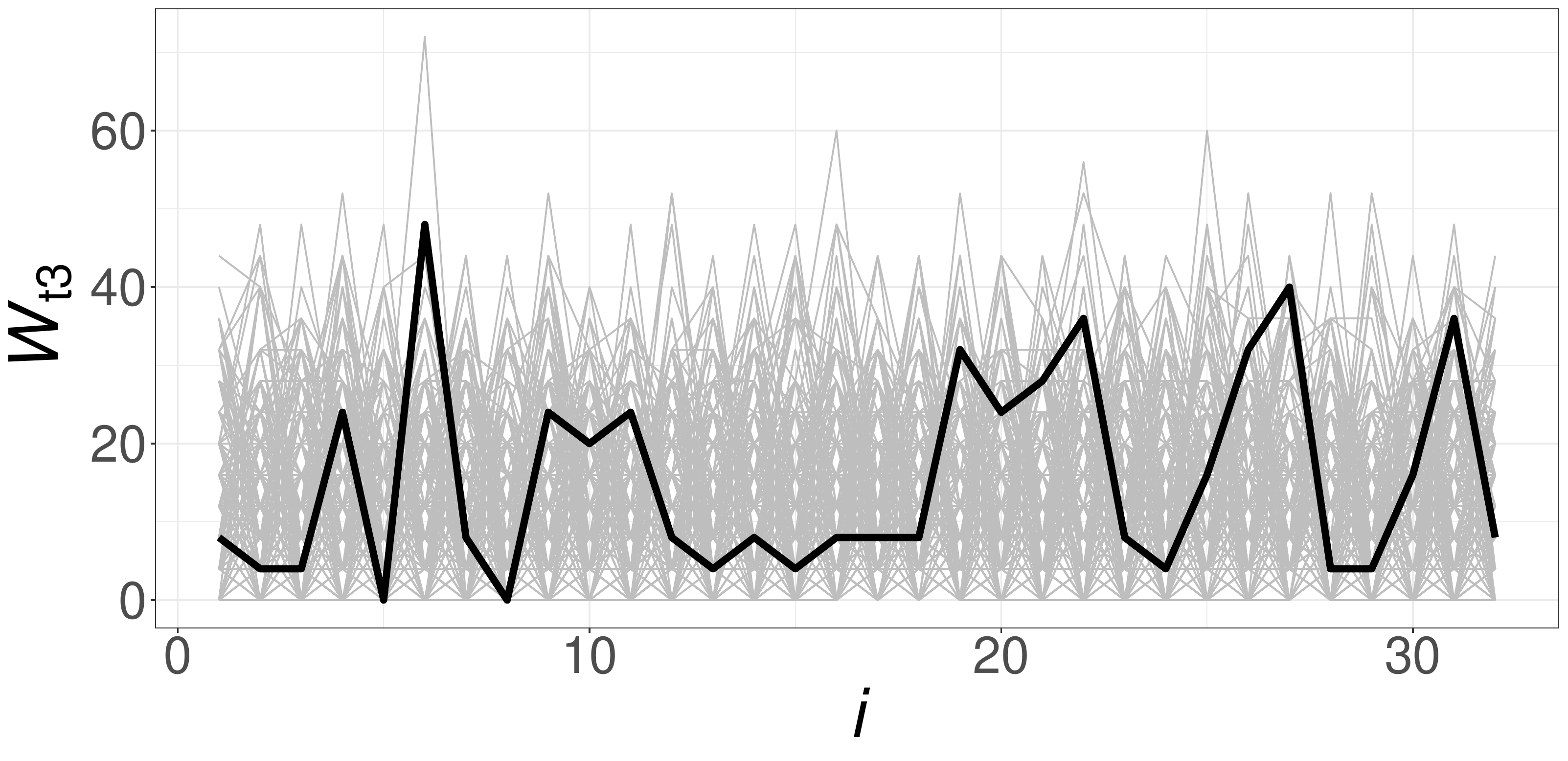}
} \\

\caption{Walsh transforms on the MixColumns multiplication obtained by $Q_0$ and $Q_1$. Black: correct subkey. Gray: wrong key candidates.}
\label{fig:Q0_Q1_UT_Walsh_bit1}
\end{figure}

The encoding imbalance shown in the results of  the Walsh transforms above improves the protection of keys against DCA attacks searching for either the highest- or lowest-ranking subkey candidate.  
While the correct subkeys are in the lowest rank in Table~\ref{tab:Q0_UT_DCA}, 
those are not in the highest or lowest ranks as shown in Table~\ref{tab:Q0_Q1_UT_DCA} when a plaintext was encrypted by looking up $Q_0$ or $Q_1$ with a 1/2 probability.
Compared to Table~\ref{tab:Q0_UT_DCA_coefficients}, the gap between correlation coefficients computed by the correct and wrong subkeys are reduced as shown in Table~\ref{tab:Q0_Q1_UT_DCA_coefficients}.
The tendency of correlation coefficients depending on the number of computational traces can be found in Appendix~\ref{sec:appendix_coff_tendency}.

\begin{table*}
\centering
\scriptsize
\caption{DCA ranking on the AES encryption using $Q_0$ or $Q_1$ with a 1/2 probability when conducting mono-bit CPA on the SubBytes output with 10,000 computational traces.}
\label{tab:Q0_Q1_UT_DCA}
\begin{tabular}{c|*{16}{{c}@{\hskip 2mm}}}
\\
\backslashbox{TargetBit}{Subkey}   &       1   &       2   &    3   &    4   &   5    &    6   &      7   &    8   &   9   &  10   & 11   & 12   & 13  & 14  & 15  & 16  \\\hline
1 & 48   & 51	& 41	& 199	& 145	& 53	& 64	& 102	& 28	& 131	& 67	& 15	& 245	& 42	& 19	& 218	\\	 
2 & 172	& 62      & 34	& 54	& 71	& 211	& 62	& 132	& 94	& 239	& 213	& 99	& 53	& 21	& 18	& 135	\\	 
3 & 4	& 65	& 57	& 66	& 76	& 230	& 244	& 153	& 95	& 62	& 3	& 71	& 252	& 219	& 112	& 125	\\	 
4 & 133	& 146	& 210	& 214	& 74	& 94	& 254	& 96	& 86	& 200	& 181	& 232	& 211	& 250	& 201	& 191	\\	  
5 & 51	& 234	& 124	& 192	& 180	& 144	& 186	& 27	& 63	& 36	& 212	& 78	& 237	& 147	& 238	& 197	\\	 
6 & 18	& 57	& 126	& 117	& 199	& 38	& 54	& 67	& 37	& 93	& 234	& 93	& 38	& 52	& 100	& 187	\\	 
7 & 4	& 66	& 10	& 70	& 88	& 38	& 177	& 143	& 64	& 85	& 188	& 128	& 161	& 145	& 15	& 111	\\	 
8 & 150	& 120	& 136	& 27	& 110	& 49	& 5	& 238	& 252	& 135	& 212	& 100	& 182	& 174	& 46	& 127	
\normalsize
\end{tabular}
\end{table*}

\begin{table*}
\centering
\scriptsize
\caption{Highest correlation of all key candidates vs. highest correlation of the correct subkeys when conducting the DCA attacks on the UT output obtained by $Q_0$ or $Q_1$ with a 1/2 probability. }
\label{tab:Q0_Q1_UT_DCA_coefficients}
\begin{tabular}{c|*{16}{{c}@{\hskip 1mm}}}
\\
Subkey    &       1   &       2   &    3   &    4   &   5    &    6   &      7   &    8   &   9   &  10   & 11   & 12   & 13  & 14  & 15  & 16  \\\hline
Highest   &  0.043   &     0.043& 	0.044& 	0.043& 	0.040& 	0.043& 	0.045& 	0.038& 	0.039& 	0.042& 	0.042& 	0.042& 	0.043& 	0.046& 	0.038& 	0.042 		 \\
Key        &  0.037  &      0.027&	0.032&	0.030&	0.024&	0.027&	0.035&	0.023&	0.026&	0.028&	0.033&	0.031&	0.027&	0.029&	0.029&	0.023 		 
\normalsize
\end{tabular}
\end{table*}

\subsection{Analysis of the Round Outputs}
\label{sec:anlysis_round_output}

The encoded round output of the first round was analyzed so that the balance in the encoding applied to the round output can be investigated.
Let $p1$ and $p2$ denote the first two byte of the plaintexts. 
When the remaining 14 bytes are fixed to 0, the first byte of the first round output can be expressed as follows.
\begin{equation*}
\delta(p1, p2) = S^2(p1 \oplus \hat{k}^0_{0,0}) \oplus S^3(p2 \oplus \hat{k}^0_{1,0}) \oplus c, 
\end{equation*}
where $c$ is a constant. 
For simplicity, $\epsilon$ represents the encoding applied to $\delta$.
Then, the first subbyte of the first round output protected by the proposed encoding can be written by $\epsilon \circ \delta(p1, p2)$.
If we assume that the attacker knows $\hat{k}^0_{0,0}$, the corresponding hypothetical value is given by 
\begin{equation*}
\gamma(p1, p2) = S^2(p1 \oplus \hat{k}^0_{0,0}) \oplus S^3(p2 \oplus {k}^{\ast}),
\end{equation*}
where ${k}^{\ast}$ is the key candidates. 
Let the subscripts $i$ and $i'$ denote the $i$- and $i'$-th bits, respectively. 
For a fixed $p1$ and the correct key candidate, by Equation (\ref{eq:barR-barS-0}), we expect
\begin{equation*}
\sum^{255}_{p2=0}{(-1)^{\epsilon \circ \delta(p1, p2)_i \oplus \gamma(p1, p2)_{i'}}} = 0.
\end{equation*}
Thus, the Walsh transform $W_{\epsilon\gamma}$ is defined:  
\begin{equation*}
W_{\epsilon\gamma}(i') = \sum^{255}_{p1=0}{\left| \sum^{255}_{p2=0}{(-1)^{\epsilon \circ \delta(p1, p2)_i \oplus \gamma(p1, p2)_{i'}}}\right|}. 
\end{equation*}

\begin{figure}
\centering
\includegraphics[width=0.9\linewidth]{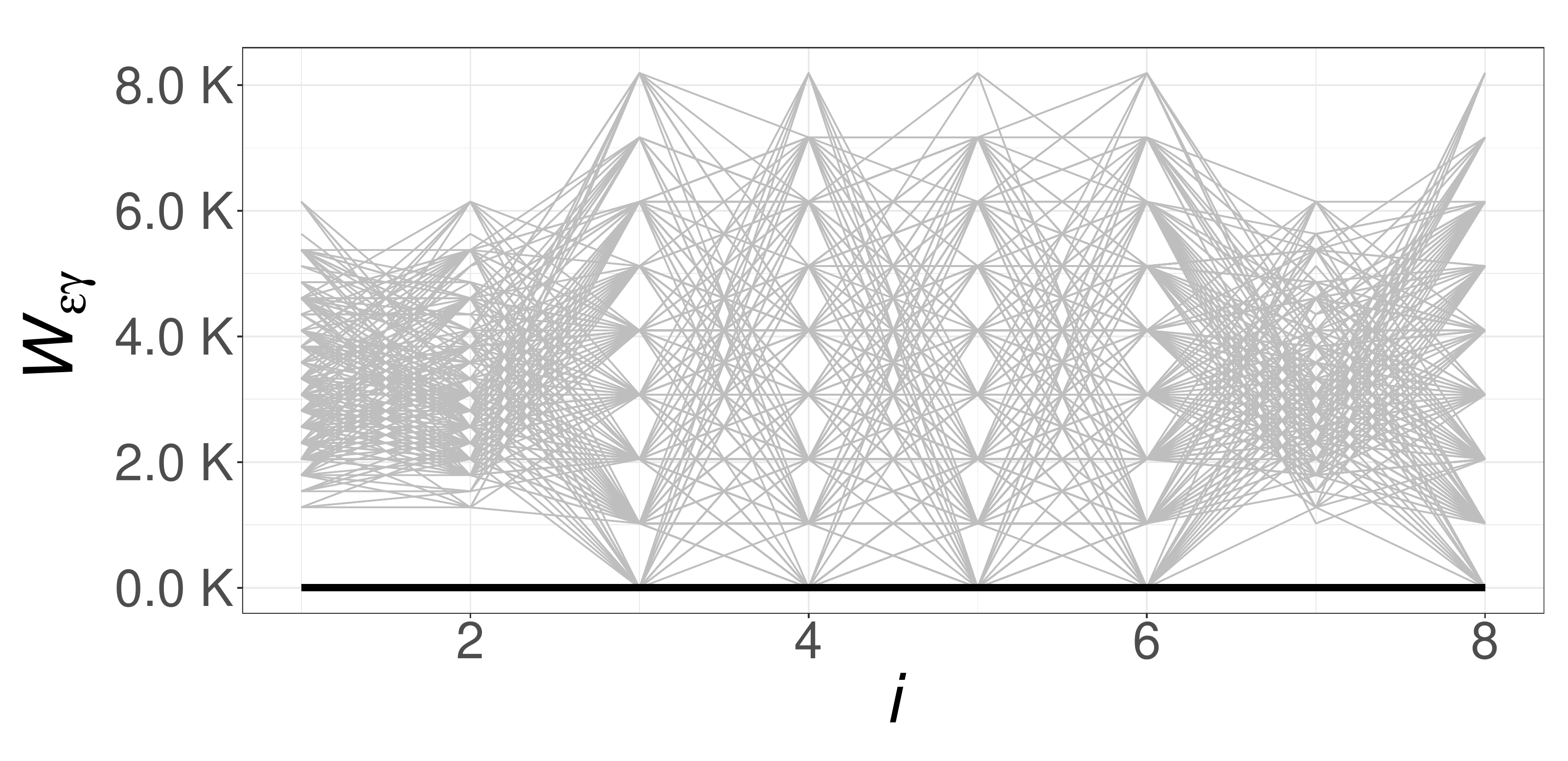}
\caption{Walsh transforms on the round output obtained by $Q_0$. Black: correct subkey. Gray: wrong key candidates.}
\label{fig:Q0_RO_Walsh_bit1}
\end{figure}

\begin{figure}
\centering
\includegraphics[width=0.9\linewidth]{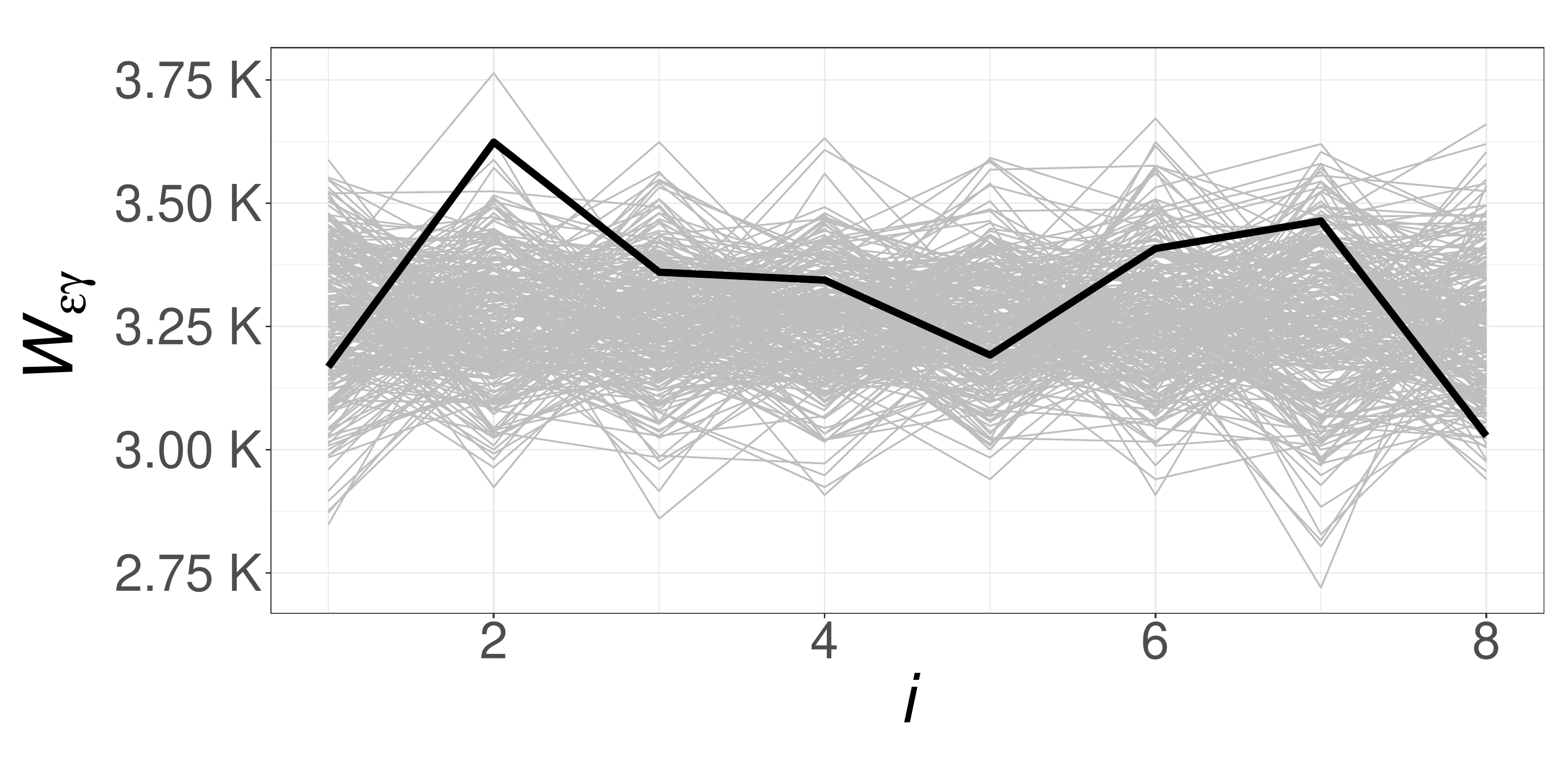}
\caption{Walsh transforms on the round output obtained by $Q_0$ and $Q_1$. Black: correct subkey. Gray: wrong key candidates.}
\label{fig:Q0_Q1_RO_Walsh_bit1}
\end{figure}

To see the effect of the balance in the encoded round output, we calculated $W_{\epsilon \gamma}$ for all $i, i' \in [1, 8]$.
As a result, the correct subkey always produced zeros on the Walsh transforms. 
The full experimental results can be found in \cite{Lee:Balanced:full} and Fig.~\ref{fig:Q0_RO_Walsh_bit1} shows, due to limited space,  one of the results in the case of $i' = 1$. 
Only the correct subkey (0x55) scores 0 for all $i \in [1, 8]$.
We also collected 10,000 computational traces while performing encryption using plaintexts consisting of the first two random bytes $(p1, p2)$ followed by the 14 zeros.
Assuming that the first subkey $\hat{k}^0_{0,0}$ is known, we mounted a customized DCA attack to recover the second subkey $\hat{k}^0_{1,0}$. 
Table~\ref{tab:Q0_RO_DCA} shows the result.
It is noteworthy that the DCA ranking of the correct subkey is relatively high compared to Table~\ref{tab:Q0_UT_DCA} because there is a decrease in the correlation coefficients computed by each of all key candidates. 
For this reason, the round output protected by the proposed encoding is not an attractive attack target for CPA attacks.
The Walsh transforms and the DCA results with respect to the round outputs obtained by looking up $Q_0$ and $Q_1$ with a 1/2 probability are provided in Fig.~\ref{fig:Q0_Q1_RO_Walsh_bit1} and Table~\ref{tab:Q0_Q1_RO_DCA}, respectively. 
There is no noticeable point of key leakage.


\begin{table}
\centering
\scriptsize
\caption{Highest correlation of all key candidates vs. highest correlation/ranking of the correct subkey when conducting the customized DCA on the first round output obtained by $Q_0$. }
\label{tab:Q0_RO_DCA}
\begin{tabular}{c|*{9}{{c}@{\hskip 1mm}}}
\\
TargetBit     &       1   &       2   &    3   &    4   &   5    &    6   &      7   &    8     \\\hline
Highest       & 0.031 & 0.029 & 0.025 & 0.032 & 0.028 & 0.033 & 0.032 & 0.034 \\
Correct subkey           & 0.0009 &	0.0006 &	0.009 &	0.008 &	0.004 &	0.003 &	0.017 & 0.0005  \\ 	
Key ranking   & 241    & 240    & 84   & 114   &  116  &   189 &  26   &  250 
\normalsize
\end{tabular}
\end{table}

\begin{table}
\centering
\scriptsize
\caption{Highest correlation of all key candidates vs. highest correlation/ranking of the correct subkey when conducting the customized DCA on the first round output obtained by $Q_0$ and $Q_1$. }
\label{tab:Q0_Q1_RO_DCA}
\begin{tabular}{c|*{9}{{c}@{\hskip 1mm}}}
\\
TargetBit       &       1   &       2   &    3   &    4   &   5    &    6   &      7   &    8     \\\hline
Highest         & 0.030  & 0.027  & 0.030 & 0.034 & 0.028 & 0.030 & 0.029  & 0.030 \\
Correct subkey  & 0.009  & 0.007  &	0.011 &	0.004 &	0.01  &	0.003 &	0.004  & 0.003  \\ 	
Key ranking     & 88     & 117    & 75    & 180   &  74   &   178 &  171   & 179 
\normalsize
\end{tabular}
\end{table}

\begin{figure}
\centering
\subfloat[$\Delta^{coll}_{k^{\ast}}$ for all subkey candidates.]{
\centering
\includegraphics[width=0.9\linewidth]{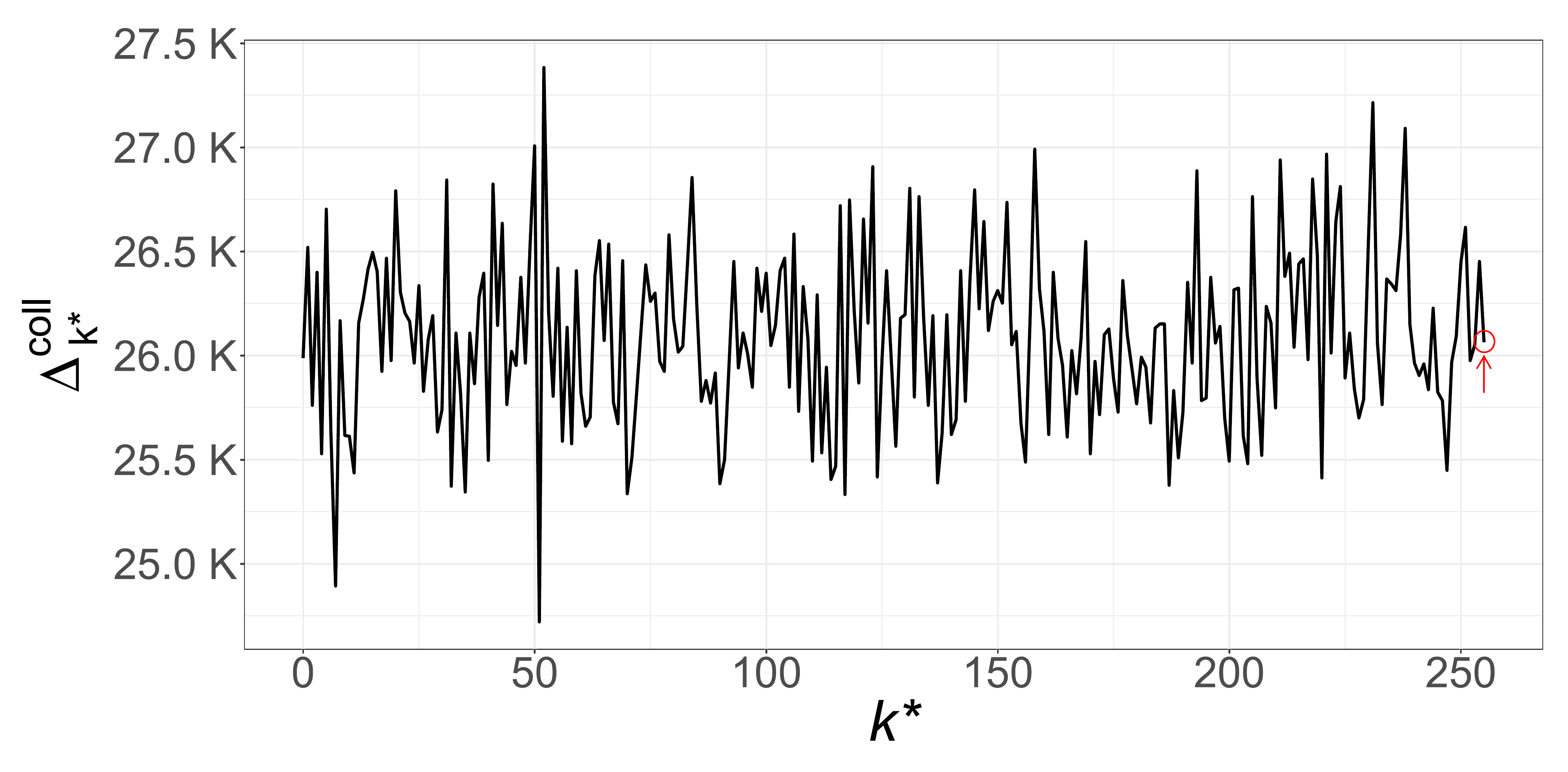}
\label{fig:delta_coll}
} \\
\subfloat[$\Delta^{sse}_{k^{\ast}}$ for all subkey candidates.]{
\includegraphics[width=0.9\linewidth]{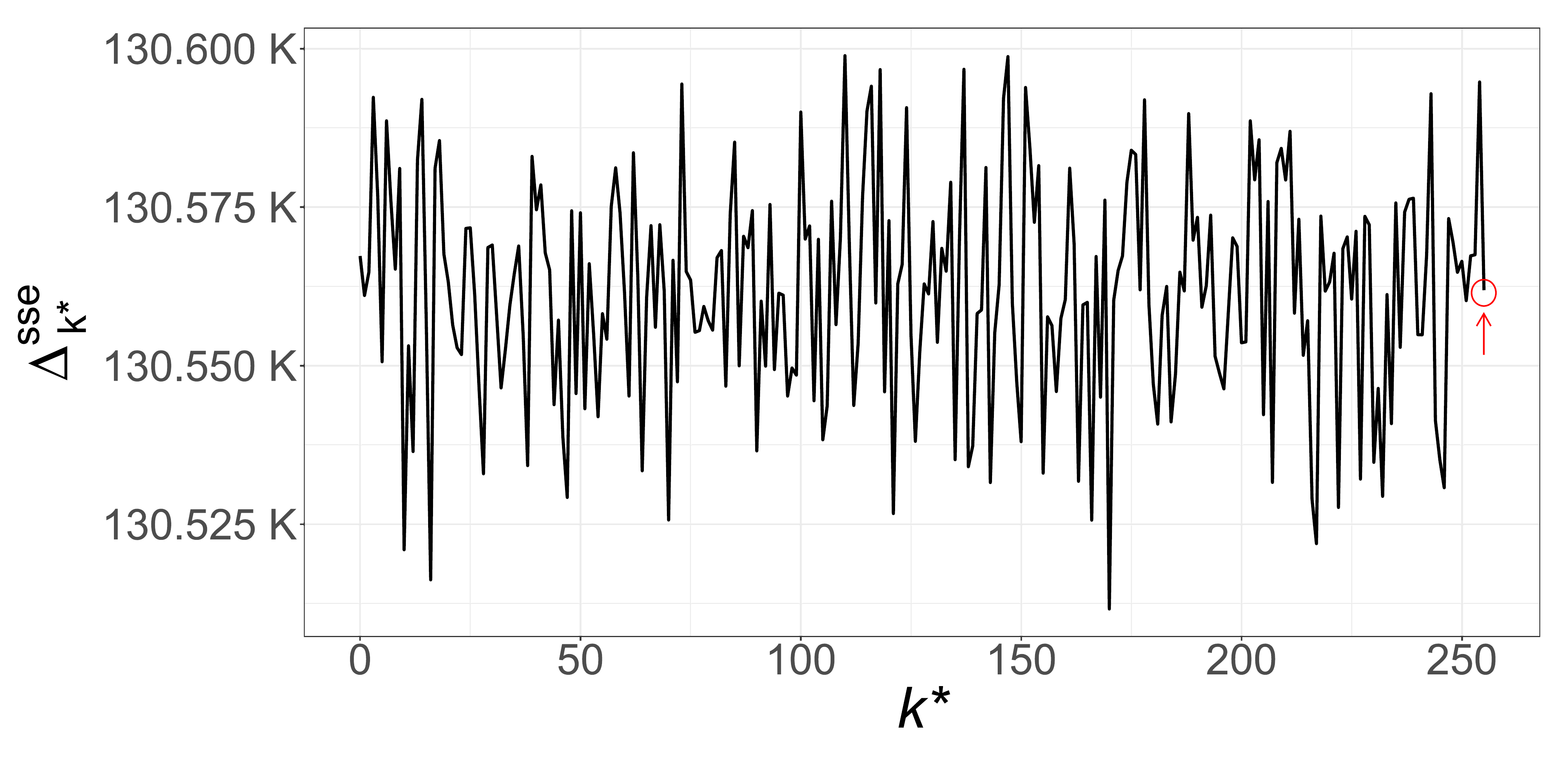}
\label{fig:delta_sse}
} \\
\caption{Disturbed collision and clustering with randomly selected $Q_0$ and $Q_1$ for each encryption. The score of the correct subkey (\texttt{0xFF}, 255) is highlighted with a red circle and arrow.}
\label{fig:deltas}
\end{figure}

In the context of collision-based attacks, as discussed in Section~\ref{sec:preliminaries}, an attacker exploits the property that if two or more distinct inputs produce the same output from a function, the corresponding encoded variable in the trace also experiences a collision.
To demonstrate the disturbed collision, we collected the following set of pairs:
$$
\mathcal{I}_{v} = \{ (a, b): a, b \in \{0, 1\}^8 | \gamma(a, b) = v, \mbox{~for~} v \in \{0, 1\}^8\}.
$$
By abuse of notation in Equation~(\ref{eq:delta}), we define:
\begin{equation*}
\label{eq:delta_collision}
    \Delta^{coll}_{k^{\ast} \in \{0,1\}^8} = \sum_{v = 0}^{255}\sum_{i = 1}^8{\left| \sum^{\ell_v}_{j=1}{(-1)^{c^j_i \oplus v_{i}}}\right|},
\end{equation*}
where $\ell_v = |\mathcal{I}_v|$, and $c^j = \epsilon \circ \delta(a^j, b^j)$ for all $(a^j, b^j) \in \mathcal{I}_v$.
With a perfect collision, the correct hypothetical subkey $k^{\ast} = k$ produces the highest score $\Delta^{coll}_{k}$, which is distinguishable from the others. However, Fig.~\ref{fig:delta_coll} demonstrates that this does not hold when $Q_0$ and $Q_1$ are randomly selected for each encryption.

Similarly, the protection of cluster analysis can be demonstrated as follows. After clustering the collected traces based on the hypothetical value $v$, we can utilize a cluster criterion function such as the sum-of-squared-error. Without loss of generality, we define:

\begin{equation*}
\label{eq:delta_sse}
    \Delta^{sse}_{k^{\ast} \in \{0,1\}^8} = \sum_{v = 0}^{255}\sum_{i = 1}^8{\sum^{\ell_v}_{j=1}{\left| {c^j_i - m_{i}}\right|}^2},
\end{equation*}
where $m_i$ = $(\sum^{\ell_v}_{j=1}{c^j_i})/\ell_v$.
The optimal partition minimizes $\Delta^{sse}_{k}$. However, Fig.~\ref{fig:delta_sse} shows that cluster analysis is also not effective in extracting the key.

\subsection{Other Considerations}

We have demonstrated that when conducting correlation-based statistical analyses on AES encryption with our lookup table employing balanced encoding, key analysis attacks prove ineffective. Now, we aim to assess whether our proposed method can defend the key against attacks such as Mutual Information Analysis (MIA) and test vector leakage assessment (TVLA). An overview of MIA and TVLA, accompanied by our evaluation,  is provided below. \\

\noindent \textbf{Mutual Information Analysis (MIA)} was introduced to address scenarios where the adversary has limited knowledge about the distribution of the leakage and its relationship with computed data~\cite{Gierlichs:MIA:2008}.  Let $X$ be the attacker's hypothetical value and $Y$ be the actual observed side-channel information such as computation traces. These are random variables on the discrete spaces $\mathcal{X}$ and $\mathcal{Y}$ with probability distributions $P_X$ and $P_Y$ respectively. The reduction in uncertainty on $X$ that is obtained by having observed $Y$ is exactly equal to the information that one has obtained on $X$ by having observed $Y$. Hence the formula for the Mutual Information $I(X; Y)$ is given by:
\begin{equation*}
I(X; Y) = H(X) - H(X|Y) = H(X) + H(Y) - H(X, Y) = I(Y; X).
\end{equation*}
The mutual information satisfies $0 \leq I(X; Y) \leq H(X)$. The lower bound is reached if and only if $X$ and $Y$ are independent. The upper bound is achieved when $Y$ fully determines $X$. Hence, the larger the mutual information, the closer the relation between $X$ and $Y$ is to a one-to-one relation. 

\begin{figure}
\centering
\subfloat[MIA on the UT outputs]{
\centering
\includegraphics[width=0.9\linewidth]{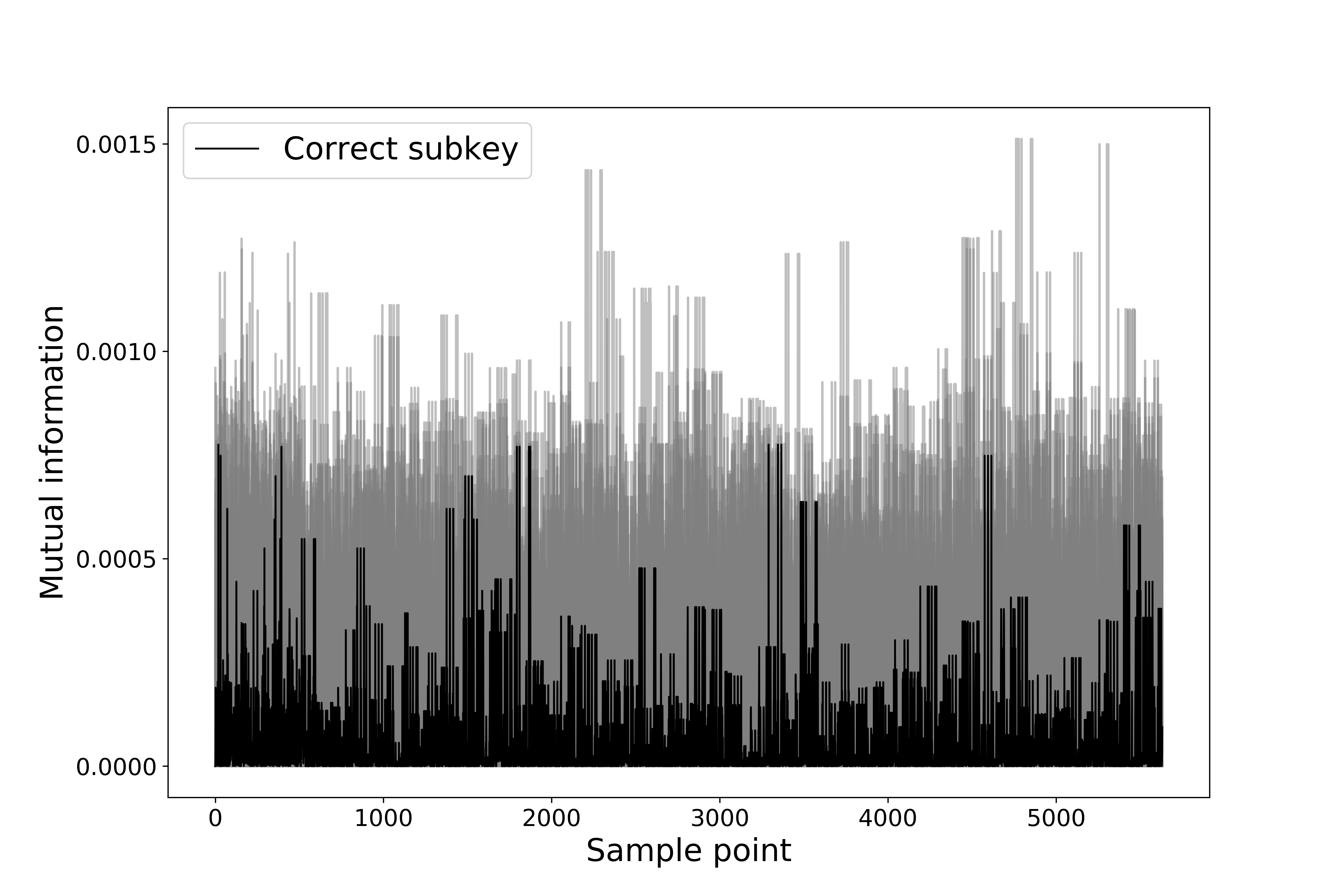}
\label{fig:Q0_Q1_UT_MIA}
} \\
\subfloat[MIA on the round outputs ]{
\includegraphics[width=0.9\linewidth]{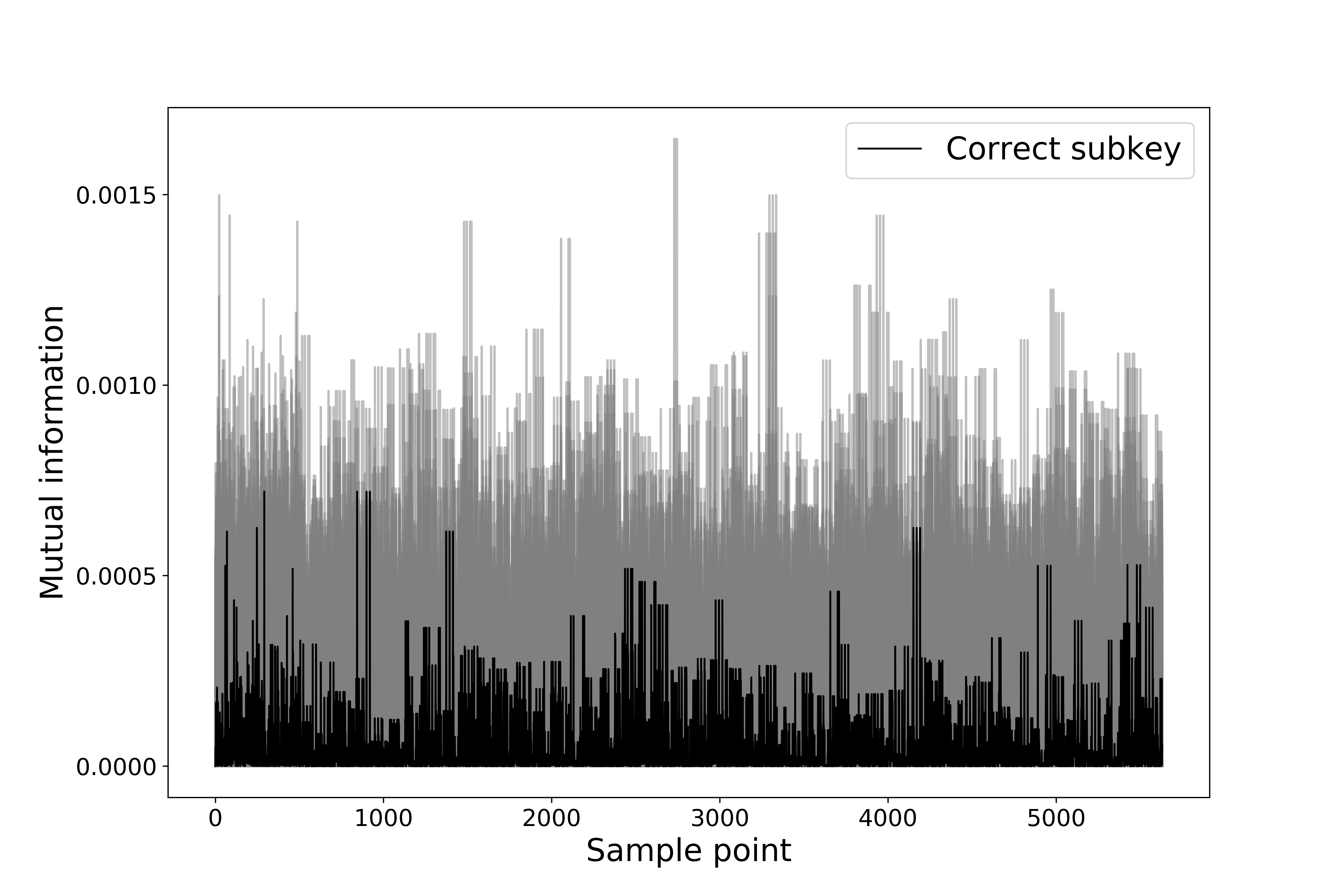}
\label{fig:Q0_Q1_UT_MIA}
} \\
\caption{MIA results on the protected AES encryption with 10,000 computational traces collected by selecting $Q_0$ or $Q_1$ with a 1/2 probability. Black: correct subkey. Gray: wrong key candidates.}
\label{fig:Q0_Q1_MIA}
\end{figure}

To assess the key's protection against MIA, we generated a pair of $Q_0$ and $Q_1$ and gathered 10,000 computational traces during encryption processes using random plaintexts. The attacker's hypothetical value derived from both the first round SubBytes' output and the first byte of that same round's output. We executed MIA attacks on each bit of these hypothetical values, and Fig.~\ref{fig:Q0_Q1_MIA} presents the attack results. Similarly to the CPA results, which were unsuccessful in key analysis, the MIA attack also showed no evidence of key leakage.
The MIA results for the remaining hypothetical bits can be found in the Appendix. \\

\noindent \textbf{Test Vector Leakage Assessment (TVLA)} is a method used to detect vulnerabilities in cryptographic systems, particularly for spotting unintended information leaks. 
A central technique in TVLA is the ``fixed-versus-random'' test. This entails comparing two trace sets: one where the input to the cryptographic algorithm is constant, and another where the input changes randomly. Notably, the cryptographic key remains consistent in both scenarios.

The distinction between these sets is assessed using Welch's \textit{t}-test. The computation is:
\[
t = \frac{\mu_F - \mu_R}{\sqrt{\frac{s^2_F}{n_F} + \frac{s^2_R}{n_R}}},
\]
where:
\begin{itemize}
    \item \( \mu_F \) and \( \mu_R \) represent the sample means of the fixed and random trace sets, respectively.
    \item \( s^2_F \) and \( s^2_R \) denote their respective sample variances.
    \item \( n_F \) and \( n_R \) indicate the number of traces in each set.
\end{itemize}
The \textit{p}-value, derived from the Student's \textit{t}-distribution, serves as an indicator of potential information leakage. Typically, a threshold of \( t = \pm 4.5 \) suggests significant evidence of a leak, especially with a large number of traces. 

In essence, one can determine whether a device's behavior is data-dependent by comparing the signals between fixed and random plaintexts. The primary objective is to identify potential points of leakage in the cryptographic implementation. If the differential signal exhibits significant variations at specific points, it may indicate measurable leakage, which could be exploited in CPA attacks. 

\begin{figure}
\centering
\subfloat[TVLA result]{
\centering
\includegraphics[width=0.9\linewidth]{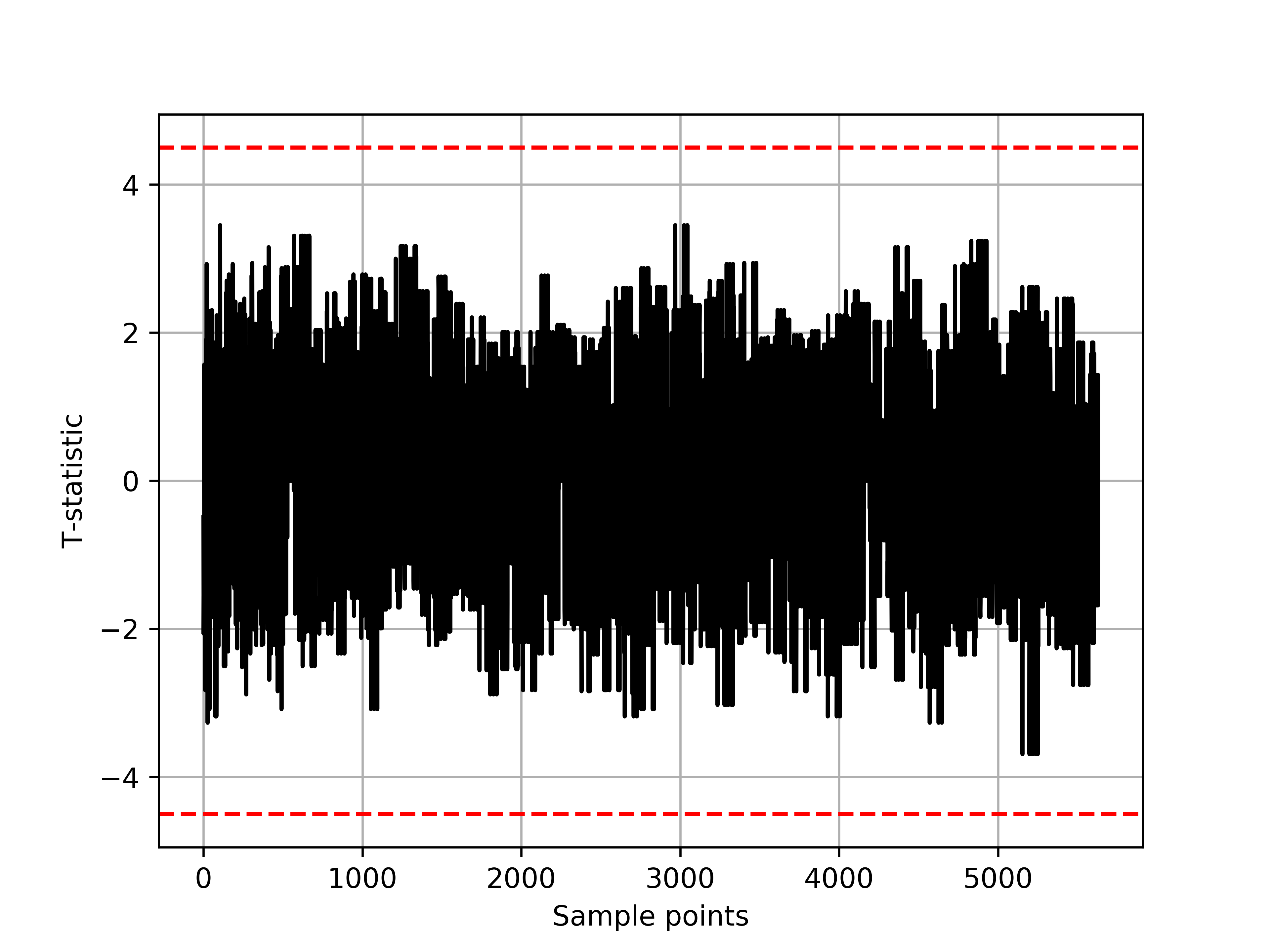}
\label{fig:Q0_Q1_UT_TVLA}
} \\
\subfloat[CPA result using the first bit of hypothetical SubBytes output]{
\includegraphics[width=0.9\linewidth]{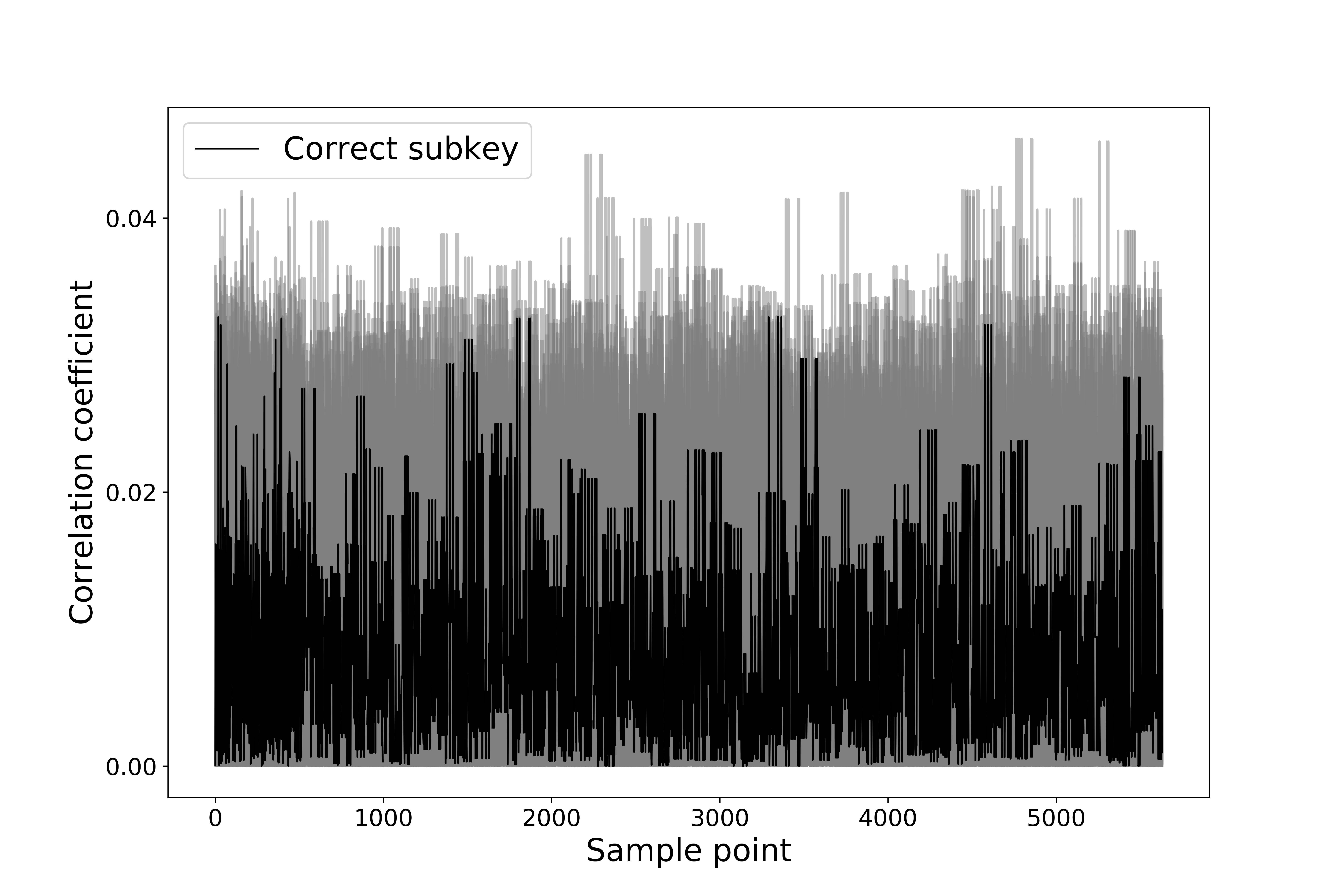}
\label{fig:Q0_Q1_UT_TVLA_CPA_bit0}
} \\
\caption{TVLA and CPA results on the protected AES encryption. Black: correct subkey. Gray: wrong key candidates.}
\label{fig:Q0_Q1_TVLA_CPA}
\end{figure}

We gathered two sets of 10,000 computational traces, each generated through the random application of $Q_0$ and $Q_1$. One set resulted from encrypting a fixed plaintext, while the other set was created by encrypting random plaintexts. During the TVLA analysis, it was observed that the peak values did not surpass the $\pm 4.5$ threshold at any point in the first round. To further confirm the absence of any potential key leakage, we performed CPA attacks using the 10,000 computational traces obtained from the encryption of random plaintexts. In these attacks, the attacker's hypothetical target was the SubBytes outputs in the first round. As depicted in Fig.~\ref{fig:Q0_Q1_TVLA_CPA}, the results clearly indicate the absence of key leakage. Please refer to the complete CPA results provided in the Appendix.

\section{Conclusion}
\label{sec:conclusion}

Power analysis has become a significant concern, especially for low-cost devices like IC cards. In many cases, cryptographic operations have been significantly slowed down to conceal key-dependent intermediate values, or memory requirements have sharply increased to enable fast and secure cryptographic operations.
In this paper, we propose an enhancement to the internal encoding of the table-based AES implementation, aiming to protect the key hidden in the tables. Our approach involves using a perfectly balanced encoding, which makes it challenging for power analysis attackers to guess the key with the lowest correlation to the correct hypothetical values. To address this issue, we suggest generating a complement set of balanced tables and randomly selecting one of the table sets for encrypting the plaintext. This modification ensures that our AES implementation remains secure against DPA-like attacks as well as collision-based attacks.
One notable advantage of our scheme is that it requires only 512KB of memory space and involves 1,024 table lookups. Additionally, it does not rely on run-time random number generators, making it more practical for low-cost devices.

As future work, we believe it is feasible to combine our implementation with countermeasures against fault attacks~\cite{Boneh:FA:1997,Biham:DFA:1997,Giraud:DFA:2004}. Specifically, our table-based implementation can be easily integrated with a table redundancy method for protecting against fault attacks~\cite{Lee:DFA-Tab:2021}. This combination would enable defense against both power analysis and fault attacks simultaneously. Notably, the table redundancy method, like our implementation, is independent of a runtime random source, making the combination suitable for low-cost devices. However, it is important to note that the proposed implementation remains secure under the gray-box model, meaning that if a white-box attacker could manipulate the execution flow to consistently select a particular set of tables, the key would still lead to the lowest correlation.

\appendix

\section{The experimental results of the Walsh Transforms}
\label{sec:appendix_walsh}

\begin{figure*}
\centering
\subfloat[$i'$ = 1.]{ \includegraphics[width=0.5\linewidth]{fig/Wt1_Q0_bit1.pdf}
\label{fig:mx_output_wfi_w1}}
\subfloat[$i'$ = 2.]{ \includegraphics[width=0.5\linewidth]{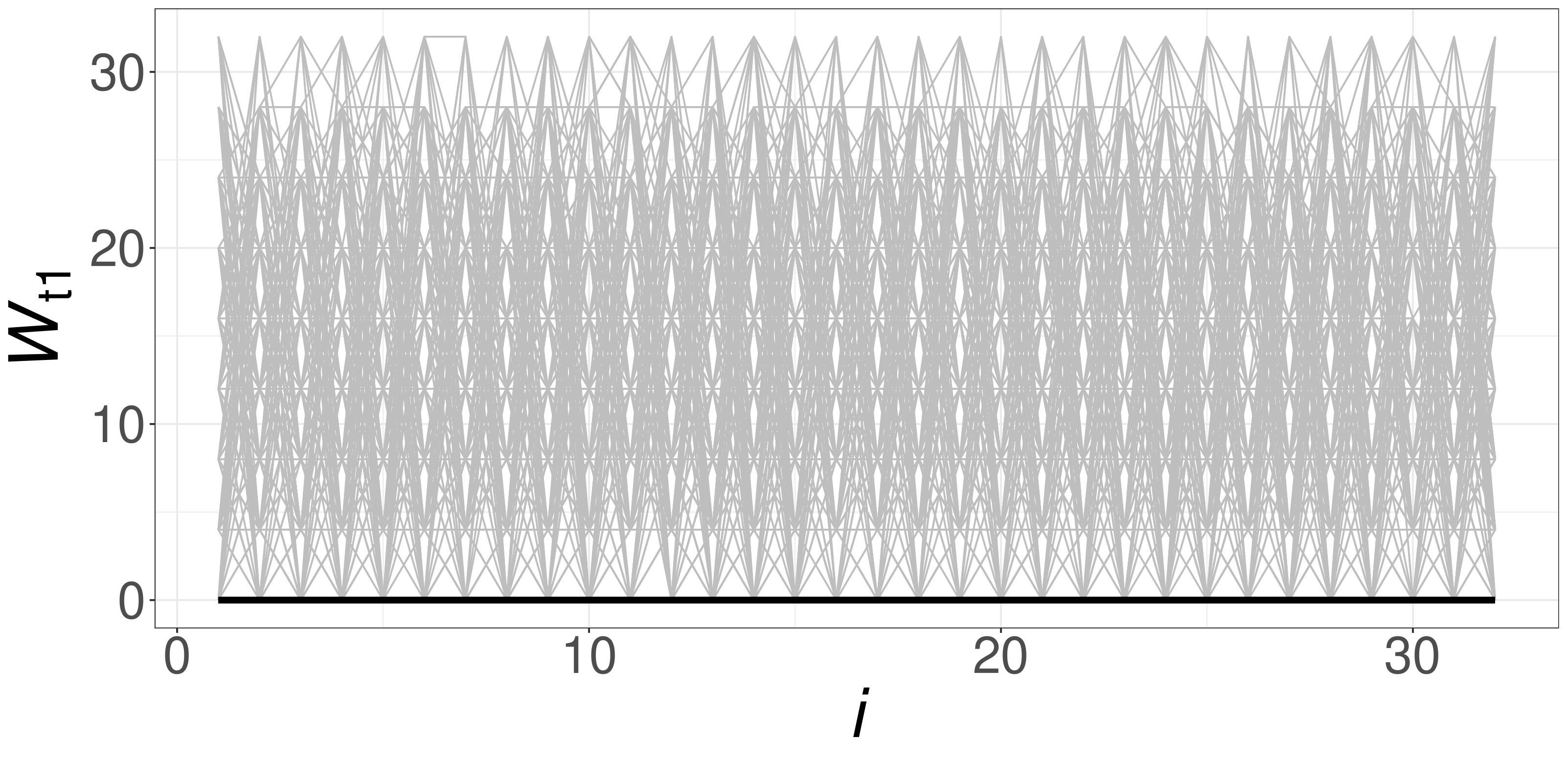}
\label{fig:mx_output_wfi_w2}}\\
\subfloat[$i'$ = 3.]{ \includegraphics[width=0.5\linewidth]{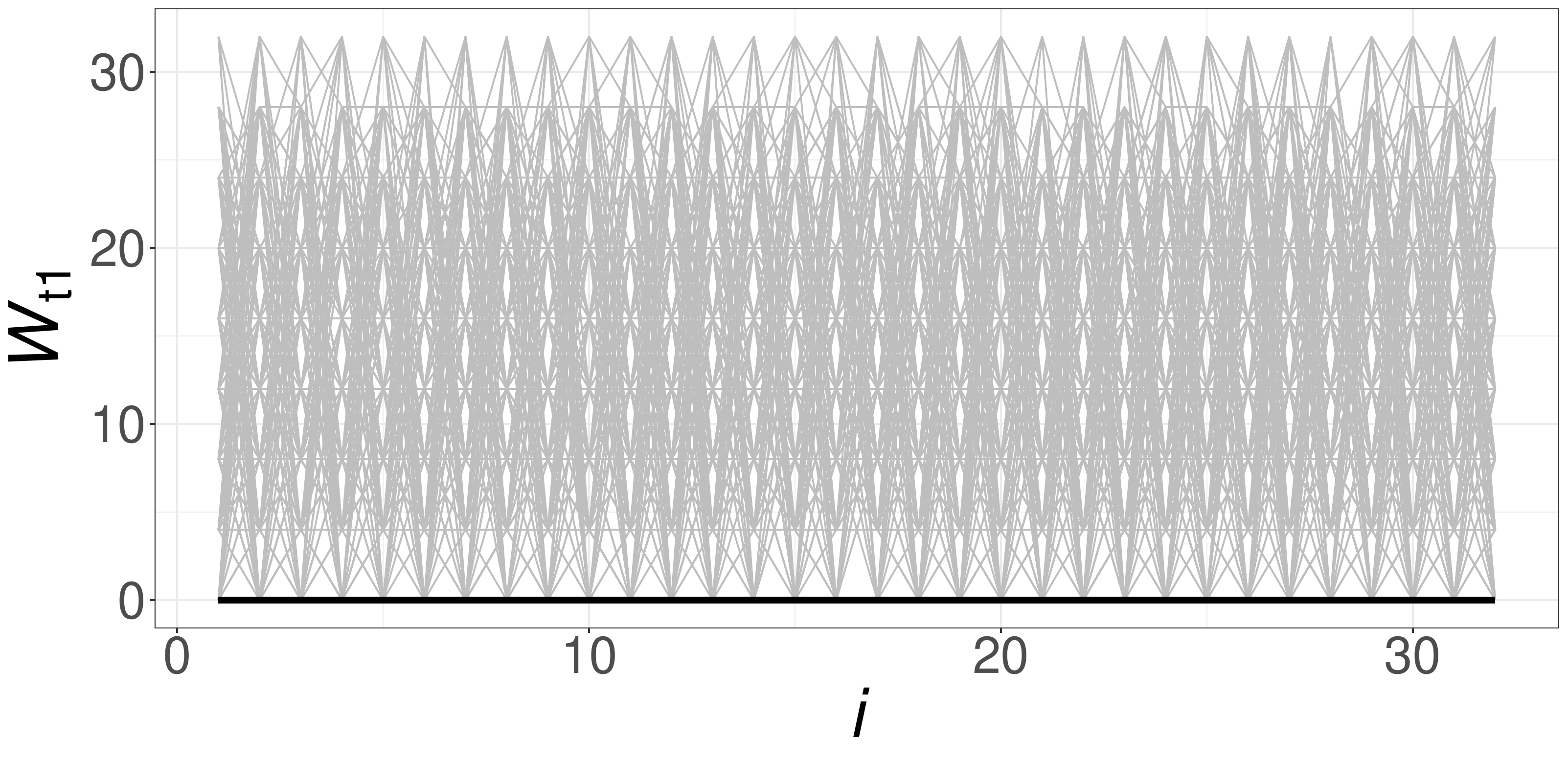}
\label{fig:mx_output_wfi_w3}}
\subfloat[$i'$ = 4.]{ \includegraphics[width=0.5\linewidth]{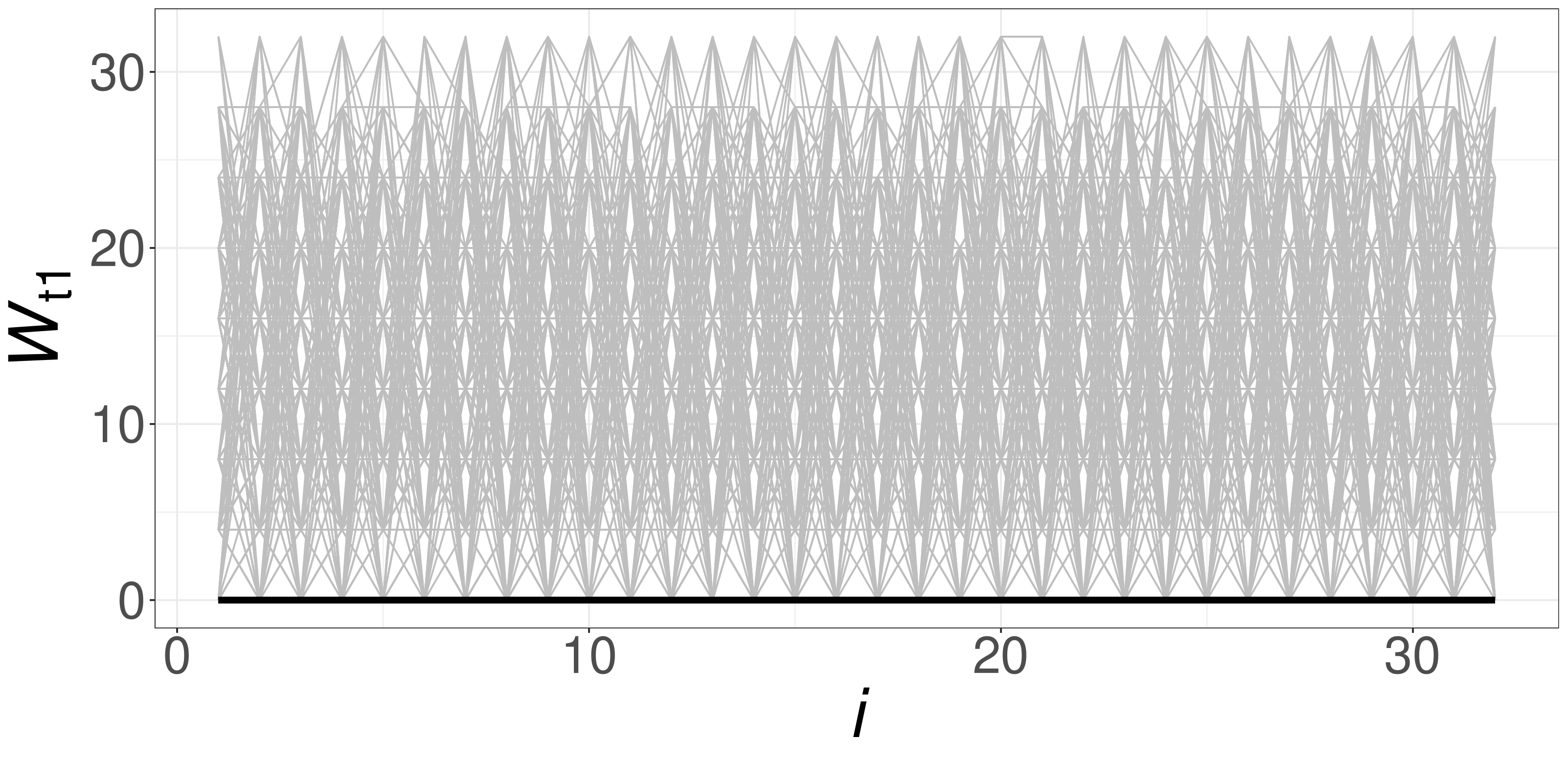}
\label{fig:mx_output_wfi_w4}}\\
\subfloat[$i'$ = 5.]{\includegraphics[width=0.5\linewidth]{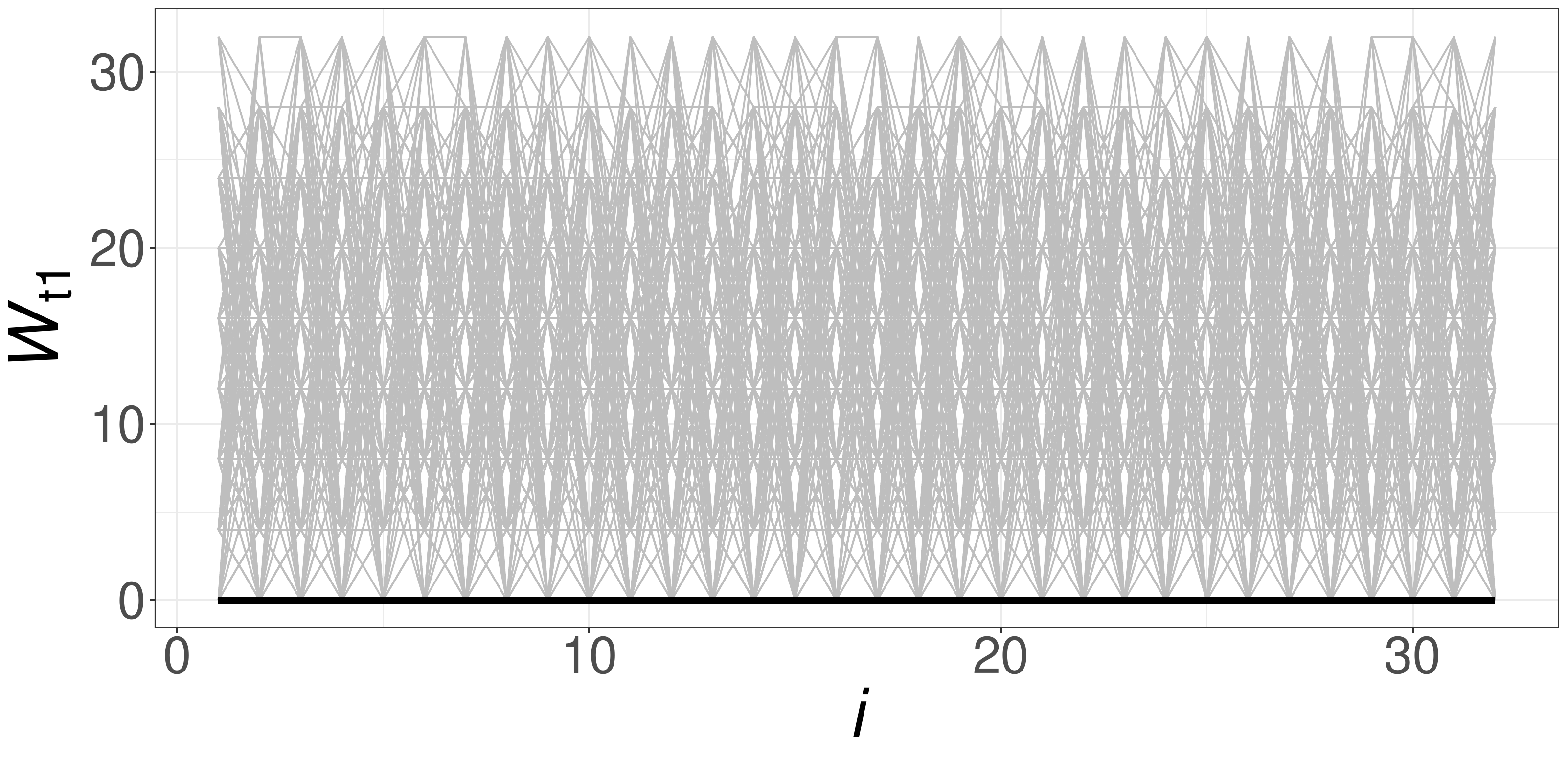}
\label{fig:mx_output_wfi_w5}}
\subfloat[$i'$ = 6.]{ \includegraphics[width=0.5\linewidth]{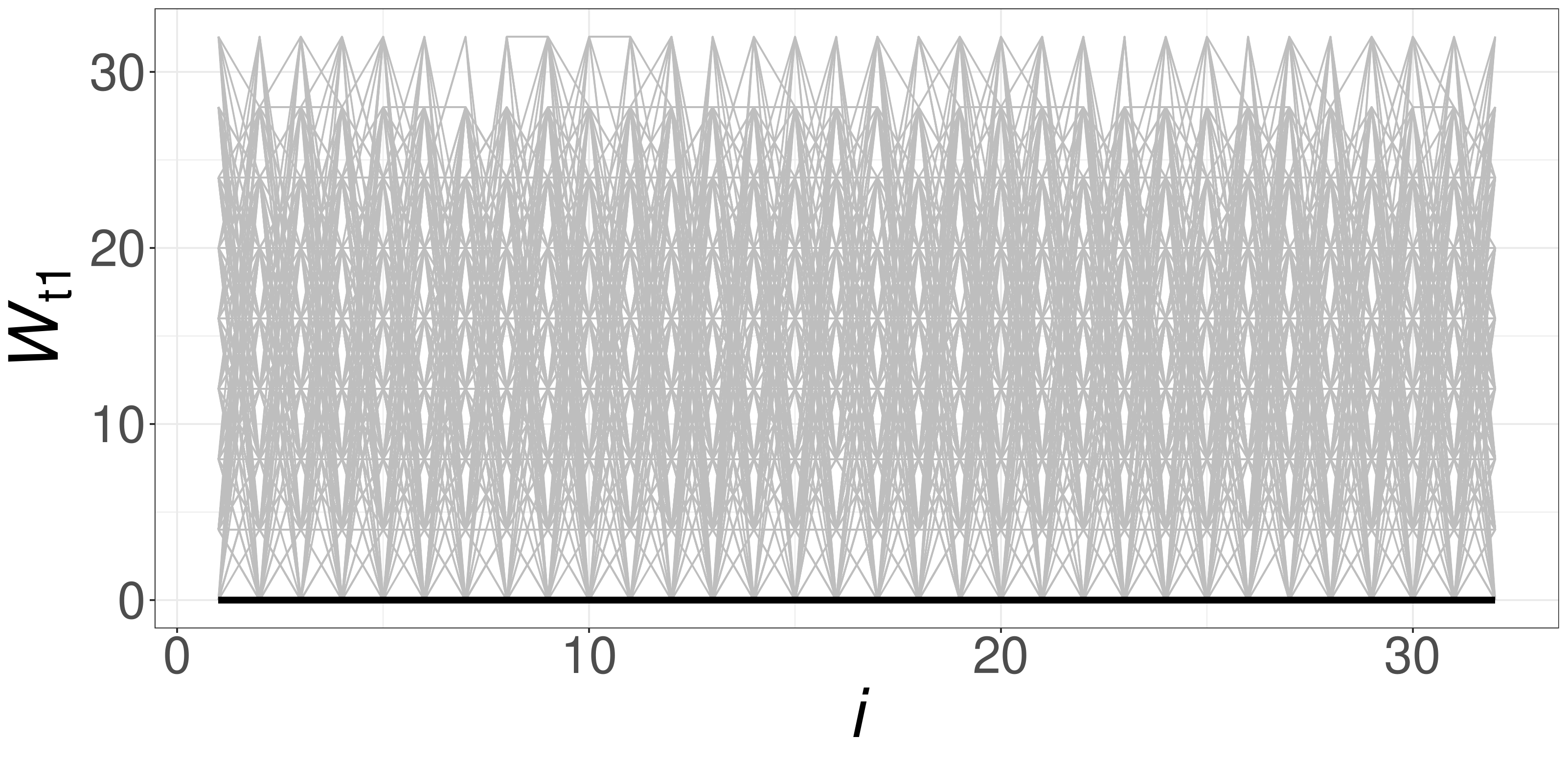}
\label{fig:mx_output_wfi_w6}}\\
\subfloat[$i'$ = 7.]{ \includegraphics[width=0.5\linewidth]{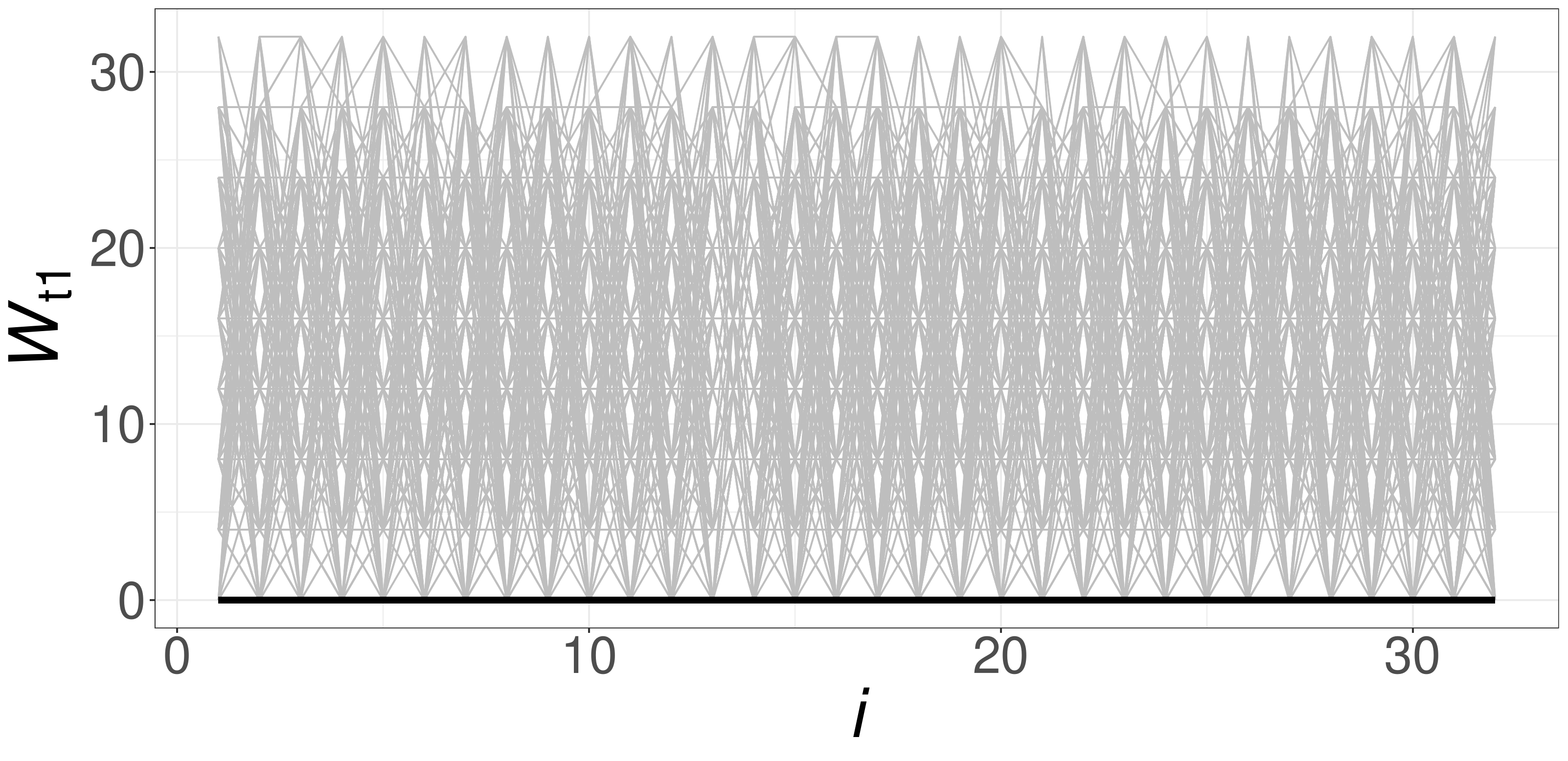}
\label{fig:mx_output_wfi_w7}}
\subfloat[$i'$ = 8.]{\includegraphics[width=0.5\linewidth]{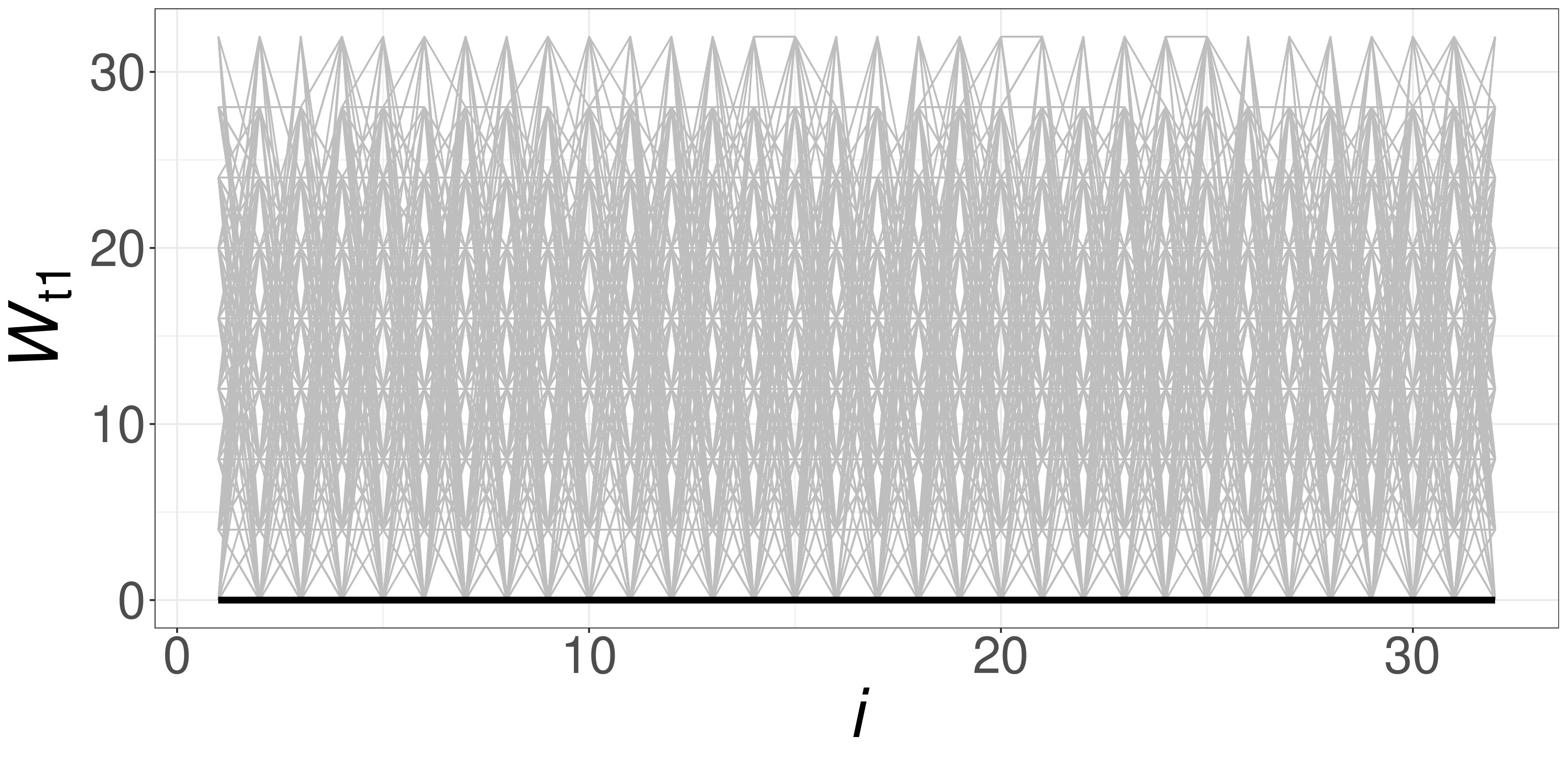}
\label{fig:mx_output_wfi_w8}}
\caption{The Walsh transforms on the $UT^{1}_{0,0}$ outputs obtained by $Q_0$ and $\textbf{S}^{1}$ in the first round. Black: correct key; gray: wrong key.}
\label{fig:Appendix_Walsh_UT_Q0_S1}
\end{figure*}

\begin{figure*}
\centering
\subfloat[$i'$ = 1.]{ \includegraphics[width=0.5\linewidth]{fig/Wt2_Q0_bit1.pdf}
\label{fig:mx_output_wfi_w1}}
\subfloat[$i'$ = 2.]{ \includegraphics[width=0.5\linewidth]{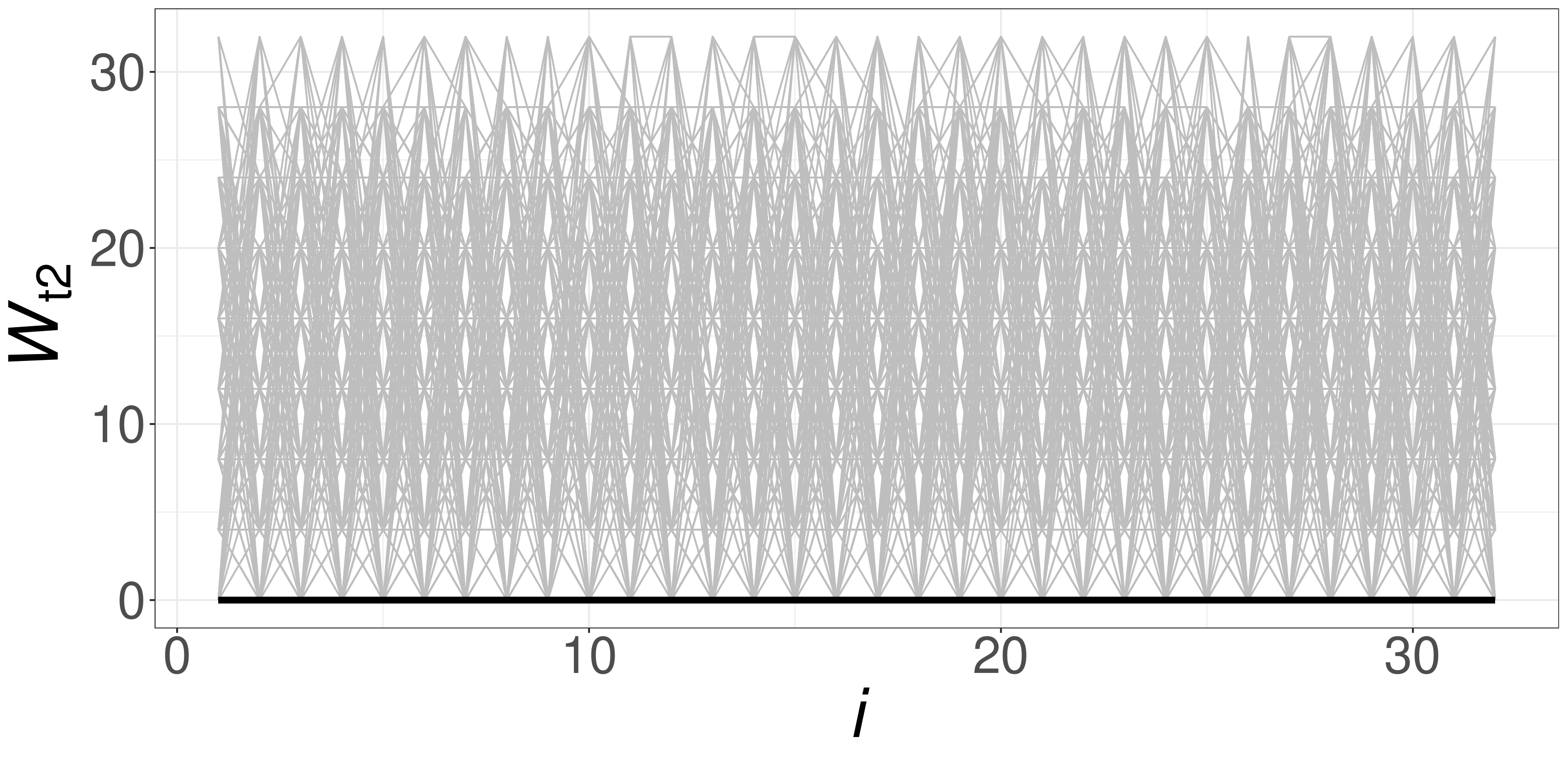}
\label{fig:mx_output_wfi_w2}}\\
\subfloat[$i'$ = 3.]{ \includegraphics[width=0.5\linewidth]{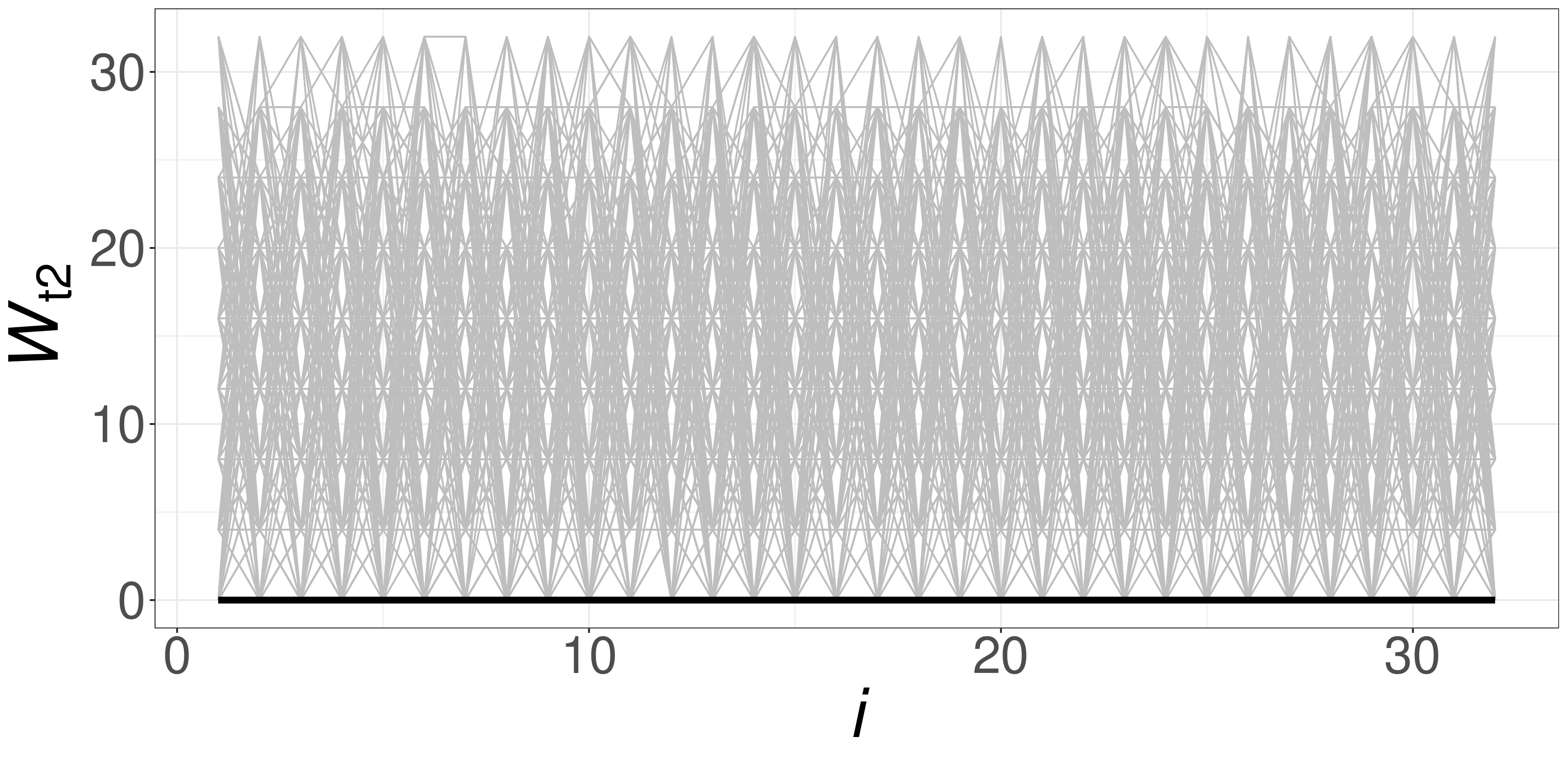}
\label{fig:mx_output_wfi_w3}}
\subfloat[$i'$ = 4.]{ \includegraphics[width=0.5\linewidth]{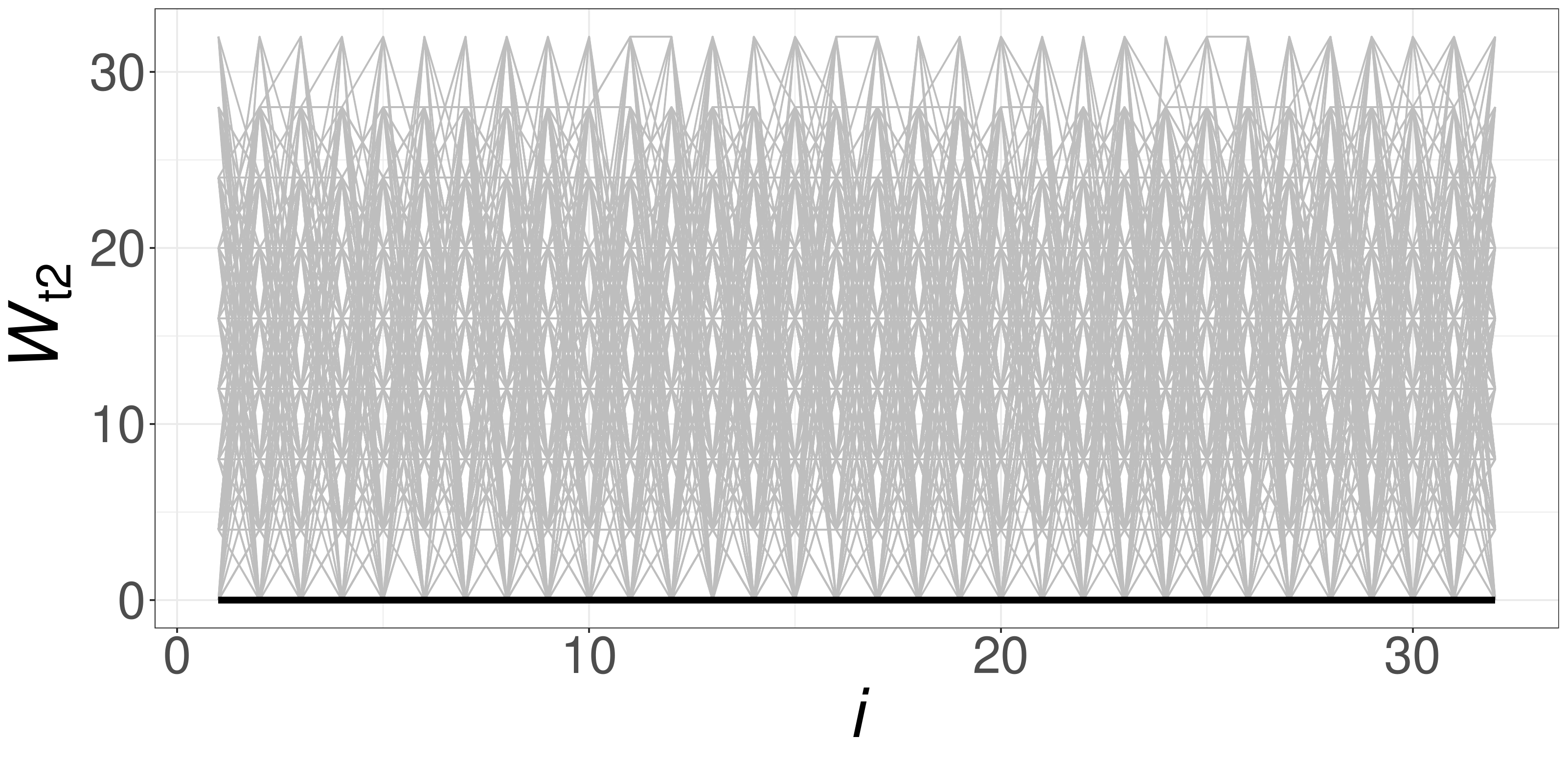}
\label{fig:mx_output_wfi_w4}}\\
\subfloat[$i'$ = 5.]{\includegraphics[width=0.5\linewidth]{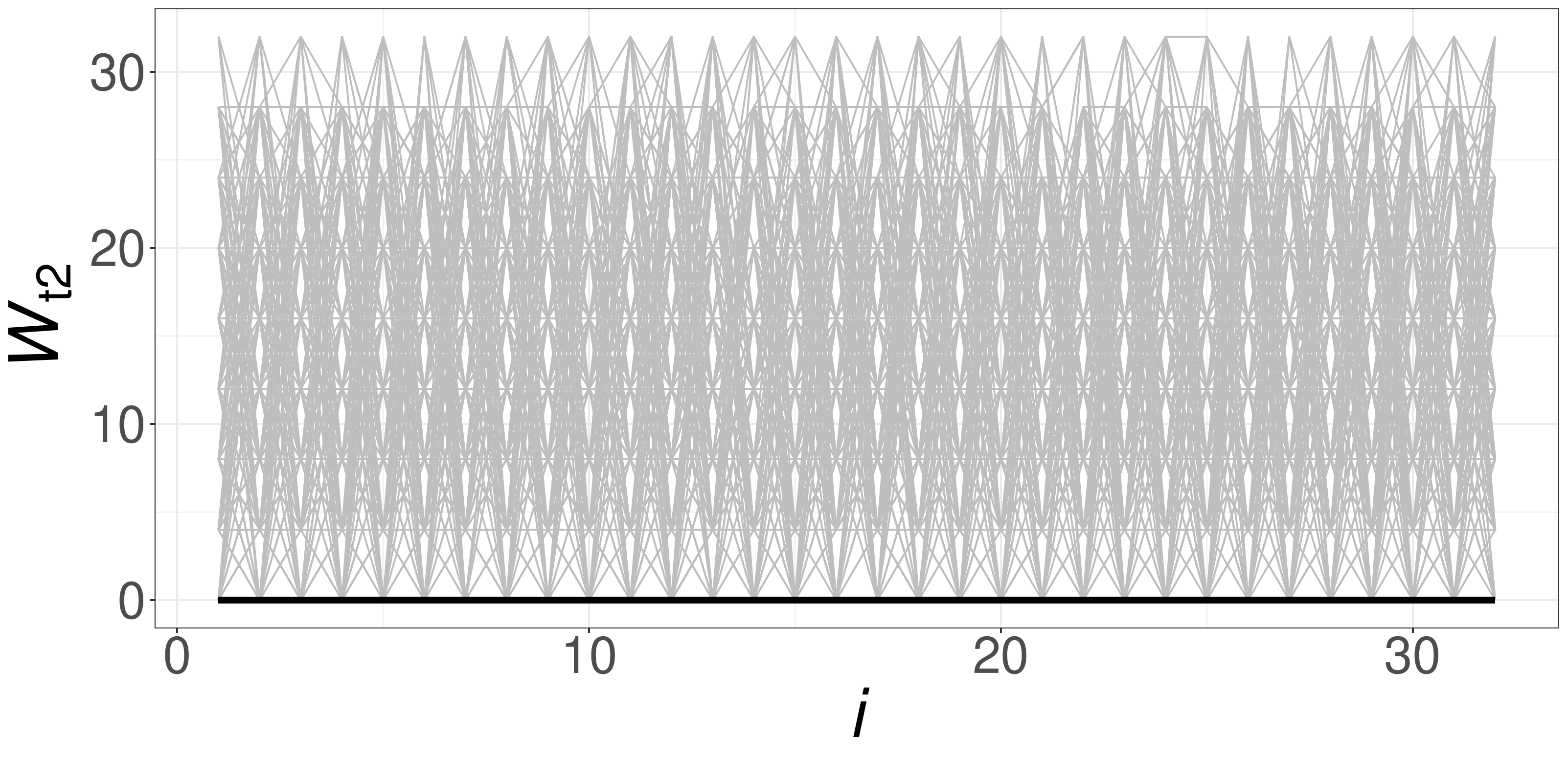}
\label{fig:mx_output_wfi_w5}}
\subfloat[$i'$ = 6.]{ \includegraphics[width=0.5\linewidth]{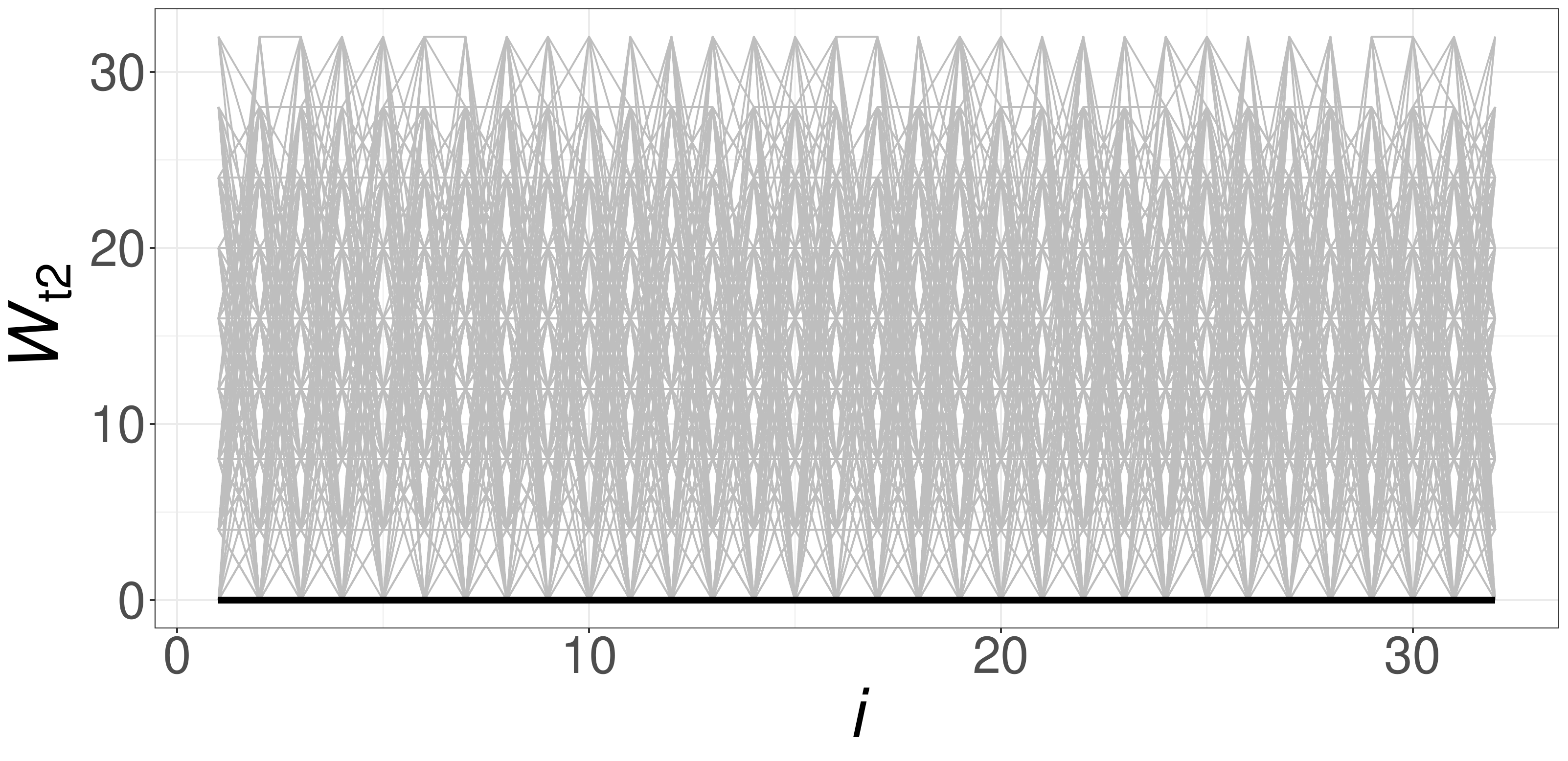}
\label{fig:mx_output_wfi_w6}}\\
\subfloat[$i'$ = 7.]{ \includegraphics[width=0.5\linewidth]{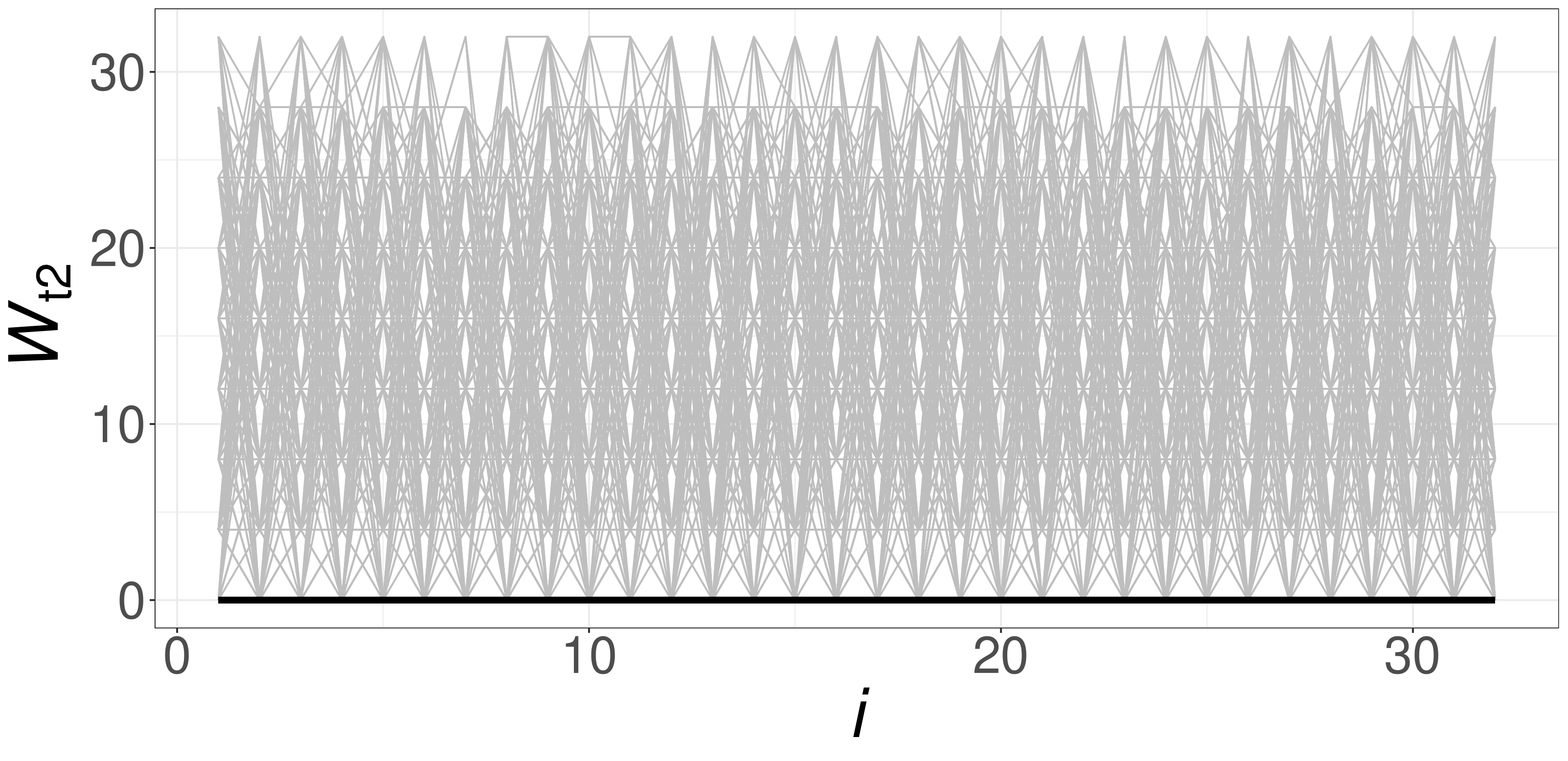}
\label{fig:mx_output_wfi_w7}}
\subfloat[$i'$ = 8.]{\includegraphics[width=0.5\linewidth]{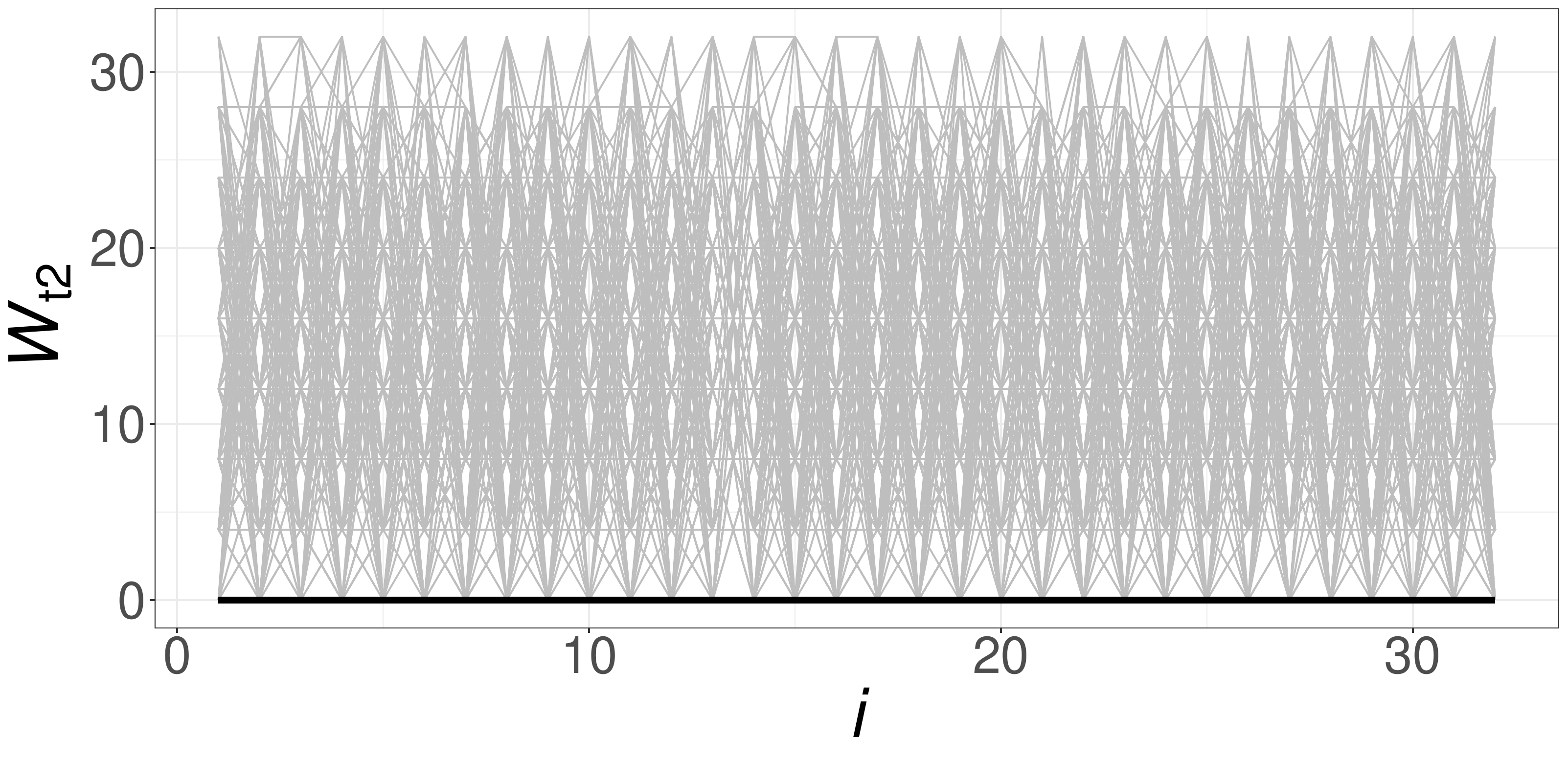}
\label{fig:mx_output_wfi_w8}}
\caption{The Walsh transforms on the $UT^{1}_{0,0}$ outputs obtained by $Q_0$ and $\textbf{S}^{2}$ in the first round. Black: correct key; gray: wrong key.}
\label{fig:Appendix_Walsh_UT_Q0_S2}
\end{figure*}

\begin{figure*}
\centering
\subfloat[$i'$ = 1.]{ \includegraphics[width=0.5\linewidth]{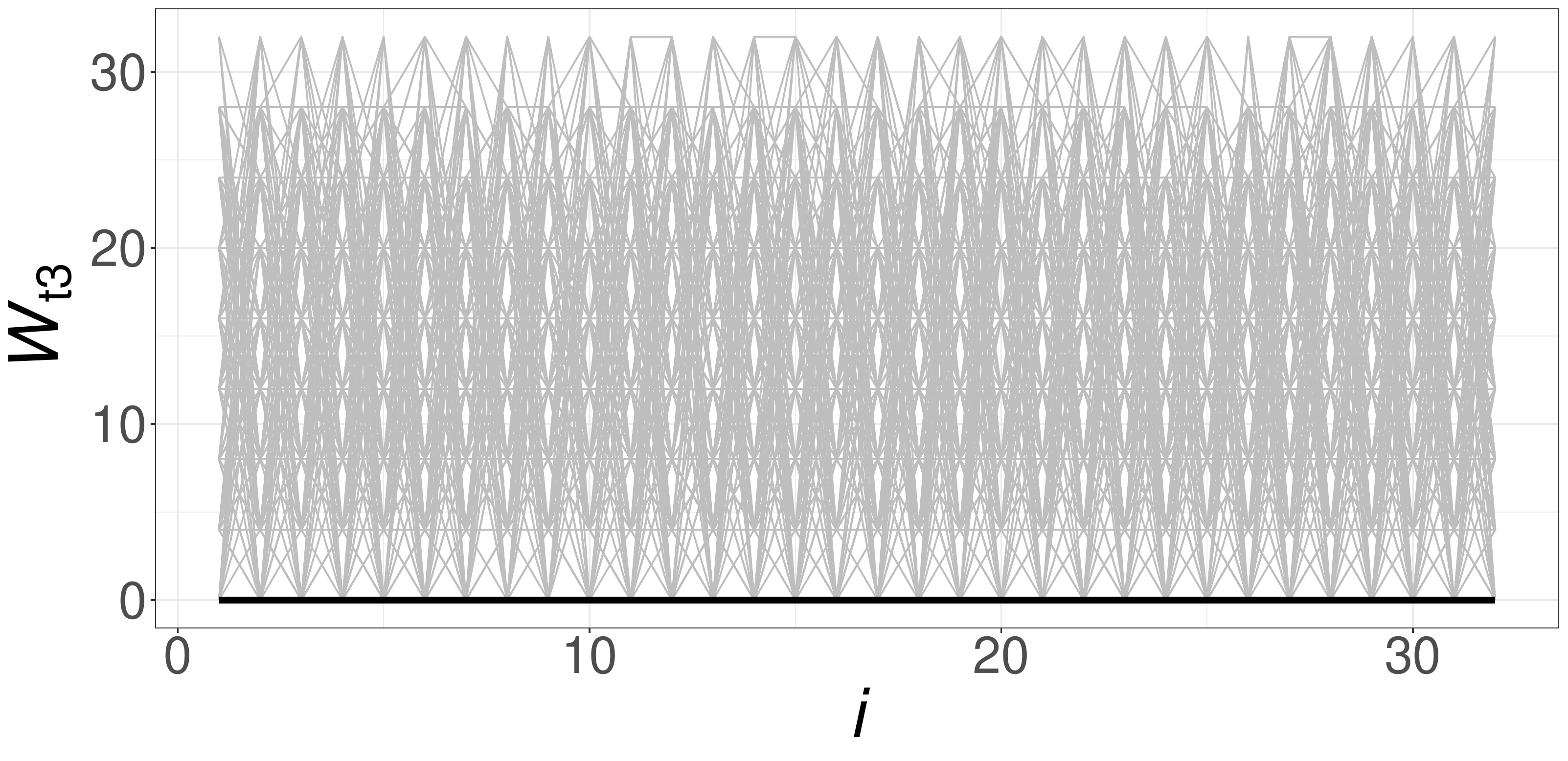}
\label{fig:mx_output_wfi_w1}}
\subfloat[$i'$ = 2.]{ \includegraphics[width=0.5\linewidth]{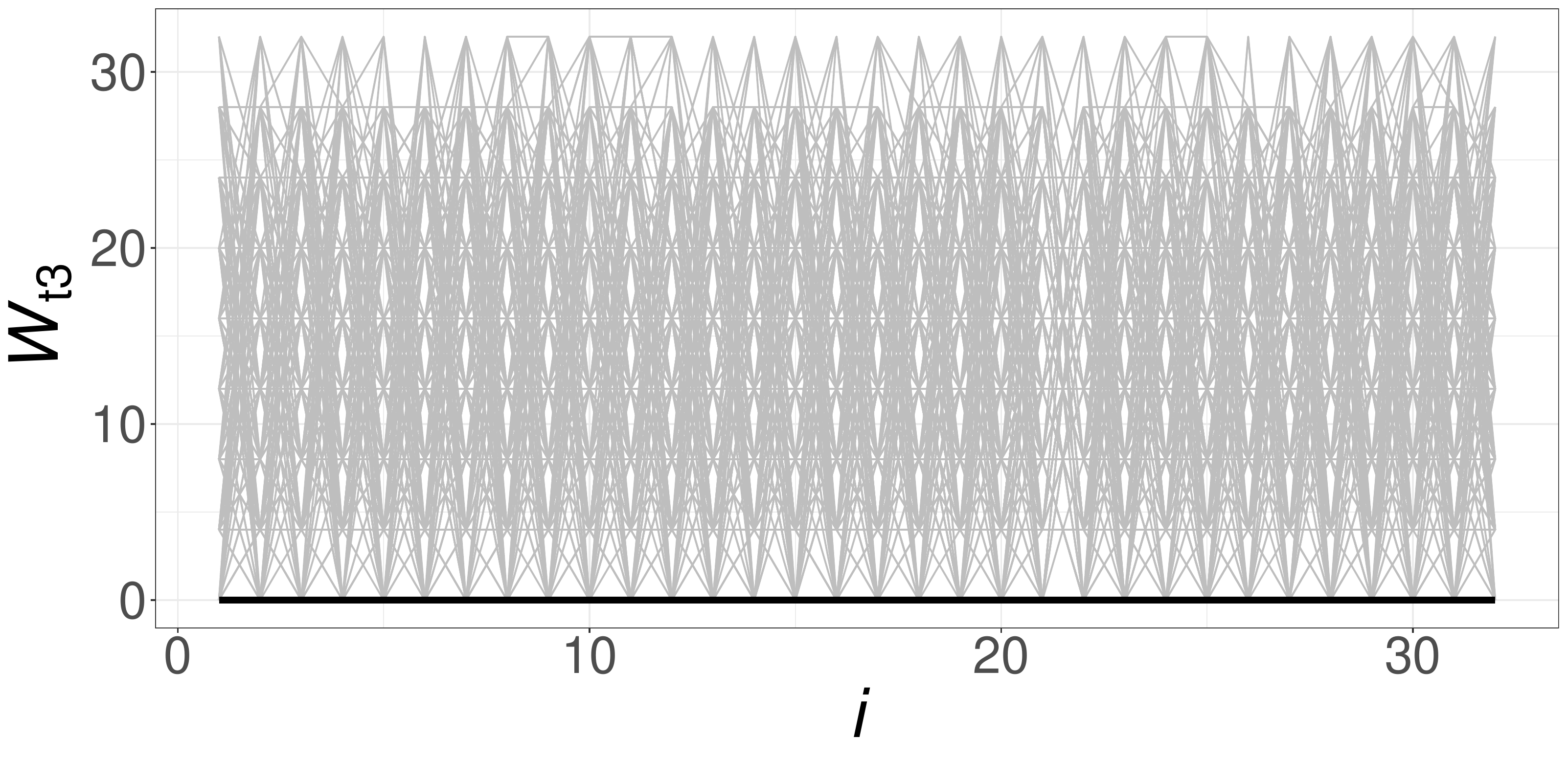}
\label{fig:mx_output_wfi_w2}}\\
\subfloat[$i'$ = 3.]{ \includegraphics[width=0.5\linewidth]{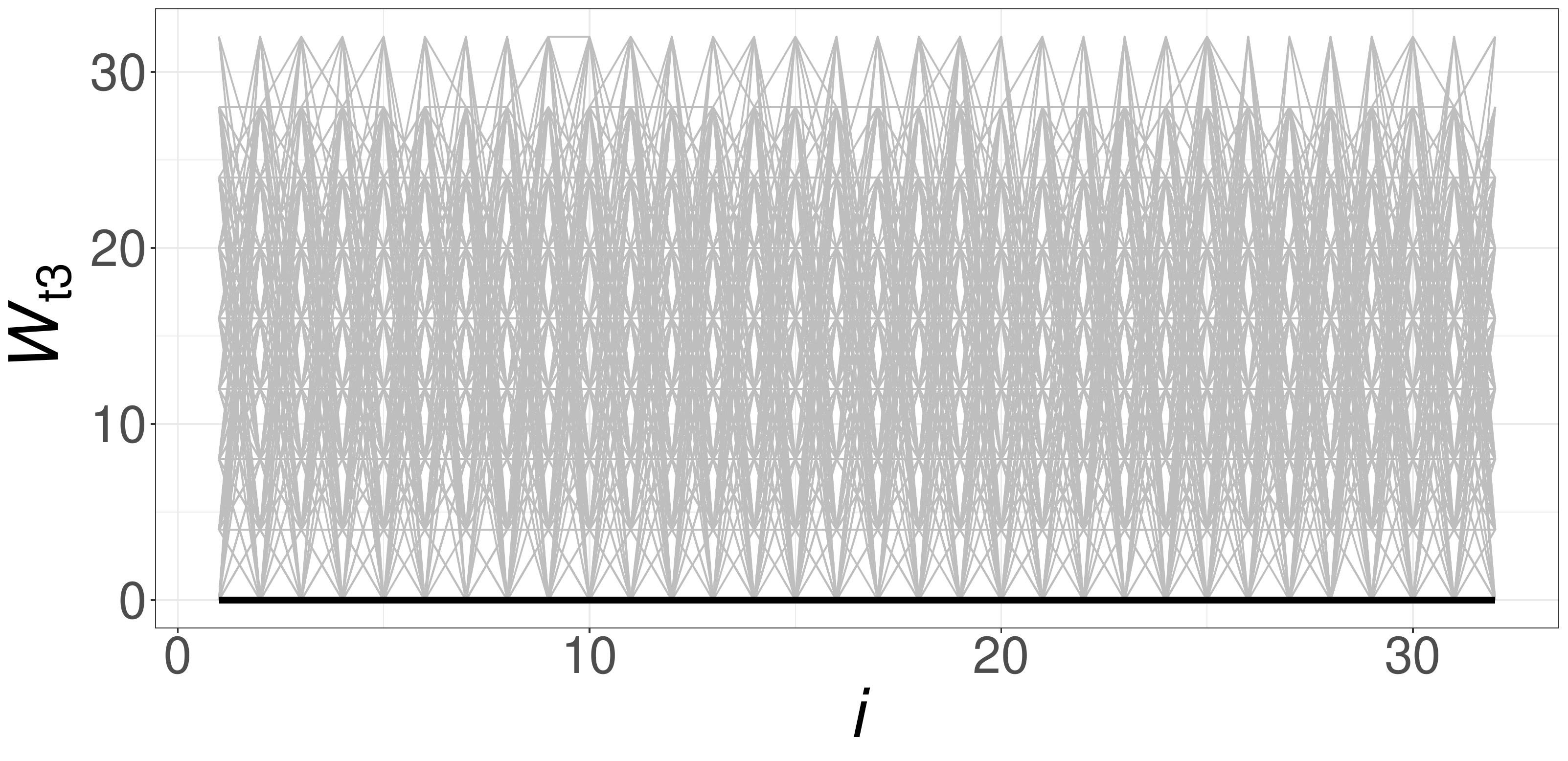}
\label{fig:mx_output_wfi_w3}}
\subfloat[$i'$ = 4.]{ \includegraphics[width=0.5\linewidth]{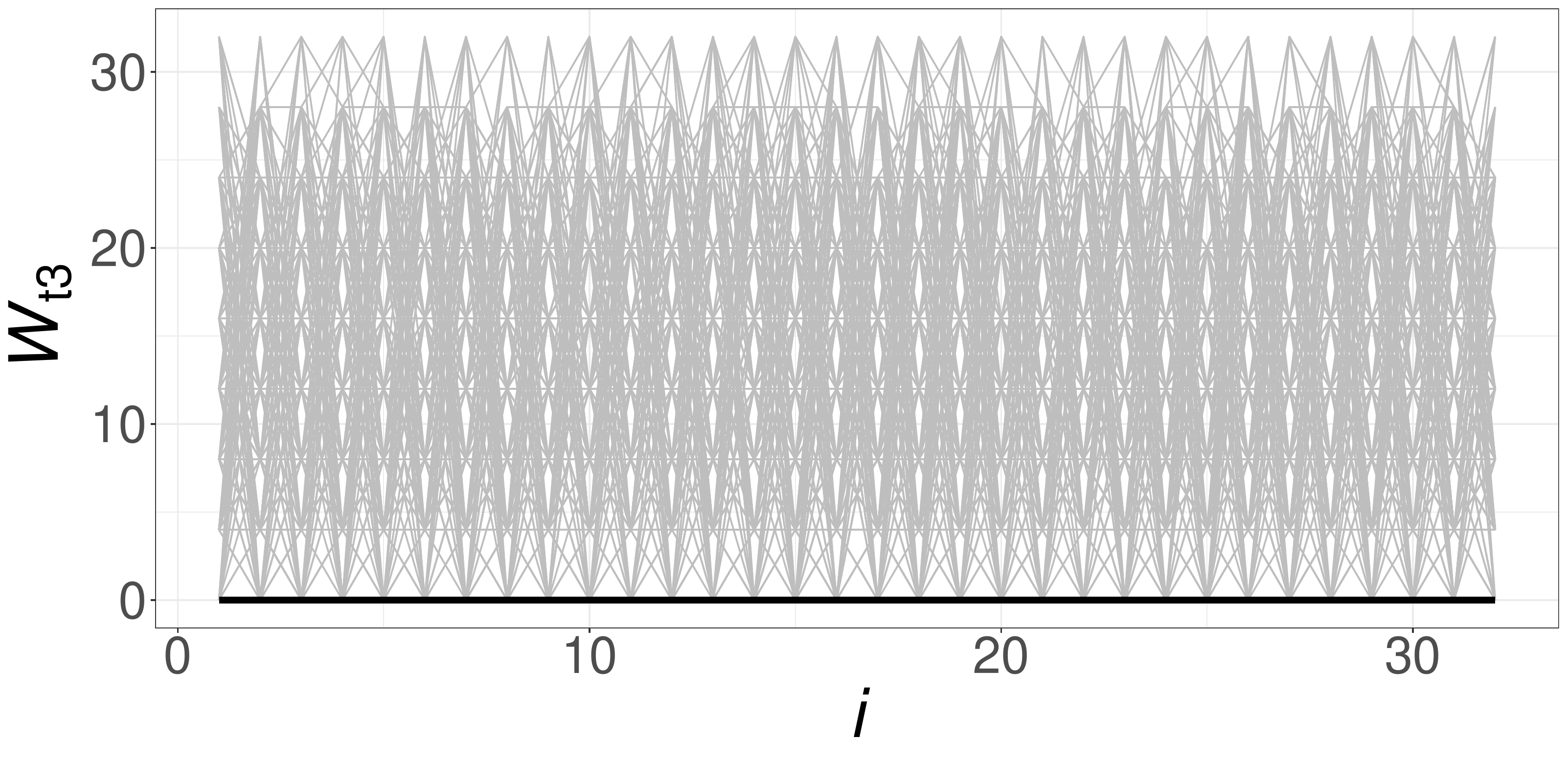}
\label{fig:mx_output_wfi_w4}}\\
\subfloat[$i'$ = 5.]{\includegraphics[width=0.5\linewidth]{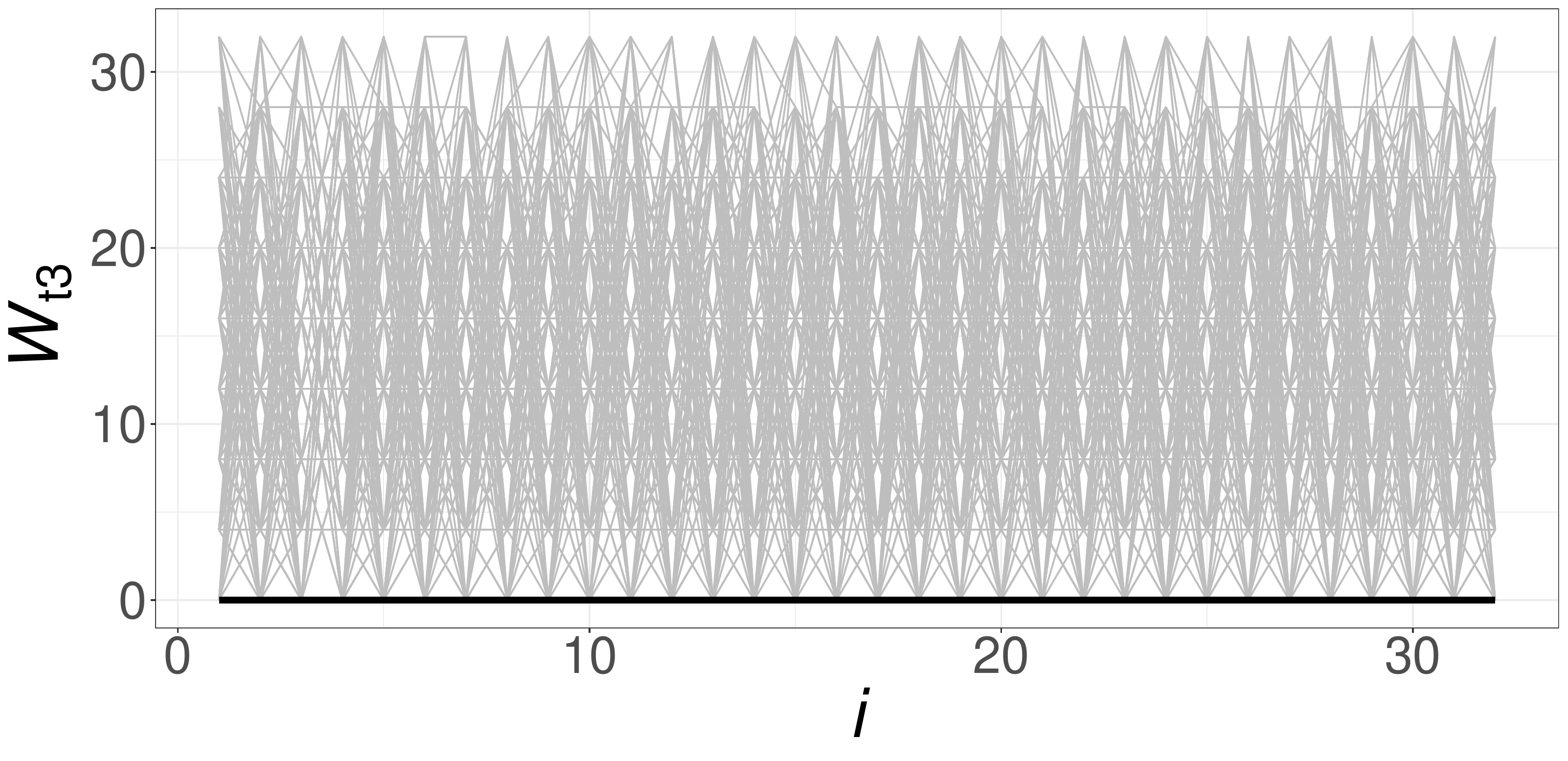}
\label{fig:mx_output_wfi_w5}}
\subfloat[$i'$ = 6.]{ \includegraphics[width=0.5\linewidth]{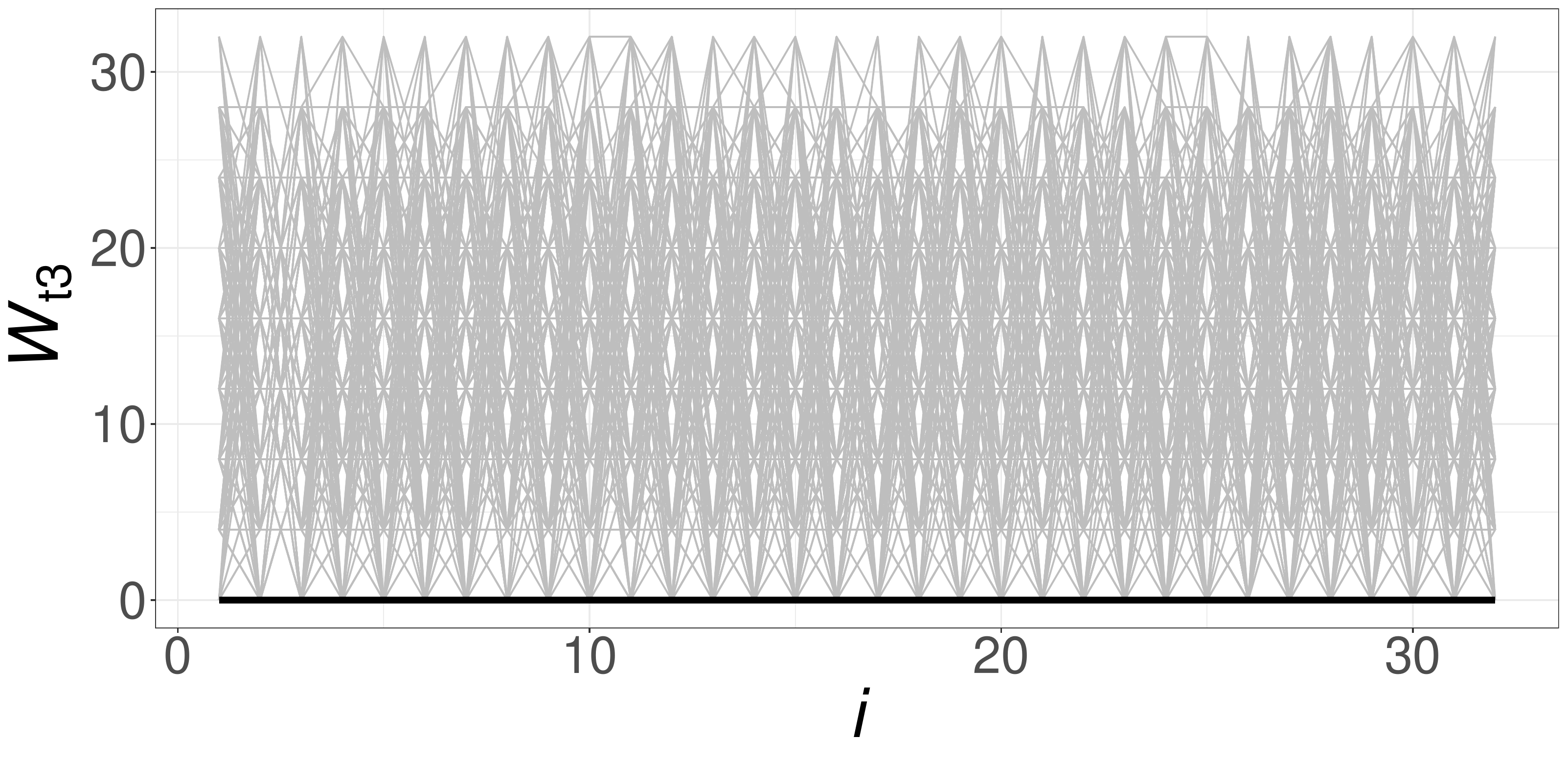}
\label{fig:mx_output_wfi_w6}}\\
\subfloat[$i'$ = 7.]{ \includegraphics[width=0.5\linewidth]{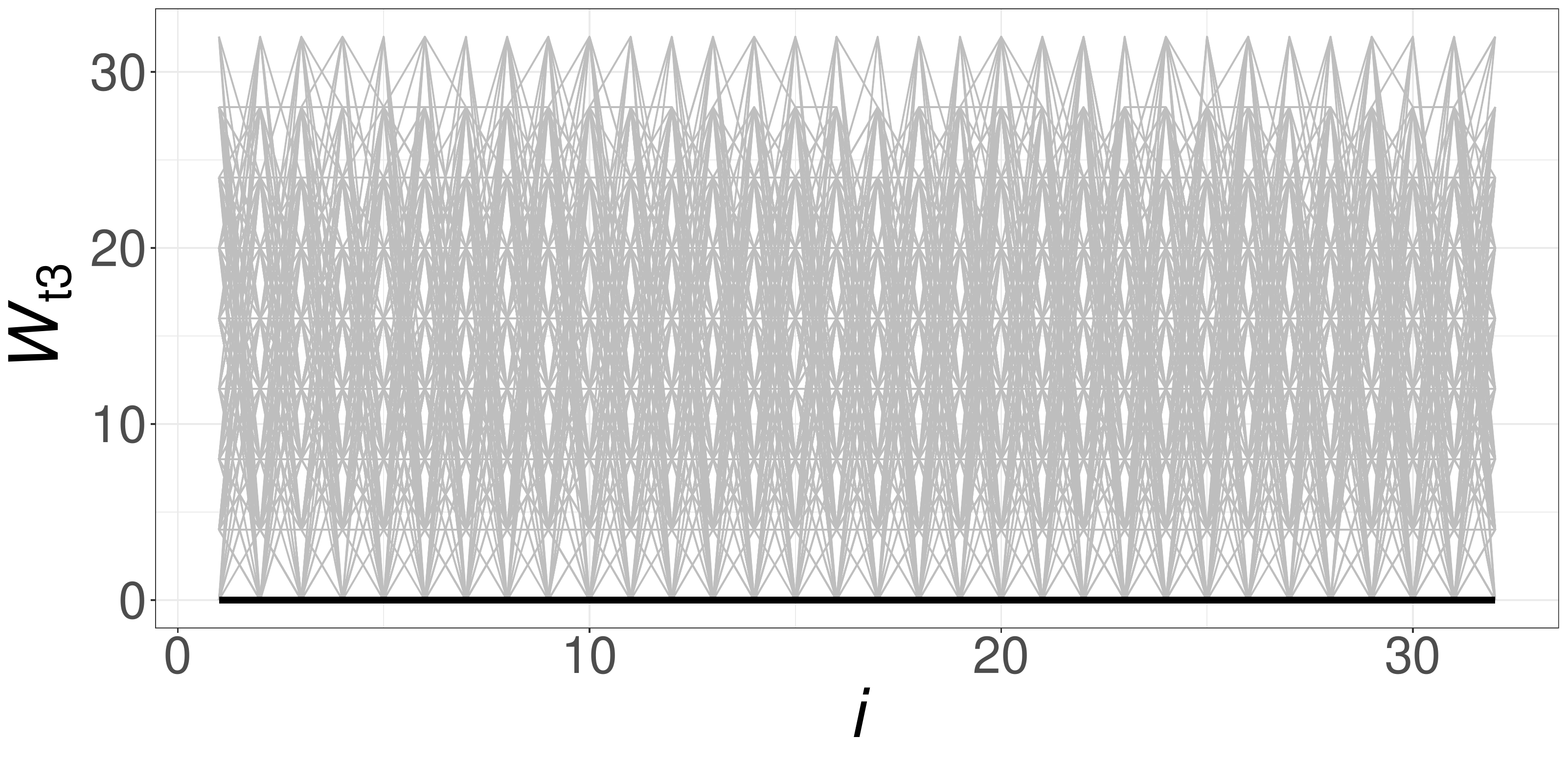}
\label{fig:mx_output_wfi_w7}}
\subfloat[$i'$ = 8.]{\includegraphics[width=0.5\linewidth]{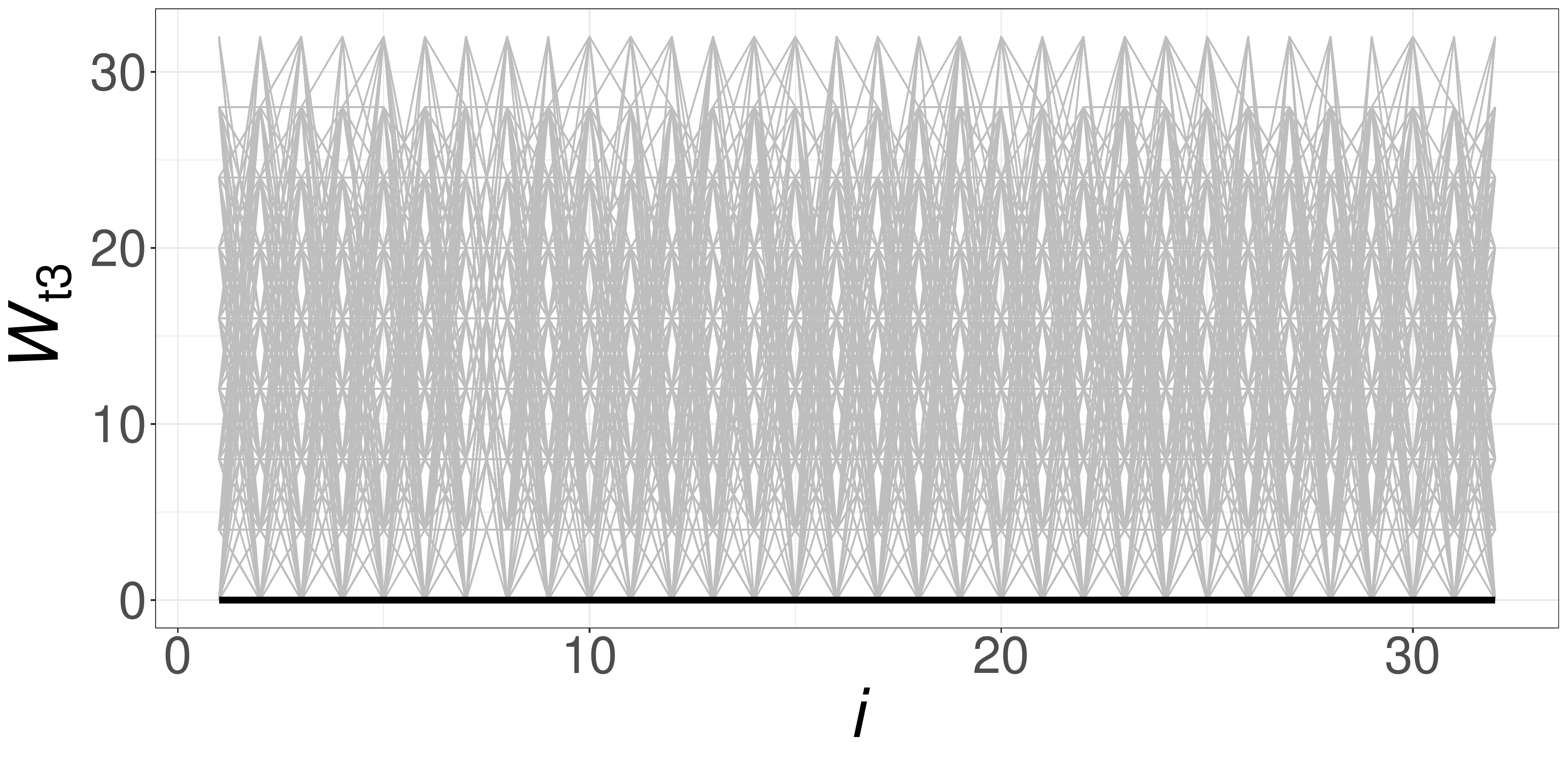}
\label{fig:mx_output_wfi_w8}}
\caption{The Walsh transforms on the $UT^{1}_{0,0}$ outputs obtained by $Q_0$ and $\textbf{S}^{3}$ in the first round. Black: correct key; gray: wrong key.}
\label{fig:Appendix_Walsh_UT_Q0_S3}
\end{figure*}

\begin{figure*}
\centering
\subfloat[$i'$ = 1.]{ \includegraphics[width=0.5\linewidth]{fig/Wer_Q0_bit1.pdf}
\label{fig:mx_output_wfi_w1}}
\subfloat[$i'$ = 2.]{ \includegraphics[width=0.5\linewidth]{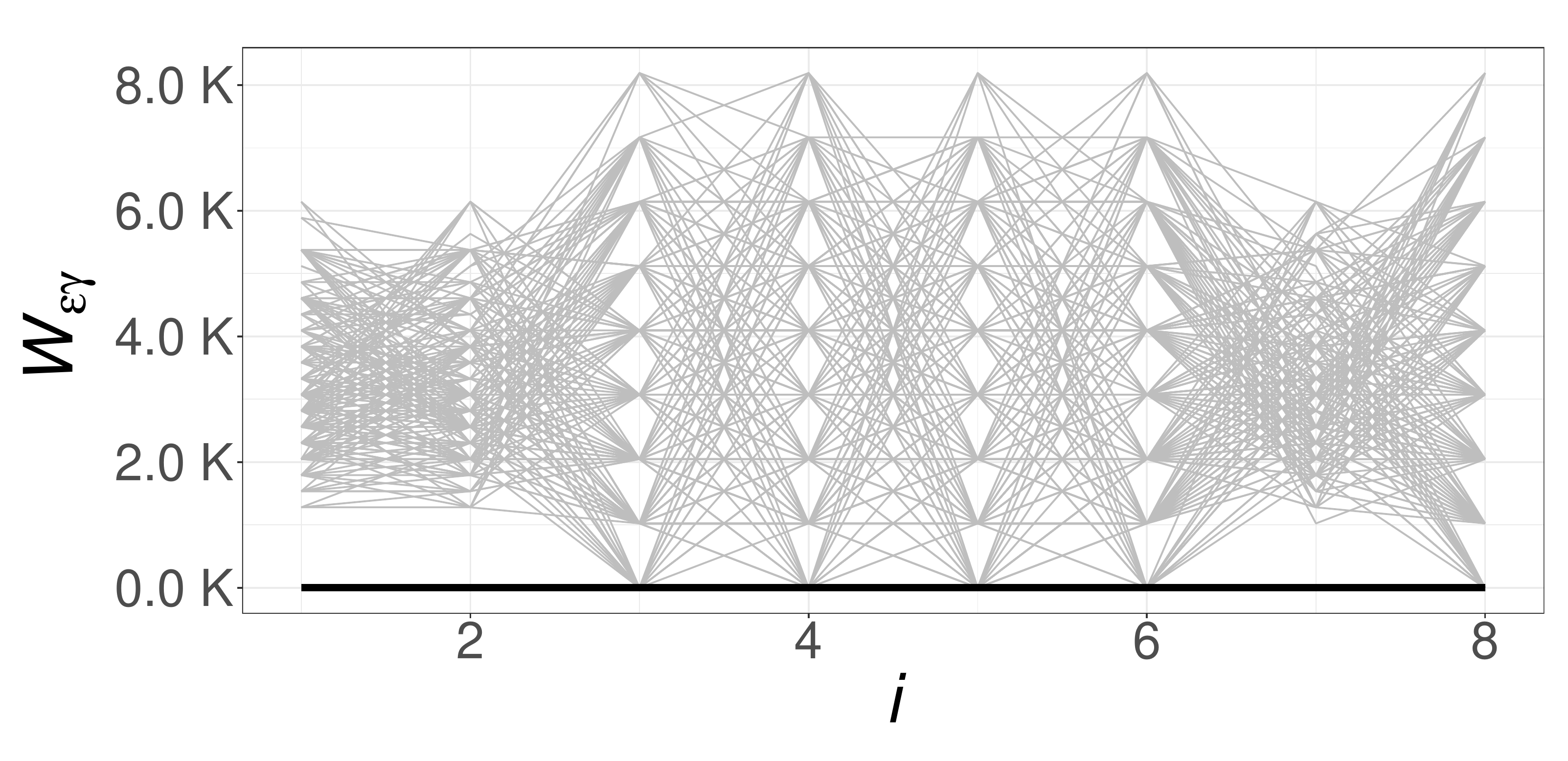}
\label{fig:mx_output_wfi_w2}}\\
\subfloat[$i'$ = 3.]{ \includegraphics[width=0.5\linewidth]{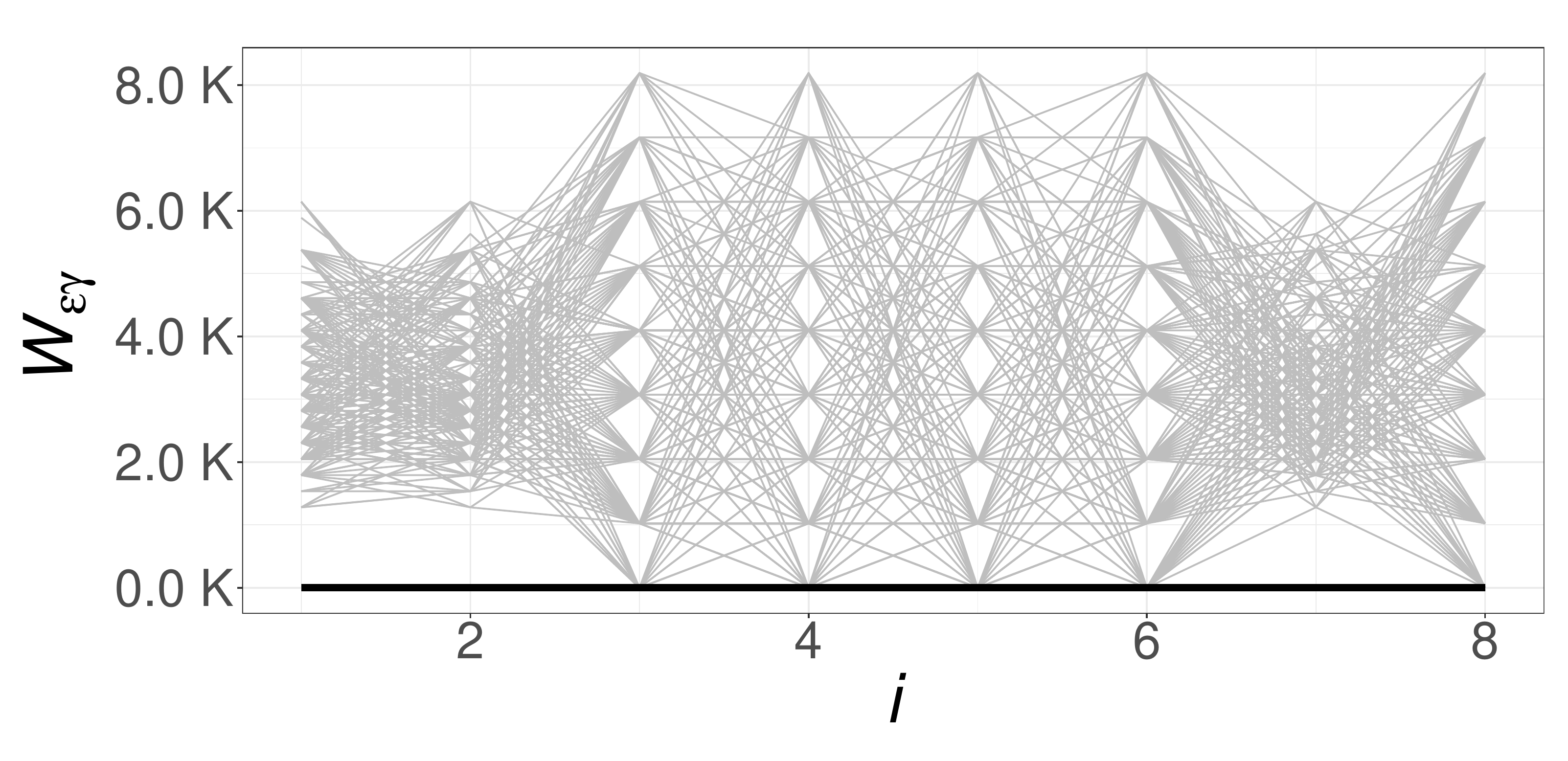}
\label{fig:mx_output_wfi_w3}}
\subfloat[$i'$ = 4.]{ \includegraphics[width=0.5\linewidth]{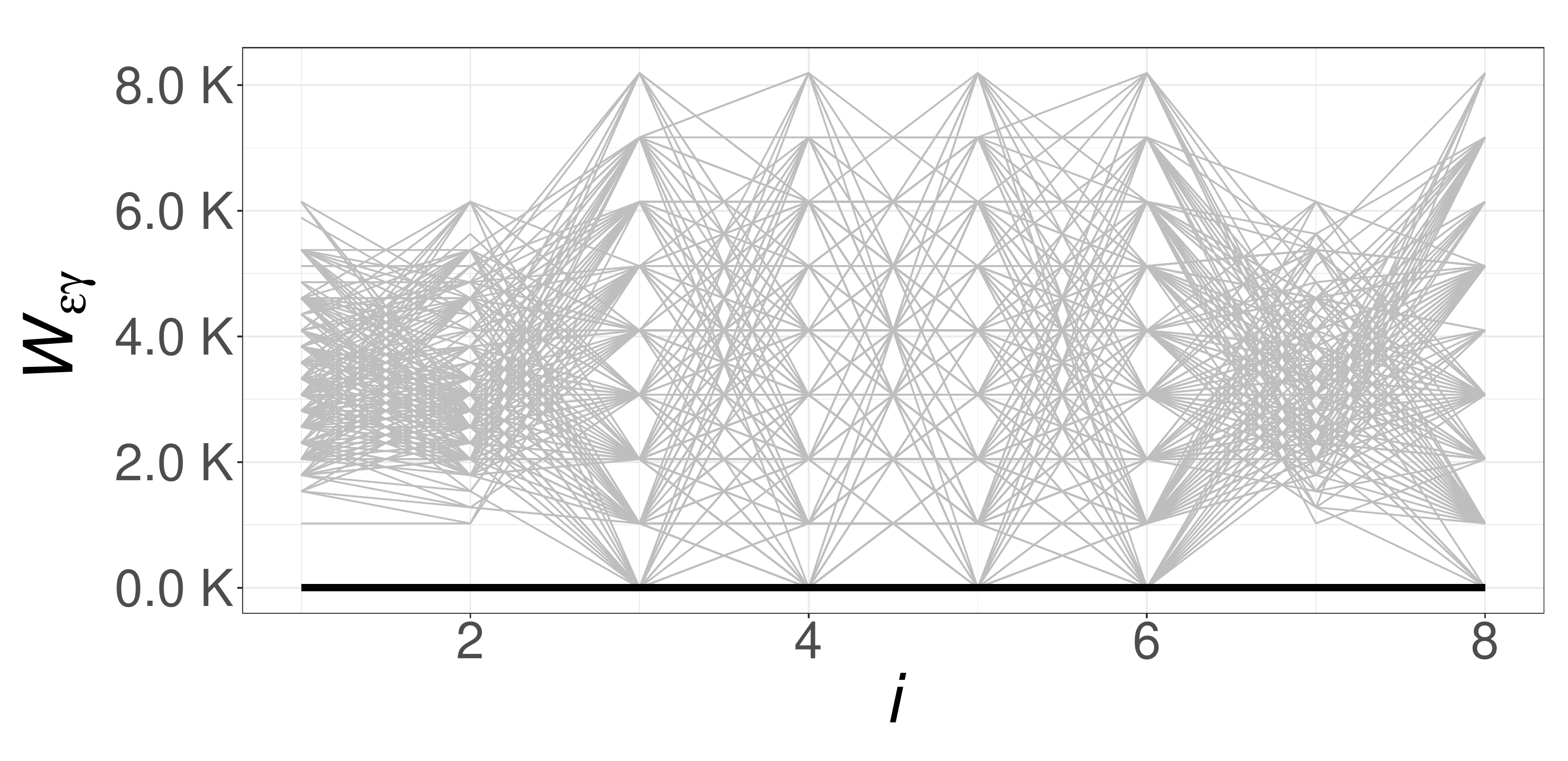}
\label{fig:mx_output_wfi_w4}}\\
\subfloat[$i'$ = 5.]{\includegraphics[width=0.5\linewidth]{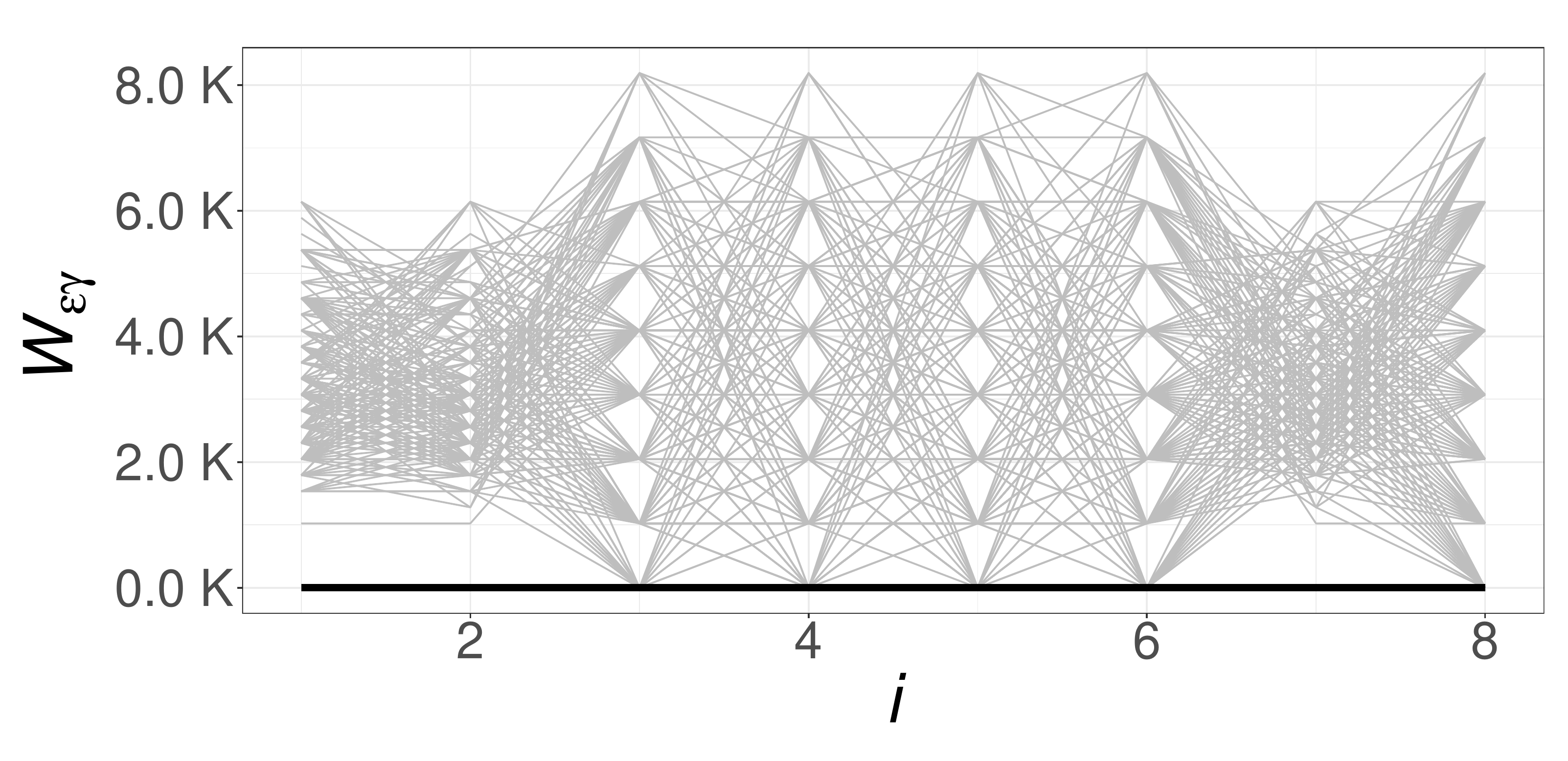}
\label{fig:mx_output_wfi_w5}}
\subfloat[$i'$ = 6.]{ \includegraphics[width=0.5\linewidth]{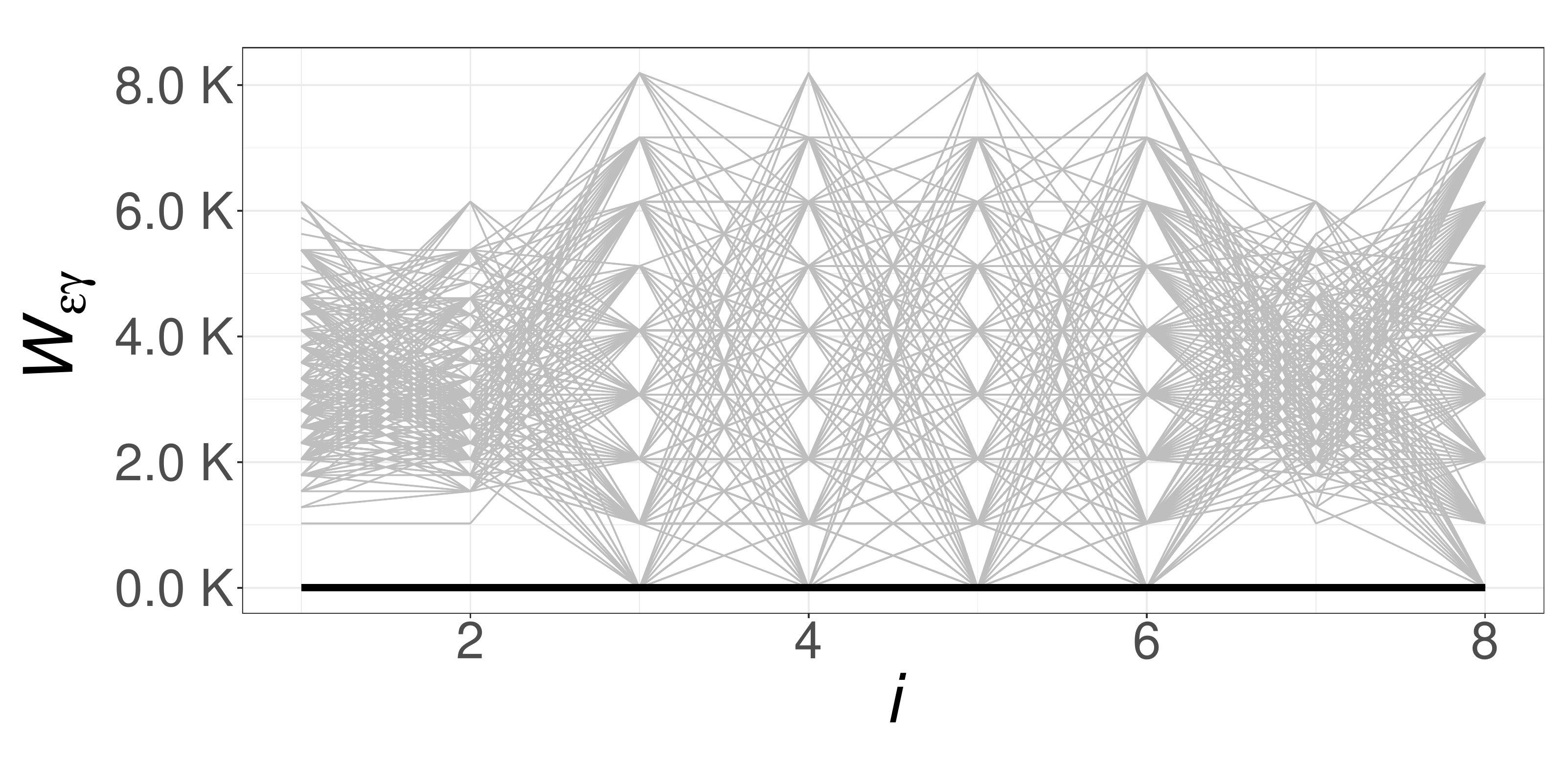}
\label{fig:mx_output_wfi_w6}}\\
\subfloat[$i'$ = 7.]{ \includegraphics[width=0.5\linewidth]{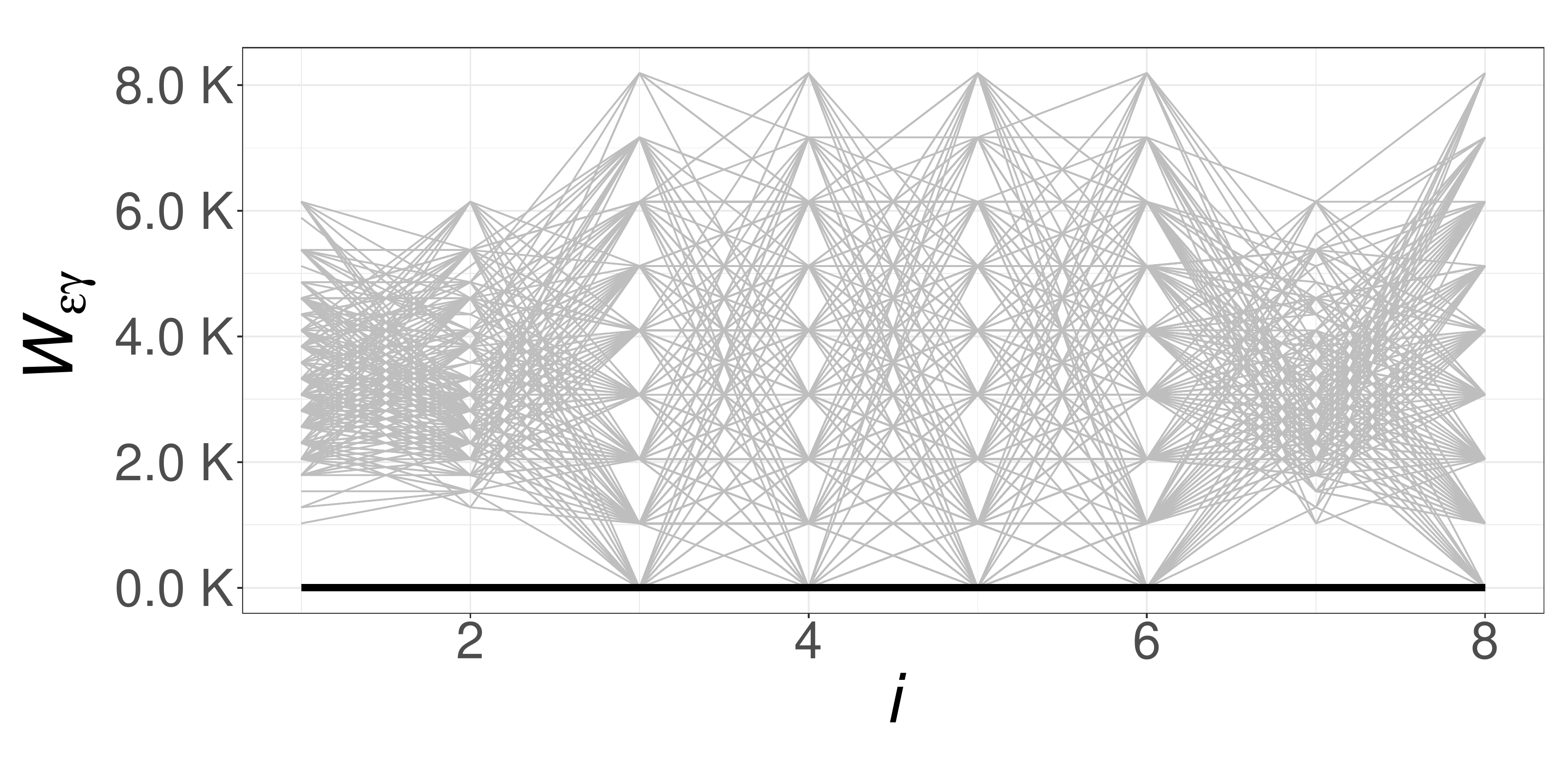}
\label{fig:mx_output_wfi_w7}}
\subfloat[$i'$ = 8.]{\includegraphics[width=0.5\linewidth]{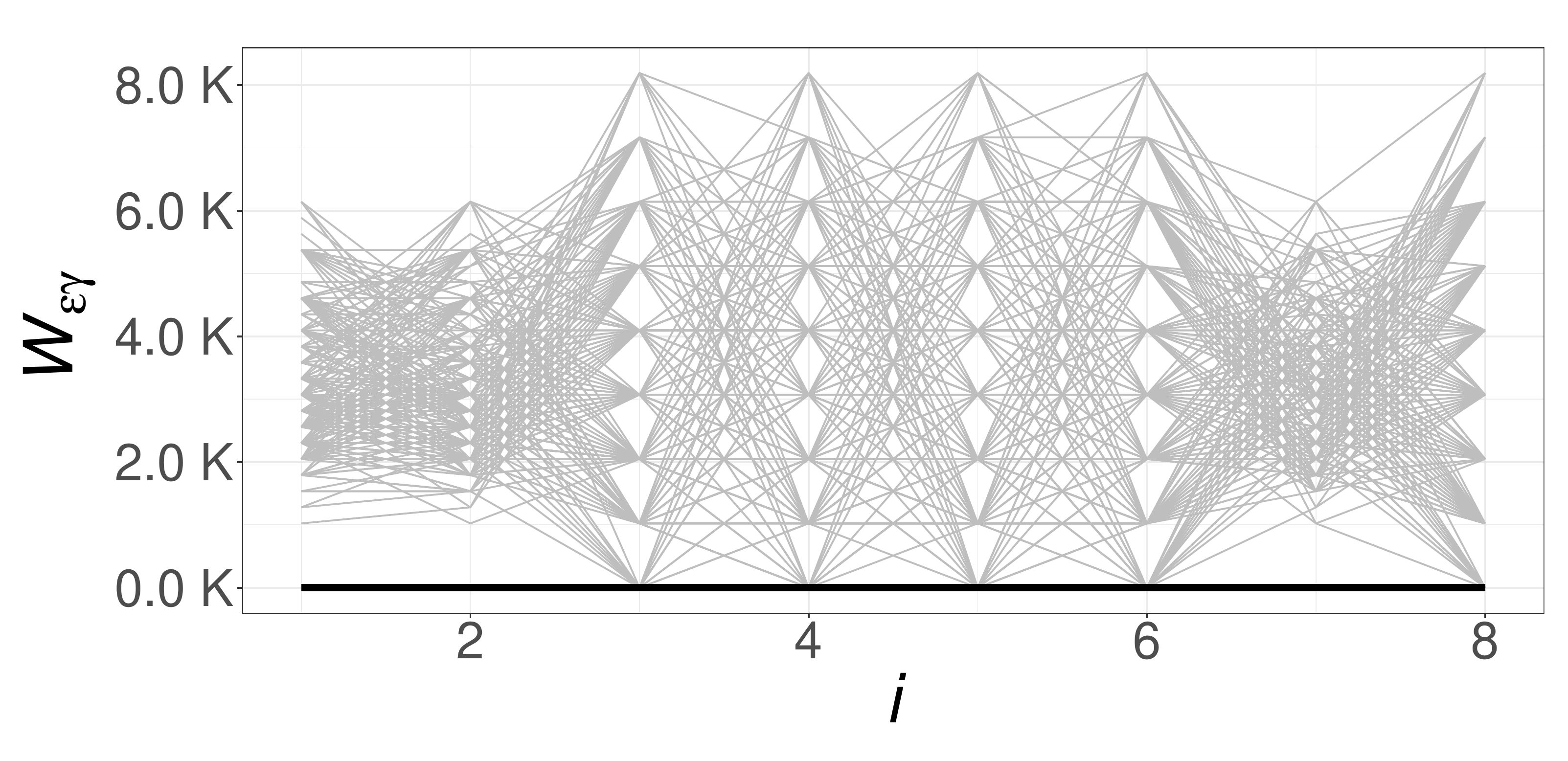}
\label{fig:mx_output_wfi_w8}}
\caption{The Walsh transforms on the round outputs obtained by $Q_0$ and the hypothetical round outputs in the first round. Black: correct key; gray: wrong key.}
\label{fig:Appendix_Walsh_RO_Q0}
\end{figure*}

\begin{figure*}
\centering
\subfloat[$i'$ = 1.]{ \includegraphics[width=0.5\linewidth]{fig/Wt1_Q0_Q1_bit1.pdf}
\label{fig:mx_output_wfi_w1}}
\subfloat[$i'$ = 2.]{ \includegraphics[width=0.5\linewidth]{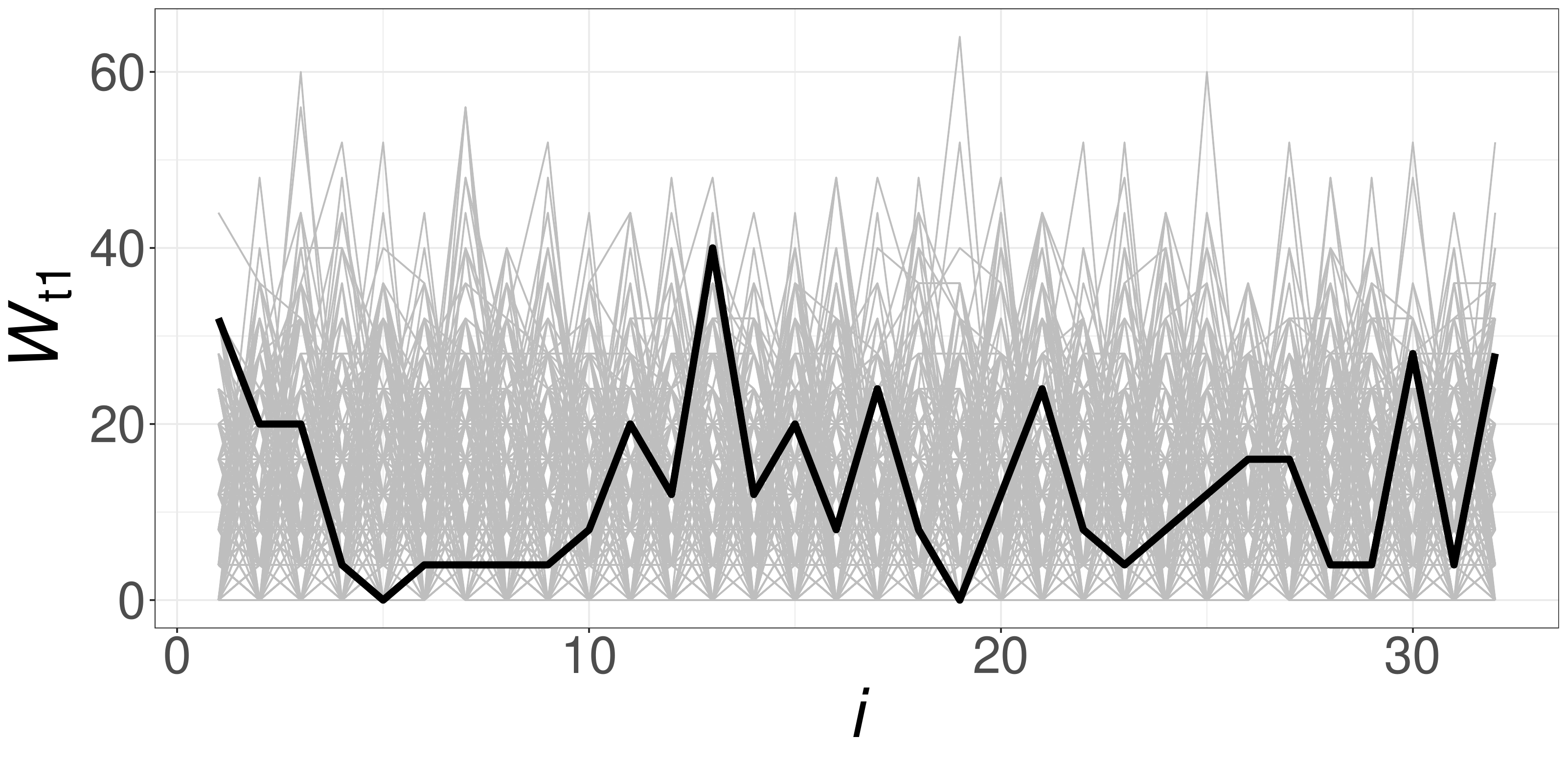}
\label{fig:mx_output_wfi_w2}}\\
\subfloat[$i'$ = 3.]{ \includegraphics[width=0.5\linewidth]{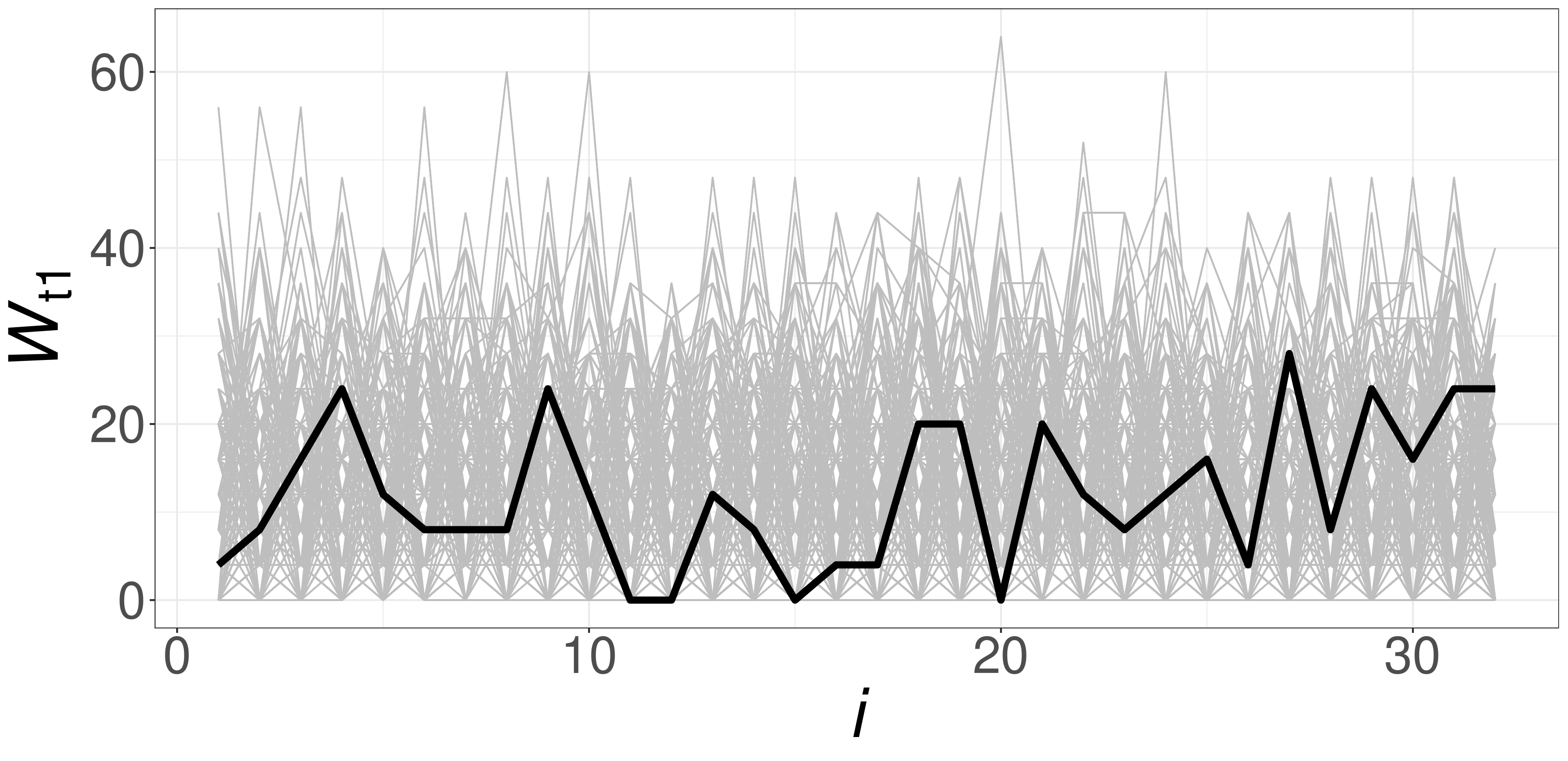}
\label{fig:mx_output_wfi_w3}}
\subfloat[$i'$ = 4.]{ \includegraphics[width=0.5\linewidth]{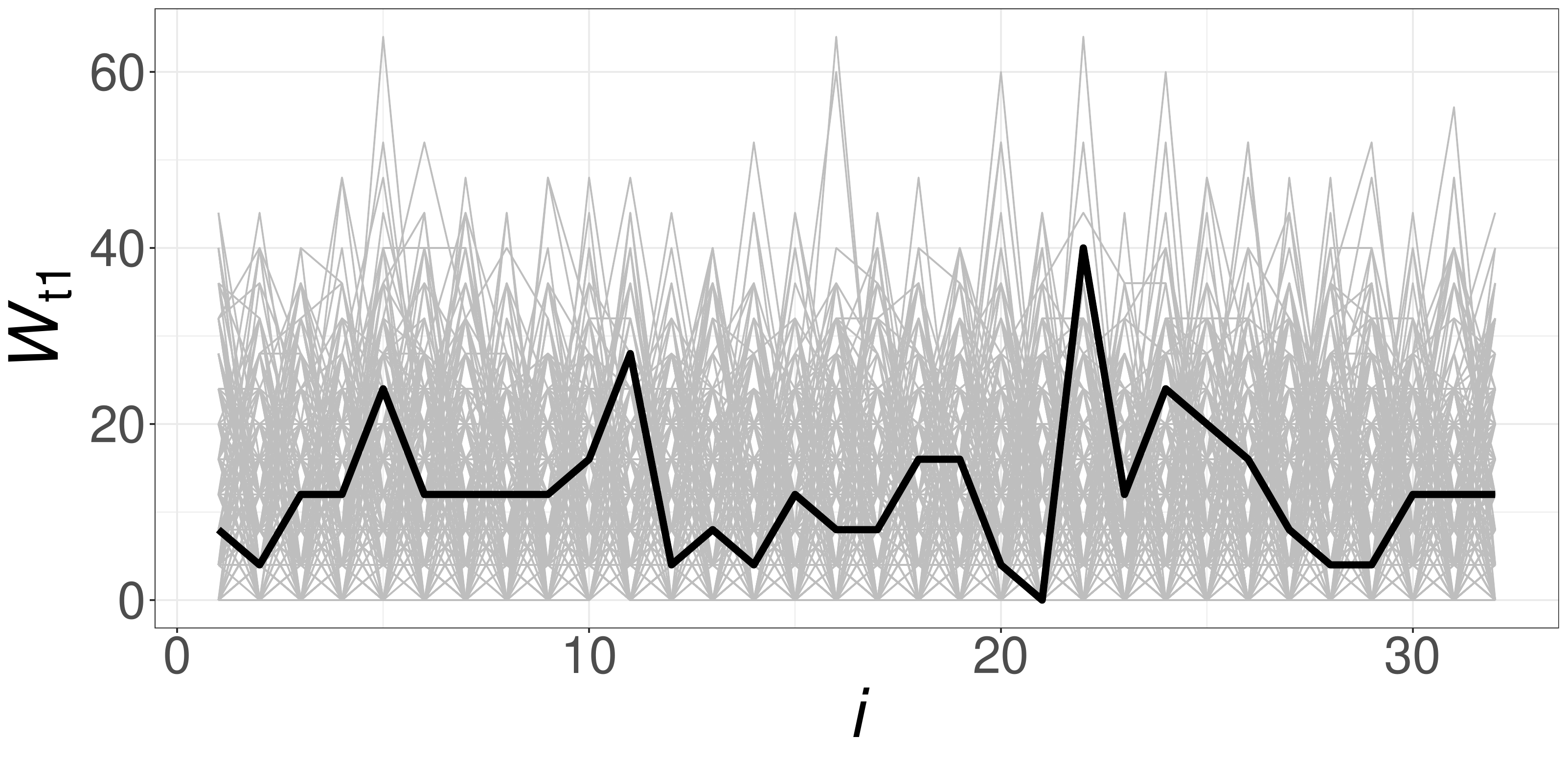}
\label{fig:mx_output_wfi_w4}}\\
\subfloat[$i'$ = 5.]{\includegraphics[width=0.5\linewidth]{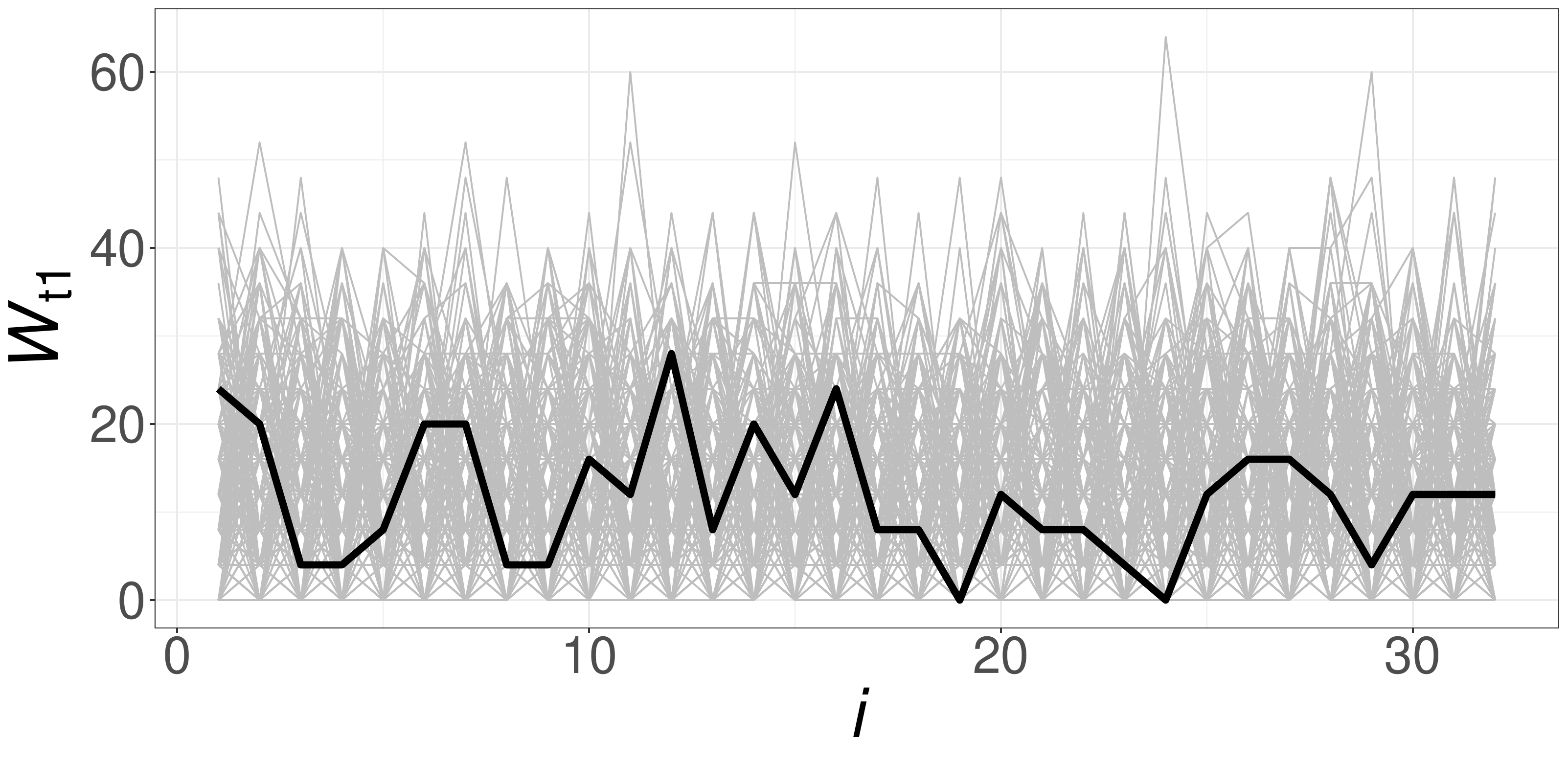}
\label{fig:mx_output_wfi_w5}}
\subfloat[$i'$ = 6.]{ \includegraphics[width=0.5\linewidth]{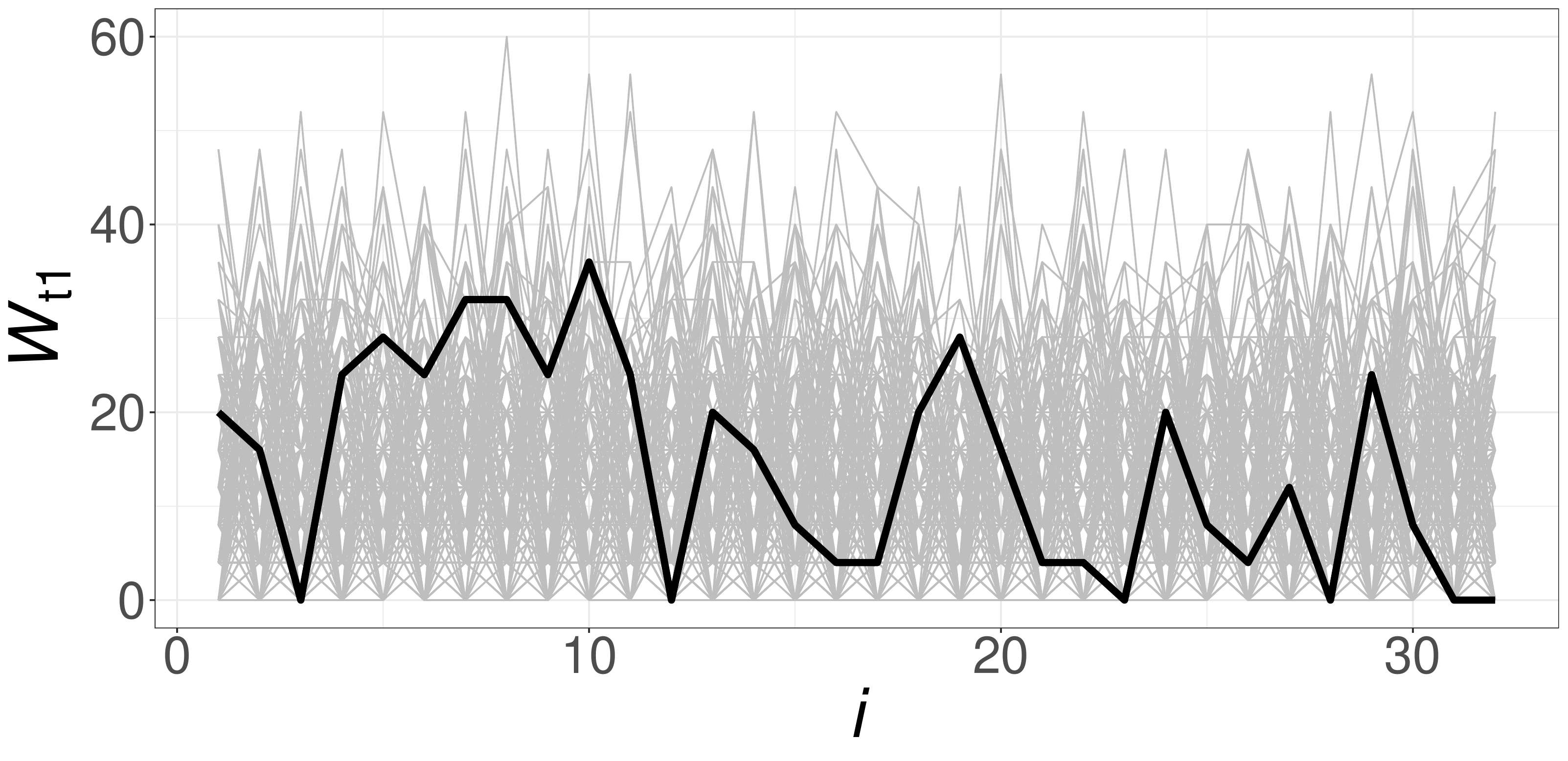}
\label{fig:mx_output_wfi_w6}}\\
\subfloat[$i'$ = 7.]{ \includegraphics[width=0.5\linewidth]{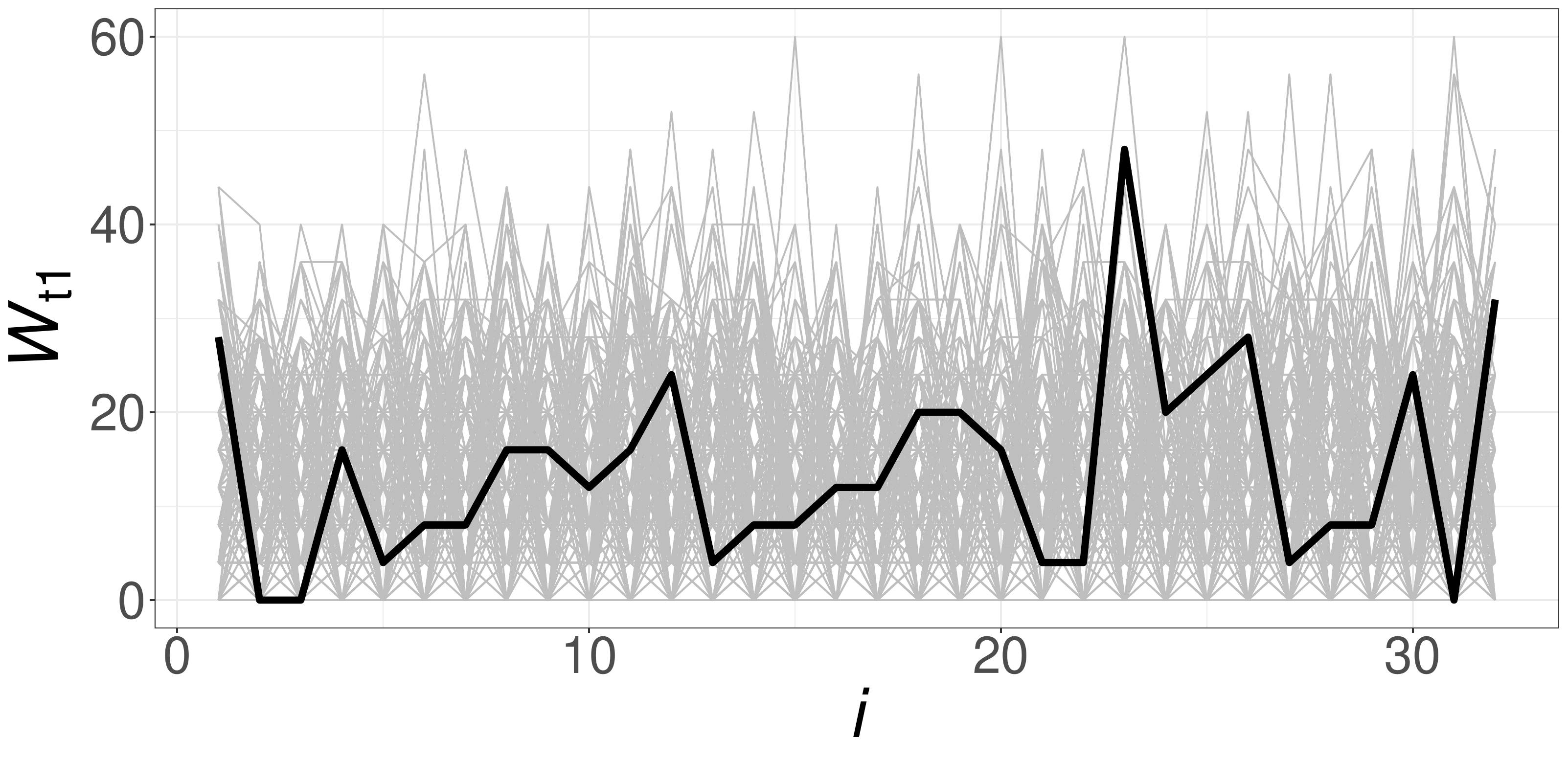}
\label{fig:mx_output_wfi_w7}}
\subfloat[$i'$ = 8.]{\includegraphics[width=0.5\linewidth]{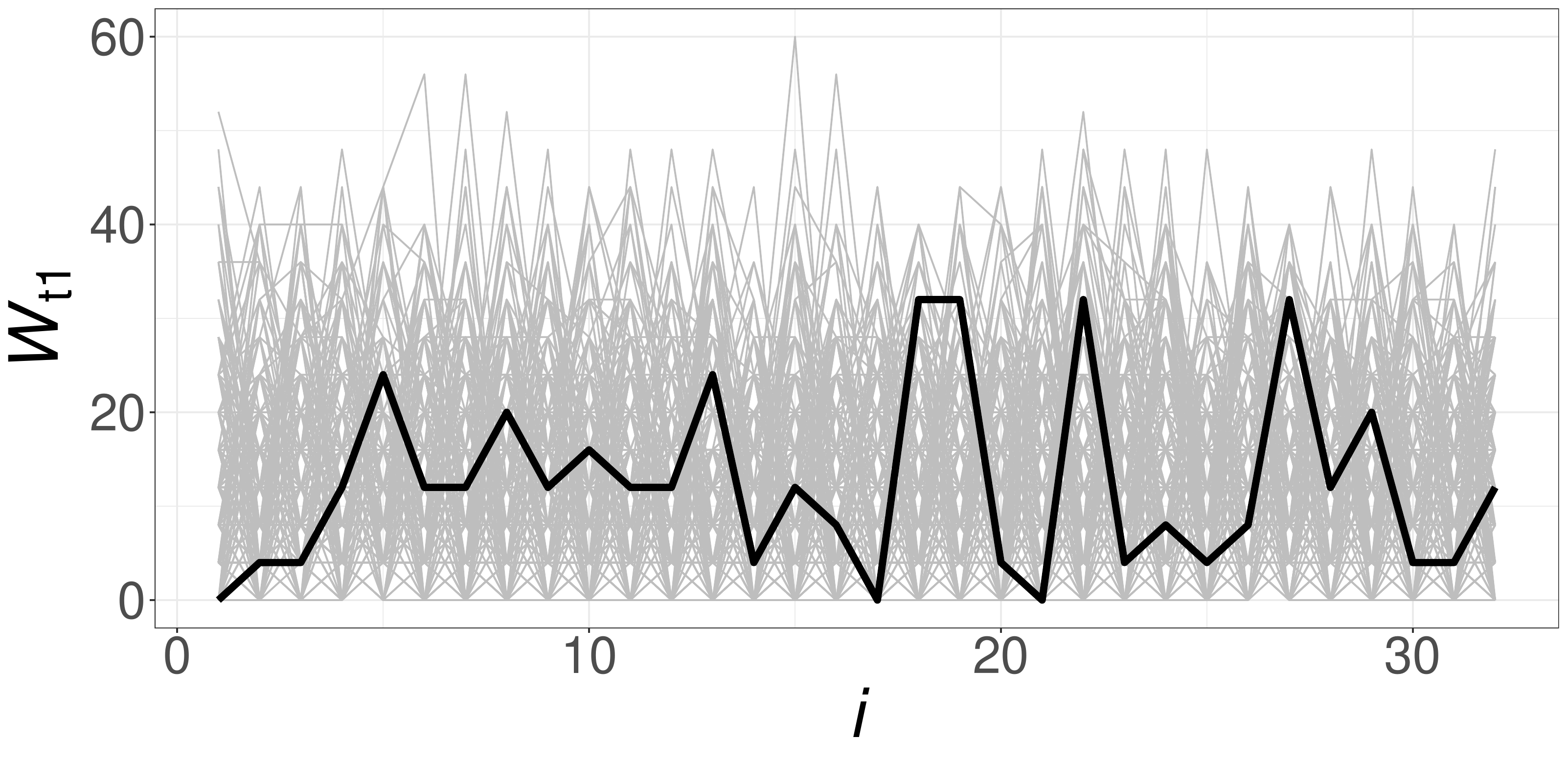}
\label{fig:mx_output_wfi_w8}}
\caption{The Walsh transforms on the $UT^{1}_{0,0}$ outputs obtained by $Q_0$ and $Q_1$ and $\textbf{S}^{1}$ in the first round. Black: correct key; gray: wrong key.}
\label{fig:Appendix_Walsh_UT_Q0_Q1_S1}
\end{figure*}

\begin{figure*}
\centering
\subfloat[$i'$ = 1.]{ \includegraphics[width=0.5\linewidth]{fig/Wt2_Q0_Q1_bit1.pdf}
\label{fig:mx_output_wfi_w1}}
\subfloat[$i'$ = 2.]{ \includegraphics[width=0.5\linewidth]{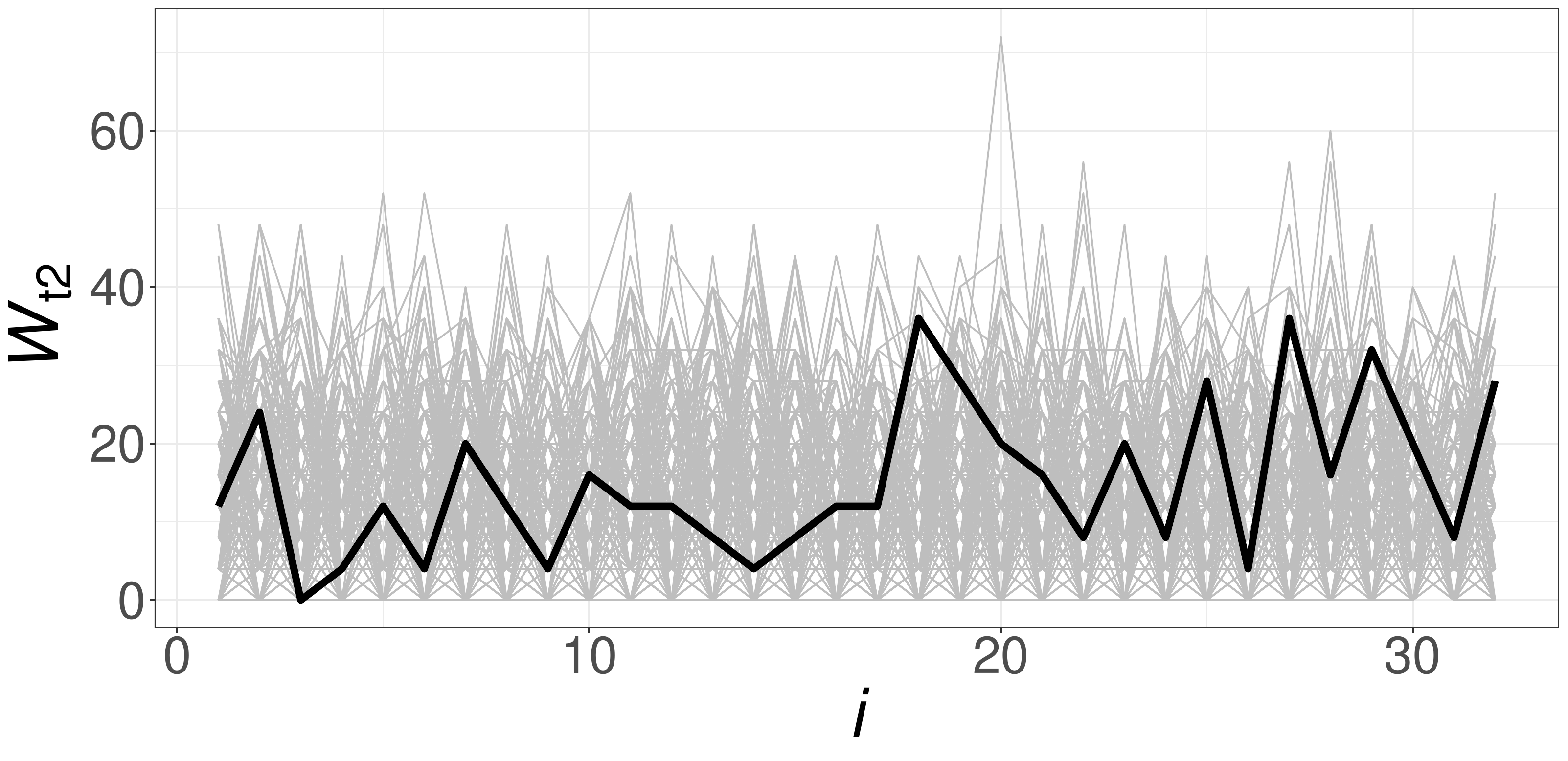}
\label{fig:mx_output_wfi_w2}}\\
\subfloat[$i'$ = 3.]{ \includegraphics[width=0.5\linewidth]{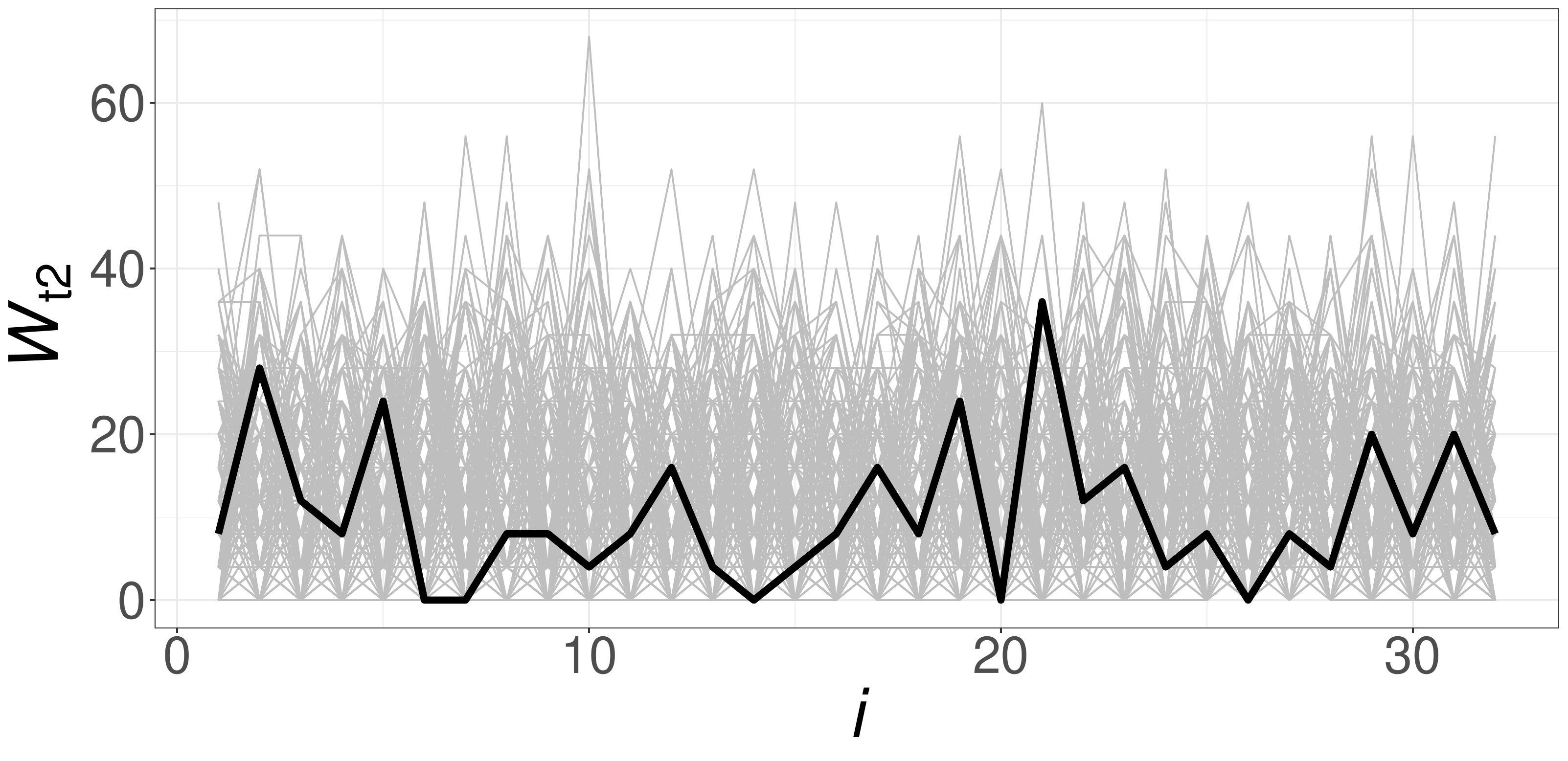}
\label{fig:mx_output_wfi_w3}}
\subfloat[$i'$ = 4.]{ \includegraphics[width=0.5\linewidth]{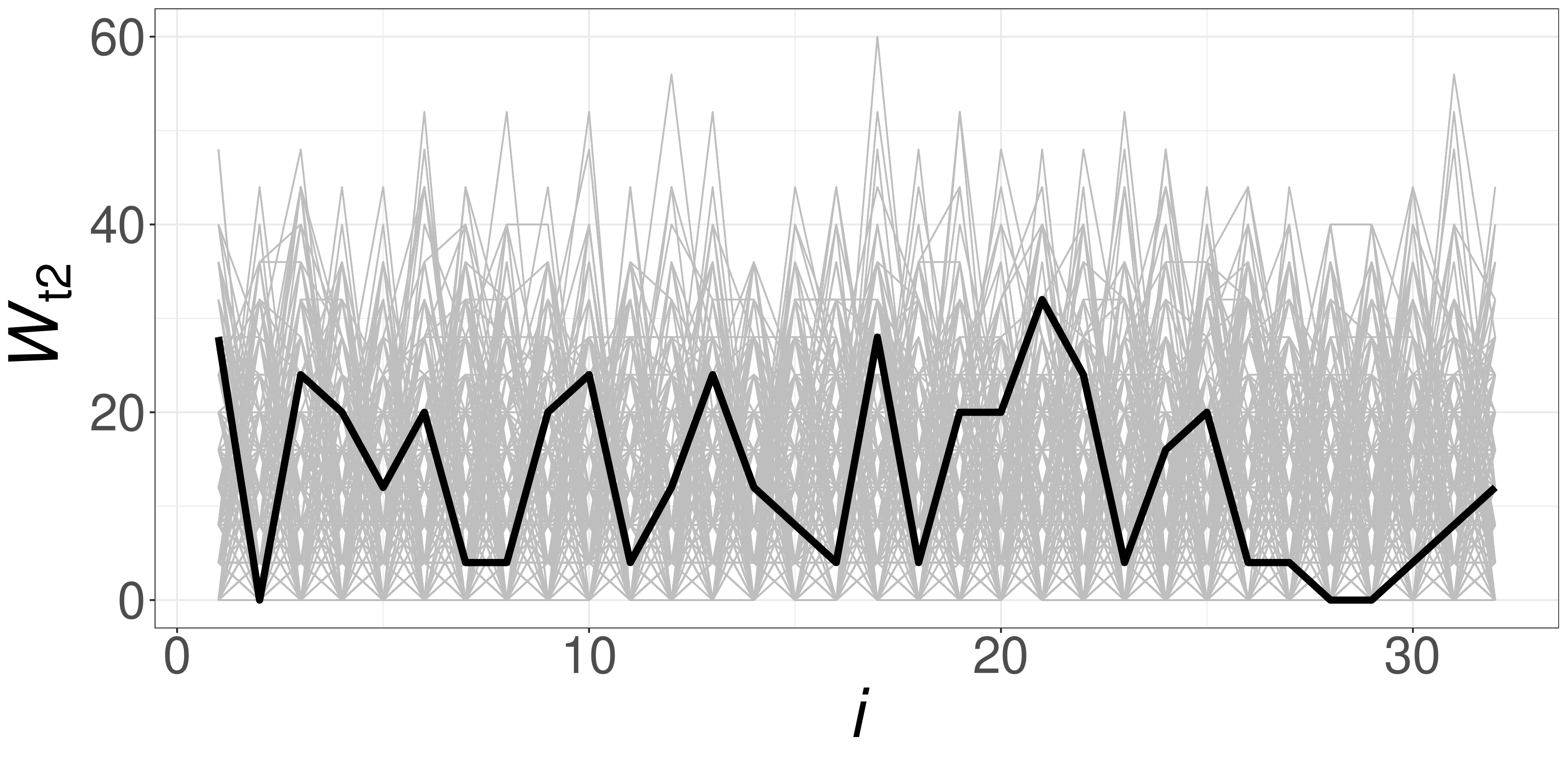}
\label{fig:mx_output_wfi_w4}}\\
\subfloat[$i'$ = 5.]{\includegraphics[width=0.5\linewidth]{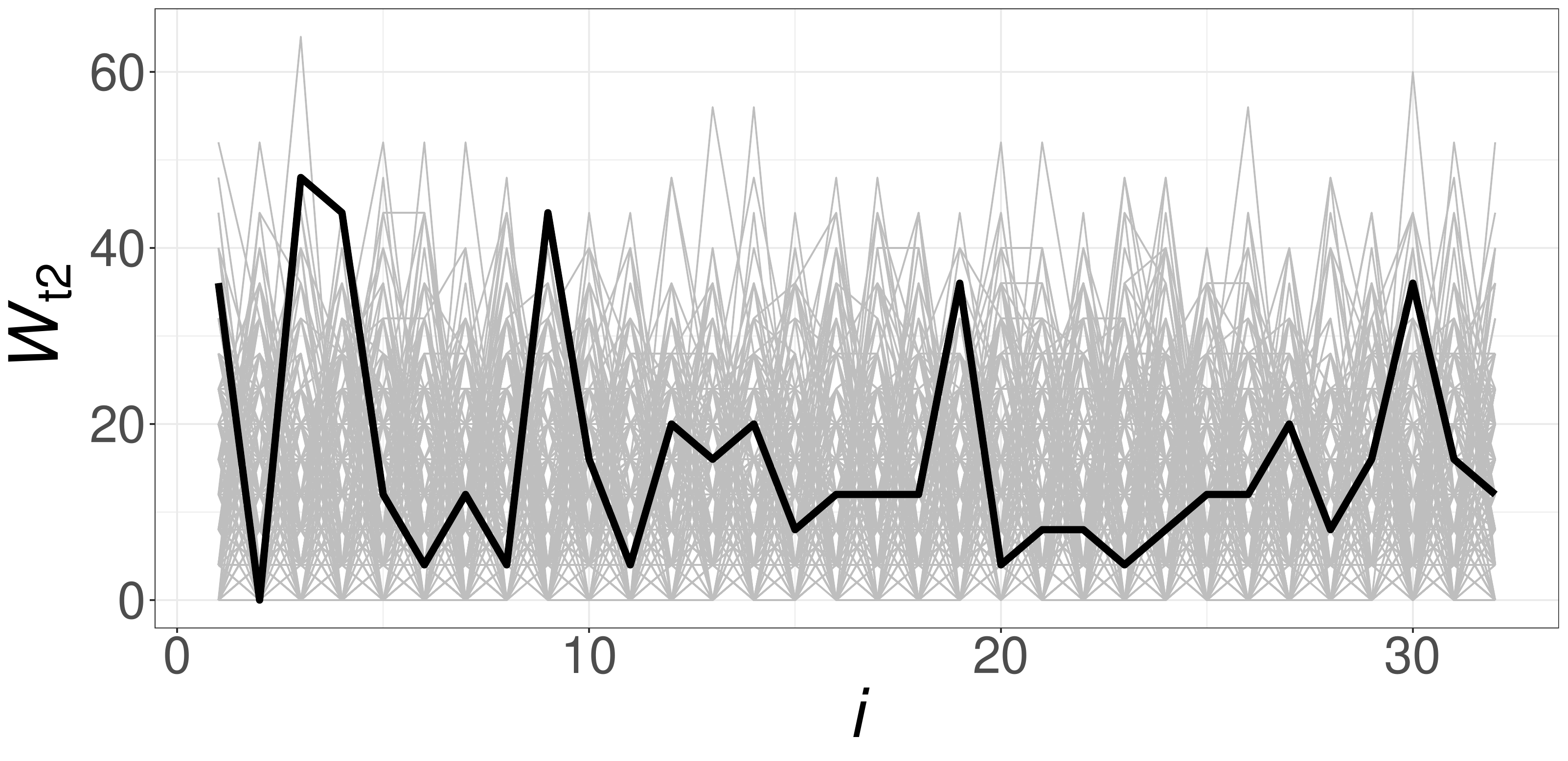}
\label{fig:mx_output_wfi_w5}}
\subfloat[$i'$ = 6.]{ \includegraphics[width=0.5\linewidth]{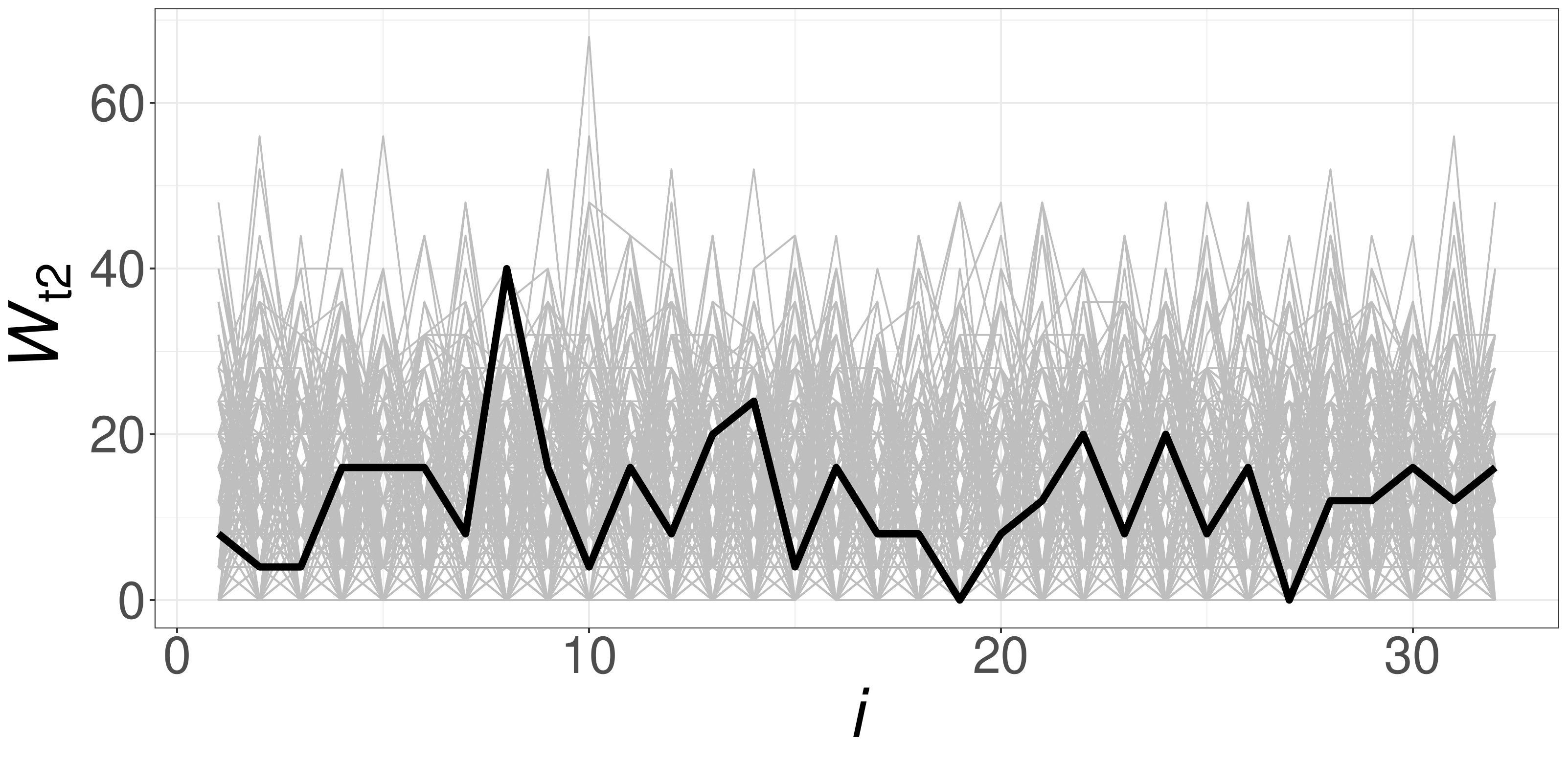}
\label{fig:mx_output_wfi_w6}}\\
\subfloat[$i'$ = 7.]{ \includegraphics[width=0.5\linewidth]{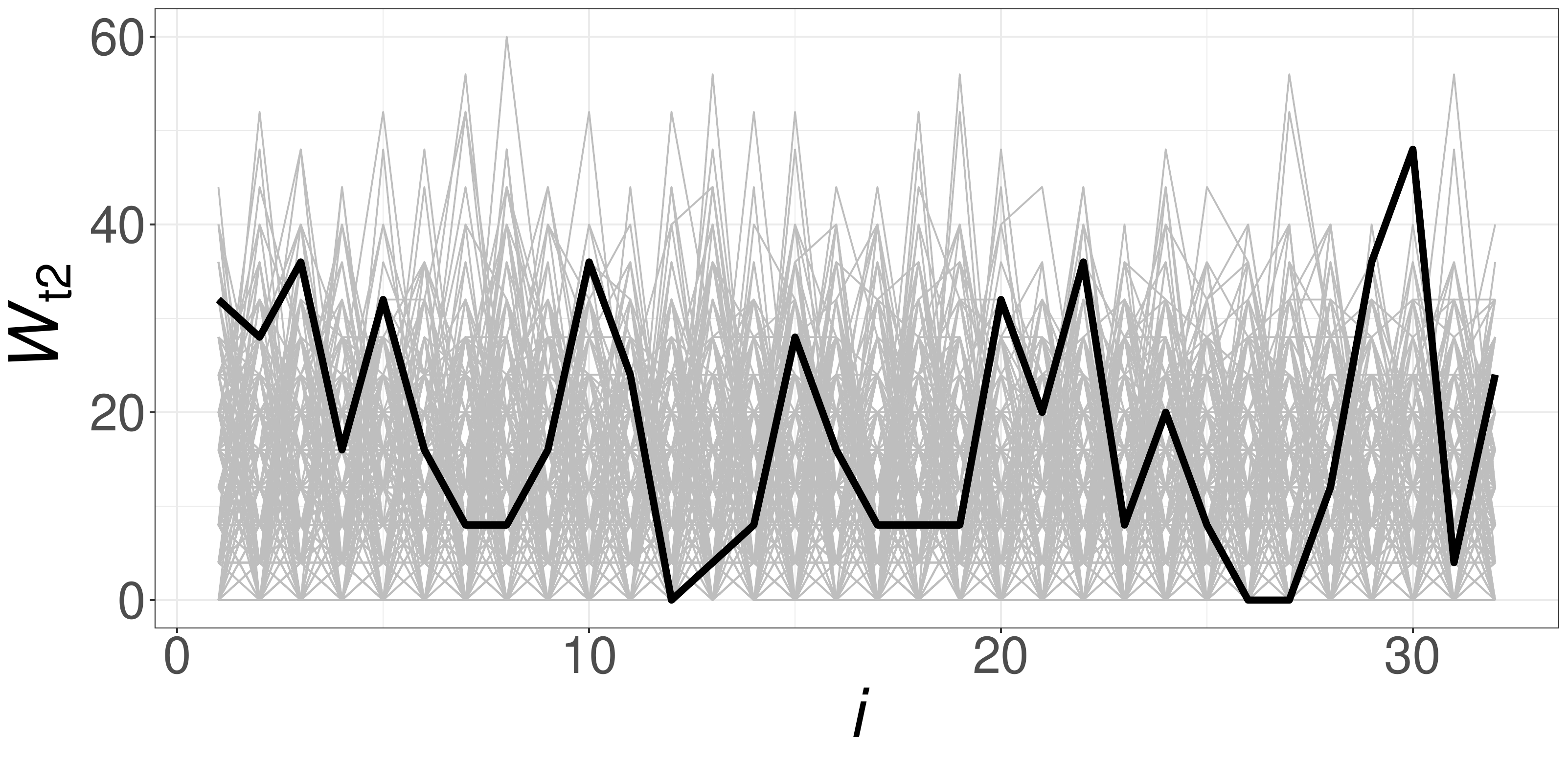}
\label{fig:mx_output_wfi_w7}}
\subfloat[$i'$ = 8.]{\includegraphics[width=0.5\linewidth]{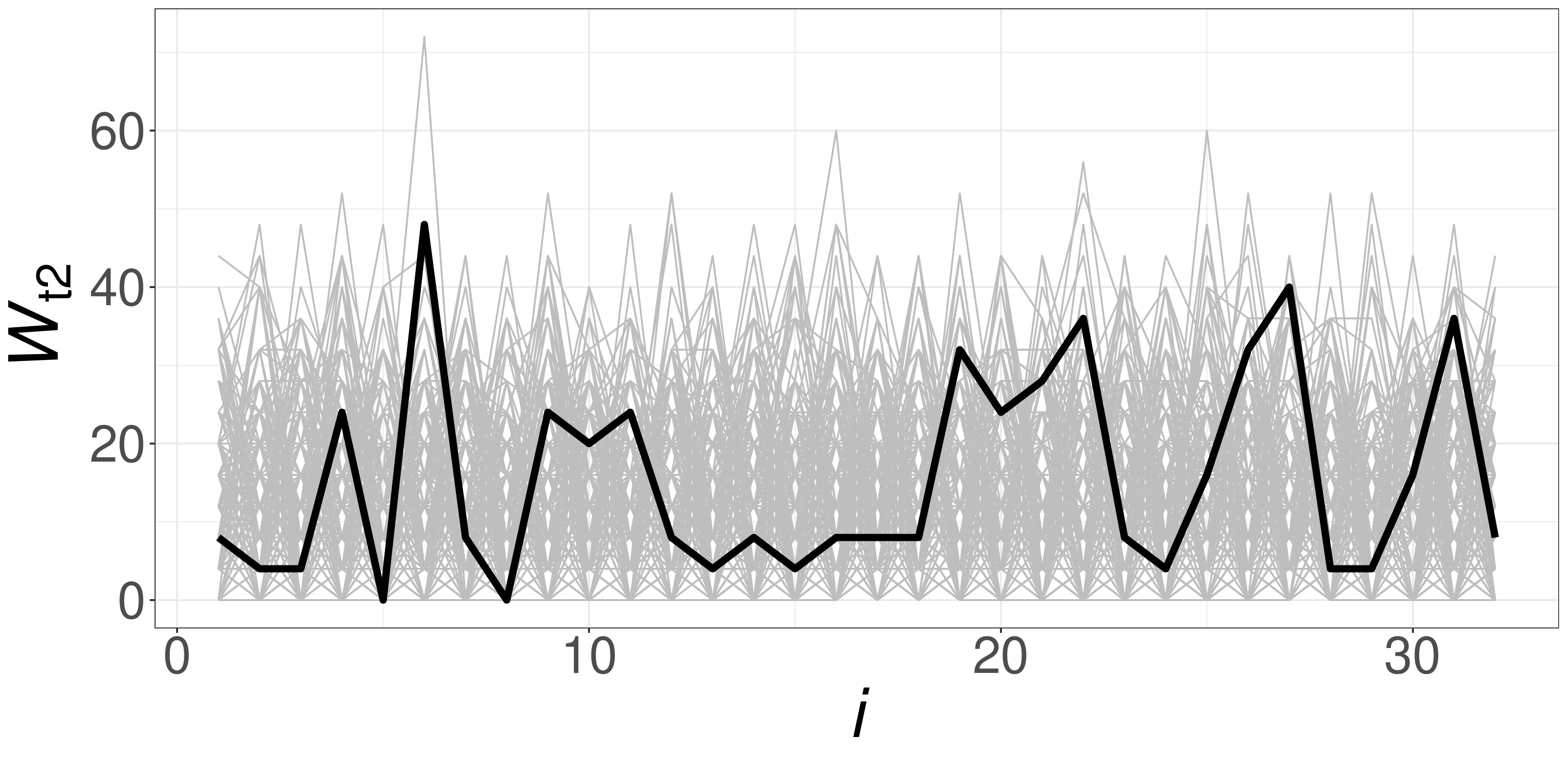}
\label{fig:mx_output_wfi_w8}}
\caption{The Walsh transforms on the $UT^{1}_{0,0}$ outputs obtained by $Q_0$ and $Q_1$  and $\textbf{S}^{2}$ in the first round. Black: correct key; gray: wrong key.}
\label{fig:Appendix_Walsh_UT_Q0_Q1_S2}
\end{figure*}

\begin{figure*}
\centering
\subfloat[$i'$ = 1.]{ \includegraphics[width=0.5\linewidth]{fig/Wt3_Q0_Q1_bit1.pdf}
\label{fig:mx_output_wfi_w1}}
\subfloat[$i'$ = 2.]{ \includegraphics[width=0.5\linewidth]{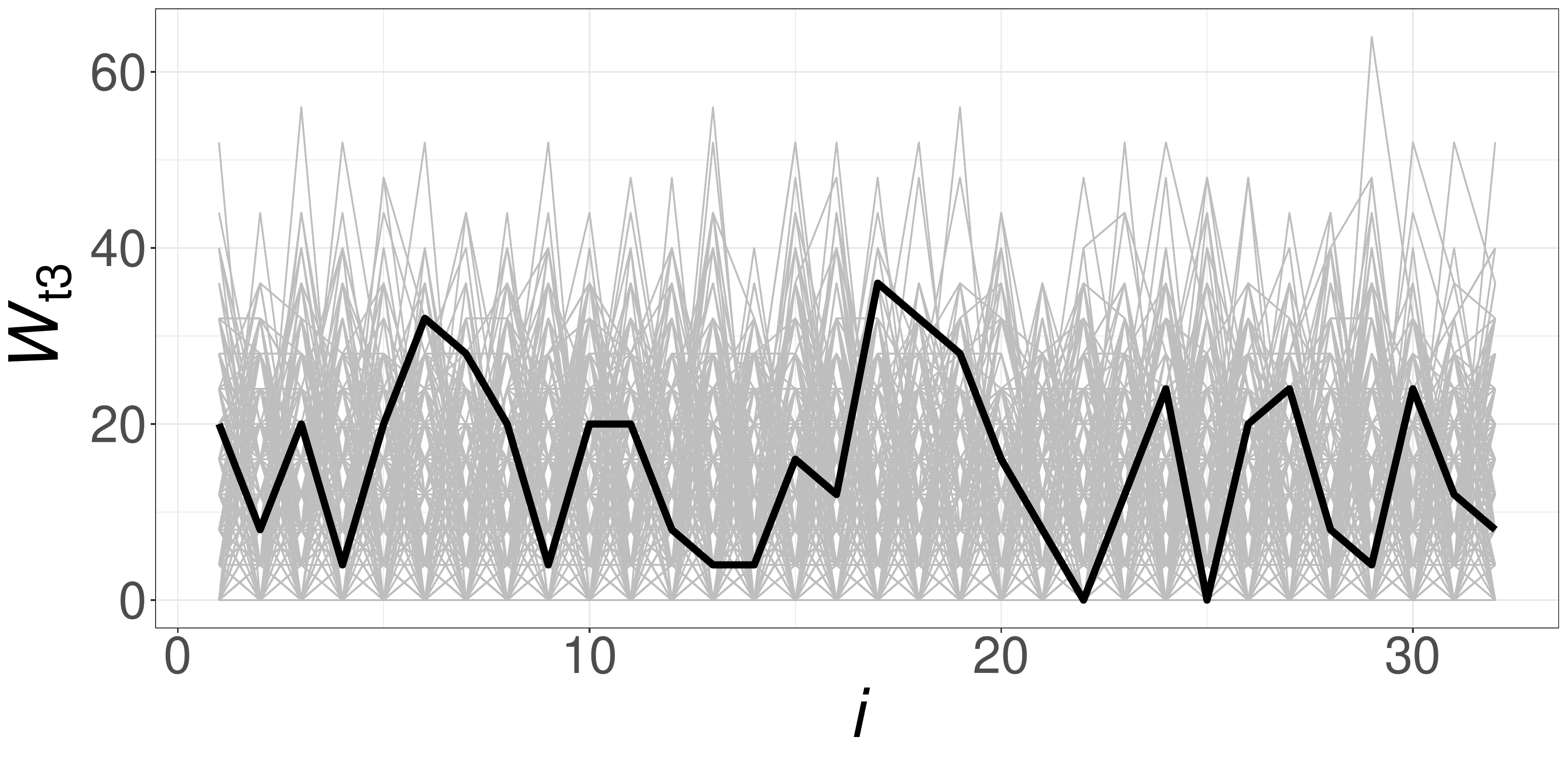}
\label{fig:mx_output_wfi_w2}}\\
\subfloat[$i'$ = 3.]{ \includegraphics[width=0.5\linewidth]{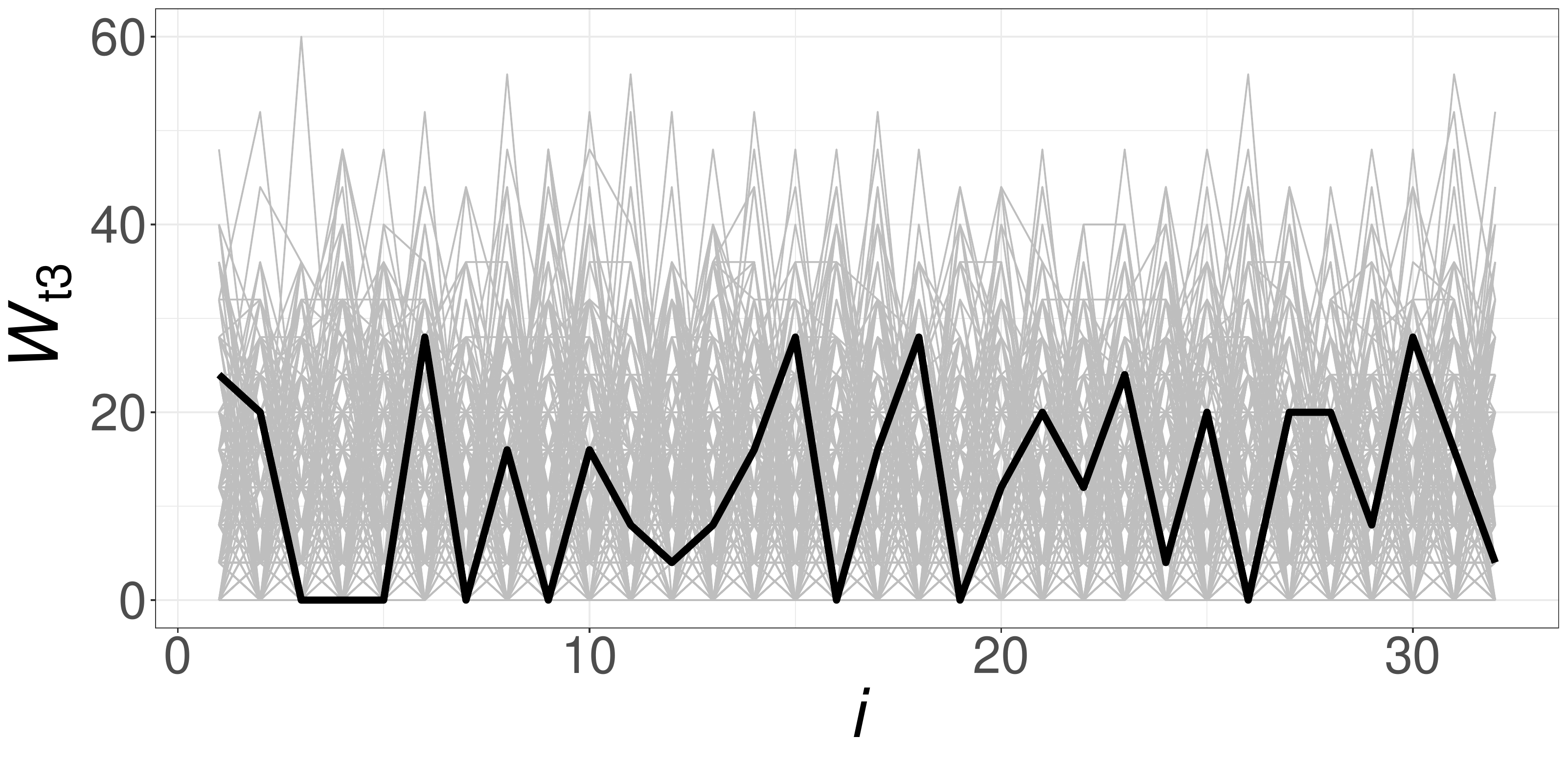}
\label{fig:mx_output_wfi_w3}}
\subfloat[$i'$ = 4.]{ \includegraphics[width=0.5\linewidth]{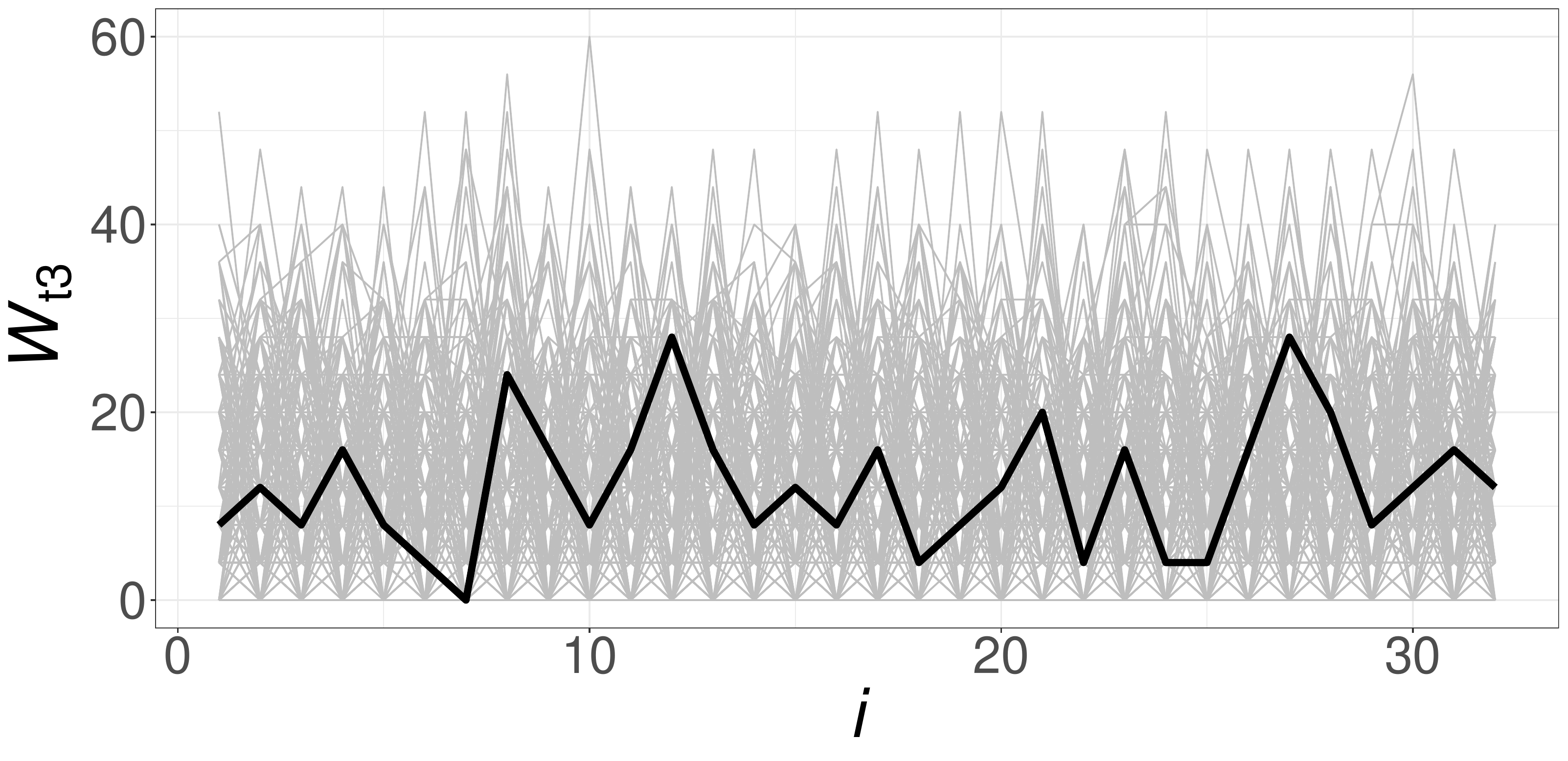}
\label{fig:mx_output_wfi_w4}}\\
\subfloat[$i'$ = 5.]{\includegraphics[width=0.5\linewidth]{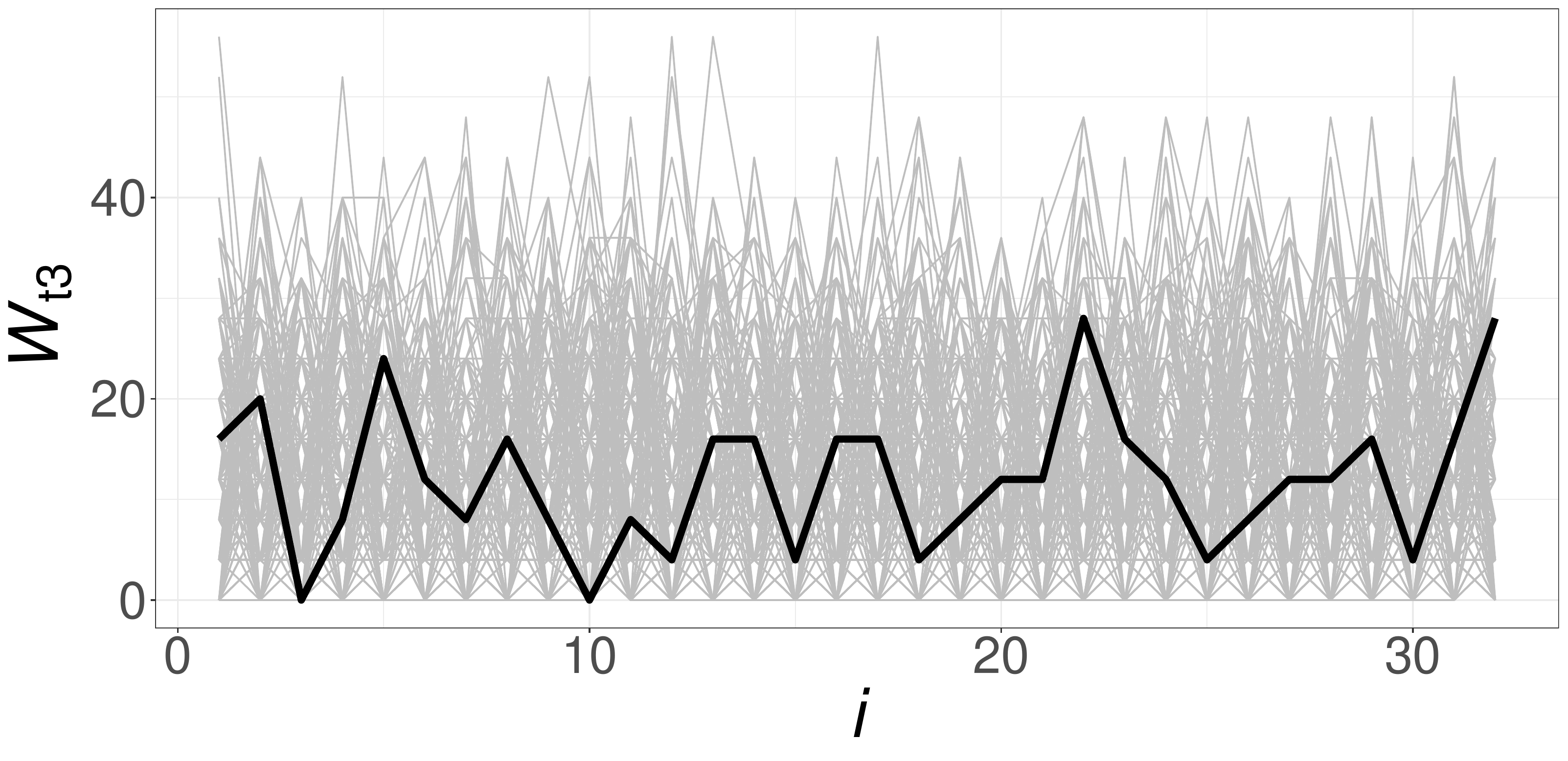}
\label{fig:mx_output_wfi_w5}}
\subfloat[$i'$ = 6.]{ \includegraphics[width=0.5\linewidth]{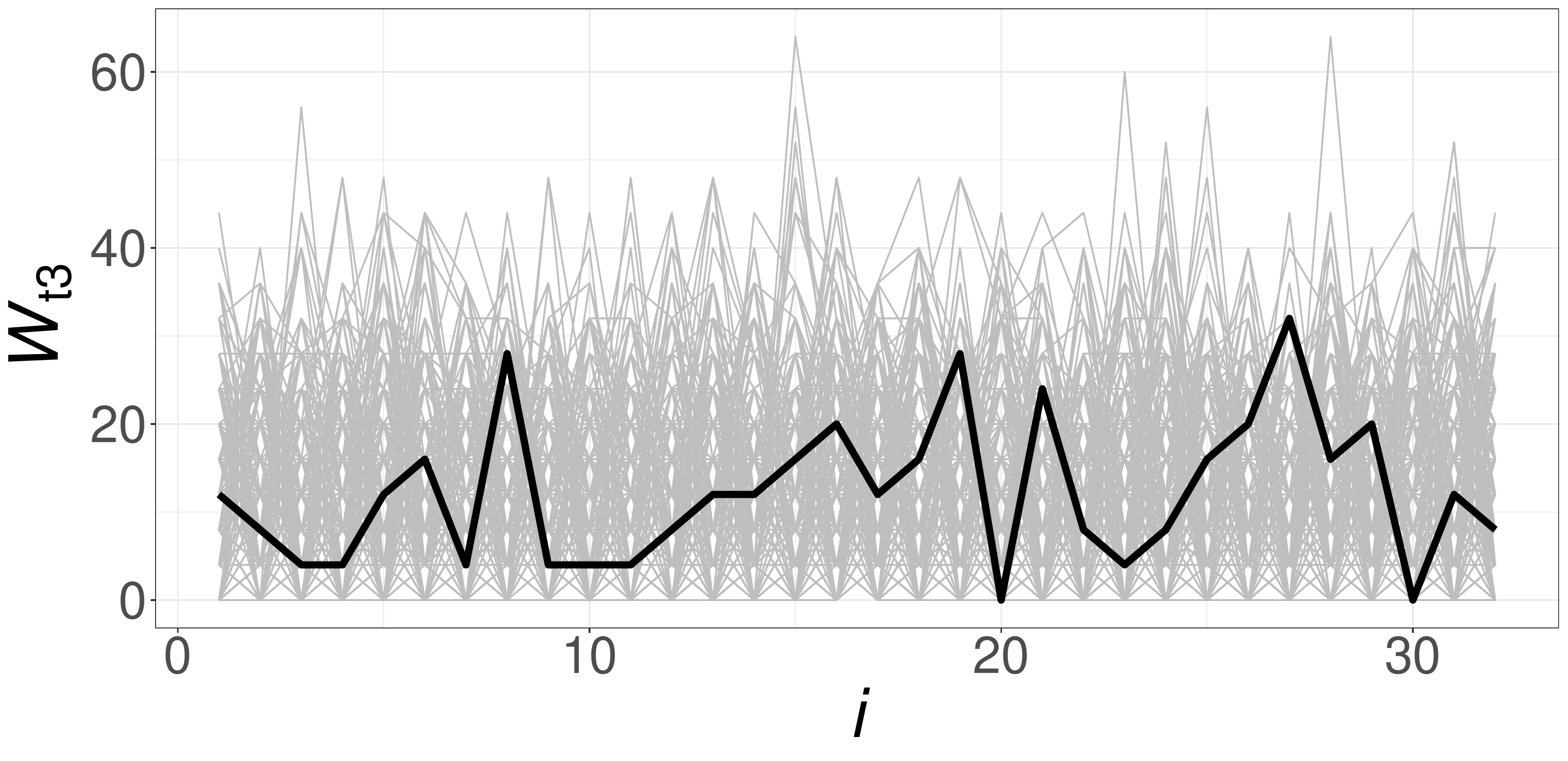}
\label{fig:mx_output_wfi_w6}}\\
\subfloat[$i'$ = 7.]{ \includegraphics[width=0.5\linewidth]{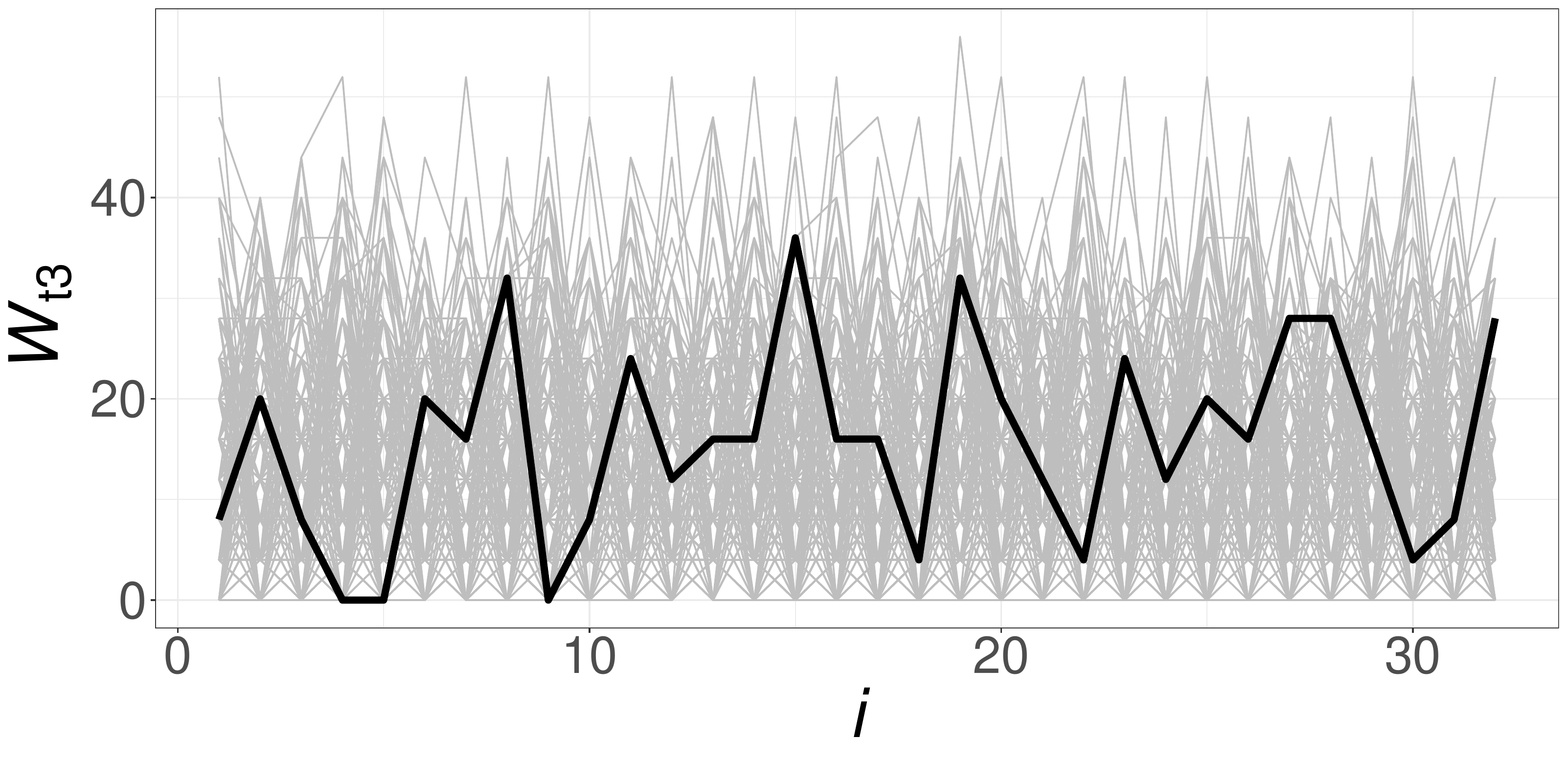}
\label{fig:mx_output_wfi_w7}}
\subfloat[$i'$ = 8.]{\includegraphics[width=0.5\linewidth]{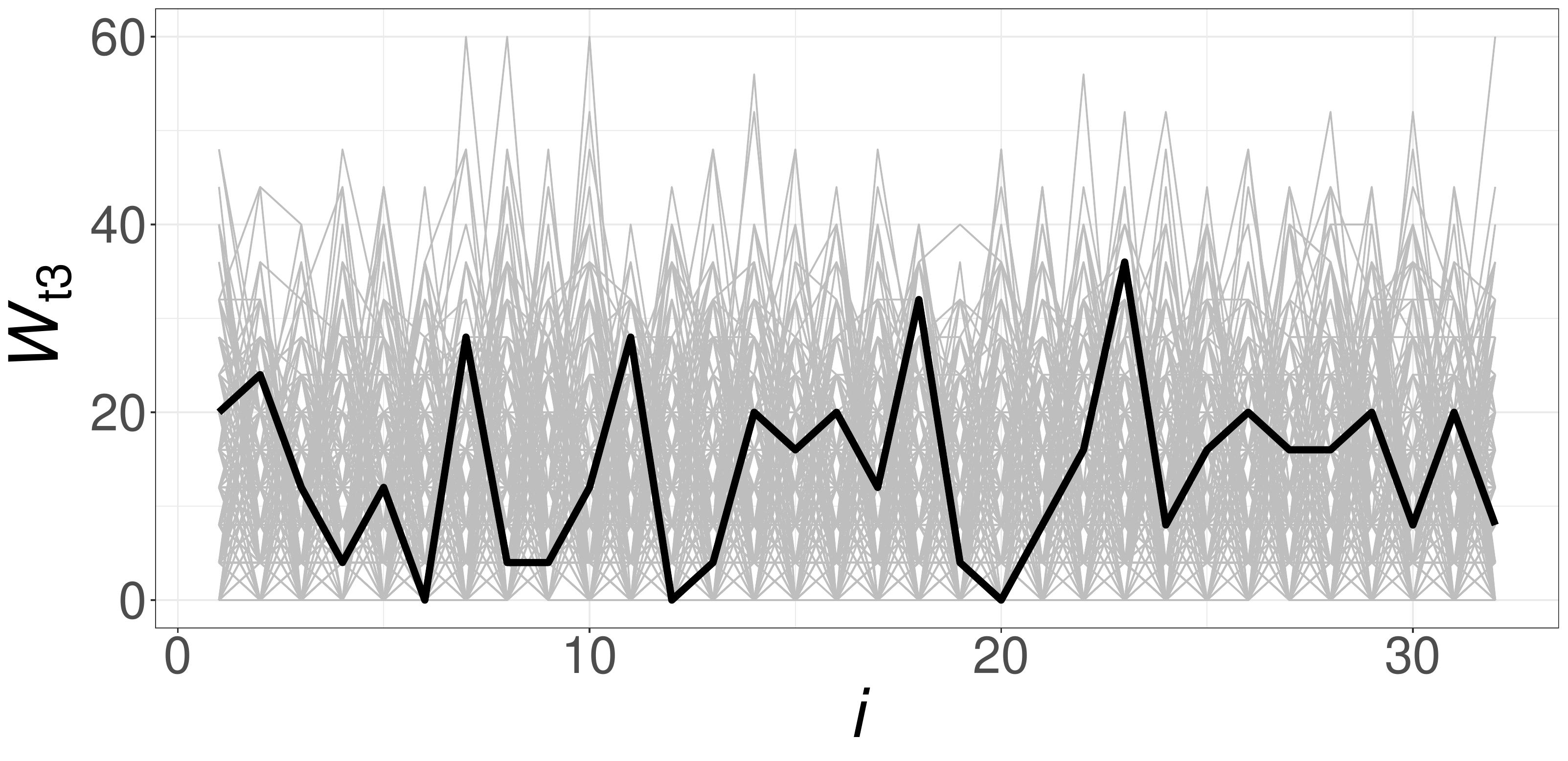}
\label{fig:mx_output_wfi_w8}}
\caption{The Walsh transforms on the $UT^{1}_{0,0}$ outputs obtained by $Q_0$ and $Q_1$  and $\textbf{S}^{3}$ in the first round. Black: correct key; gray: wrong key.}
\label{fig:Appendix_Walsh_UT_Q0_Q1_S3}
\end{figure*}

\begin{figure*}
\centering
\subfloat[$i'$ = 1.]{ \includegraphics[width=0.5\linewidth]{fig/Wer_Q0_Q1_bit1.pdf}
\label{fig:mx_output_wfi_w1}}
\subfloat[$i'$ = 2.]{ \includegraphics[width=0.5\linewidth]{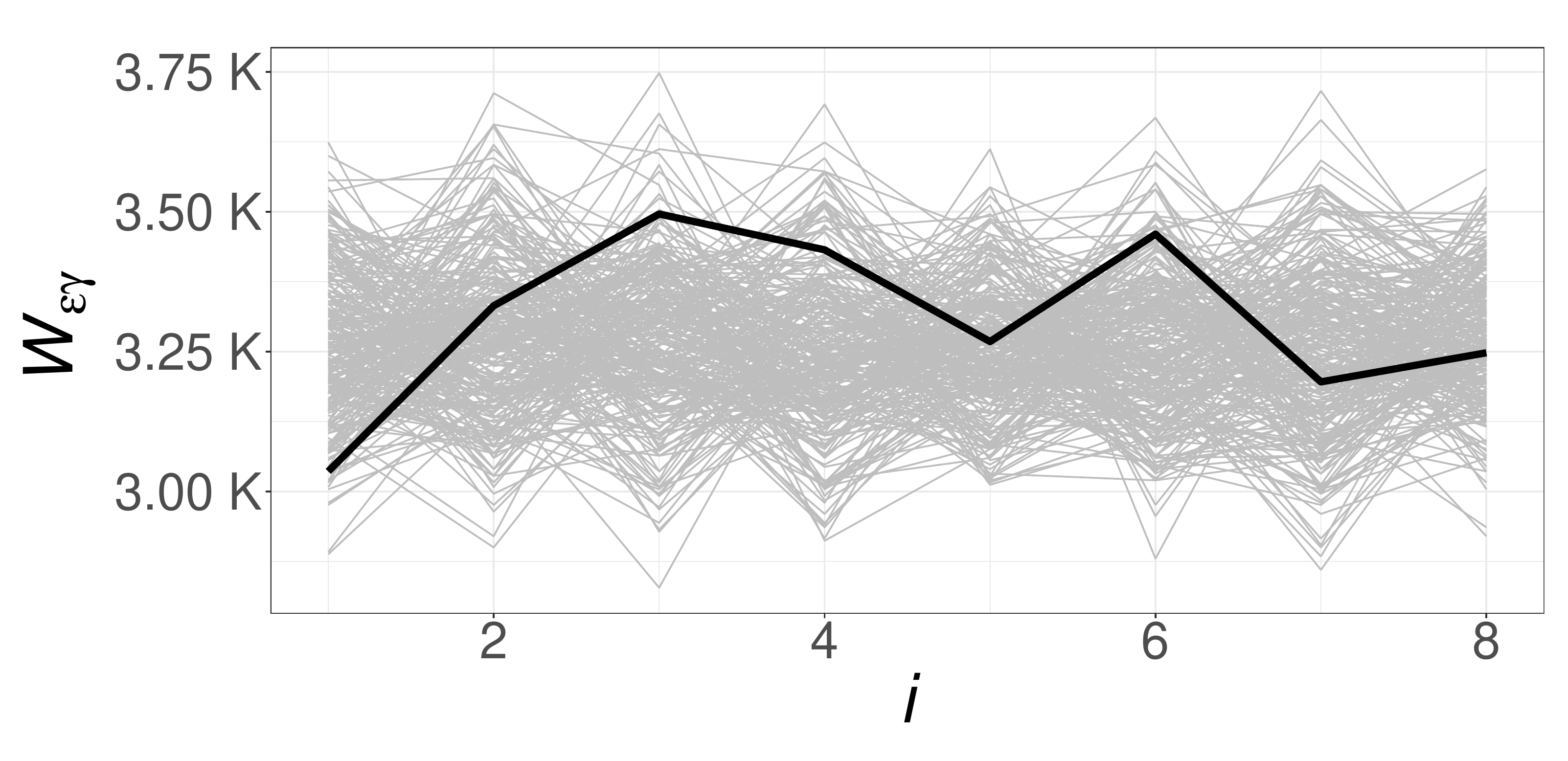}
\label{fig:mx_output_wfi_w2}}\\
\subfloat[$i'$ = 3.]{ \includegraphics[width=0.5\linewidth]{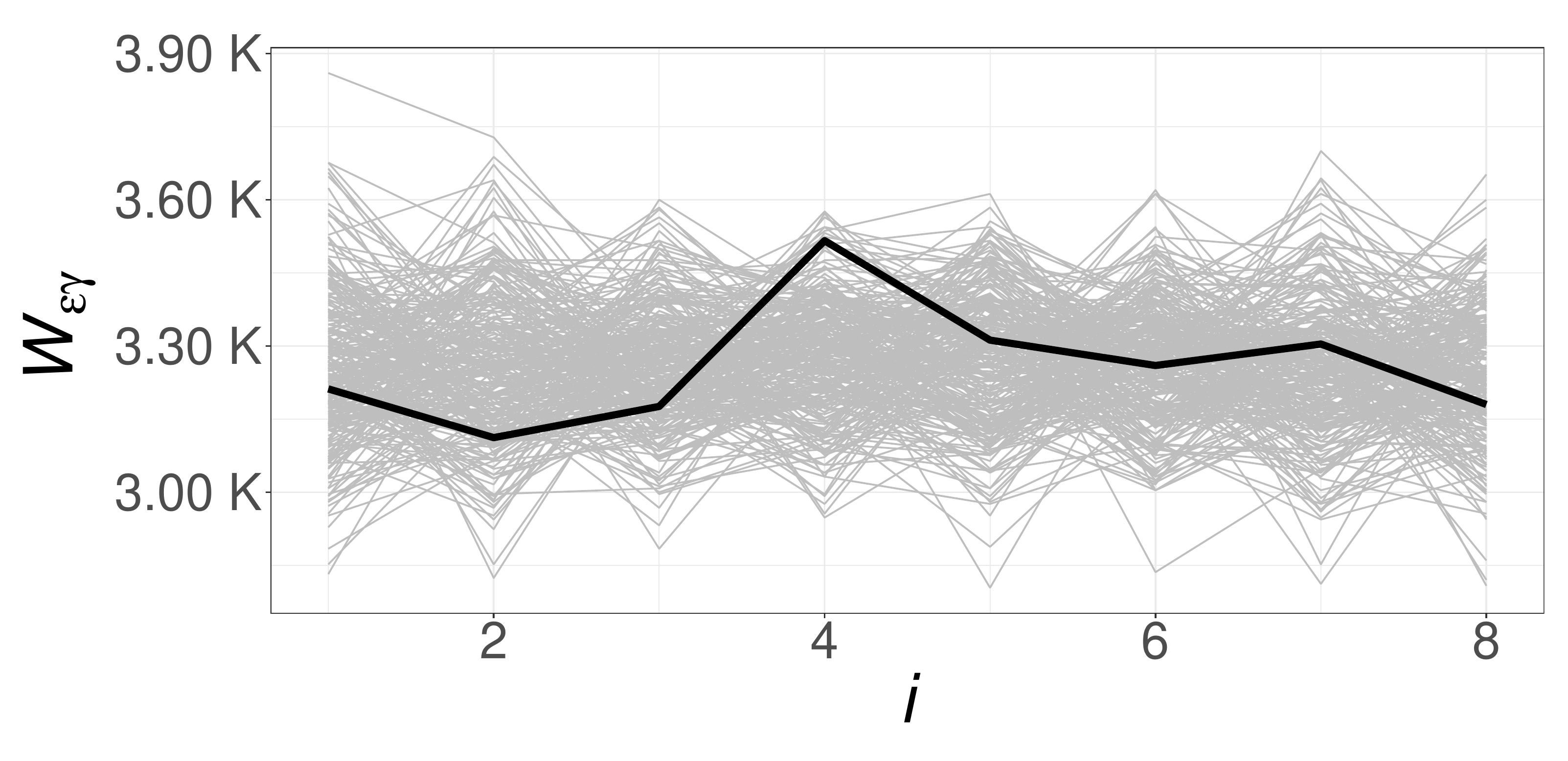}
\label{fig:mx_output_wfi_w3}}
\subfloat[$i'$ = 4.]{ \includegraphics[width=0.5\linewidth]{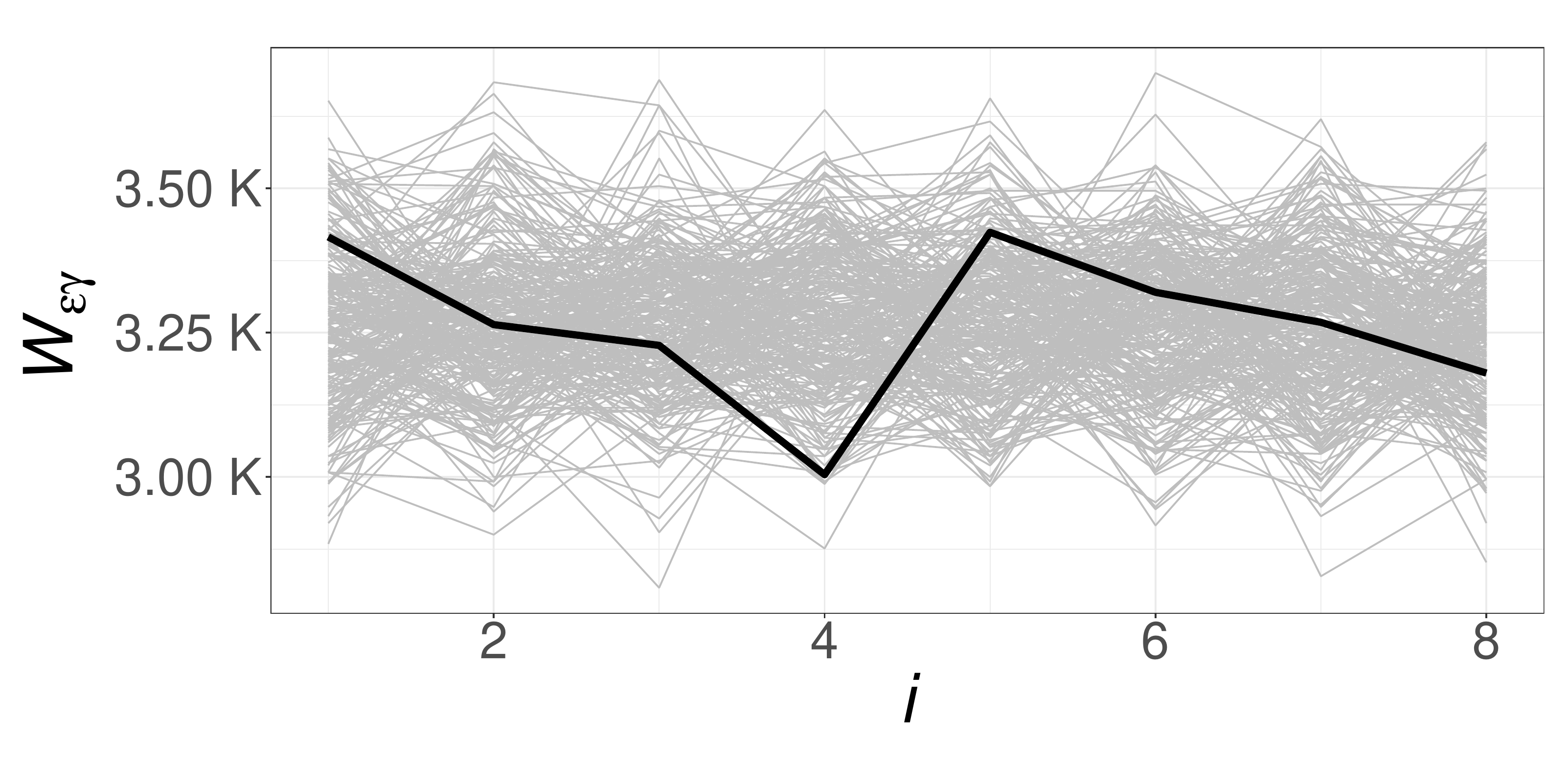}
\label{fig:mx_output_wfi_w4}}\\
\subfloat[$i'$ = 5.]{\includegraphics[width=0.5\linewidth]{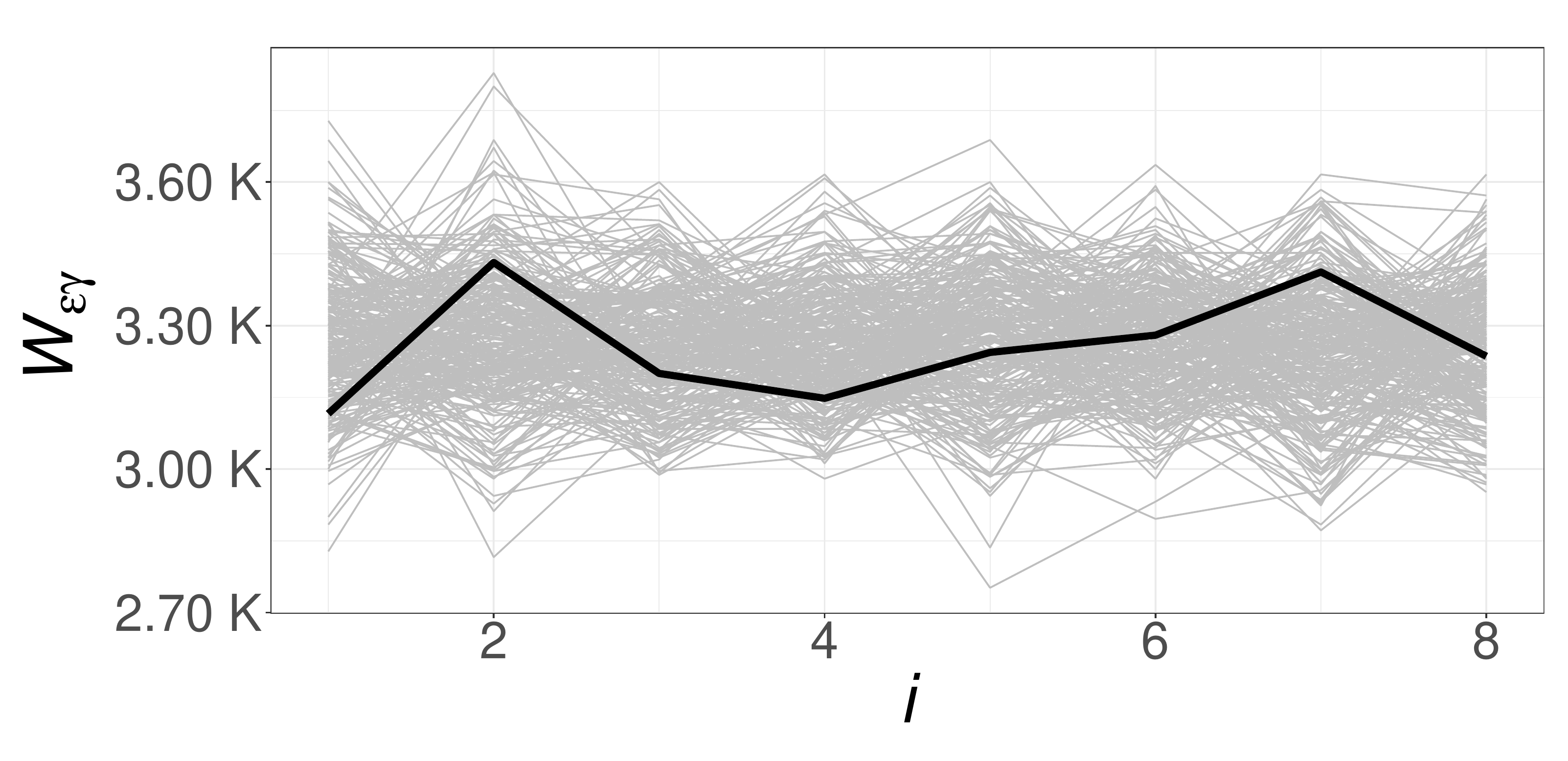}
\label{fig:mx_output_wfi_w5}}
\subfloat[$i'$ = 6.]{ \includegraphics[width=0.5\linewidth]{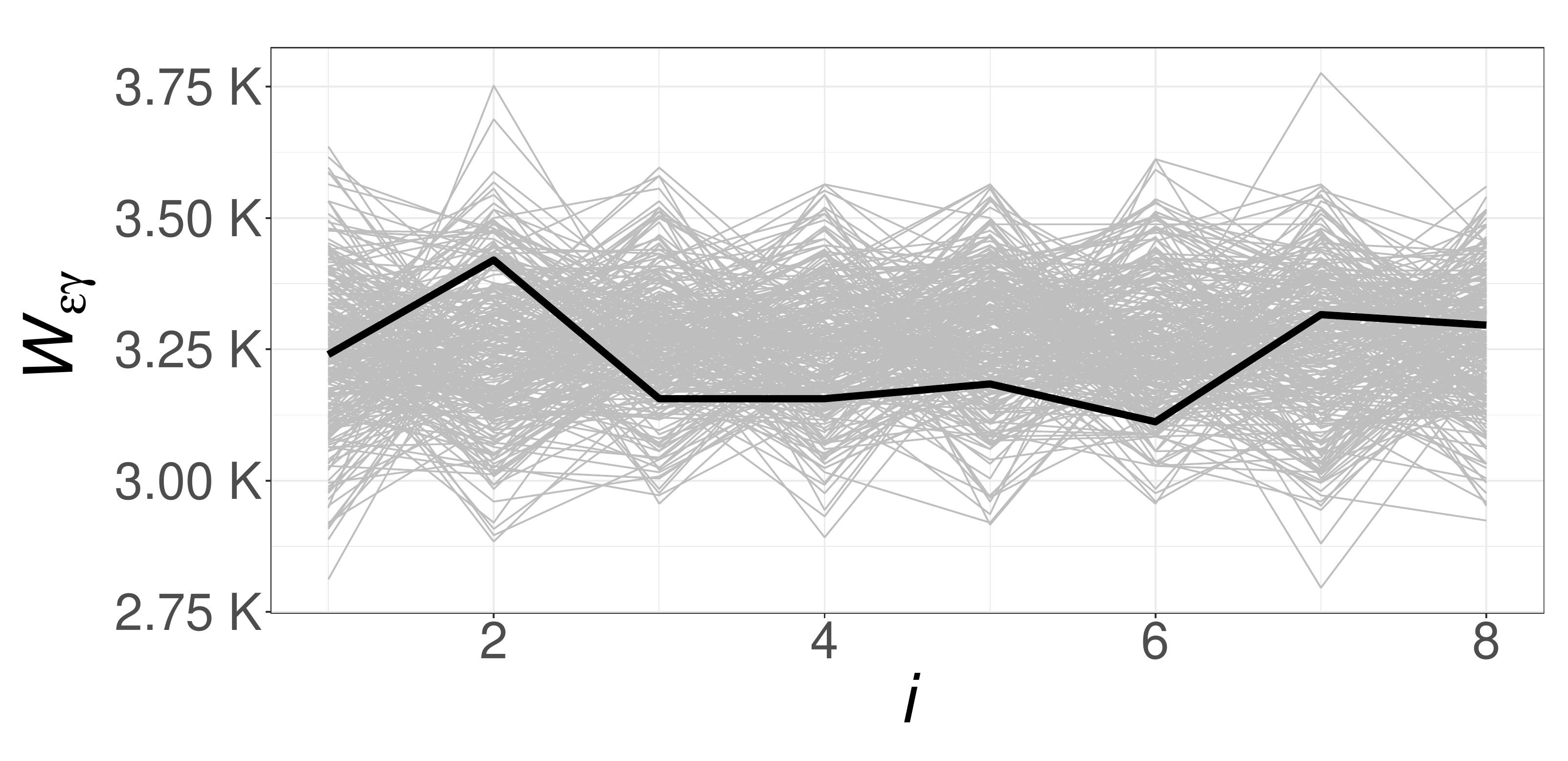}
\label{fig:mx_output_wfi_w6}}\\
\subfloat[$i'$ = 7.]{ \includegraphics[width=0.5\linewidth]{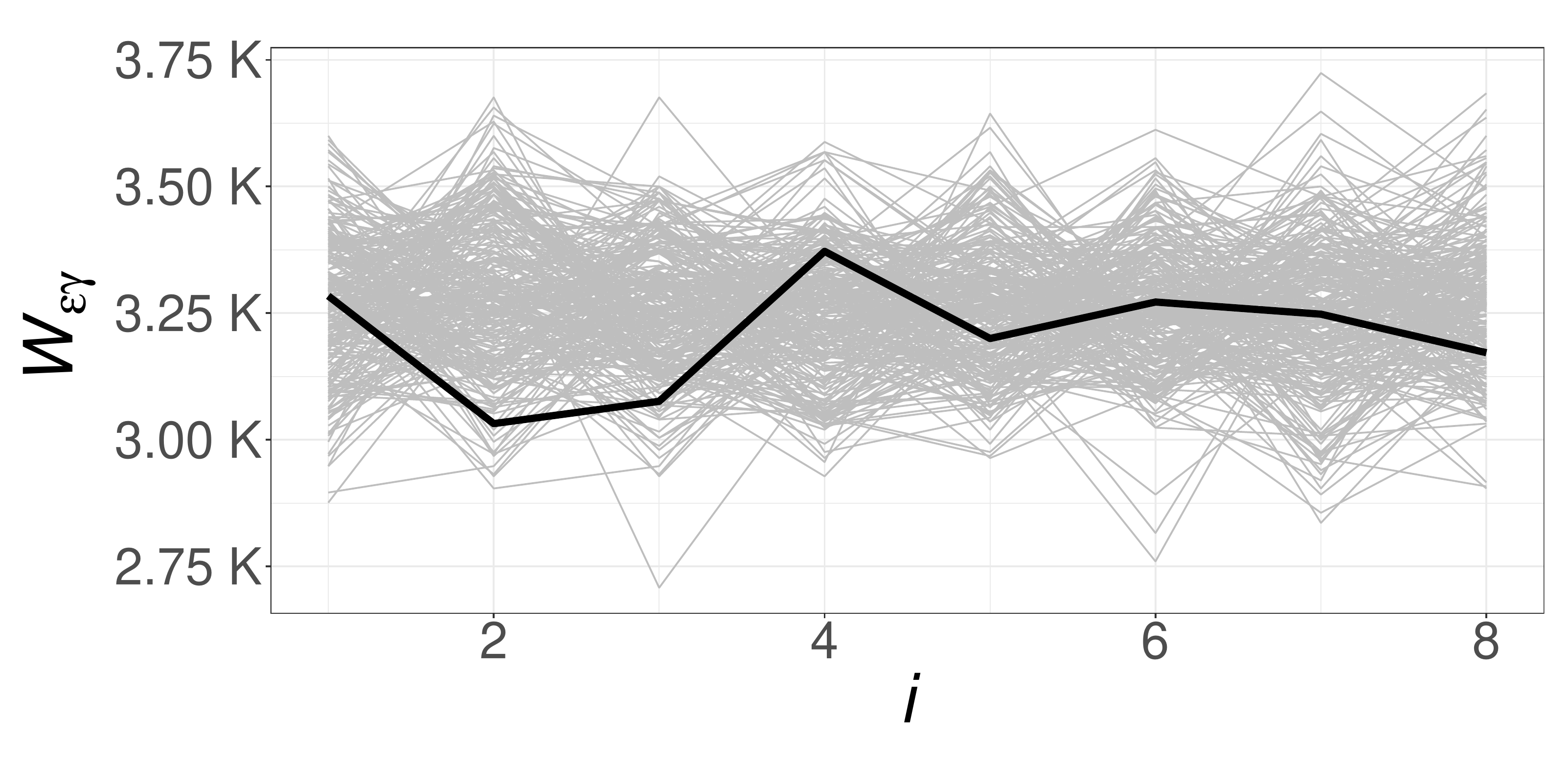}
\label{fig:mx_output_wfi_w7}}
\subfloat[$i'$ = 8.]{\includegraphics[width=0.5\linewidth]{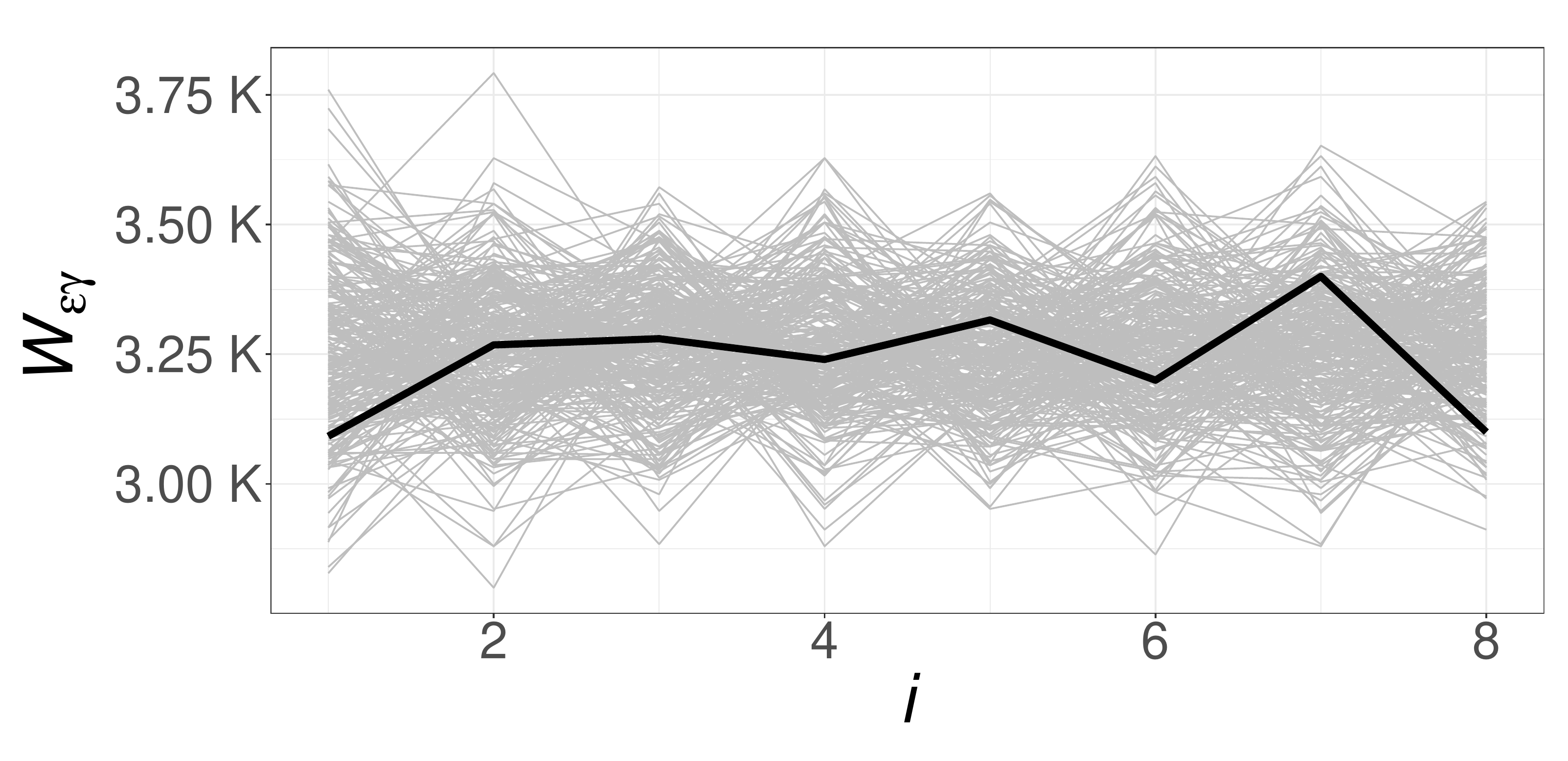}
\label{fig:mx_output_wfi_w8}}
\caption{The Walsh transforms on the round outputs obtained by $Q_0$ and $Q_1$ and the hypothetical round outputs in the first round. Black: correct key; gray: wrong key.}
\label{fig:Appendix_Walsh_RO_Q0_Q1}
\end{figure*}

All of the Walsh transform results explained in Section~\ref{sec:experiment} are provided. 
First, Fig.~\ref{fig:Appendix_Walsh_UT_Q0_S1} - Fig.~\ref{fig:Appendix_Walsh_UT_Q0_S3} show $W_{t1}$ - $W_{t3}$, respectively, using only $Q_0$.
Next, Fig.~\ref{fig:Appendix_Walsh_RO_Q0} depicts $W_{\epsilon\gamma}$.
Every Walsh transform with respect to the correct subkeys and $Q_0$ results in the value of 0.   
In contrary, Fig.~\ref{fig:Appendix_Walsh_UT_Q0_Q1_S1} - Fig.~\ref{fig:Appendix_Walsh_UT_Q0_Q1_S3} illustrate $W_{t1}$ - $W_{t3}$, respectively, with an increase in the encoding imbalance when using $Q_0$ and $Q_1$.
Fig.~\ref{fig:Appendix_Walsh_RO_Q0_Q1} also demonstrate it with respect to $W_{\epsilon\gamma}$.

\section{The experimental results of the DCA attacks on the UT outputs}
\label{sec:appendix_only_ut}

Table~\ref{tab:Q0_UT_DCA} and Table~\ref{tab:Q0_UT_DCA_coefficients} in Section~\ref{sec:experiment} show the results of DCA using the computational traces that contain every computation in the encryption. 
For this reason, the DCA ranking of the correct subkey was not always the lowest.
To show the effectiveness of our balanced encoding in more detail, we would like to provide DCA results using the 10,000 computational traces consisting of only the UT output values.
This will help us understand more accurately how the Walsh transform score 0 for the UT output appears in the correlation coefficients.
As shown in Table~\ref{tab:Appendix_Q0_UT_only_DCA}, all correct subkeys were ranked lowest in all cases as a result of the DCA attacks.
In this regard, it can be seen that the correlation coefficients calculated by the correct subkeys shown in Table~\ref{tab:Appendix_Q0_UT_only_DCA_coefficients} is much lower than the coefficients provided in Table~\ref{tab:Q0_UT_DCA_coefficients}.

\begin{table*}
\centering
\scriptsize
\caption{DCA ranking on the AES encryption using only $Q_0$ when conducting mono-bit CPA on the SubBytes output with 10,000 computational traces consisting of only the UT outputs.}
\label{tab:Appendix_Q0_UT_only_DCA}
\begin{tabular}{c|*{16}{{c}@{\hskip 2mm}}}
\\
\backslashbox{TargetBit}{Subkey}   &       1   &       2   &    3   &    4   &   5    &    6   &      7   &    8   &   9   &  10   & 11   & 12   & 13  & 14  & 15  & 16  \\\hline
1 & 256 & 256 & 256 & 256 & 256 & 256 & 256 & 256 & 256 & 256 & 256 & 256 & 256 & 256 & 256 & 256 \\
2 & 256 & 256 & 256 & 256 & 256 & 256 & 256 & 256 & 256 & 256 & 256 & 256 & 256 & 256 & 256	& 256 \\
3 & 256 & 256 & 256 & 256 & 256 & 256 & 256 & 256 & 256 & 256 & 256 & 256 & 256 & 256 & 256 & 256 \\
4 & 256 & 256 & 256 & 256 & 256 & 256 & 256 & 256 & 256 & 256 & 256 & 256 & 256 & 256 & 256 & 256 \\
5 & 256 & 256 & 256 & 256 & 256 & 256 & 256 & 256 & 256 & 256 & 256 & 256 & 256 & 256 & 256 & 256 \\
6 & 256 & 256 & 256 & 256 & 256 & 256 & 256 & 256 & 256 & 256 & 256 & 256 & 256 & 256 & 256 & 256 \\
7 & 256 & 256 & 256 & 256 & 256 & 256 & 256 & 256 & 256 & 256 & 256 & 256 & 256 & 256 & 256 & 256 \\
8 & 256 & 256 & 256 & 256 & 256 & 256 & 256 & 256 & 256 & 256 & 256 & 256 & 256 & 256 & 256 & 256 
\normalsize
\end{tabular}
\end{table*}

\begin{table*}
\centering
\scriptsize
\caption{Highest correlation of all key candidates vs. highest correlation of the correct subkeys when conducting the DCA attacks above. }
\label{tab:Appendix_Q0_UT_only_DCA_coefficients}
\begin{tabular}{c|*{16}{{c}@{\hskip 1mm}}}
\\
Subkey    &       1   &       2   &    3   &    4   &   5    &    6   &      7   &    8   &   9   &  10   & 11   & 12   & 13  & 14  & 15  & 16  \\\hline
Highest   &  0.156	& 0.154 & 0.157 & 0.155 & 0.153 & 0.161 & 0.162 & 0.158 & 0.161 & 0.158 & 0.157 & 0.158	& 0.152  & 0.155 & 0.153 & 0.159	 		 \\
Key        &  0.030	& 0.024 & 0.024 & 0.025 & 0.028 & 0.029 & 0.038 & 0.034 & 0.029 & 0.030 & 0.021 & 0.026	& 0.032  & 0.028 & 0.028 & 0.033 		 
\normalsize
\end{tabular}
\end{table*}

\section{Correlation coefficients and the number of traces}
\label{sec:appendix_coff_tendency}

Table~\ref{tab:Appendix_DCA_ranking_2500} - Table~\ref{tab:Appendix_DCA_ranking_7500} show the DCA ranking results obtained using $Q_0$ or $Q_1$ with a 1/2 probability during mono-bit CPA on the SubBytes output with 2,500, 5,000, and 7,500 computational traces, respectively. Additionally, Table~\ref{tab:Appendix_DCA_coefficients_2500} - Table~\ref{tab:Appendix_DCA_coefficients_7500} illustrate the correlation coefficient trends as the number of computational traces increases. It can be observed that the correlation coefficients computed by incorrect key candidates and the correct key candidate decrease as the number of computational traces increases.

\begin{table*}
\centering
\scriptsize
\caption{DCA ranking on the AES encryption using $Q_0$ or $Q_1$ with a 1/2 probability when conducting mono-bit CPA on the SubBytes output with 2,500 computational traces.}
\label{tab:Appendix_DCA_ranking_2500}
\begin{tabular}{c|*{16}{{c}@{\hskip 2mm}}}
\\
\backslashbox{TargetBit}{Subkey}   &       1   &       2   &    3   &    4   &   5    &    6   &      7   &    8   &   9   &  10   & 11   & 12   & 13  & 14  & 15  & 16  \\\hline
1&80  & 176 & 150 & 127 & 37  & 34  & 139 & 242 & 152 & 196 & 202  & 109 & 56  & 169 & 122 & 126 \\
2&26  & 148 & 159 & 13  & 156 & 144 & 161 & 237 & 164 & 135 & 59    & 210 & 150 & 89  & 89  & 167 \\
3&179 & 212 & 188 & 57  & 46  & 145 & 79  & 245 & 217 & 134 & 103  & 143 & 171 & 47  & 105 & 72  \\
4&157 & 178 & 172 & 115 & 151 & 72  & 216 & 49  & 117 & 178 & 193  & 148 & 22  & 210 & 132 & 119 \\
5&33  & 177 & 186 & 21  & 253 & 176 & 53  & 129 & 253 & 4   & 68    & 5   & 85  & 103 & 117 & 126 \\
6&183 & 153 & 92  & 52  & 55  & 38  & 178 & 85  & 80  & 216 & 205  & 63  & 161 & 254 & 121 & 12  \\
7&183 & 174 & 2   & 34  & 65  & 222 & 65  & 211 & 111 & 36  & 58    & 256 & 51  & 149 & 111 & 160 \\
8&208 & 212 & 203 & 113 & 26  & 146 & 138 & 140 & 185 & 202 & 30    & 146 & 80  & 179 & 127 & 77 
\normalsize
\end{tabular}
\end{table*}

\begin{table*}
\centering
\scriptsize
\caption{Highest correlation of all key candidates vs. highest correlation of the correct subkeys when conducting the DCA attacks with 2,500 computational traces on the UT output obtained by $Q_0$ or $Q_1$ with a 1/2 probability. }
\label{tab:Appendix_DCA_coefficients_2500}
\begin{tabular}{c|*{16}{{c}@{\hskip 1mm}}}
\\
Subkey    &       1   &       2   &    3   &    4   &   5    &    6   &      7   &    8   &   9   &  10   & 11   & 12   & 13  & 14  & 15  & 16  \\\hline
Highest   &  0.098	& 0.079 & 0.086 & 0.087 & 0.078 & 0.084  & 0.089 & 0.086 & 0.086 & 0.104 & 0.081 & 0.086	& 0.081  & 0.083 & 0.081 & 0.082	 		 \\
Key        &  0.058	& 0.043 & 0.076 & 0.058 & 0.060 & 0.055 & 0.059 & 0.054 & 0.048 & 0.069 & 0.057 & 0.067	& 0.057  & 0.056 & 0.050 & 0.060		 
\normalsize
\end{tabular}
\end{table*}

\begin{table*}
\centering
\scriptsize
\caption{DCA ranking on the AES encryption using $Q_0$ or $Q_1$ with a 1/2 probability when conducting mono-bit CPA on the SubBytes output with 5,000 computational traces.}
\label{tab:Appendix_DCA_ranking_5000}
\begin{tabular}{c|*{16}{{c}@{\hskip 2mm}}}
\\
\backslashbox{TargetBit}{Subkey}   &       1   &       2   &    3   &    4   &   5    &    6   &      7   &    8   &   9   &  10   & 11   & 12   & 13  & 14  & 15  & 16  \\\hline
1 & 245 & 80  & 131 & 104 & 178 & 8   & 148 & 220 & 171 & 167 & 112 & 208 & 89  & 118 & 100 & 209 \\
2 & 248 & 230 & 189 & 75  & 159 & 113 & 97  & 83  & 210 & 235 & 206 & 7   & 104 & 127 & 189 & 237 \\
3 & 36  & 136 & 7   & 75  & 146 & 157 & 184 & 119 & 210 & 119 & 181 & 188 & 68  & 84  & 116 & 164 \\
4 & 11  & 88  & 197 & 99  & 84  & 93  & 108 & 189 & 218 & 122 & 218 & 5   & 158 & 105 & 191 & 11  \\
5 & 36  & 233 & 136 & 207 & 191 & 170 & 218 & 172 & 23  & 214 & 124 & 98  & 221 & 208 & 192 & 93  \\
6 & 106 & 79  & 213 & 96  & 164 & 8   & 54  & 121 & 222 & 190 & 123 & 252 & 95  & 238 & 36  & 120 \\
7 & 42  & 136 & 53  & 153 & 115 & 3   & 99  & 39  & 179 & 159 & 7   & 12  & 207 & 147 & 96  & 66  \\
8 & 43  & 80  & 207 & 112 & 111 & 59  & 164 & 137 & 193 & 115 & 118 & 70  & 58  & 180 & 78  & 108
\normalsize
\end{tabular}
\end{table*}

\begin{table*}
\centering
\scriptsize
\caption{Highest correlation of all key candidates vs. highest correlation of the correct subkeys when conducting the DCA attacks with 5,000 computational traces on the UT output obtained by $Q_0$ or $Q_1$ with a 1/2 probability. }
\label{tab:Appendix_DCA_coefficients_5000}
\begin{tabular}{c|*{16}{{c}@{\hskip 1mm}}}
\\
Subkey    &       1   &       2   &    3   &    4   &   5    &    6   &      7   &    8   &   9   &  10   & 11   & 12   & 13  & 14  & 15  & 16  \\\hline
Highest   &  0.064	& 0.050 & 0.061 & 0.064 & 0.061 & 0.056 & 0.060 & 0.064 & 0.059 & 0.068 & 0.053 & 0.068 & 0.072	& 0.060  & 0.065 & 0.063 \\
Key        &  0.043	& 0.030 & 0.049 & 0.038 & 0.036 & 0.047 & 0.036 & 0.039 & 0.038 & 0.032 & 0.045 & 0.045	& 0.041  & 0.032 & 0.041 & 0.046 		 
\normalsize
\end{tabular}
\end{table*}

\begin{table*}
\centering
\scriptsize
\caption{DCA ranking on the AES encryption using $Q_0$ or $Q_1$ with a 1/2 probability when conducting mono-bit CPA on the SubBytes output with 7,500 computational traces.}
\label{tab:Appendix_DCA_ranking_7500}
\begin{tabular}{c|*{16}{{c}@{\hskip 2mm}}}
\\
\backslashbox{TargetBit}{Subkey}   &       1   &       2   &    3   &    4   &   5    &    6   &      7   &    8   &   9   &  10   & 11   & 12   & 13  & 14  & 15  & 16  \\\hline
1 & 35  & 152 & 79  & 4   & 224 & 112 & 21  & 134 & 122 & 43  & 213 & 148 & 243 & 170 & 75  & 206 \\
2 & 97  & 99  & 151 & 186 & 61  & 198 & 154 & 255 & 82  & 131 & 79  & 167 & 68  & 243 & 75  & 244 \\
3 & 14  & 155 & 228 & 45  & 106 & 59  & 160 & 103 & 80  & 171 & 243 & 123 & 72  & 234 & 199 & 115 \\
4 & 168 & 243 & 129 & 48  & 97  & 62  & 161 & 114 & 74  & 192 & 73  & 144 & 185 & 34  & 73  & 226 \\
5 & 252 & 54  & 35  & 190 & 123 & 154 & 23  & 101 & 137 & 205 & 253 & 223 & 98  & 77  & 195 & 7   \\
6 & 117 & 212 & 189 & 216 & 105 & 72  & 97  & 206 & 85  & 133 & 134 & 15  & 20  & 137 & 181 & 7   \\
7 & 101 & 149 & 60  & 146 & 204 & 67  & 80  & 142 & 156 & 208 & 246 & 190 & 20  & 176 & 196 & 156 \\
8 & 190 & 179 & 149 & 151 & 220 & 191 & 7   & 81  & 243 & 100 & 17  & 19  & 223 & 87  & 252 & 194
\normalsize
\end{tabular}
\end{table*}

\begin{table*}
\centering
\scriptsize
\caption{Highest correlation of all key candidates vs. highest correlation of the correct subkeys when conducting the DCA attacks with 7,500 computational traces on the UT output obtained by $Q_0$ or $Q_1$ with a 1/2 probability. }
\label{tab:Appendix_DCA_coefficients_7500}
\begin{tabular}{c|*{16}{{c}@{\hskip 1mm}}}
\\
Subkey    &       1   &       2   &    3   &    4   &   5    &    6   &      7   &    8   &   9   &  10   & 11   & 12   & 13  & 14  & 15  & 16  \\\hline
Highest   &  0.054	& 0.051 & 0.050 & 0.058 & 0.051 & 0.052 & 0.050 & 0.046 & 0.055 & 0.048 & 0.052 & 0.044	& 0.054  & 0.050 & 0.047 & 0.047	 		 \\
Key        &  0.038	& 0.031 & 0.032 & 0.044 & 0.029 & 0.029 & 0.043 & 0.024 & 0.029 & 0.030 & 0.035 & 0.033	& 0.031  & 0.031 & 0.027 & 0.038 		 
\normalsize
\end{tabular}
\end{table*}

\section{The experimental results of the MIA attacks}
\label{sec:appendix_mia}

\begin{figure*}
\centering
\subfloat[Target Bit 1.]{ \includegraphics[width=0.5\linewidth]{fig/mia/mia_bit0.png}
\label{fig:mia_ut_bit1}}
\subfloat[Target Bit 2.]{ \includegraphics[width=0.5\linewidth]{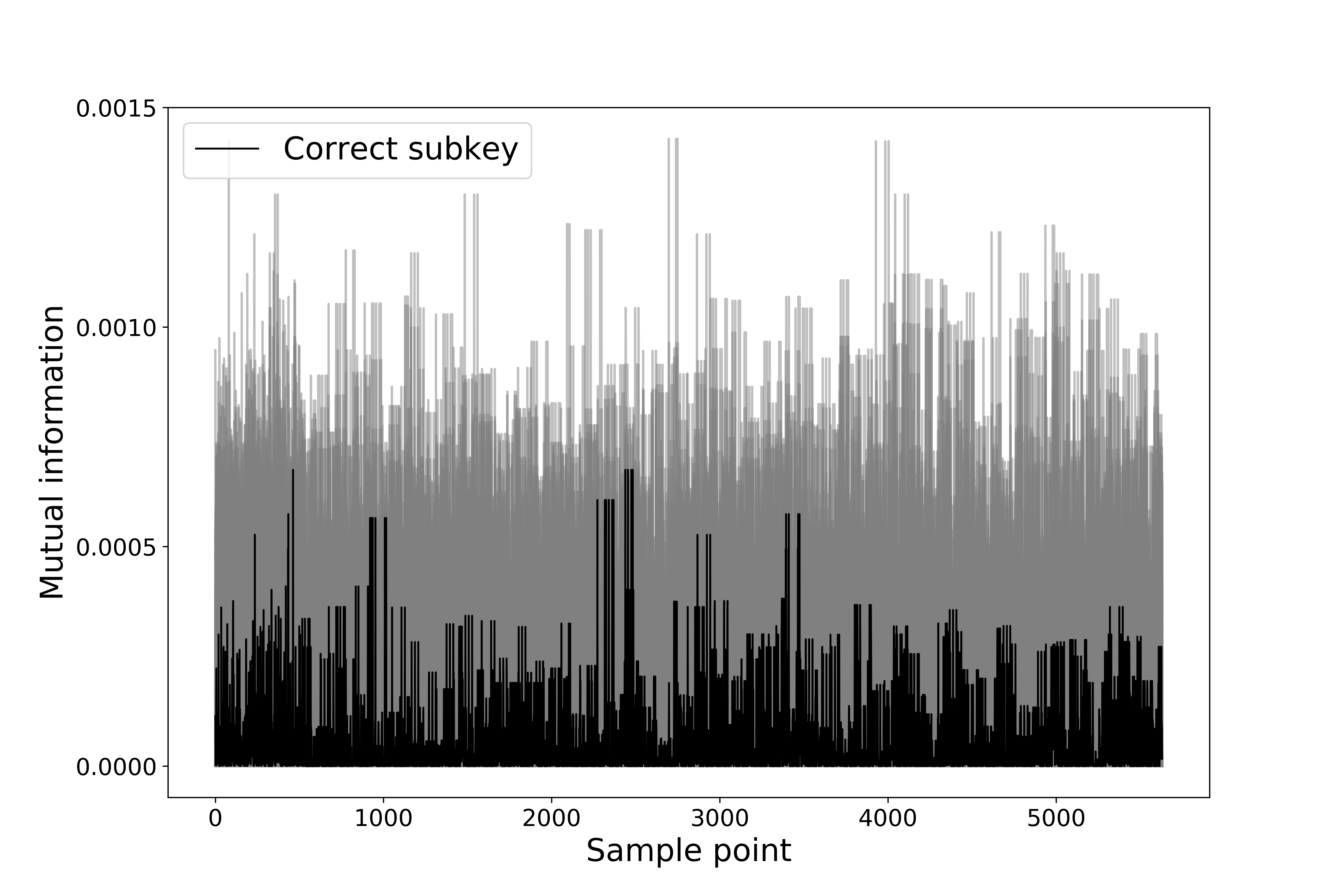}
\label{fig:mia_ut_bit2}}\\
\subfloat[Target Bit 3.]{ \includegraphics[width=0.5\linewidth]{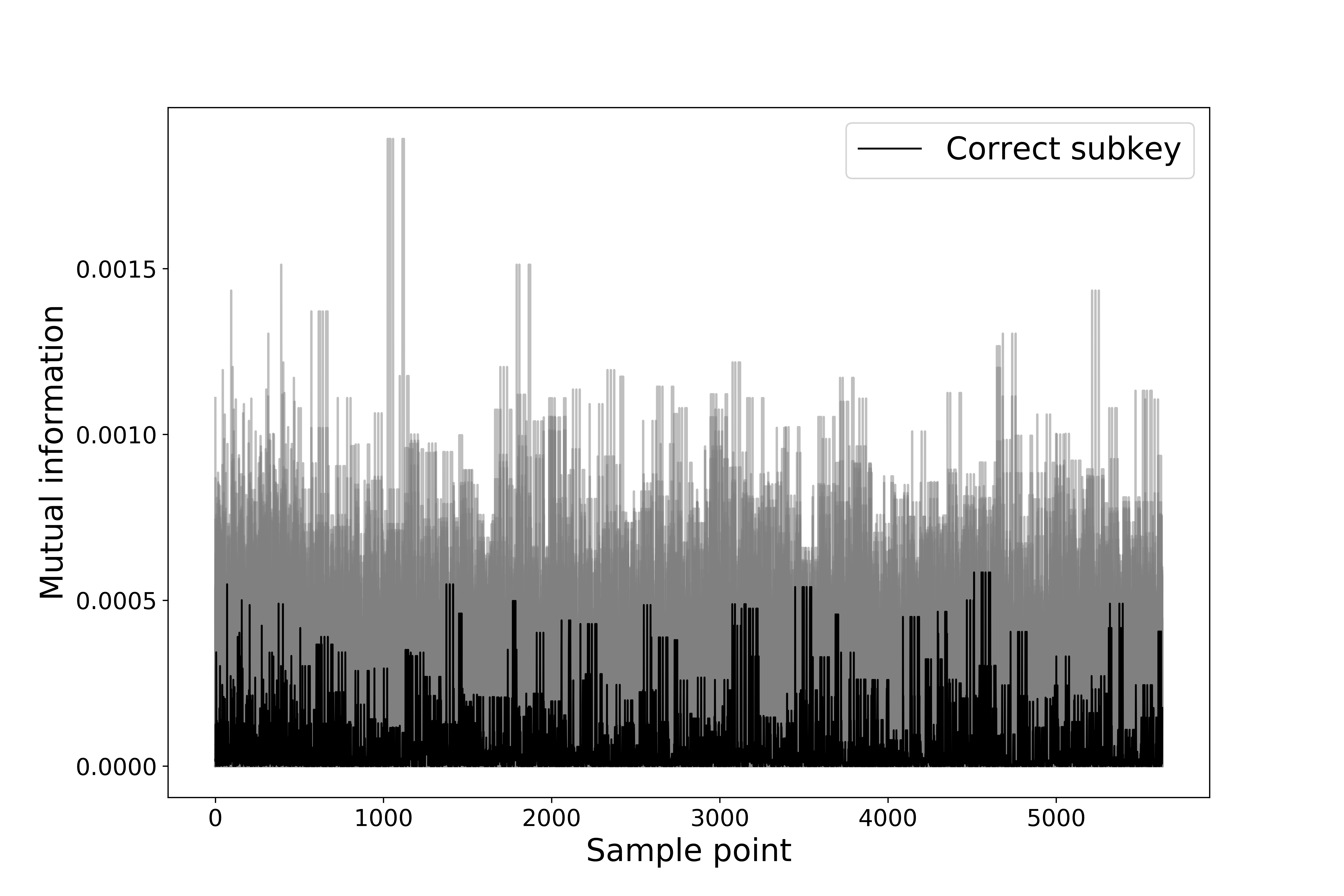}
\label{fig:mia_ut_bit3}}
\subfloat[Target Bit 4.]{ \includegraphics[width=0.5\linewidth]{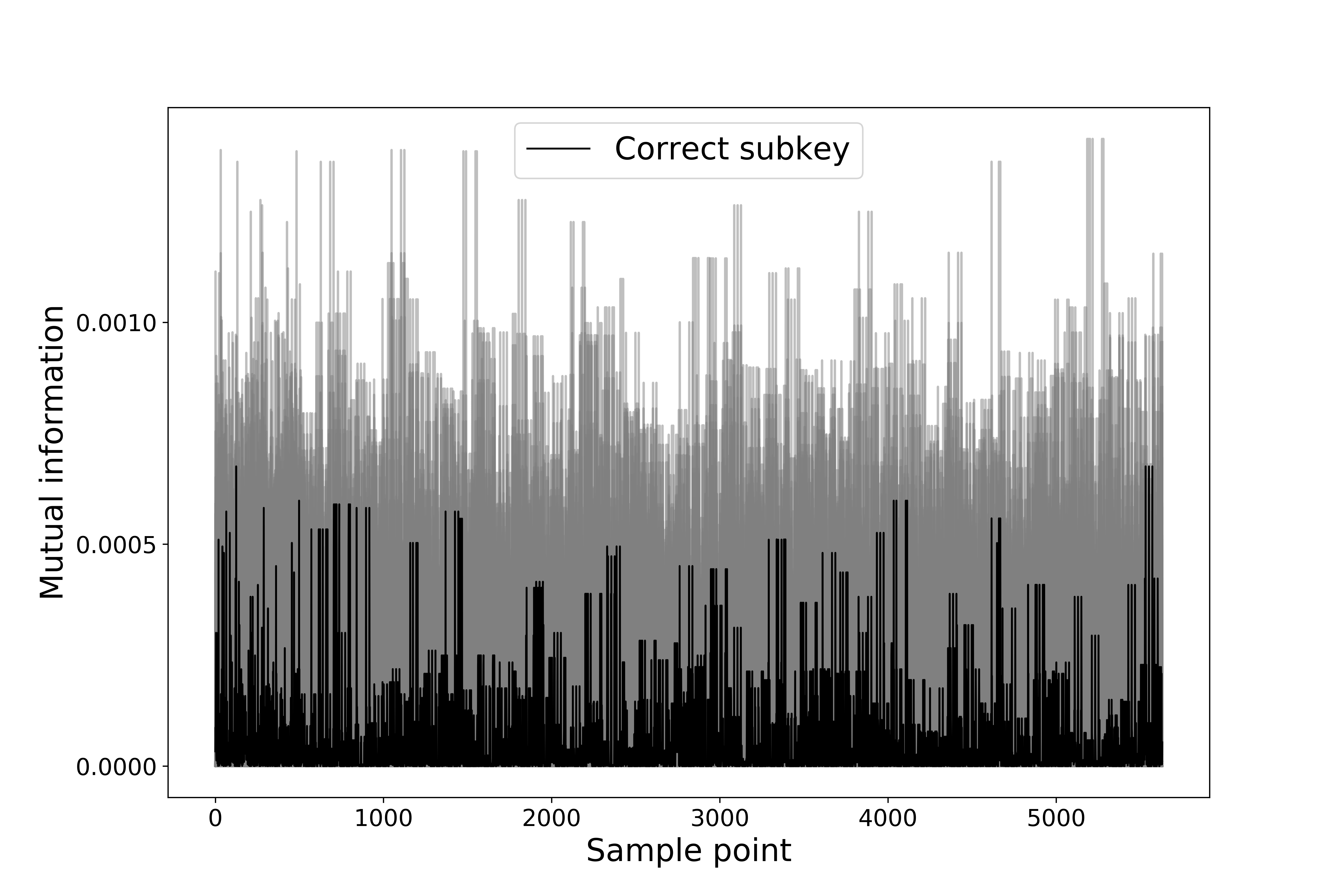}
\label{fig:mia_ut_bit4}}\\
\subfloat[Target Bit 5.]{\includegraphics[width=0.5\linewidth]{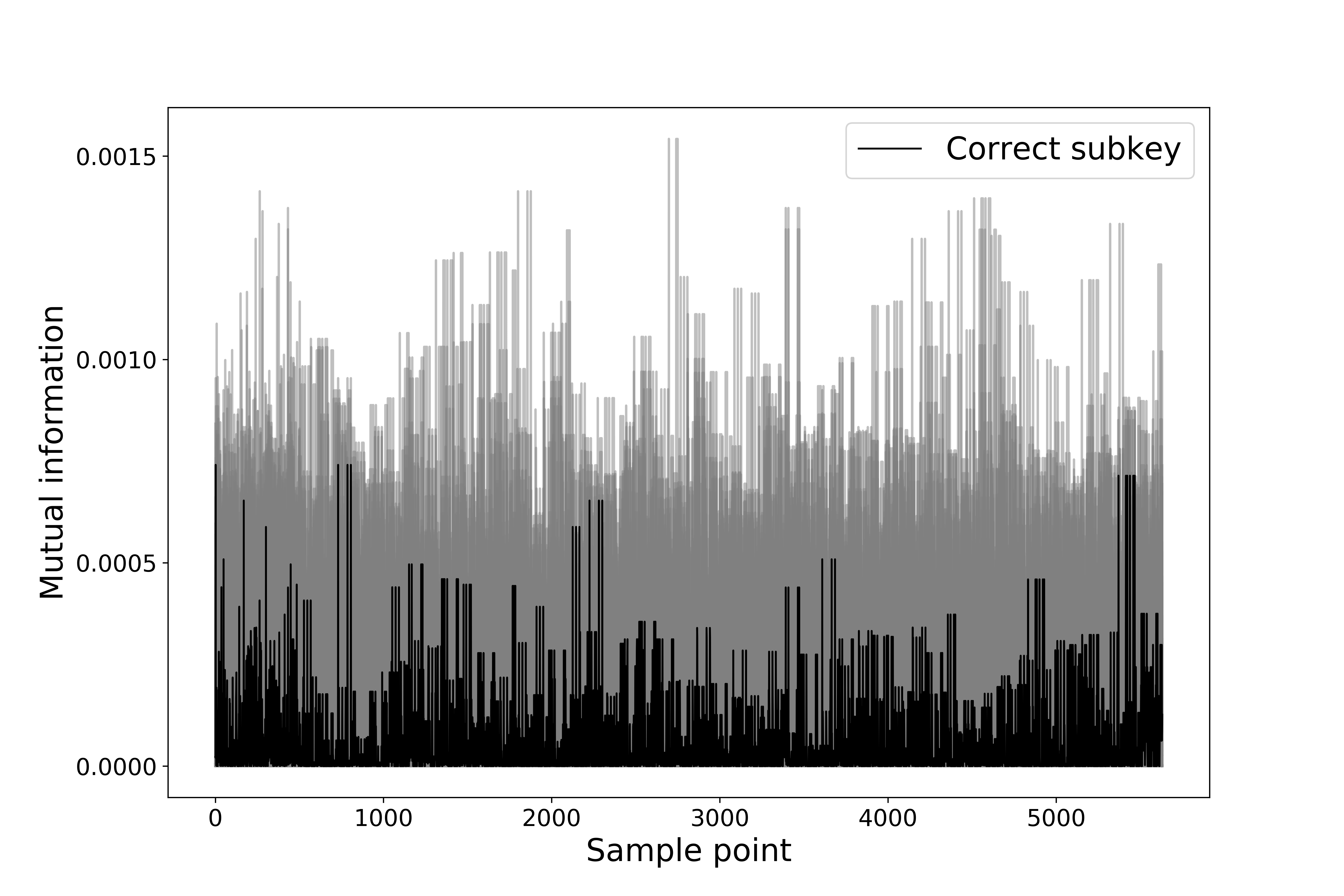}
\label{fig:mia_ut_bit5}}
\subfloat[Target Bit 6.]{ \includegraphics[width=0.5\linewidth]{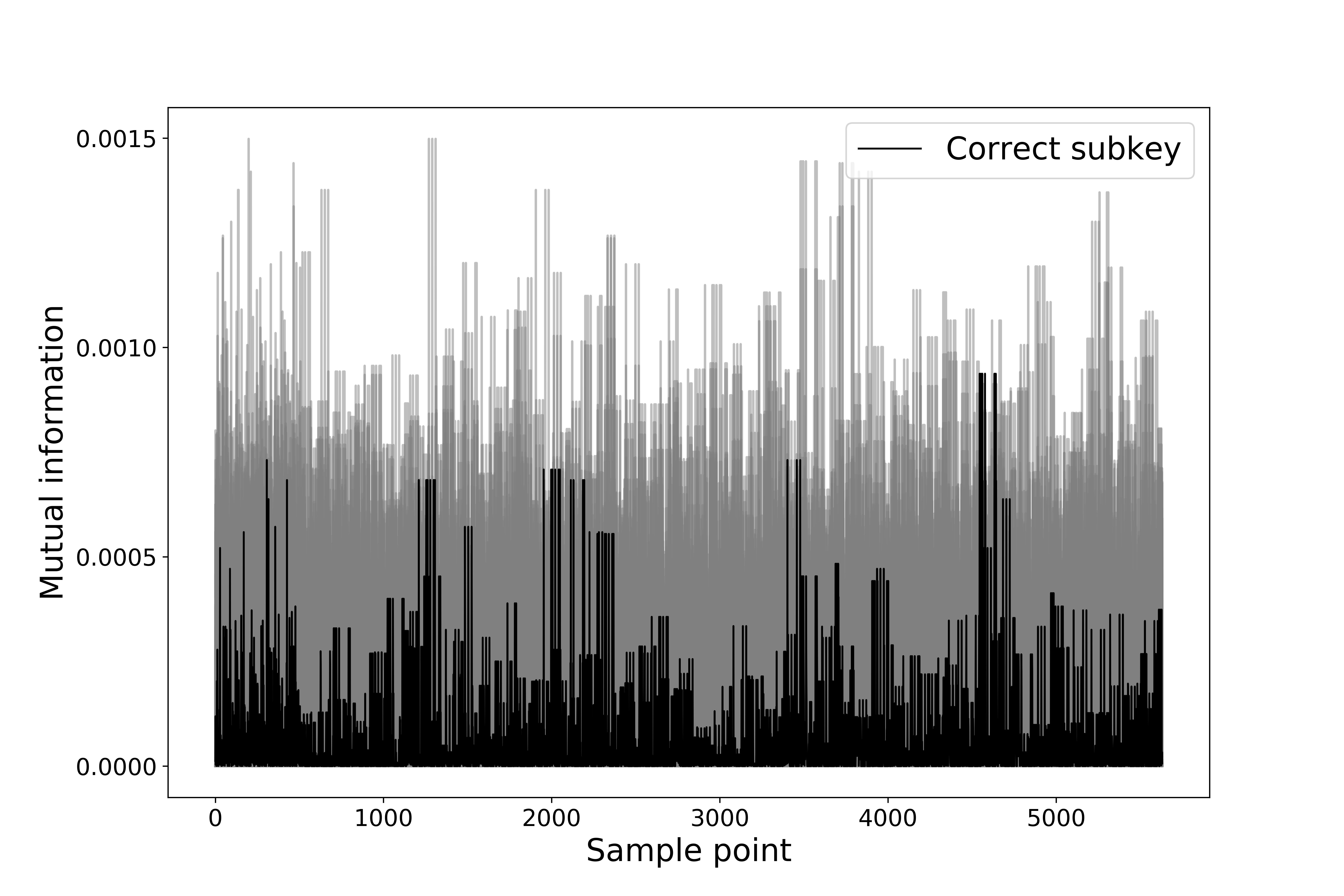}
\label{fig:mia_ut_bit6}}\\
\subfloat[Target Bit 7.]{ \includegraphics[width=0.5\linewidth]{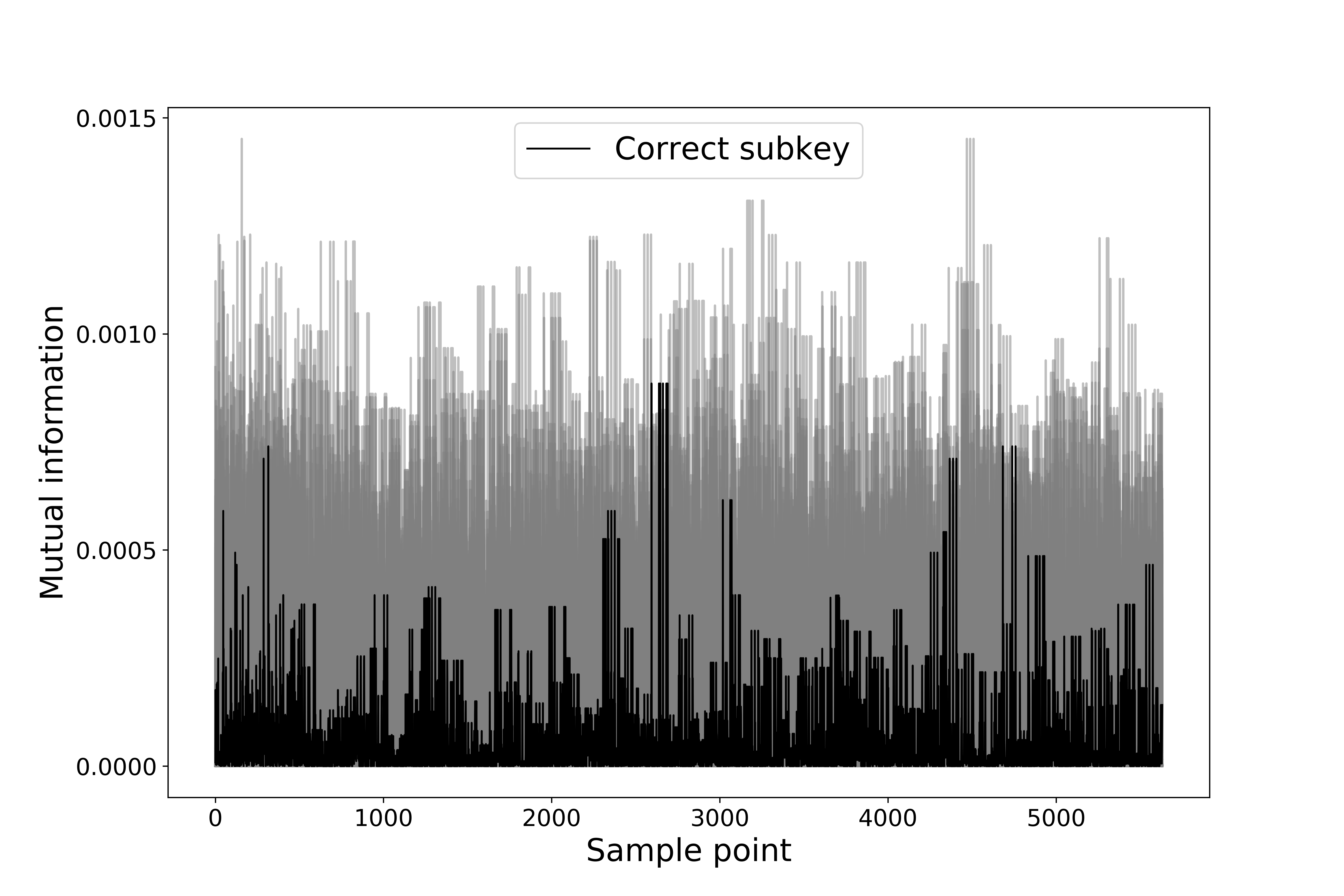}
\label{fig:mia_ut_bit7}}
\subfloat[Target Bit 8.]{\includegraphics[width=0.5\linewidth]{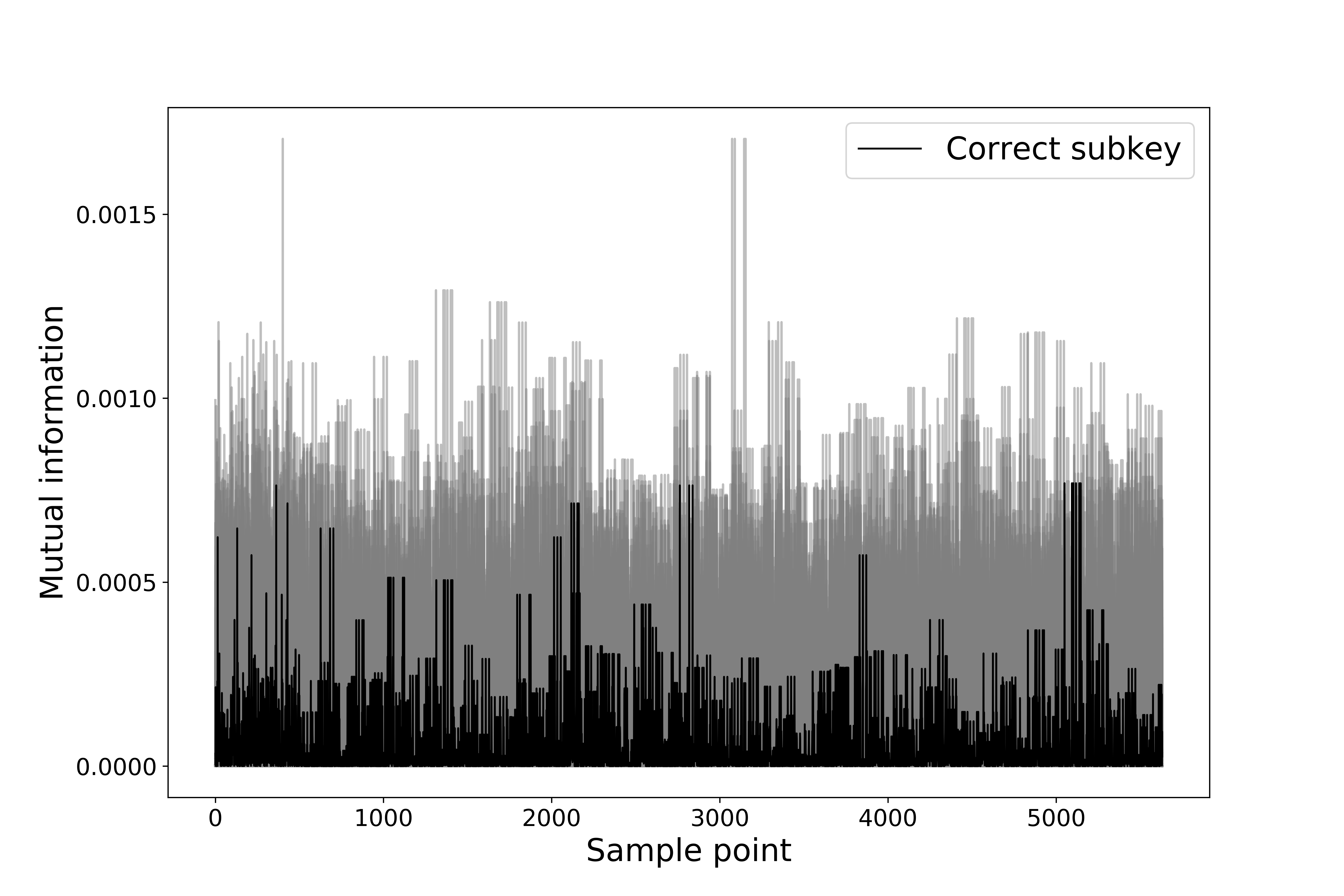}
\label{fig:mia_ut_bit8}}
\caption{MIA results on the $UT^{1}_{0,0}$ obtained by $Q_0$ and $Q_1$ and each bit of the hypothetical SubBytes outputs in the first round. Black: correct key; gray: wrong key.}
\label{fig:Appendix_MIA_UT_Q0_Q1}
\end{figure*}

To analyze the security of the MIA (Mutual Information Analysis) attack, we generated pairs of Q0 and Q1 while collecting computational traces during the encryption of 10,000 random plaintexts. The hypothetical value for the attacker in measuring mutual information is the output of the first round's SubBytes operation. Fig.~\ref{fig:Appendix_MIA_UT_Q0_Q1} displays the results of mutual information for each bit of the SubBytes output. As in previous experimental results, the black results represent mutual information for the correct subkey, while the results obtained from other wrong subkey candidates are depicted in gray.

\begin{figure*}
\centering
\subfloat[Target Bit 1.]{ \includegraphics[width=0.5\linewidth]{fig/mia/mia_ro_bit0.png}
\label{fig:mia_ro_bit1}}
\subfloat[Target Bit 2.]{ \includegraphics[width=0.5\linewidth]{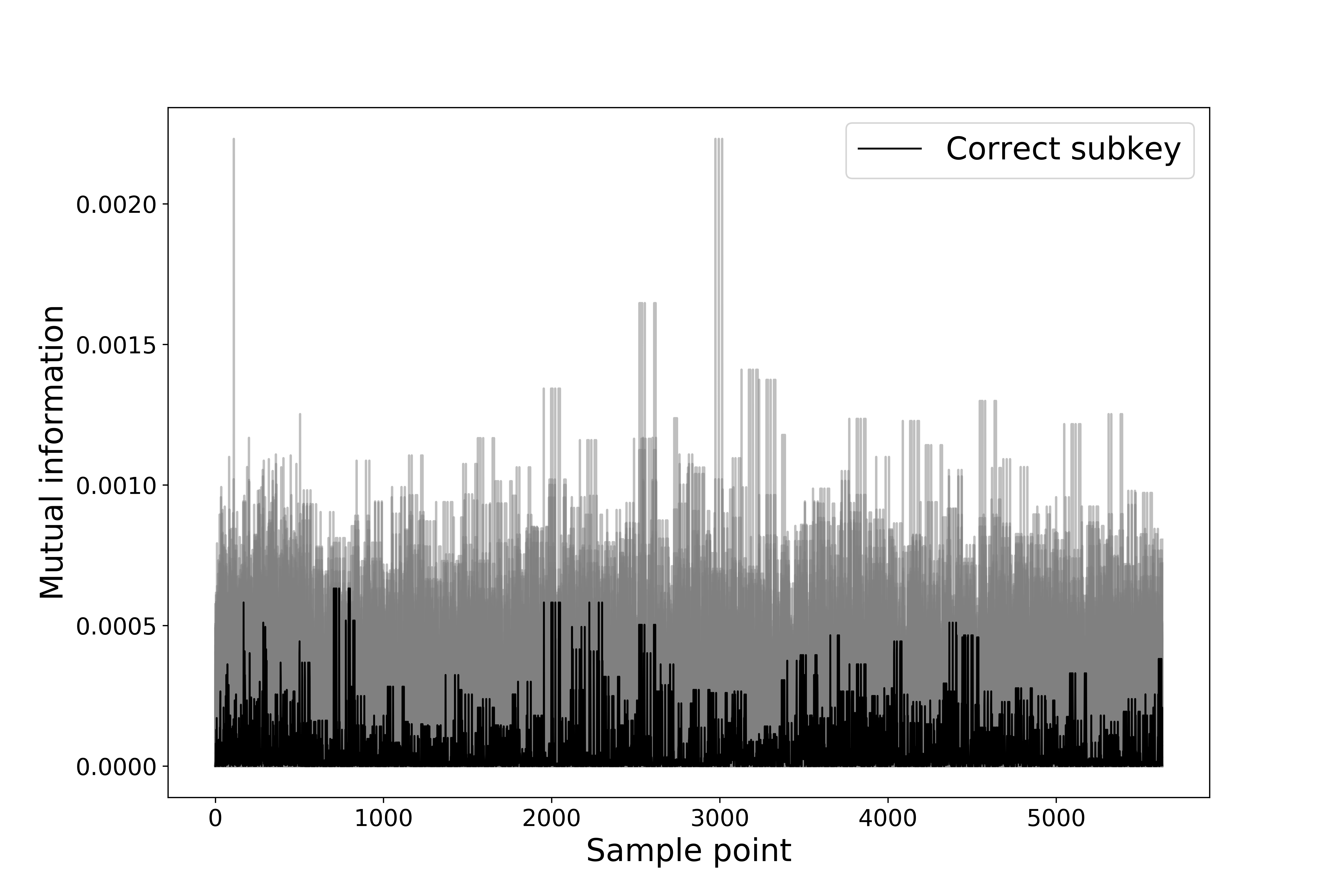}
\label{fig:mia_ro_bit2}}\\
\subfloat[Target Bit 3.]{ \includegraphics[width=0.5\linewidth]{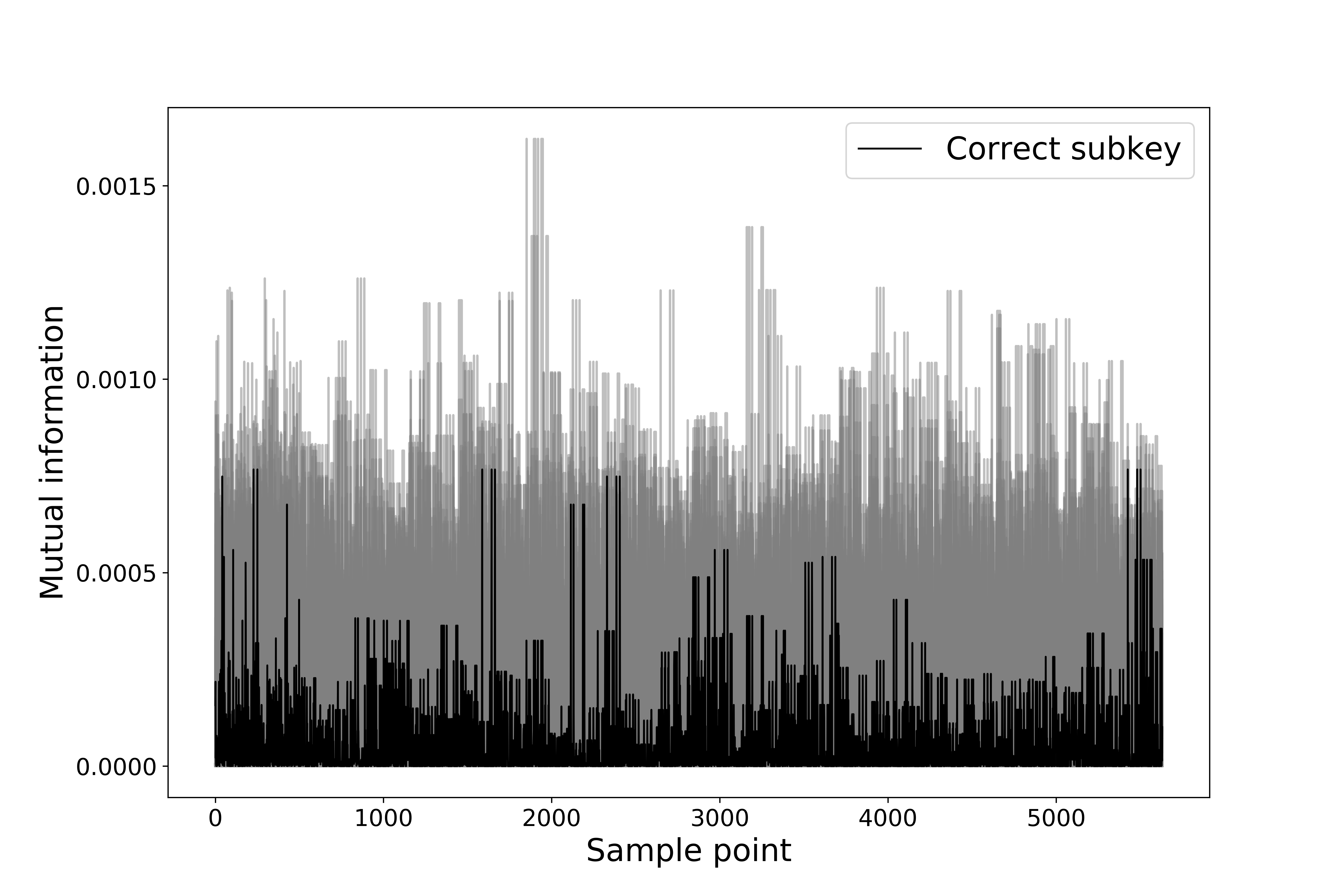}
\label{fig:mia_ro_bit3}}
\subfloat[Target Bit 4.]{ \includegraphics[width=0.5\linewidth]{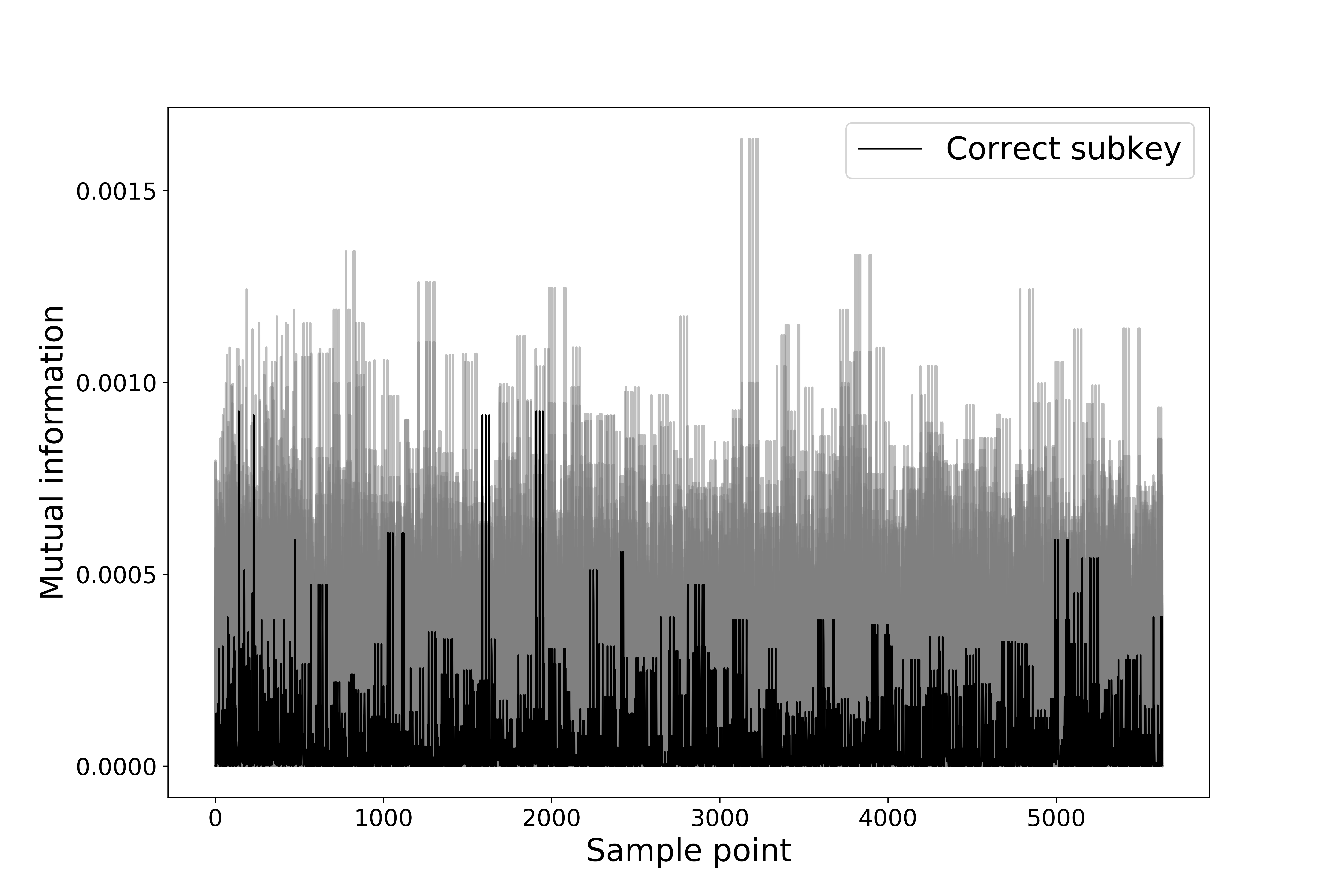}
\label{fig:mia_ro_bit4}}\\
\subfloat[Target Bit 5.]{\includegraphics[width=0.5\linewidth]{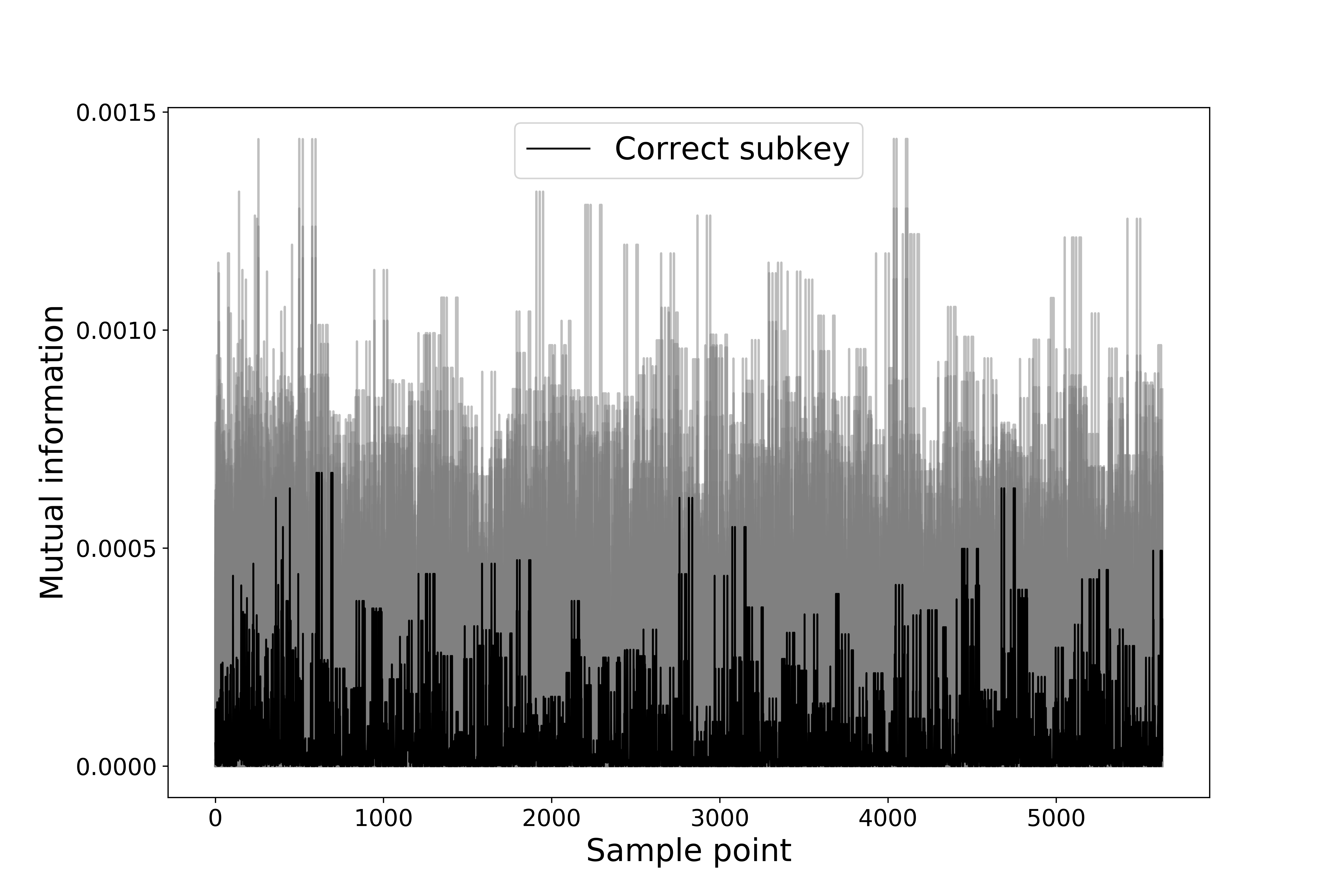}
\label{fig:mia_ro_bit5}}
\subfloat[Target Bit 6.]{ \includegraphics[width=0.5\linewidth]{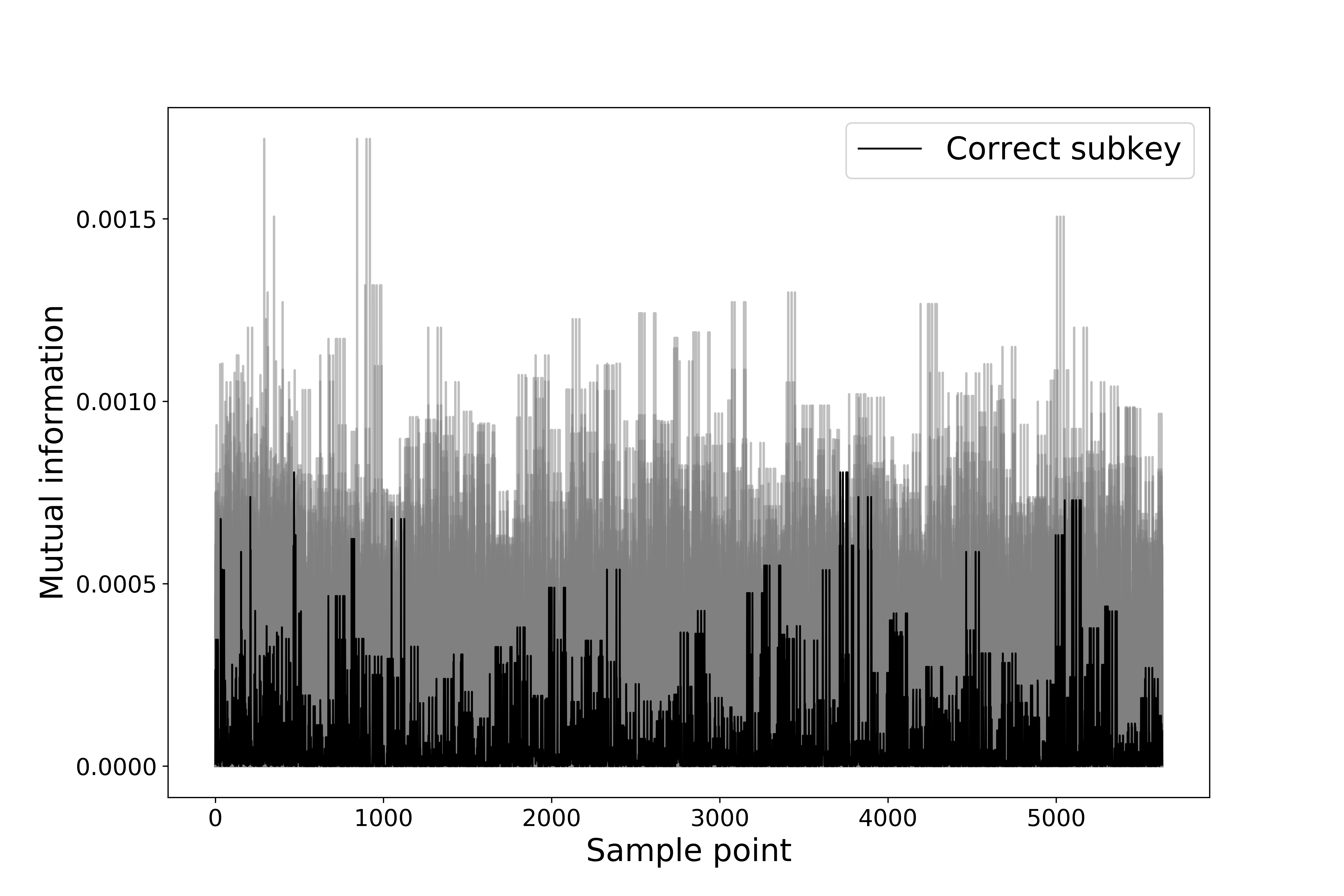}
\label{fig:mia_ro_bit6}}\\
\subfloat[Target Bit 7.]{ \includegraphics[width=0.5\linewidth]{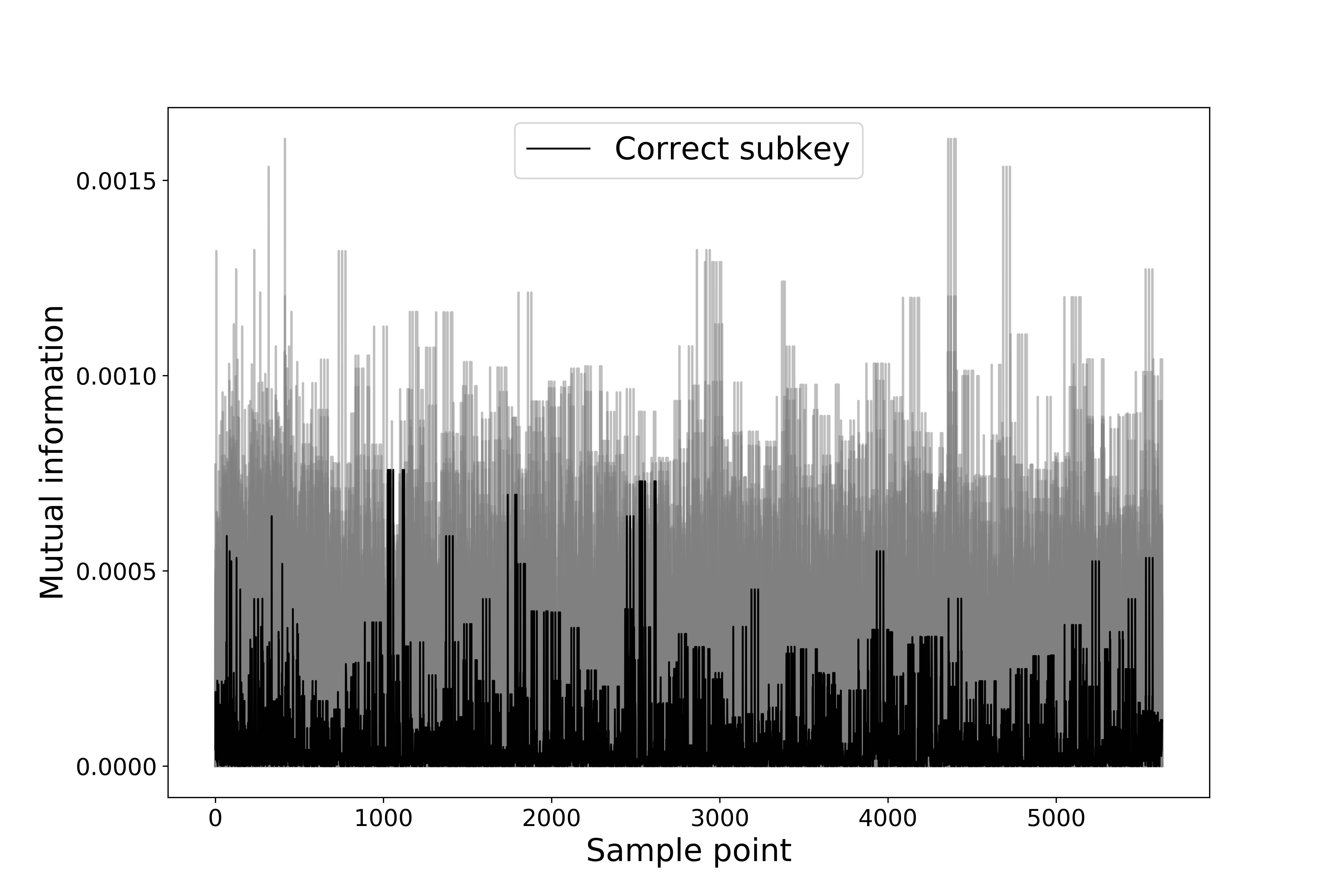}
\label{fig:mia_ro_bit7}}
\subfloat[Target Bit 8.]{\includegraphics[width=0.5\linewidth]{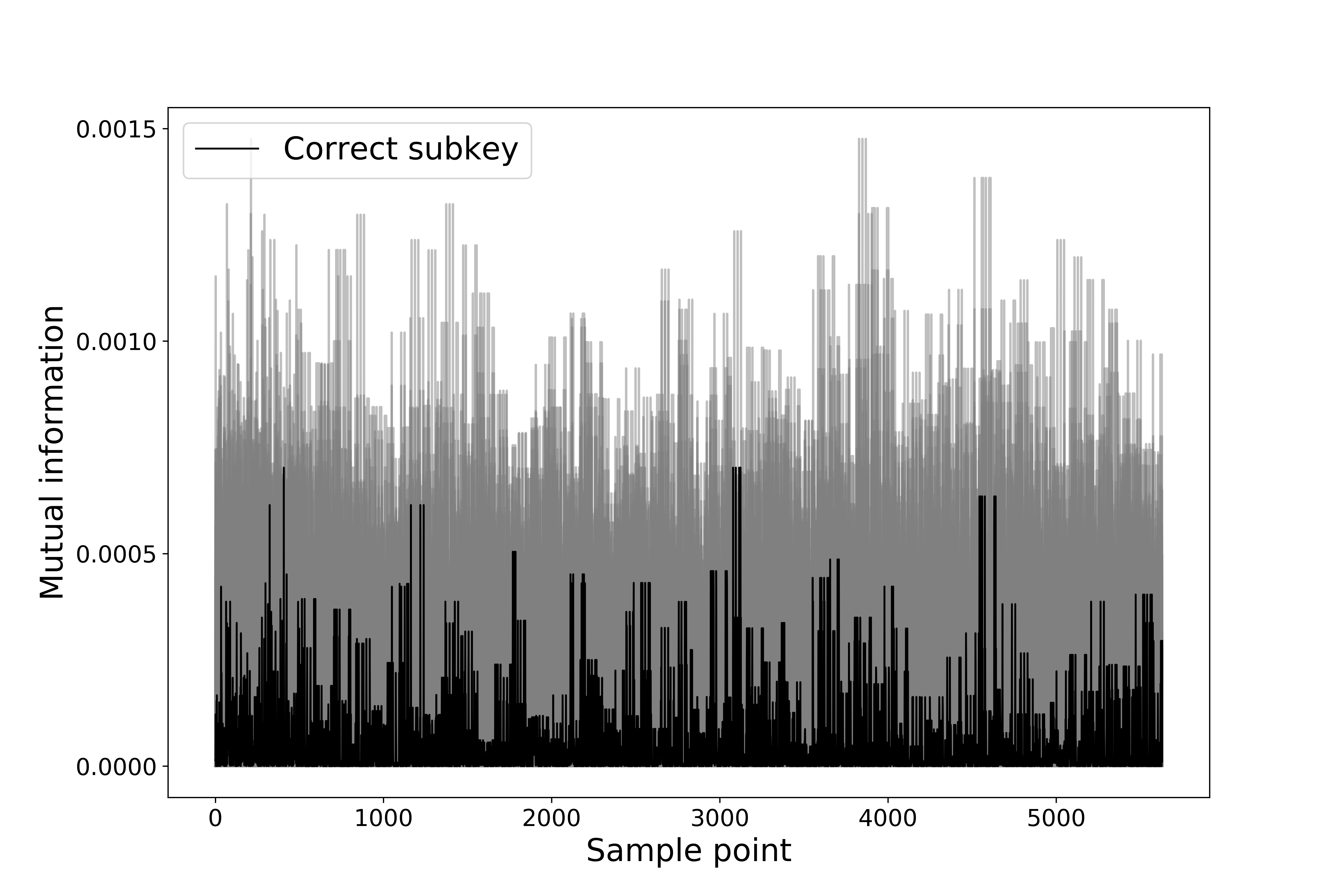}
\label{fig:mia_ro_bit8}}
\caption{MIA results on the round output obtained by $Q_0$ and $Q_1$ and each bit of the hypothetical first round outputs. Black: correct key; gray: wrong key.}
\label{fig:Appendix_MIA_RO_Q0_Q1}
\end{figure*}

Furthermore, we measured mutual information for the round output values. AES requires four subkeys to calculate each byte of the first-round output. For the sake of attack efficiency, we assumed that the attacker knows three out of the four required subkeys to compute the first byte of the round output. The results for the correct subkey associated with the unknown fourth subkey are in black, and the results for the other wrong subkey candidates are in gray. As evident from the results shown in Fig.~\ref{fig:Appendix_MIA_RO_Q0_Q1}, our proposed method demonstrates the ability to protect the secret key against mutual information analysis.

\section{The experimental results of TVLA and CPA attacks}
\label{sec:appendix_tvla}

\begin{figure*}
\centering
\subfloat[TVLA result.]{ \includegraphics[width=0.5\linewidth]{fig/tvla/tvla_plot.png}
\label{fig:tvla}}
\subfloat[CPA Target Bit 2.]{ \includegraphics[width=0.5\linewidth]{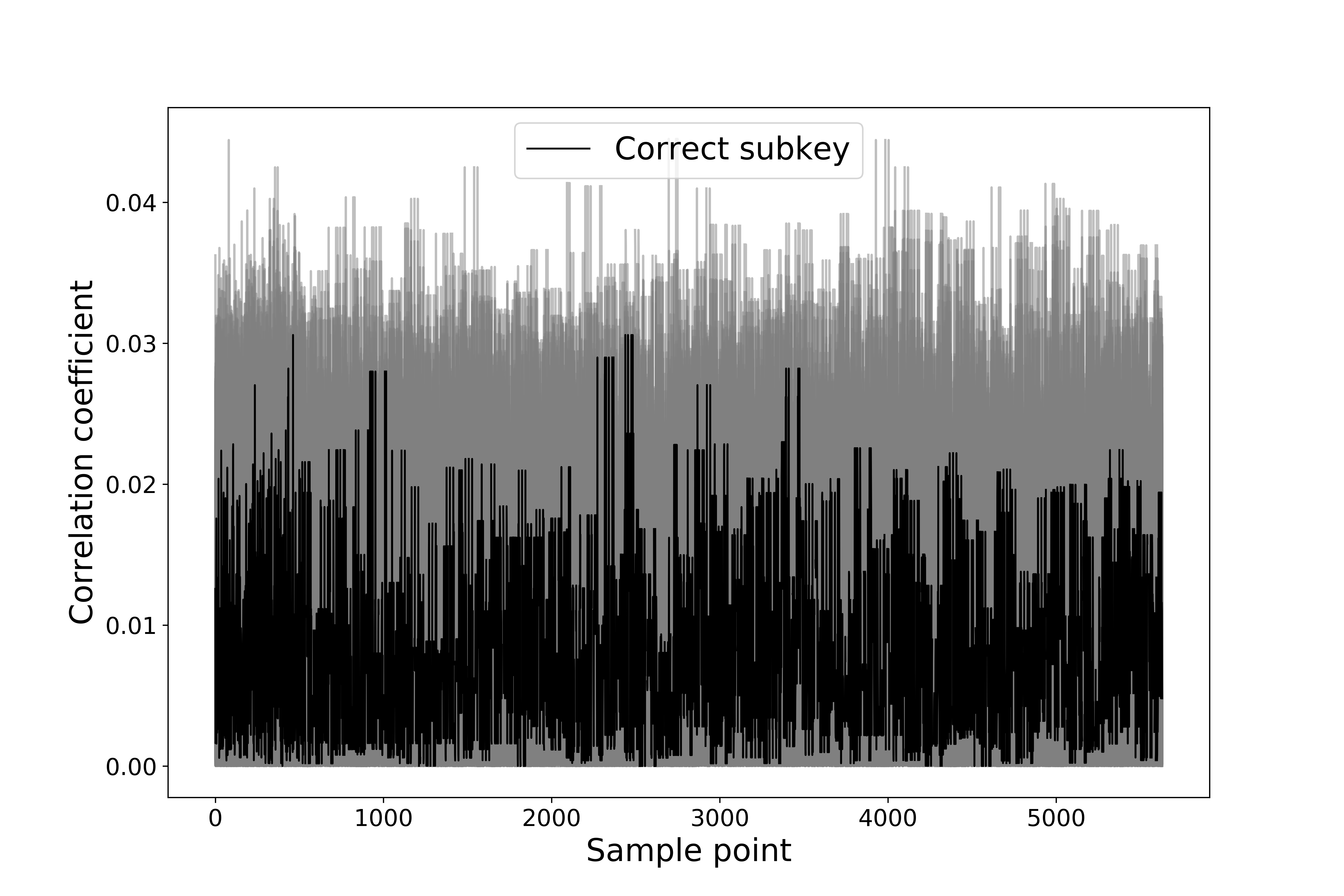}
\label{fig:tvla_cpa_bit2}}\\
\subfloat[CPA Target Bit 3.]{ \includegraphics[width=0.5\linewidth]{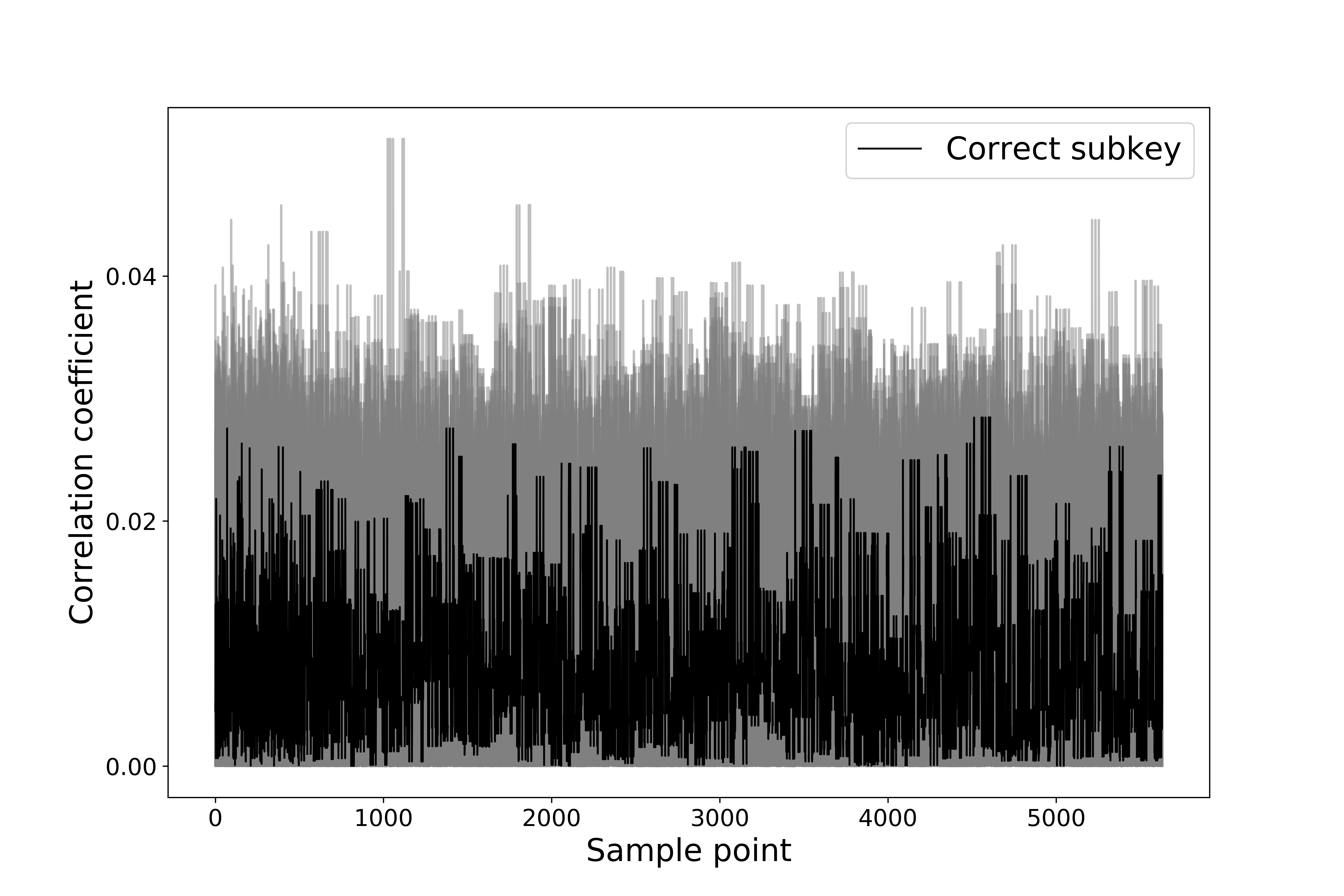}
\label{fig:tvla_cpa_bit3}}
\subfloat[CPA Target Bit 4.]{ \includegraphics[width=0.5\linewidth]{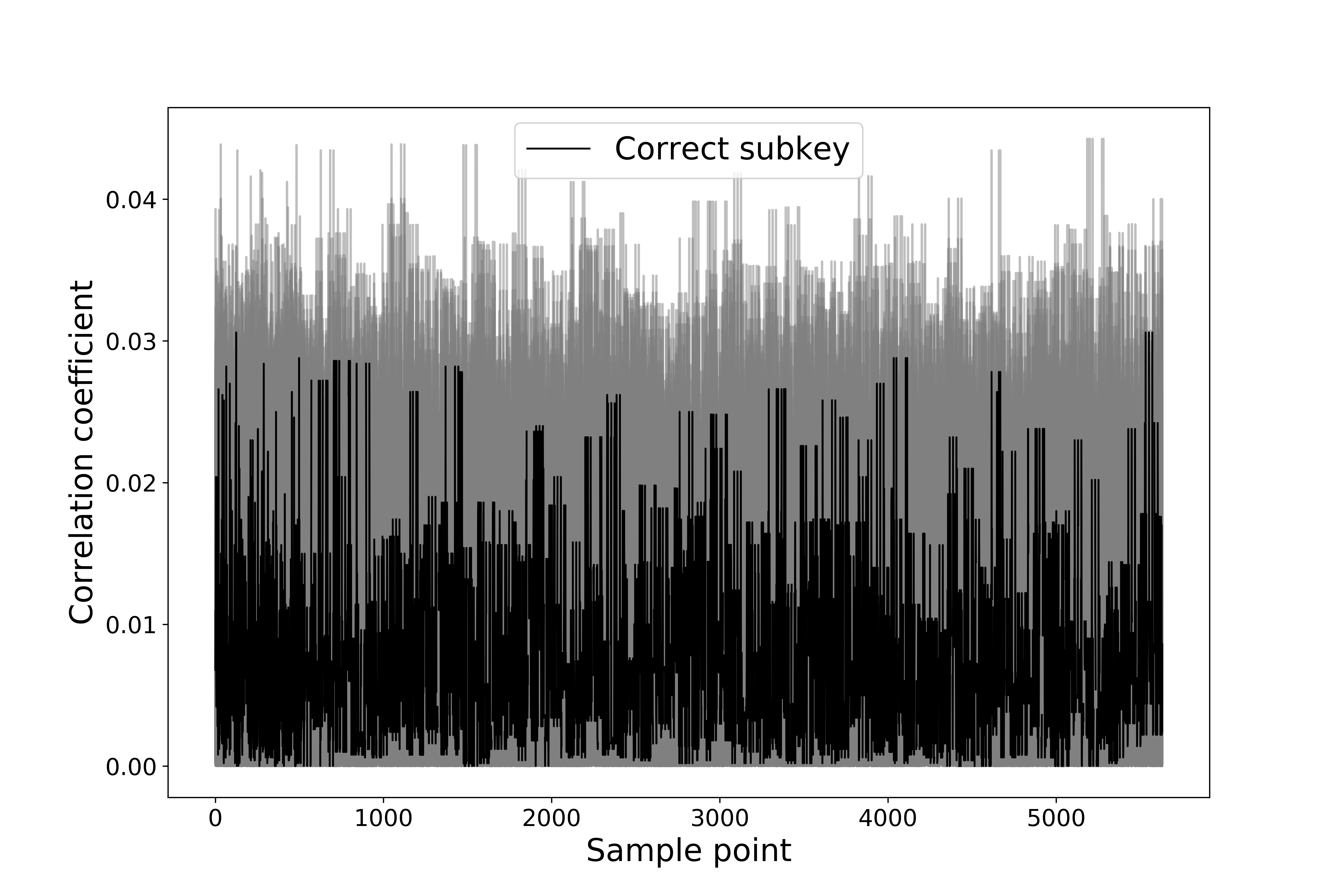}
\label{fig:tvla_cpa_bit4}}\\
\subfloat[CPA Target Bit 5.]{\includegraphics[width=0.5\linewidth]{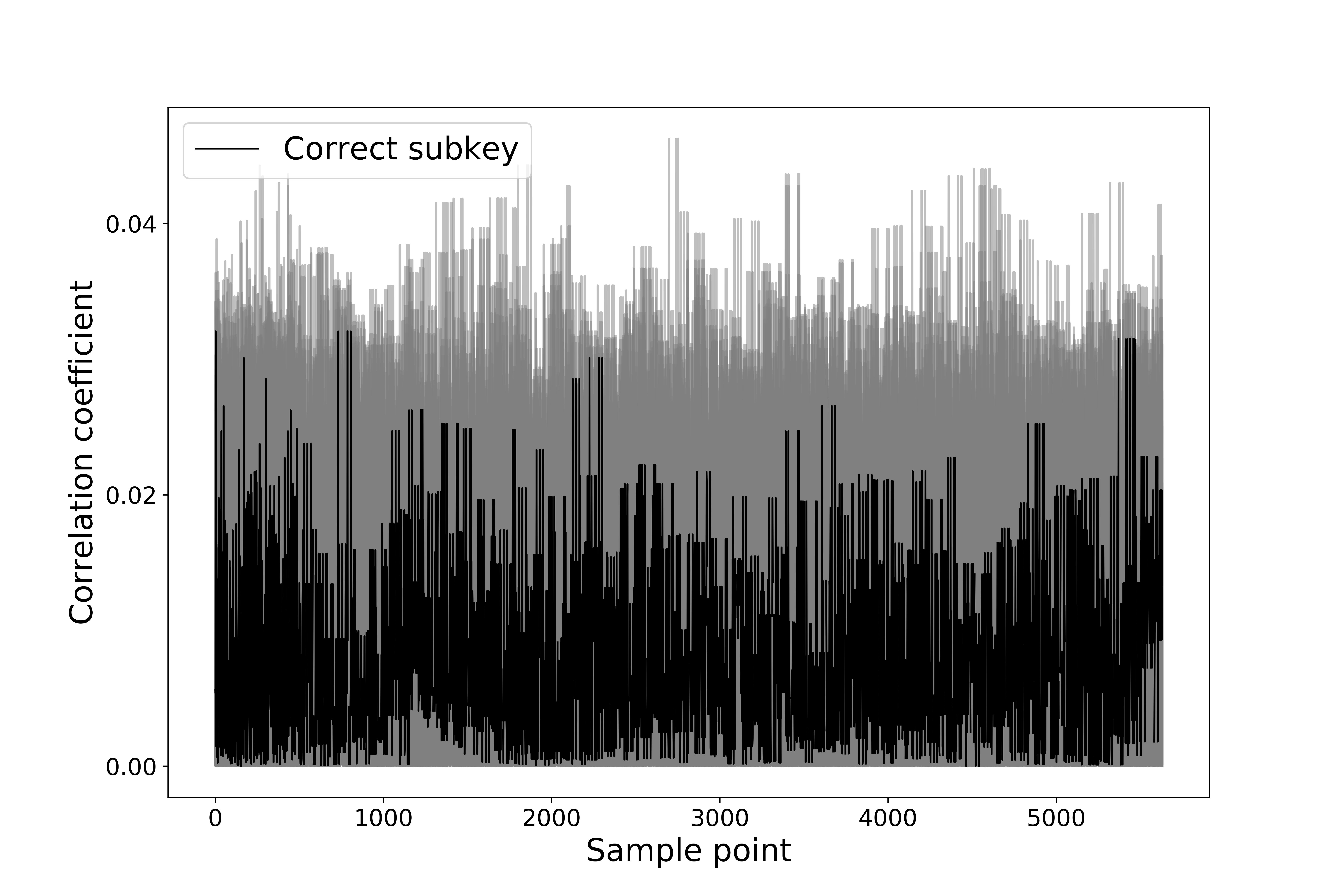}
\label{fig:tvla_cpa_bit5}}
\subfloat[CPA Target Bit 6.]{ \includegraphics[width=0.5\linewidth]{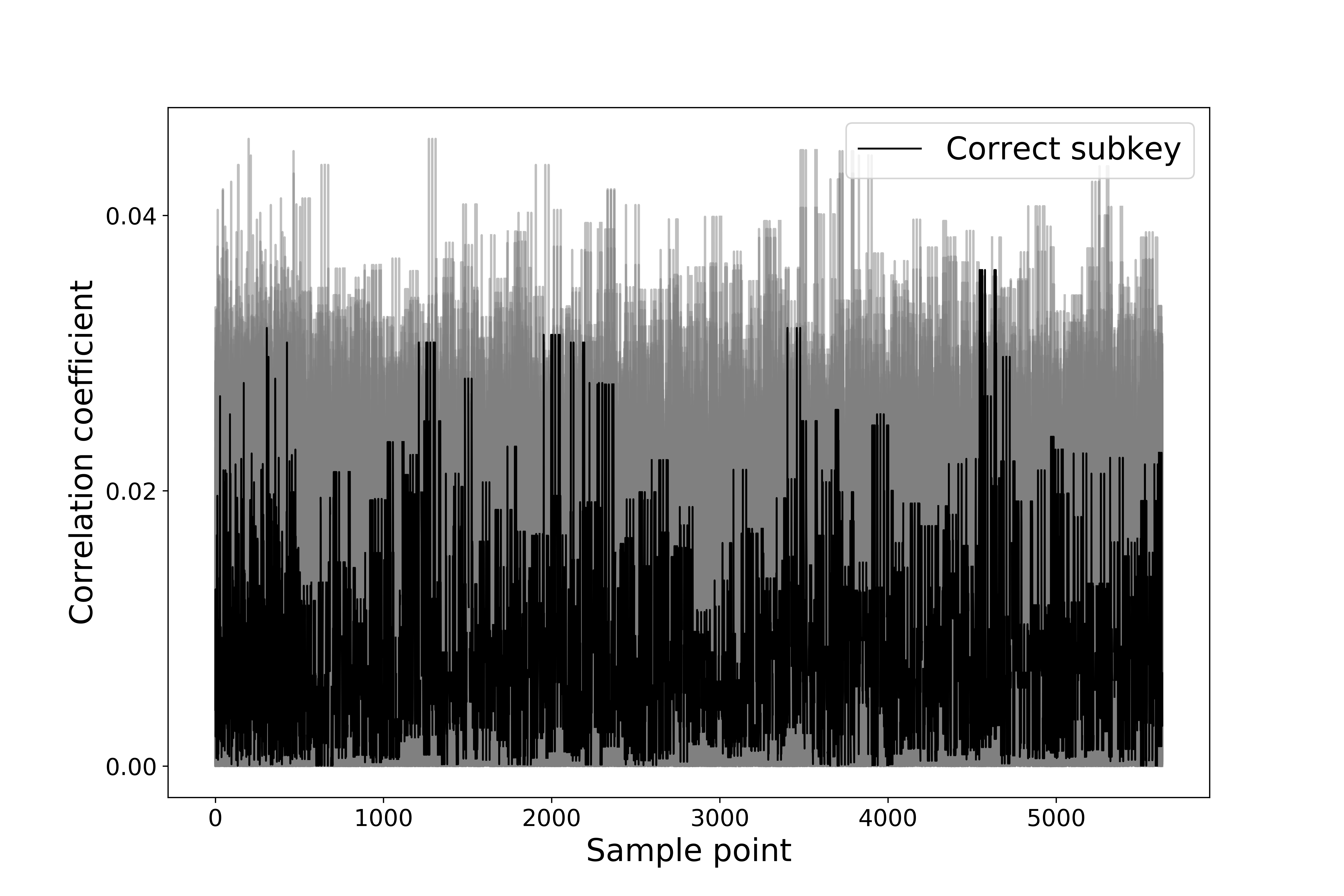}
\label{fig:tvla_cpa_bit6}}\\
\subfloat[CPA Target Bit 7.]{ \includegraphics[width=0.5\linewidth]{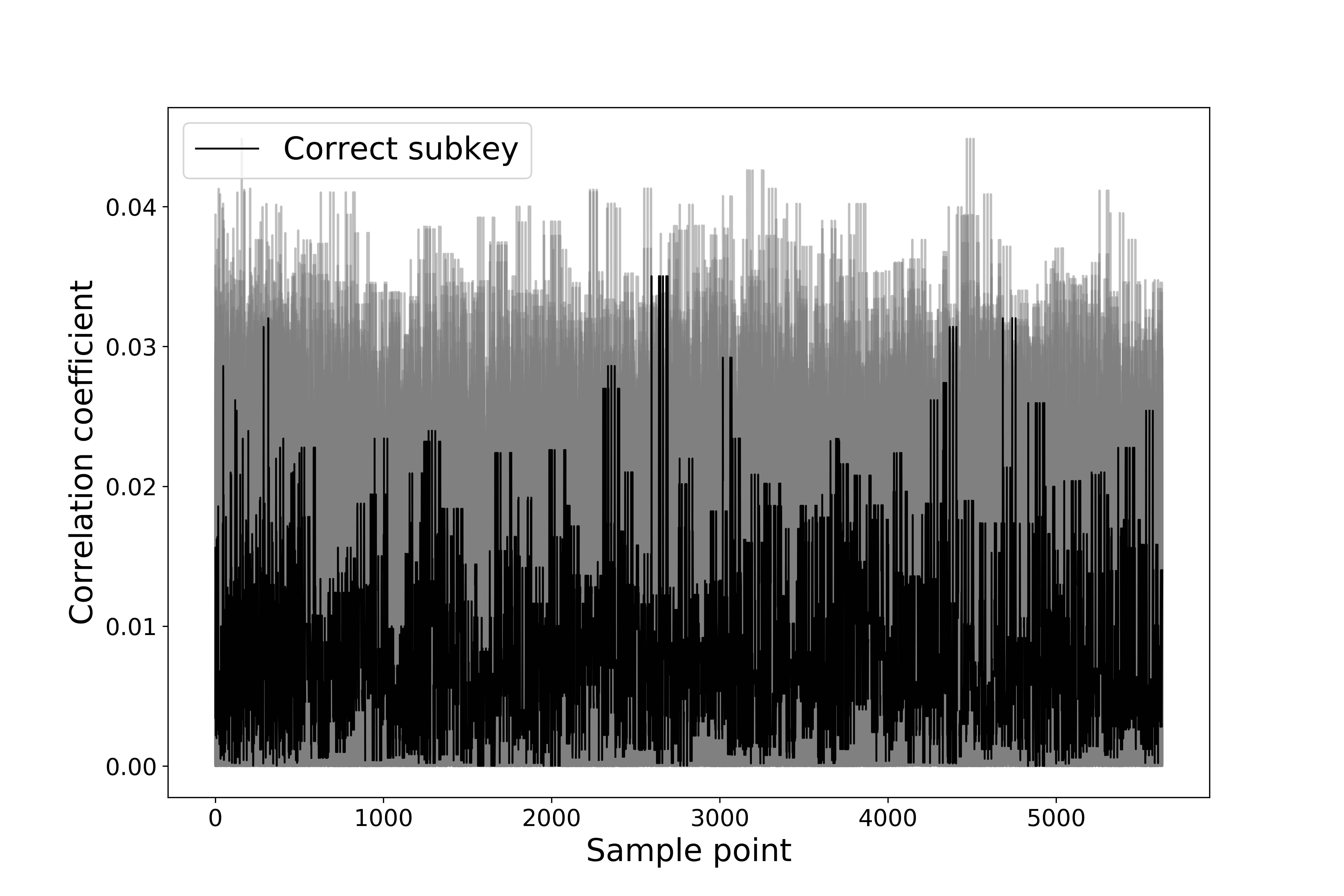}
\label{fig:tvla_cpa_bit7}}
\subfloat[CPA Target Bit 8.]{\includegraphics[width=0.5\linewidth]{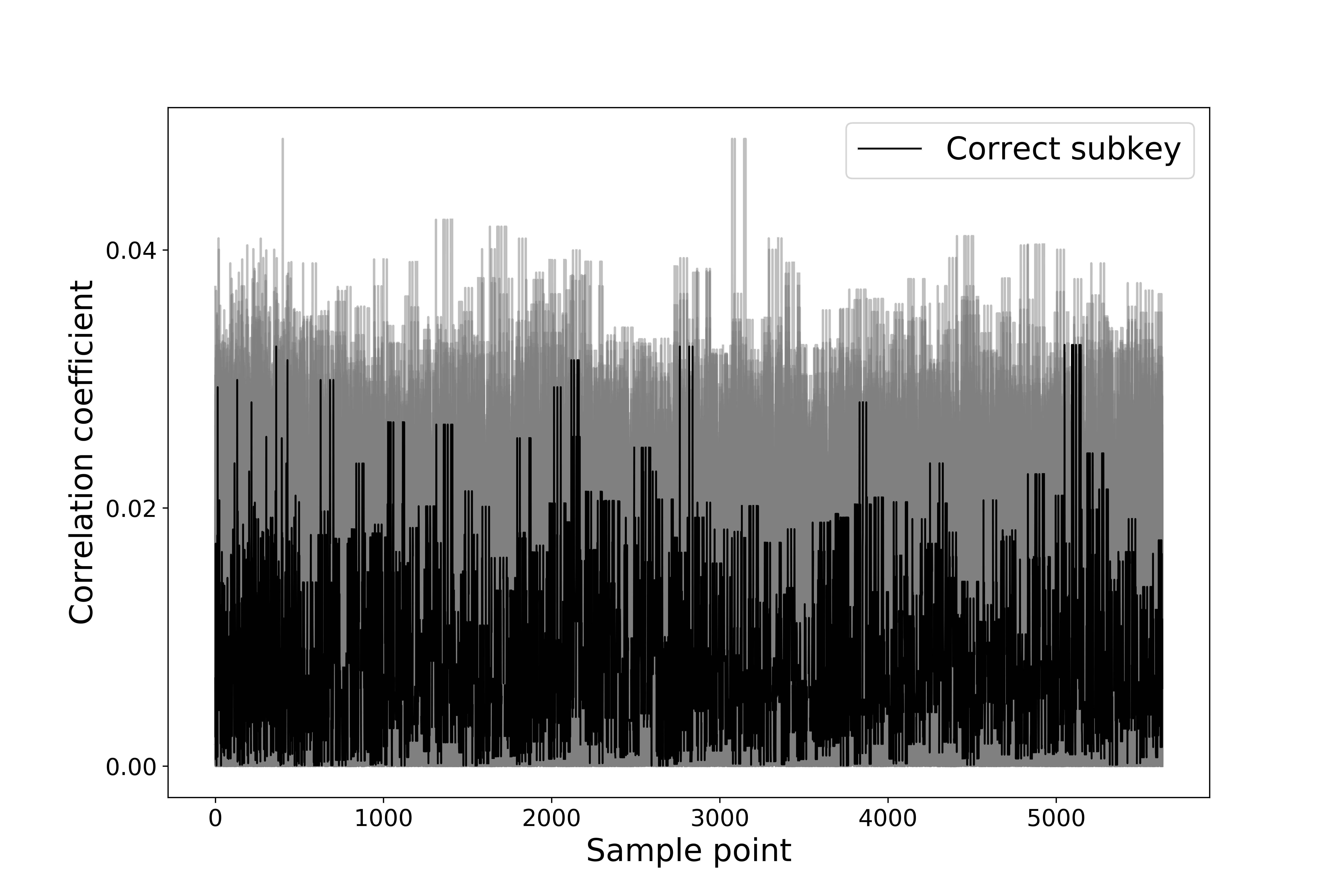}
\label{fig:tvla_cpa_bit8}}
\caption{TVLA and CPA results on the SubBytes output in the first round. Black: correct key; gray: wrong key.}
\label{fig:Appendix_TVLA_CPA_Q0_Q1}
\end{figure*}

We previously provided the results of the TVLA  and the CPA attacks to demonstrate that key leakage did not occur. The attacker's hypothetical value was each bit of the first-round SubBytes operation, and we had provided the results of the CPA attack for the first bit only. Here, we aim to provide the CPA results for the remaining 7 bits as depicted in Fig.~\ref{fig:Appendix_TVLA_CPA_Q0_Q1}.

\clearpage

\bibliographystyle{splncs03}
\bibliography{reference}

\end{document}